\documentclass[seceq,supplement]{ptptex}
\usepackage{epsfig}
\usepackage{color}
\newcommand{\e}{\mathrm{e}}
\newcommand{\vev}[1]{\left\langle #1 \right\rangle}
\newcommand{\SL}{S_\Lambda}
\newcommand{\SIL}{S_{I,\Lambda}}
\newcommand{\SFL}{S_{F,\Lambda}}

\newcommand{\Si}{S_I}
\newcommand{\bSL}{\bar{S}_\Lambda}
\newcommand{\bSIL}{\bar{S}_{I,\Lambda}}
\newcommand{\bSinf}{\bar{S}_\infty}
\newcommand{\Scl}{S_{cl}}
\newcommand{\bScl}{\bar{S}_{cl}}

\newcommand{\tSL}{\tilde{S}_\Lambda}
\newcommand{\hSL}{\hat{S}_\Lambda}

\newcommand{\K}[1]{K\left( #1/\Lambda \right)}
\newcommand{\Kz}[1]{K\left( #1/\Lambda_0 \right)}

\newcommand{\Kb}[1]{K_b \left( #1/\Lambda \right)}
\newcommand{\Kf}[1]{K_f \left( #1/\Lambda \right)}

\newcommand{\fmslash}[1]{\hbox{$#1$\kern-0.5em\raise0.3ex\hbox{/}}}
\newcommand{\Ld}[1]{\frac{\overrightarrow{\delta}}{\delta #1}}
\newcommand{\Rd}[1]{\frac{\overleftarrow{\delta}}{\delta #1}}

\newcommand{\Op}{\mathcal{O}}
\newcommand{\ep}{\epsilon}
\newcommand{\nn}{\nonumber}
\newcommand{\lb}{\left\lbrace}
\newcommand{\rb}{\right\rbrace}

\newcommand{\Tr}{\mathrm{Tr}\,}

\newcommand{\V}{\mathcal{V}}
\newcommand{\asym}{\stackrel{\Lambda \to \infty}{\longrightarrow}}
\newcommand{\N}{\mathcal{N}}
\newcommand{\de}{\delta_\ep}
\newcommand{\A}{\mathcal{A}}
\newcommand{\C}{\mathcal{C}}
\newcommand{\PN}{\left[ \Phi_N \right]_\Lambda}
\newcommand{\Slash}[1]{\ooalign{\hfil/\hfil\crcr$#1$}}
\newcommand{\pa}{\partial}

\newcommand{\bref}[1]{(\ref{#1})}
\newcommand{\cbar}{{\bar c}}
\newcommand{\Cbar}{{\bar C}}

\newcommand{\ds}{{\displaystyle}}

\title{
Realization of symmetry in the ERG approach \\to quantum field theory
}

\author{
Yuji \textsc{Igarashi}${}^1$,
Katsumi \textsc{Itoh}${}^1$,
and Hidenori \textsc{Sonoda}${}^2$
}

\inst{
${}^1$ Department of Education, Niigata University, Japan\\
${}^2$ Physics Department, Kobe University, Japan
}

\abst{ We review the use of the exact renormalization group for
  realization of symmetry in renormalizable field theories.  The
  review consists of three parts.  In part I (\S\S 2,3,4), we start
  with the perturbative construction of a renormalizable field theory
  as a solution of the exact renormalization group (ERG) differential
  equation.  We show how to characterize renormalizability by an
  appropriate asymptotic behavior of the solution for a large momentum
  cutoff.  Renormalized parameters are introduced to control the
  asymptotic behavior.  In part II (\S\S 5--9), we introduce two
  formalisms to incorporate symmetry: one by imposing the
  Ward-Takahashi identity, and another by imposing the generalized
  Ward-Takahashi identity via sources that generate symmetry
  transformations.  We apply the two formalisms to concrete models
  such as QED, YM theories, and the Wess-Zumino model in four
  dimensions, and the O(N) non-linear sigma model in two dimensions.
  We end this part with calculations of the abelian axial and chiral
  anomalies.  In part III (\S\S 10,11), we overview the
  Batalin-Vilkovisky formalism adapted to the Wilson action of a bare
  theory with a UV cutoff.  We provide a few appendices to give
  details and extensions that can be omitted for the understanding of
  the main text. The last appendix is a quick summary for the reader's
  convenience.}

\begin{document}

\maketitle
\tableofcontents

\newpage

\section{Introduction\label{intro}}

This is a review of the \textbf{exact renormalization group}
(\textbf{ERG}) with an emphasis on the use of ERG for realization of
symmetry in renormalizable field theories.  In writing this review, we
have set two goals.  The first is to popularize the ERG formalism among
the practitioners of quantum field theory.  We believe ERG has enough
merits to become part of the shared knowledge of all those who use
quantum field theory to think and calculate.  The second is to convince
the reader of the universal applicability of ERG: if a field theoretical
model can have a continuous symmetry, we can formulate it exactly using
the ERG formalism.  ERG is the second best after a formalism that
realizes the symmetry manifestly with no fine-tuning of parameters.

Modern development of the renormalization group started with the work of
K.~G.~Wilson, whose main results on the subject, summarized below, are
reviewed in the well known lecture notes \citen{Wilson:1973jj}:
\begin{enumerate}
\item He introduced the exact renormalization group as a
      non-perturbative framework in which we can define the continuum
      limit of a field theory.  Given a theory near criticality, he
      showed how to construct a massive field theory as a double limit
      of criticality and infinite distances.
\item He derived an approximate recursion formula that embodies the
      renormalization group for the real scalar theory in $D$
      dimensions, and computed a critical exponent (the anomalous
      dimension of the squared mass) by solving the formula numerically.
\item He, together with M.~E.~Fisher, devised the epsilon expansions to
      compute non-trivial critical exponents in powers of $\ep$, the
      difference of the space dimension from $4$ (or the dimension in
      which the corresponding critical theory is gaussian).
\end{enumerate}
In addition to these main results Wilson introduced the exact
renormalization group equations in differential form, the principal
tools for the symmetry realization in this review.  Many of the results
we give in \S 2 can already be found in \S 11 of \citen{Wilson:1973jj}.
\footnote{Wilson also predicted that his ERG differential equations
would become the basis for most future work on the renormalization
group.}\footnote{Wegner and Houghton introduced an exact renormalization
group differential equation independently of
Wilson.\cite{Wegner:1972ih}.  The one-particle-irreducible version of
the Wegner-Houghton equation was subsequently derived in
\citen{Nicoll:1977hi}.}

It took almost ten years before Wilson's ERG differential equation was
taken up by Polchinski in \citen{Polchinski:1983gv}.  After modifying
the equation to a form suitable for perturbation theory, Polchinski
applied the equation to prove the perturbative renormalizability of the
$\phi^4$ theory in $4$ dimensions.  We base all our discussions of
symmetry realization on Polchinski's differential equation, since we are
primarily interested in perturbation theory.  For those interested in
Wilson's original ERG differential equation, we describe it briefly in
Appendix \ref{WilsonRG} for the case of a real scalar field.

Let us now recall briefly the basic idea behind Wilson's exact
renormalization group.\footnote{The equations in this section are given
for the purpose of illustration, and they require modifications to be
strictly correct.}  The starting point is a bare action $S_B$, a
functional of field variables with an ultraviolet cutoff $\Lambda_0$.
The correlation of field variables is given by a functional integral
with the weight $\e^{S_B}$:
\begin{equation}
\vev{\phi (p_1) \cdots \phi (p_n)} = \int [d\phi]\,
\phi (p_1) \cdots \phi (p_n) \, \e^{S_B [\phi]}\,.
\end{equation}
By integrating over the field variables with momentum between
$\Lambda_0$ and $\Lambda < \Lambda_0$, we obtain an equivalent action
$\SL$, called the Wilson action, that gives the same correlation
functions:
\begin{equation}
\vev{\phi (p_1) \cdots \phi (p_n)} = \int [d\phi]\,
\phi (p_1) \cdots \phi (p_n) \, \e^{\SL [\phi]}\,.
\end{equation}
The Wilson action $\SL$ has the ultraviolet cutoff $\Lambda$.  As we
lower $\Lambda$, we generate a flow of equivalent Wilson actions that
provide the same correlation functions.\footnote{The precise definition
of the Wilson action is given by (\ref{derivation-WilsonI}), and the
$\Lambda$ independence of the correlation functions are given by
(\ref{derivation-vevSBSL}).}

In defining the continuum limit of a field theory, we usually look at
the correlation functions as we raise $\Lambda_0 \to
\infty$.\footnote{Here, we describe the more traditional perturbative
definition of renormalizability, not the non-perturbative definition
given in \citen{Wilson:1973jj}.  Hence, no rescaling of space is
introduced.} If the theory is renormalizable, we can give appropriate
$\Lambda_0$ dependence to the parameters of the bare action $S_B$ so
that the correlation functions have a limit as $\Lambda_0 \to \infty$:
\begin{equation}
\lim_{\Lambda_0 \to \infty} \vev{\phi (p_1) \cdots \phi (p_n)}\,.
\end{equation}
Alternatively, we can look at the Wilson action $\SL$ at a fixed
$\Lambda$.  $S_B$ and $\SL$ give the same correlation functions, and if
$\SL$ has a limit as $\Lambda_0 \to \infty$, the correlation functions
must also have a limit.  Polchinski used this idea to simplify the proof
of the perturbative renormalizability of the $\phi^4$
theory.\cite{Polchinski:1983gv}

Polchinski's work was subsequently developed further.  We just mention a
few as examples.  An early attempt was made in
\citen{Warr:1986we,Warr:1986ux} to apply Polchinski's differential
equation to construct Yang-Mills, chiral, and supersymmetric Yang-Mills
theories.  The beta functions of renormalized parameters were derived
from the Wilson action,\cite{Hughes:1987rf} and the proof of
renormalizability in \citen{Polchinski:1983gv} was
simplified\cite{Keller:1990ej},
extended\cite{Keller:1991bz,Keller:1992by}, and given more
rigor.\cite{Ball:1993zy}.

Turning now to the subject of symmetry realization, we call a symmetry
\textbf{manifest} if it exists in the bare action $S_B$.  An example is
a continuous linear symmetry, such as O($N$) of $N$ real scalar fields,
for which the transformation is linear in the elementary fields.
Another example is the gauge symmetry of the lattice gauge theory, for
which the field transformation is highly non-linear.  Not all symmetry
is manifest, though: even if the symmetry cannot be realized manifestly,
it may exist in the continuum limit of the theory.

Let us recall that the continuum limit is described fully by the Wilson
action.  If a theory has symmetry in its continuum limit, the symmetry
must be present in the Wilson action.  This simple reasoning suggests
the possibility of formulating symmetry using the Wilson action in the
continuum limit.  It is the second goal of this review to explain this
formulation.  We introduce two identities:
\begin{enumerate}
\item the \textbf{Ward-Takahashi identity} (\textbf{WT} identity) (to be
      discussed in \S\S \ref{WT}, \ref{WTexamples}) --- this familiar
      identity is usually given for the bare action in the case of
      manifest symmetry.  In general any continuous symmetry, whether
      linear or non-linear, can be formulated as the invariance of the
      Wilson action under an infinitesimal transformation of field
      variables.  The infinitesimal change is highly non-linear due to
      integration of higher momentum modes, giving rise to a non-trivial
      jacobian of the transformation.  Hence, the Wilson action is
      invariant only if the jacobian is taken into account.  The WT
      identity for the Wilson action was discussed earlier by Becchi in
      his seminal work on the ERG approach to YM theories.\cite{Becchi:1996an}
\item the \textbf{quantum master equation} (to be discussed in \S\S
      \ref{AF}, \ref{AFexamples}, and again in \S\S \ref{BV},
      \ref{BVexamples}) --- we introduce classical external sources that
      generate the symmetry transformation of fields.  For gauge theory,
      this is quite familiar from the Zinn-Justin equation satisfied by
      the effective action.\cite{ZinnJustin:1993wc} The quantum master
      equation results from an adaptation of the general
      Batalin-Vilkovisky formalism\cite{Batalin:1981jr}\footnote{We
      recommend \citen{Henneaux-Teitelboim,BVreview-Gomis} as reviews on
      the BV formalism.} to the Wilson action.  It is not hard to expect
      the presence of a rich algebraic structure in this formulation.
      In fact, for YM theories we must generalize the WT identity to the
      quantum master equation in order to complete the perturbative
      proof of the theory's existence.  (\S\S \ref{WTexamples-YM} and
      \ref{AFexamples-YM})
\end{enumerate}

\vspace{0.2cm}
In formulating either the WT identity or the quantum master equation, we
must introduce renormalized parameters to parameterize the radiative
corrections to the field transformation.  We must tune not only the
parameters of the Wilson action but also these parameters to satisfy the
WT identity or the quantum master equation.  The necessity of giving
cutoff dependence on the symmetry transformation is reminiscent of
chiral symmetry on the lattice.  There, the symmetry transformation is
not simply specified by the standard $\gamma_{5}$ matrix, but by its
non-trivial extension that depends on the Dirac operator as well as
lattice spacing.  Ginsparg and Wilson\cite{Ginsparg:1981bj} found a
specific form of the Ward-Takahashi (WT) identity for the presence of a
lattice chiral symmetry.\cite{Luscher:1998pqa} In \S
\ref{WTexamples-axial} we derive an analogous identity in the ERG
framework\cite{Igarashi:1999rm}.

As we will explain in \S \ref{derivation-Wetterich}, there is an
alternative way of introducing the Wilson action $\SL$.  Instead of
$S_B$, we start with a bare action $S_{B,\Lambda}$ not only with an UV
cutoff $\Lambda_0$ but also an IR cutoff $\Lambda$.  The Wilson action
can be defined as the generating functional of the connected correlation
functions of $S_{B,\Lambda}$.  We can then introduce the effective
average action $\Gamma_{B,\Lambda}$ as the Legendre transform of $\SL$,
or the generating functional of 1PI correlation functions.  Symmetry of
the continuum limit can be formulated using $\Gamma_{B,\Lambda}$ instead
of $\SL$.  For YM theories, this approach was initiated by
Ellwanger.\cite{Ellwanger:1994iz,Ellwanger:1995qf} The corresponding WT
identity is called the \textbf{modified Slavnov-Taylor (ST) identity},
because this identity approaches the Slavnov-Taylor identity or the
Jinn-Zustin equation as $\Lambda \to 0$.  In the modified ST identity,
the jacobian of the symmetry transformation contains the inverse of the
second-order differential of $\Gamma_{B,\Lambda}$.  This makes the
perturbative analysis of the modified ST identity a little more
complicated than that of the WT identity.  For this reason we use the WT
identity and the corresponding quantum master equation in this review;
we explain the modified ST identity only briefly in \S\ref{BV-ST} and
Appendix \ref{appgamma}.  For a review of the modified ST identity, we
refer the reader to one of the references we will cite toward the end of
this introduction.

This review consists of three main parts.  Assuming that most of the
readers are not familiar with the exact renormalization group or
Polchinski's differential equations, we give a detailed introduction to
the technique before starting its application to symmetry realization.
Except in \S\S \ref{derivation}, \ref{BV}, we treat theories
perturbatively.  The review is organized as follows.

Part I consists of \S\S \ref{derivation}, \ref{cl}, and \ref{comp}.  In
section 2, we define the Wilson action $\SL$ and derive the ERG
differential equation (Polchinski equation) that gives the $\Lambda$
dependence of $\SL$.  Primarily for notational simplicity, we take a
real scalar field theory as a generic example.  We then introduce the
effective average action as the Legendre transform of the Wilson action.
The $\Lambda$ dependence of the effective average action is given by the
Wetterich equation, widely used for non-perturbative applications of
ERG.  The diagrammatic interpretation of these two flow equations are
given for more insights.  In \S \ref{cl}, we define perturbative
renormalizability in terms of a Wilson action.  A renormalizable theory
is characterized by the behavior of the Wilson action for the cutoff
$\Lambda$, large compared with the momenta carried by the field
variables.  In \S \ref{comp}, we introduce ``composite operators.''
Following Becchi,\cite{Becchi:1996an} we give a specific meaning to the
composite operators, more than an arbitrary functional of field
variables.  The composite operators play essential roles in our later
discussions of symmetry realization: the presence of symmetry amounts to
the vanishing of a certain composite operator.

Part II consists of \S\S \ref{WT} through \ref{anomaly}.  In the
ordinary treatment of a field theory, the presence of continuous
symmetry is expressed as the Ward-Takahashi (WT) identity.  In \S
\ref{WT}, we derive the corresponding relation satisfied by the Wilson
action.  This relation is expressed as the vanishing of the WT composite
operator, given as the sum of the change of the Wilson action under an
infinitesimal transformation and the jacobian of the transformation.  We
provide several examples of the formalism in \S \ref{WTexamples}: QED,
YM theories, the Wess-Zumino model, the O(N) non-linear sigma model in
two dimensions, and fermionic systems with an axial symmetry.  The
discussions in \S\S\ref{WT} \& \ref{WTexamples} have a serious
shortcoming: the field transformations for the WT identity are applied
only once, and they lack an algebraic structure. Particularly for YM
theories, the BRST transformation for the Wilson action lacks
nilpotency, and as a result we cannot prove the possibility of
satisfying the WT identity by fine-tuning the parameters of the theory.
To circumvent this difficulty, we extend the WT identity by introducing
classical sources that generate the symmetry transformations.  The
discussion in \S \ref{AF} emphasizes the practical aspect of the
formalism, and it is developed only to the extent necessary for the
concrete application to YM theories in \S \ref{AFexamples-YM}.  In \S
\ref{anomaly} we apply the formalism of \S \ref{WT} to the axial and
chiral anomalies.  The discussions are quite limited: we discuss only
abelian gauge theories up to 1-loop.

Part III consists of \S\S \ref{BV} and \ref{BVexamples}.  We adapt the
general formalism of Batalin and Vilkovisky to an arbitrary bare action
with an UV cutoff and its Wilson action.  To emphasize the generality of
the framework, we adopt a matrix notation that handles bosonic and
fermionic fields equally.  We show how naturally the Batalin-Vilkovisky
formalism applies to the Wilson action.  Contrary to \S\S \ref{AF} \&
\ref{AFexamples}, we do not base our discussions on loop expansions, but
we use a general functional method of \S \ref{derivation}.

Throughout this review we work on the euclidean space with $D$
dimensions, where $D$ may be specified in concrete models.  We use the
notation
\begin{equation}
\int_q \equiv \int \frac{d^D q}{(2 \pi)^D}
\end{equation}
for integrals over momenta.  

Please be warned that this review is by no means a comprehensive review
of the exact renormalization group approach to field theory.  In the
main text we cite only those references that are relevant to symmetry
realization and that happen to be familiar to us.  We owe an apology to
those whose contributions deserve citations in a better researched
review article.

Before closing the introductory section, however, we would like to give
references on the aspects of ERG that may not be directly relevant to
this review; this may help the reader broadening his perspective.  The
exact renormalization group equation of Wilson and that of Wegner \&
Houghton were introduced originally for non-perturbative studies, and
accordingly there have been many such applications.  Most
non-perturbative studies use an approximation called the \textbf{local
potential approximation (LPA)}, where only the local potential part of
the Wilson action transforms under
ERG.\cite{Nicoll:1974zz}\cite{Hasenfratz:1985dm}\cite{Margaritis:1987hv}
\cite{Ball:1994ji}\cite{Aoki:1996fn,Aoki:1996gx,Aoki:1998um} The
analytic properties of the LPA have only recently been studied
\cite{Bervillier:2008kh,Bervillier:2008pt}.  General classes of
approximations to the ERG differential equation have been introduced in
\citen{Golner:fp}.

The ERG differential equation for the effective average action
$\Gamma_{B,\Lambda}$ is now commonly called the Wetterich
equation,\cite{Nicoll:1977hi}\cite{Wetterich:1992yh}\cite{Wetterich:1993ne}
\cite{Morris:1993qb}\cite{Bonini:1992vh}
which we will derive in \S\ref{derivation-Wetterich}. The added note to
\citen{Morris:1993qb}, where the relation between the Polchinski and
Wetterich equations was clarified, has a short but careful analysis of
the history of the Wetterich equation.  For numerical works, Wetterich's
equation is known to have better convergence properties than
Polchinski's.\cite{Tetradis:1993ts}\cite{Alford:1994fa} See
Refs.~\citen{Benitez:2009xg, Benitez:2007mk} for the state-of-the-art
calculations of the critical exponents for the O(N) linear sigma model.
The equivalence between the LPA for Polchinski's and that for
Wetterich's has been shown in \citen{Morris:2005ck}.

As for applications of ERG to gauge theories, there have been two other
approaches besides the WT identity and its extension explained in this
review:
\begin{enumerate}
\item Manifestly gauge invariant ERG formalism for YM theories have been
      introduced by Morris, Rosten, and their collaborators.
      \cite{Morris:1999px,Morris:2000fs}\cite{Arnone:2001iy}
      \cite{Morris:2005tv,Arnone:2005fb}\cite{Morris:2006in}
      \cite{Rosten:2006qx}\cite{Rosten:2006tk}\cite{Rosten:2005ka}
\item The background field method has been incorporated within the ERG
      framework.\cite{Reuter:1993kw}\cite{Freire:1996db}
      \cite{Freire:2000mn,Litim:2002hj}
      \cite{Freire:2000bq,Bonini:2001xm}
\end{enumerate}
ERG has also been applied to quantum gravity, and a possible scenario
for asymptotic safety has been
proposed\cite{Reuter:2007rv,Lauscher:2007zz}\cite{Litim:2008tt}.

Finally, we give references to the review articles that complement the
present review: Morris\cite{Morris:1998da}, Litim \cite{Litim:1998nf},
Aoki\cite{Aoki:2000wm}, Bagnuls \& Bervillier \cite{Bagnuls:2001mw},
Berges, Tetradis, and Wetterich \cite{Berges:2000ew},
Polonyi\cite{Polonyi:2001se}, and Pawlowski \cite{Pawlowski:2005xe}.
Especially, we recommend the pedagogical review on the functional RG
and gauge theories by Gies\cite{Gies:2006wv}.  For applications of ERG
to condensed matter physics, we recommend Shankar
\cite{Shankar:1993pf}, Fisher\cite{Fisher:1998kv} (on RG in general),
Salmhofer \cite{Salmhofer:2001tr}, and
Delamotte\cite{Delamotte:2007pf}.

\newpage

\section{Derivation of the ERG differential
  equations\label{derivation}}

\subsection{Splitting fields into high and low momentum
  modes\label{derivation-split}} 

\begin{figure}[t]
\begin{center}
\epsfig{file=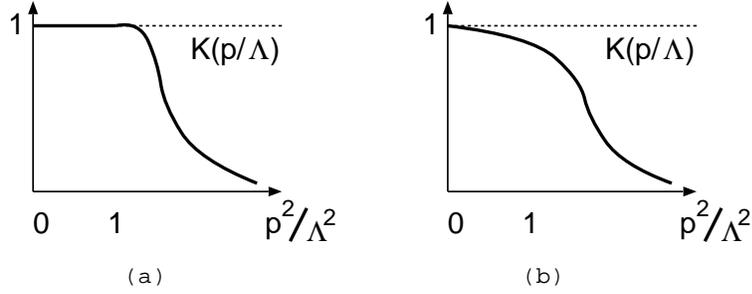, width=10cm}
\caption{Cutoff function --- (a) satisfying 2, (b) satisfying $2'$ but not
 2.}
\label{derivation-K}
\end{center}
\end{figure}

As a generic case, we consider a real scalar theory defined by a bare
action
\begin{equation}
S_B \equiv - \frac{1}{2} \int_p \frac{p^2 + m^2}{\Kz{p}} \phi (p) \phi
(-p) + S_{I,B} [\phi]\,,\label{derivation-SB}
\end{equation}
where $K$ is a cutoff function, for which we demand the following
properties:
\begin{enumerate}
\item $K (p/\Lambda)$ is a smooth non-increasing positive function of
    $p^2/\Lambda^2$. 
\item $K (p/\Lambda)$ is $1$ for $p^2/\Lambda^2 < 1$. 
\item $K (p/\Lambda)$ damps sufficiently fast as $p^2/\Lambda^2 \to \infty$. 
\end{enumerate}
(See Fig.~\ref{derivation-K}(a).)  Hence, $\Kz{p}$ is small for $p^2 >
\Lambda_0^2$, and the high momentum modes are suppressed by the
large gaussian term.    
$\Lambda_0$ is the UV cutoff of the theory, but the high momentum modes
of $\phi$ are not cut off abruptly at $\Lambda_0$: the smoothness of
$\Kz{p}$ for $p^2/\Lambda_0^2 > 1$ gives rise to smooth damping of the
modes $p^2 > \Lambda_0^2$.

The regularization in terms of a smooth cutoff function $K$
generalizes regularization in terms of higher order derivatives which
corresponds to a choice of a finite order polynomial:
\begin{equation}
\frac{1}{\Kz{p}} = 1 + \sum_{n=1}^N a_n
\left(\frac{p^2}{\Lambda_0^2}\right)^n\,.
\end{equation}
With $N$ finite, this cannot satisfy the second property
\begin{equation}
- p^2 \frac{d}{dp^2} \K{p} = 0 \quad (p^2 < \Lambda^2)
\end{equation}
that assures the locality of the differential equation for the exact
renormalization group.  We usually assume 2 for locality. (See
sect.~\ref{derivation-polchinski}.) But it is sometimes necessary to
keep $K$ strictly less than $1$ except at $p=0$; for example in
\S\ref{derivation-Wetterich}, we need that division by $1-\K{p}$ make
sense.  We can replace 2 by
\begin{itemize}
\item[2${}^\prime$.] $K(0) = 1$.
\end{itemize}
(See Fig.~\ref{derivation-K}(b).)  For practical calculations the
condition 2 is often replaced by 2${}^\prime$.

Let us give some concrete examples of the cutoff function.  One popular
choice for numerical calculations is that of Litim\cite{Litim:2001up}:
\begin{equation}
K(p/\Lambda) = \lb\begin{array}{c@{\quad}l}
1 - p^2/\Lambda^2& (p^2 < \Lambda^2)\,,\\
0 & (p^2 > \Lambda^2)\,.
\end{array}\right.
\end{equation}
This satisfies 1, 2${}'$, and 3, but not 2.  The gaussian function
\begin{equation}
\K{p} = \exp ( - p^2/\Lambda^2 )
\end{equation}
also satisfies only 1, 2${}^\prime$, and 3.  One example, satisfying 1,
2, and 3, is
\begin{equation}
\K{p} = \lb\begin{array}{c@{\quad}l}
1& (p^2 < \Lambda^2)\,,\\
1 - \exp\left[ - \left(\frac{\Lambda^2}{p^2 -
		  \Lambda^2}\right)^n\right]& (p^2 > \Lambda^2)\,,
\end{array}\right.
\end{equation}
where $n$ is a big enough positive number for UV convergence.  (For the
scalar theory in $D=4$, we can take $n = 2$, for example.)  The step
function, first used in \citen{Wegner:1972ih},
\begin{equation}
\K{p} = \theta (1-p^2/\Lambda^2) \equiv \lb \begin{array}{l@{\quad}l}
1& (p^2 < \Lambda^2)\\
0& (p^2 > \Lambda^2)\end{array}\right.
\end{equation}
is not smooth, but is often convenient for practical calculations,
especially for analytic calculations.

We should remark that the choice of $K$ is unimportant theoretically; in
Appendix \ref{appcomp} on universality we will show that a different
choice of $K$ merely amounts to reparameterization and renormalization of
fields.

\subsubsection{Functional integrals without a source}

All the momentum modes $p^2 < \Lambda_0^2$ contribute to the
vacuum functional integral
\begin{eqnarray}
Z_B &\equiv& \int [d\phi] \, \exp \left[ S_B [\phi] \right]\nn\\
&=& \int [d\phi]\, \exp \left[ - \frac{1}{2} \int_p \frac{p^2 + m^2}{\Kz{p}}
    \phi (p) \phi (-p) + S_{I,B} [\phi] \right]\,,
\end{eqnarray}
where the contribution of the modes $p^2 > \Lambda_0^2$ are suppressed
by the cutoff function.  We would like to split $\phi (p)$ into the
high and low momentum modes
\begin{equation}
\phi (p) = \phi_h (p) + \phi_l (p)\,,
\end{equation}
so that, roughly speaking, $\phi_h$ contains the momentum modes
$\Lambda^2 < p^2 < \Lambda_0^2$, where $\Lambda$ is an arbitrarily
chosen momentum scale, and $\phi_l$ contains the low momentum
modes $p^2 < \Lambda^2$.  To be more precise, in what follows we will
obtain the splitting formula
\begin{eqnarray}
Z_B &=& \int [d\phi_l] [d\phi_h] \exp \left[ - \frac{1}{2} \int_p
    \frac{p^2 + m^2}{K(p/\Lambda)} \phi_l (p) \phi_l (-p)\right.\nn\\
&&\left. - \frac{1}{2} \int_p \frac{p^2 + m^2}{K(p/\Lambda_0) -
  K(p/\Lambda)} \phi_h (p) \phi_h (-p) + S_{I,B} \left[\phi_h +
    \phi_l\right] \right]\,.\label{derivation-splitformula}
\end{eqnarray}
The propagator of $\phi_l$ is the low momentum propagator
\begin{equation}
\frac{\K{p}}{p^2 + m^2}\,,
\end{equation}
while that of $\phi_h$ is the high momentum one
\begin{equation}
\frac{\Kz{p} - \K{p}}{p^2 + m^2}\,.
\end{equation}
The original propagator 
\begin{equation}
\frac{\Kz{p}}{p^2 + m^2}
\end{equation}
of $\phi = \phi_h + \phi_l$ is reproduced as their sum.

The splitting of $\phi$ into high and low momentum modes is an example
of the following general formula:
\begin{eqnarray}
&& \int [d\phi] \exp \left[ - \frac{1}{2} \int_p \frac{1}{A(p) + B(p)}
		      \phi (-p) \phi (p) + S_{I,B} [\phi]\right]\nn\\
&=& \int [d\phi_1][d\phi_2] \exp \Bigg[ - \frac{1}{2} \int_p
				  \frac{1}{A(p)} \phi_1 (-p)
\phi_1 (p) - \frac{1}{2} \int_p \frac{1}{B(p)} \phi_2 (-p) \phi_2 (p) \nn\\
&&\qquad \qquad \qquad +
S_{I,B} [\phi_1 +\phi_2] \Bigg]\,,\label{derivation-splitformulaAB}
\end{eqnarray}
where $A$ and $B$ are both non-negative functions of $p^2$.  By choosing
\begin{equation}
A(p) = \Kz{p} - \K{p}\,,\quad B(p) = \K{p}\,,
\end{equation}
we obtain (\ref{derivation-splitformula}) from
(\ref{derivation-splitformulaAB}).  By applying
(\ref{derivation-splitformulaAB}) multiple times, we obtain
\begin{eqnarray}
&&\int [d\phi] \exp \left[ - \frac{1}{2} \int_p \frac{1}{\sum_{i=1}^n
		     A_i (p)} \phi (-p) \phi (p) + S_{I,B}
		     [\phi]\right]\nn\\
&=& \int [d\phi_1] \cdots [d\phi_n] \exp \left[ - \frac{1}{2}
					  \sum_{i=1}^n
\int_p \frac{1}{A_i (p)} \phi_i (-p) \phi_i (p) + S_{I,B} \left[
\sum_{i=1}^n \phi_i \right] \right]\,.
\end{eqnarray}
For example, we can choose
\begin{equation}
A_i (p) = K\left(p/\Lambda_{i-1}\right) -
 K\left(p/\Lambda_i\right)\,,
\end{equation}
where
\begin{equation}
\Lambda_0 > \Lambda_1 > \cdots > \Lambda_{n-1} > \Lambda_n = 0\,.
\end{equation}

Let us now prove (\ref{derivation-splitformulaAB}).  It is easier to go
from the right-hand side to the left.  Writing
\begin{equation}
\phi = \phi_1 + \phi_2\,,\quad
\phi' = \phi_2 - \frac{B}{A+B} \phi\,,
\end{equation}
we obtain
\begin{eqnarray}
&&- \frac{1}{2} \int_p \frac{1}{A(p)} \phi_1 (-p)
\phi_1 (p) - \frac{1}{2} \int_p \frac{1}{B(p)} \phi_2 (-p) \phi_2 (p)
+ S_{I,B} [\phi_1 + \phi_2]\nn\\
&=& - \frac{1}{2} \int_p \frac{1}{A+B} \phi(-p) \phi (p) -
 \frac{1}{2} \int_p 
 \left( \frac{1}{A} + \frac{1}{B}\right) \phi'(-p) \phi'(p) +
 S_{I,B} [\phi]\,. 
\end{eqnarray}
Since the jacobian for the change of variables from $\phi_1, \phi_2$ to
$\phi, \phi'$ is unity, we obtain
\begin{eqnarray}
&&\int [d\phi] \exp \Bigg[ - \frac{1}{2} \int_p \frac{1}{A} \phi_1 \phi_1
 - \frac{1}{2} \int_p \frac{1}{B} \phi_2 \phi_2 + S_{I,B}[\phi_1+\phi_2]
\Bigg]\nn\\
&& = \int [d\phi] \exp \left[  - \frac{1}{2} \int_p \frac{1}{A+B}
			\phi(-p) \phi (p) + S_{I,B} [\phi]\right]\nn\\
&&\qquad\times \int [d\phi'] \exp \left[ - \frac{1}{2} \int_p 
 \left( \frac{1}{A} + \frac{1}{B}\right) \phi'(-p) \phi'(p)\right]\,.
\end{eqnarray}
The gaussian integral over $\phi'$ can be regarded as unity
\begin{equation}
\int [d\phi'] \exp \left[ - \frac{1}{2} \int_p 
 \left( \frac{1}{A} + \frac{1}{B}\right) \phi'(-p) \phi'(p)\right] =
1\,,
\end{equation}
if we disregard a constant factor, the exponential of an additive
constant proportional to the volume of the entire space.  Hence, we
obtain (\ref{derivation-splitformulaAB}), and consequently
(\ref{derivation-splitformula}).

We now define $S_{I,\Lambda}$, the interaction part of the
\textbf{Wilson action}, by
\begin{equation}
\exp \left[S_{I,\Lambda} [\phi]\right]
\equiv \int [d\phi'] \exp \left[ - \frac{1}{2} \int_p
    \frac{p^2 + m^2}{K_0 (p) -K (p)} \phi' (p) \phi' (-p) 
+ S_{I,B} [\phi + \phi'] \right] \,,\label{derivation-WilsonI}
\end{equation}
where we have used a short hand notation
\begin{equation}
K_0 (p) \equiv \Kz{p}\,,\quad
K(p) \equiv \K{p}\,.\label{derivation-shorthand}
\end{equation}
Defining the full Wilson action by
\begin{equation}
S_\Lambda [\phi] \equiv - \frac{1}{2} \int_p \frac{p^2 + m^2}{\K{p}} \phi
(-p) \phi (p) + S_{I,\Lambda} [\phi] \,,\label{derivation-Wilson}
\end{equation}
we obtain $Z_B$ as the functional integral of the full Wilson action
\begin{equation}
Z_B = \int [d\phi] \exp \left[ S_\Lambda [\phi] \right]\,.
\end{equation}
We have thus shown that as far as the vacuum functional integral $Z_B$
is concerned, the bare action $S_B$ with cutoff $\Lambda_0$ can be
replaced by a Wilson action $S_\Lambda$ with an arbitrary cutoff
$\Lambda < \Lambda_0$.  In the next subsection, we wish to show that
this equivalence goes further: $S_\Lambda$ and $S_B$ give the same
correlation functions of the fields.

The transformation from $S_B$ to $S_\Lambda$ is called an \textbf{exact
  renormalization group} (\textbf{ERG}) transformation.  The word
``\textbf{exact}'' is a mnemonic for no loss of information.

\subsection{Functional integrals with a source\label{derivation-source}}

We consider the generating functional $W_B [J]$ of connected
correlation functions:
\begin{equation}
\exp \left[W_B [J]\right] \equiv \int [d\phi] \exp \left[ S_B [\phi] +
    \int_p J(-p) \phi 
    (p) \right]\,. \label{derivation-WB}
\end{equation}
Similarly, we consider
\begin{equation}
\exp \left[W_\Lambda [J]\right] \equiv \int [d\phi] \exp \left[
    S_\Lambda [\phi] + 
    \int_p J(-p) \phi (p) \right] \label{derivation-WL}
\end{equation}
for the Wilson action.  In the following we wish to show that
$W_B$ and $W_\Lambda$ are related as
\begin{equation}
W_B [J] = W_\Lambda \left[ \frac{K_0}{K} J \right] - \frac{1}{2}
\int_p \frac{K_0 - K}{p^2 + m^2} \frac{K_0}{K} J(-p) J(p)\,,
\label{derivation-WBWL} 
\end{equation}
where we have used the short hand notation \bref{derivation-shorthand},
omitting the momentum variable $p$.  This relation implies that the
correlation functions of the bare theory can be calculated using the
Wilson action.

To derive (\ref{derivation-WBWL}), we first define 
\begin{equation}
S_B [\phi; J] \equiv - \frac{1}{2}\int_p \frac{p^2 + m^2}{K_0} \phi
(p) \phi (-p) + S_{I,B} [\phi; J]\,,
\end{equation}
where
\begin{equation}
S_{I,B} [\phi; J] \equiv S_{I, B} [\phi] + \int_p J(-p) \phi (p)
\end{equation}
is a bare action that has a linear coupling to the classical external
field $J$.  By definition we obtain
\begin{equation}
\exp \left[ W_B [J] \right] = \int [d\phi] \exp \left[ S_B [\phi; J]
\right]\,.
\end{equation}
Let us now call the corresponding Wilson action by $S_\Lambda [\phi;
J]$ and its interaction part by $S_{I,\Lambda} [\phi; J]$.  Then, from
the result of the previous section, we obtain
\begin{equation}
\exp \left[ W_B [J] \right] = \int [d\phi] \exp \left[ S_\Lambda [\phi; J]
\right]\,.
\end{equation}
We obtain (\ref{derivation-WBWL}) by working out $S_\Lambda [\phi; J]$,
as we will show below.

From the definition (\ref{derivation-WilsonI}) of a Wilson action, we
obtain
\begin{eqnarray}
&&\exp \left[ S_{I,\Lambda} [\phi; J] \right] \equiv \int [d\phi']
\exp \left[ - \frac{1}{2} \int_p \frac{p^2 + m^2}{K_0-K} \phi'  \phi'
    + S_{I,B} [\phi+\phi' ; J] \right]\nn\\
&& = \int  [d\phi']
\exp \left[ - \frac{1}{2} \int_p \frac{p^2 + m^2}{K_0-K} \phi' \phi' 
    + S_{I,B} [\phi+\phi'] + \int J (\phi + \phi')
\right]\,.\label{derivation-defSILJ} 
\end{eqnarray}
We can rewrite this as
\begin{eqnarray}
\exp \left[ S_{I,\Lambda} [\phi; J] \right] &=& \int [d\phi']
\exp  \left[ - \frac{1}{2} \int_p \frac{p^2 + m^2}{K_0-K} \left(\phi' -
        \frac{K_0-K}{p^2 + m^2} J \right) \left(\phi' -
        \frac{K_0-K}{p^2 + m^2} J \right) \right.\nn\\
&&\left.\qquad\qquad  + S_{I,B} [\phi+\phi'] + \int J \phi + \frac{1}{2} \int
\frac{K_0-K}{p^2 + m^2} J J \right]\,.
\end{eqnarray}
Now, shifting $\phi'$, we obtain
\begin{eqnarray}
&&\exp \left[ S_{I,\Lambda} [\phi; J] \right] = \int [d\phi']
\exp \left[ - \frac{1}{2} \int_p \frac{p^2 + m^2}{K_0-K} \phi'\phi'\right.\nn\\
&&\qquad \left.+ S_{I,B} \left[\phi+\frac{K_0-K}{p^2 + m^2} J +
        \phi'\right] + \int J \phi + \frac{1}{2} \int \frac{K_0-K}{p^2
      + m^2} J J \right]\nn\\
&&\quad= \exp \left[ S_{I,\Lambda} \left[ \phi+\frac{K_0-K}{p^2 + m^2} J
    \right] + \int J \phi + \frac{1}{2} \int \frac{K_0-K}{p^2 + m^2}
J J \right]\,.
\end{eqnarray}
This implies an important intermediate result
\begin{equation}
S_{I,\Lambda} [\phi; J] = S_{I,\Lambda} \left[ \phi+\frac{K_0-K}{p^2 + m^2} J
    \right] + \int_p J (-p) \phi (p) + \frac{1}{2} \int_p
    \frac{K_0-K}{p^2 + m^2} J(-p) J (p)\,,\label{derivation-Jshift}
\end{equation}
that the dependence on $J$ is given mostly by the shift of
$\phi$ proportional to $J$.  If $J$ couples non-linearly to the field
in the bare action, however, there is no simple formula like this for
the $J$ dependence of the Wilson action.

Now, using (\ref{derivation-Jshift}), we can compute $W_B[J]$ as follows:
\begin{eqnarray}
\exp \left[ W_B [J]\right] &=& \int [d\phi] \exp \left[ S_\Lambda
    [\phi; J]\right]
= \int [d\phi] \exp \left[ - \frac{1}{2} \int_p \frac{p^2 + m^2}{K}
    \phi \phi\right.\nn\\
&&\quad \left.+ S_{I,\Lambda} \left[ \phi + \frac{K_0-K}{p^2 + m^2} J
      \right] + \int J \phi + \frac{1}{2} \int \frac{K_0-K}{p^2 + m^2}
      J J \right]\,.
\end{eqnarray}
Changing the variable from $\phi (p)$ to
\begin{equation}
\phi' (p) \equiv \phi(p) + \frac{K_0-K}{p^2 + m^2} J(p)\,,
\end{equation}
we obtain the desired relation \bref{derivation-WBWL}:
\begin{eqnarray}
    \exp \left[ W_B [J]\right] 
    &=& \int [d\phi'] \exp \left[ S_\Lambda [\phi'] + \int \frac{K_0}{K} J
        \phi' - \frac{1}{2} \int \frac{K_0-K}{p^2 + m^2} \frac{K_0}{K} J J
    \right]\nn\\
    &=& \exp \left[ W_\Lambda \left[ \frac{K_0}{K} J \right] -
        \frac{1}{2} \int \frac{K_0-K}{p^2 + m^2} \frac{K_0}{K} J J
    \right]\,.
\end{eqnarray}
In fact we can go backward: given (\ref{derivation-WBWL}), we can derive
the definition (\ref{derivation-WilsonI}) of the Wilson action.  Thus,
(\ref{derivation-WilsonI}) and (\ref{derivation-WBWL}) are equivalent.
We postpone the derivation of (\ref{derivation-WilsonI}) from
(\ref{derivation-WBWL}) till \S \ref{comp}.

Let us now derive the consequences of the relation
(\ref{derivation-WBWL}).  $W_B$ and $W_\Lambda$ are the generating
functionals of connected correlation functions, which we denote by
brackets:
\begin{eqnarray}
\vev{\phi (p_1) \cdots \phi (p_n)}_{S_B} (2\pi)^D \delta^{(D)} (p_1 +
\cdots + p_n) &\equiv& \prod_{i=1}^n \frac{\delta}{\delta J(-p_i)}
\cdot W_B [J] \Big|_{J=0}\,,\\
\vev{\phi (p_1) \cdots \phi (p_n)}_{S_\Lambda} (2\pi)^D \delta^{(D)} (p_1 +
\cdots + p_n) &\equiv& \prod_{i=1}^n \frac{\delta}{\delta J(-p_i)} \cdot
 W_\Lambda [J] \Big|_{J=0}\,.
\end{eqnarray}
Then, (\ref{derivation-WBWL}) implies
\begin{equation}
\lb\begin{array}{r@{~=~}l}
\vev{\phi (p) \phi (-p)}_{S_B} & \frac{K_0^2}{K^2} \vev{\phi (p) \phi
  (-p)}_{S_\Lambda} - \frac{K_0 - K}{p^2 + m^2} \frac{K_0}{K}\,,\\
\vev{\phi (p_1) \cdots \phi (p_n)}_{S_B} & \prod_{i=1}^n \frac{K_0
  (p_i)}{K(p_i)} \cdot \vev{\phi (p_1) \cdots \phi (p_n)}_{S_\Lambda}\,,
\end{array}\right.\label{derivation-vevSBSL}
\end{equation}
where $n > 2$.  Hence, the correlation functions of the bare action
are completely determined by those of the Wilson action.  We do not
miss any physics by reducing the momentum cutoff from $\Lambda_0$ to
$\Lambda$, as long as $K(p/\Lambda)$ is a smooth non-vanishing
function so that the inverse $1/K(p/\Lambda)$ makes sense for all
momenta. 

\subsection{Polchinski equation \label{derivation-polchinski}}

The Wilson action $S_\Lambda$ is obtained from the bare action $S_B$ by
integrating over the momenta between $\Lambda$ and $\Lambda_0$.  We wish
to consider the change of the Wilson action under an infinitesimal
change of $\Lambda$, and express the result as a differential equation.

\begin{figure}[t]
\begin{center}
\epsfig{file=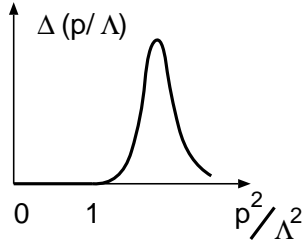, width=4cm}
\caption{$\Delta (p/\Lambda) \equiv - 2 p^2 \frac{\partial
 \K{p}}{\partial p^2}$ vanishes for $p^2 <
 \Lambda^2$ if $\K{p} = 1$ for $p^2 < \Lambda^2$.}
\label{derivation-Delta}
\end{center}
\end{figure}

We first recall the precise definition of the Wilson action
(\ref{derivation-WilsonI}):
\begin{equation}
\exp \left[S_{I,\Lambda} [\phi]\right]
\equiv \int [d\phi'] \exp \left[ - \frac{1}{2} \int_p
    \frac{p^2 + m^2}{K_0-K} \phi' (p) \phi' (-p) 
+ S_{I,B} [\phi + \phi'] \right] \,.
\end{equation}
In deriving the Polchinski equation, it is convenient to write $\phi'$
for $\phi' + \phi$ to rewrite
\begin{equation}
\exp \left[S_{I,\Lambda} [\phi]\right]
\equiv \int [d\phi'] \exp \left[ - \frac{1}{2} \int_p
    \frac{p^2 + m^2}{K_0-K} (\phi'-\phi) (p) (\phi'-\phi) (-p) 
+ S_{I,B} [\phi'] \right] \,.
\end{equation}
Differentiating the above with respect to $\Lambda$, we obtain
\begin{eqnarray}
&&- \Lambda \frac{\partial}{\partial \Lambda} \,
\exp \left[S_{I,\Lambda} [\phi]\right] = \frac{1}{2} \int [d\phi'] \int_p
\frac{p^2 + m^2}{K_0 - K} \frac{\Delta (p/\Lambda)}{K_0 - K} (\phi' -
\phi)(p) (\phi'-\phi) (-p) \nn\\
&&\qquad\qquad \times \exp \left[ - \frac{1}{2} \int_p
    \frac{p^2 + m^2}{K_0-K} (\phi'-\phi) (p) (\phi'-\phi) (-p) 
+ S_{I,B} [\phi'] \right] \,,\label{derivation-dSdL}
\end{eqnarray}
where we have defined
\begin{equation}
\Delta (p/\Lambda) \equiv - 2 p^2 \frac{\partial}{\partial p^2}
 \K{p}
= \Lambda \frac{\partial}{\partial \Lambda}
\K{p}\,.\label{derivation-Deltadef}  
\end{equation}
The sketch of $\Delta (p/\Lambda)$ is given in
Fig.~\ref{derivation-Delta}, showing a peak just above $p^2 =
\Lambda^2$, and vanishing for $p^2 < \Lambda^2$ if the property 2 is
assumed for $K$.\footnote{If only 2${}'$ is assumed for $K$, the
function $\Delta (p/\Lambda)$ vanishes at $p=0$, but is non-vanishing
for $p^2 < \Lambda^2$.}
We now compare the $\Lambda$-derivative with the second order
functional derivative with respect to $\phi$.  The first order
derivative is
\begin{eqnarray}
&&\frac{\delta}{\delta \phi (p)} \,\exp \left[S_{I,\Lambda}
    [\phi]\right]
 = \frac{p^2 + m^2}{K_0-K} \int [d\phi'] (\phi' - \phi) (-p) \nn\\
&&\qquad \times \exp \left[ - \frac{1}{2} \int_q
    \frac{q^2 + m^2}{K_0-K} (\phi'-\phi) (q) (\phi'-\phi) (-q) 
+ S_{I,B} [\phi'] \right] \,.\label{derivation-dSILdphi}
\end{eqnarray}
In differentiating this further with respect to $\phi (-p)$, we ignore
differentiating $\phi (-p)$ in front of the exponential, since it
would generate merely an additive constant to $\partial
S_{I,\Lambda}/\partial \Lambda$, independent of $\phi$.\footnote{The constant
is proportional to the volume of the entire space $(2\pi)^D
\delta^{(D)} (0)$.}  We then obtain
\begin{eqnarray}
&&\frac{\delta^2}{\delta \phi (-p) \delta \phi (p)} \,\exp \left[S_{I,\Lambda}
  [\phi]\right] = \left(\frac{p^2 
  + m^2}{K_0-K}\right)^2 \int [d\phi'] (\phi' - \phi) (-p) \left(\phi'
- \phi\right) (p) \nn\\
&&\qquad \times \exp \left[ - \frac{1}{2} \int_q
    \frac{q^2 + m^2}{K_0-K} (\phi'-\phi) (q) (\phi'-\phi) (-q) 
+ S_{I,B} [\phi'] \right] \,.
\end{eqnarray}
Hence, comparing this with (\ref{derivation-dSdL}), we obtain
the differential equation
\begin{equation}
- \Lambda \frac{\partial}{\partial \Lambda} \,\exp \left[S_{I,\Lambda}
    [\phi]\right] 
= \frac{1}{2} \int_p \frac{\Delta (p/\Lambda)}{p^2 + m^2}
\frac{\delta^2}{\delta \phi (p) \delta \phi (-p)} \,\exp \left[S_{I,\Lambda}
  [\phi]\right]\,.
\end{equation}
Rewriting this for $S_{I,\Lambda}$, we get
\begin{equation}
- \Lambda \frac{\partial}{\partial \Lambda} S_{I,\Lambda} [\phi]
= \frac{1}{2} \int_p \frac{\Delta (p/\Lambda)}{p^2 + m^2}
\lb \frac{\delta S_{I,\Lambda}[\phi]}{\delta \phi (-p)}
\frac{\delta S_{I,\Lambda}[\phi]}{\delta \phi (p)} +
\frac{\delta^2 S_{I,\Lambda}[\phi]}{\delta \phi (-p)\delta \phi (p)}
\rb\,. \label{derivation-polchinskieq}
\end{equation}
This is called the \textbf{Polchinski differential
equation}.\cite{Polchinski:1983gv} 

If we assume property 2, $\K{p} = 1$ for $p^2 < \Lambda^2$, then
\begin{equation}
\Delta (p/\Lambda) = 0\quad \textrm{for}\quad p^2 < \Lambda^2\,.
\end{equation}
This implies that the deformation of the Wilson action is restricted to
the momentum region $p^2 > \Lambda^2$.  It is in this sense that we call
the ERG transformation \textbf{local}.  If only 2${}'$ is satisfied, we
may still call the ERG transformation ``almost local.''

In the above we have derived the Polchinski differential equation
starting from the definition (\ref{derivation-WilsonI}) of the Wilson
action.  Note that (\ref{derivation-WilsonI}) in fact gives an
integral formula for the solution of the differential equation under
the initial condition
\begin{equation}
S_{I,\Lambda} \Big|_{\Lambda = \Lambda_0} = S_{I,B}\,.
\end{equation}

Finally let us rewrite the Polchinski equation for
the full action
\begin{equation}
\SL [\phi] = - \frac{1}{2} \int_p \frac{p^2 + m^2}{\K{p}} \phi (-p) \phi
 (p) + \SIL [\phi]\,.
\end{equation}
It is straightforward to obtain
\begin{eqnarray}
    &&- \Lambda \frac{\partial}{\partial \Lambda} S_\Lambda [\phi]
\label{derivation-polchinskieqfull}\\
 &=& \int_p
    \frac{\Delta (p/\Lambda)}{p^2 + m^2} \left[ \frac{p^2 + m^2}{\K{p}} \phi
        (p) \frac{\delta S_\Lambda [\phi]}{\delta \phi (p)} 
 + \frac{1}{2} \lb \frac{\delta
          S_{\Lambda}[\phi]}{\delta \phi (-p)} 
        \frac{\delta S_{\Lambda}[\phi]}{\delta \phi (p)} +
        \frac{\delta^2 S_{\Lambda}[\phi]}{\delta \phi (-p)\delta \phi
          (p)} \rb \right] \,.\nn
\end{eqnarray}

\subsection{Wetterich equation \label{derivation-Wetterich}}

The purpose of this subsection is to derive what is commonly called the
Wetterich equation,\cite{Nicoll:1977hi} \cite{Wetterich:1992yh}
\cite{Wetterich:1993ne} \cite{Morris:1993qb} \cite{Bonini:1992vh} a
counterpart of the Polchinski equation
\bref{derivation-polchinskieq}.\footnote{See the added note of
\citen{Morris:1993qb} for a historical perspective on this equation.}
Even though almost all our discussions on symmetry are done with the
Wilson action that satisfies the Polchinski equation, we would like to
dedicate a whole subsection on the Wetterich equation, because most
non-perturbative numerical works deal with the Wetterich equation
instead of the Polchinski equation.

\begin{figure}[t]
\begin{center}
\epsfig{file=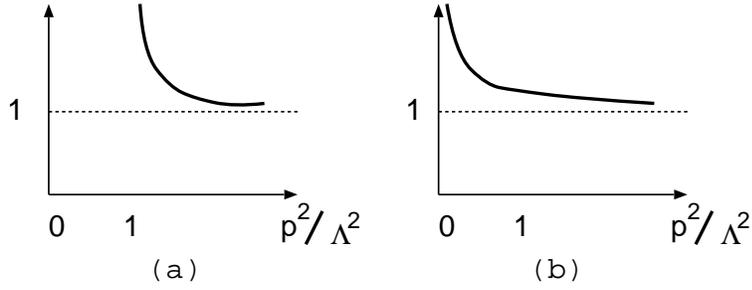, width=10cm} 
\caption{Sketch of $\frac{1}{K_0-K}$.  
(a) If $\K{p} = 1$ for $p^2 < \Lambda^2$, this is
infinite for $p^2 < \Lambda^2$.  (b) If $\K{p} < 1$ for $p^2 <
\Lambda^2$, it diverges only at $p=0$.}  
\label{derivation-IRUV}
\end{center}
\end{figure}

We introduce an action
\begin{equation}
S_{B,\Lambda} [\phi] \equiv - \frac{1}{2} \int_p \frac{p^2+m^2}{K_0 -
  K} \phi (-p) \phi (p) + S_{I,B} [\phi]\,,\label{derivation-SBL}
\end{equation}
which is obtained from the bare action $S_B$ by the following
replacement of the gaussian term:
\begin{equation}
\frac{p^2 + m^2}{\Kz{p}} \longrightarrow \frac{p^2 + m^2}{\Kz{p} - \K{p}}\,.
\end{equation}
With the weight $\exp [S_B]$, $\phi (p)$ of all momenta $p^2 <
\Lambda_0^2$ contribute to the functional integral.  With the above
replacement, however, the main contribution comes from those high
momentum modes $\phi (p)$ with $\Lambda^2 < p^2 < \Lambda_0^2$.  (See
Fig.~\ref{derivation-IRUV}.)
Hence, $S_{B,\Lambda}$ has $\Lambda$ as the IR cutoff and $\Lambda_0$ as
the UV cutoff.  Taking the difference between $S_{B,\Lambda}$ and $S_B$,
we can also write
\begin{equation}
S_{B,\Lambda} [\phi] = S_B [\phi] - \frac{1}{2} \int_p R_\Lambda (p)
\phi (p) \phi (-p)\,,
\end{equation}
where we define
\begin{equation}
    R_\Lambda (p) \equiv (p^2+m^2) \left( \frac{1}{K_0 -
          K} - \frac{1}{K_0} \right)
= (p^2 + m^2) \frac{K}{K_0 (K_0 - K)} \,.
\end{equation}
In the limit $\Lambda \to 0+$, $K$ and hence $R_\Lambda$ vanish, and we
obtain the bare action:
\begin{equation}
\lim_{\Lambda \to 0+} S_{B,\Lambda} = S_B\,.\label{derivation-SBzero}
\end{equation}

Throughout \S\ref{derivation-Wetterich}, we will assume
\begin{equation}
\K{p} < 1\quad\textrm{for}\quad 0 < p^2 < \Lambda^2\,.
\end{equation}
This is necessary so that 
\begin{equation}
\Kz{p} - \K{p} \ne 0
\end{equation}
except at $p=0$ and $p \to \infty$.  Then the division by $\Kz{p} -
\K{p}$ makes sense.  We will do the following in the remainder of
\S\ref{derivation-Wetterich}: we
\begin{enumerate}
\item introduce the generating functional $W_{B,\Lambda}$ of
    $S_{B,\Lambda}$ by (\ref{derivation-WBLdef}), and define its
    Legendre transform $\Gamma_{B,\Lambda}$ by
    (\ref{derivation-GammaBLdef}) and (\ref{derivation-PhidWdJ}),
\item show $W_{B,\Lambda}$ is essentially the Wilson action
    $S_\Lambda$ as given by (\ref{derivation-WBLvsSL}), and derive the
    relation between $S_\Lambda$ and $\Gamma_{B,\Lambda}$ as given by
    (\ref{derivation-GammaBSL}) and (\ref{derivation-Phiphi}),
\item derive the Wetterich equation (\ref{derivation-Wettericheq}),
    giving the $\Lambda$ dependence of the interaction part of
    $\Gamma_{B,\Lambda}$.
\end{enumerate}

\subsubsection{Introducing $W_{B,\Lambda}$ and $\Gamma_{B,\Lambda}$}

Let us first define the generating functional $W_{B,\Lambda}$ of
$S_{B,\Lambda}$ by
\begin{equation}
\exp \left[W_{B,\Lambda} [J]\right] \equiv \int [d\phi] \exp \left[
    S_{B,\Lambda} [\phi] + \int_p J(-p) \phi (p)
\right]\,.\label{derivation-WBLdef} 
\end{equation}
Note
\begin{equation}
\lim_{\Lambda \to 0} W_{B,\Lambda} = W_B\,.\label{derivation-WBzero}
\end{equation}
For $\Lambda > 0$, $W_{B,\Lambda}$ differs from both $W_B$ and
$W_\Lambda$.  Recall that $W_\Lambda$, the generating functional of
the Wilson action $S_\Lambda$, is essentially the same as $W_B$, as
has been shown in subsect.~\ref{derivation-source}.

We then define its Legendre transform, called the \textbf{effective average
action}, by
\begin{equation}
\Gamma_{B,\Lambda} [\Phi] \equiv W_{B,\Lambda} [J] - \int_p J(-p) \Phi
(p)\,,\label{derivation-GammaBLdef}
\end{equation}
where $J$ is determined by the condition
\begin{equation}
\frac{\delta W_{B,\Lambda} [J]}{\delta J(-p)} = \Phi (p) 
\label{derivation-PhidWdJ}
\end{equation}
as a functional of $\Phi$.  From the general properties of Legendre
transformations, we can immediately conclude the following:
\begin{itemize}
\item[(a)] We can obtain $W_{B,\Lambda}$ as the inverse Legendre
    transform of $\Gamma_{B,\Lambda}$:
\begin{equation}
W_{B,\Lambda} [J] = \Gamma_{B,\Lambda} [\Phi] + \int_p J(-p) \Phi (p)\,,
\end{equation}
where $\Phi$ is determined by
\begin{equation}
\frac{\delta \Gamma_{B,\Lambda} [\Phi]}{\delta \Phi (p)} = - J(-p)
\label{derivation-JdGammadPhi}
\end{equation}
as a functional of $J$.
\item[(b)] (\ref{derivation-PhidWdJ}) and
    (\ref{derivation-JdGammadPhi}) give
\begin{equation}
\frac{\delta \Phi (p)}{\delta J (q)} = \frac{\delta^2 W_{B,\Lambda}
  [J]}{\delta J(-p) \delta J(q)}\,,
\end{equation}
and
\begin{equation}
\frac{\delta J (q)}{\delta \Phi (p)} = - \frac{\delta^2
  \Gamma_{B,\Lambda} [\Phi]}{\delta \Phi (-q) \delta \Phi (p)}
\,.
\end{equation}
\item[(c)] The above are the inverse of each other, and we obtain
\begin{equation}
\int_r \frac{\delta^2 W_{B,\Lambda} [J]}{\delta J(-p) \delta J(r)}
\frac{\delta^2 \Gamma_{B,\Lambda} [\Phi]}{\delta \Phi (-r) \delta \Phi
  (q)} = - (2 \pi)^D \delta^{(D)} (p-q)\,,
\end{equation}
and
\begin{equation}
\int_r \frac{\delta^2 \Gamma_{B,\Lambda} [\Phi]}{\delta \Phi (-p)
  \delta \Phi (r)} \frac{\delta^2 W_{B,\Lambda} [J]}{\delta J(-r)
  \delta J(q)} =  - (2 \pi)^D \delta^{(D)} (p-q)\,.
\end{equation}
\item[(d)] Given an infinitesimal change $\Delta W_{B,\Lambda} [J]$,
    the corresponding infinitesimal change of $\Gamma_{B,\Lambda}
    [\Phi]$ satisfies
\begin{equation}
\Delta \Gamma_{B,\Lambda} [\Phi] = \Delta W_{B,\Lambda} [J]
\label{derivation-deltaGammadeltaW}
\end{equation}
where $\Phi$ is related to $J$ by (\ref{derivation-PhidWdJ}).  (Quick
proof) By definition (suppressing the subscript ${}_{B,\Lambda}$), we obtain
\begin{equation}
\left(\Gamma + \Delta \Gamma\right) [\Phi + \Delta \Phi] =
\left(W + \Delta W\right) [J] - \int J \cdot (\Phi + \Delta \Phi)\,,
\end{equation}
where 
\begin{equation}
    \frac{\delta (W+\Delta W) [J]}{\delta J (-p)} = \Phi (p) +
    \Delta \Phi (p) \,.
\end{equation}
Hence,
\begin{equation}
\Delta \Gamma [\Phi] + \int \frac{\delta \Gamma}{\delta \Phi} \Delta
\Phi = \Delta W [J] - \int J \cdot \Delta \Phi\,.
\end{equation}
From (\ref{derivation-JdGammadPhi}), the terms proportional to $\Delta
\Phi$ cancel, and we obtain (\ref{derivation-deltaGammadeltaW}).
\end{itemize}

\subsubsection{Relating $W_{B,\Lambda}$ to $S_\Lambda$, and deriving the
   Legendre transformation between 
  $\Gamma_{B,\Lambda}$ and $S_\Lambda$}

The relation between the Wilson action $\SL$ and the effective average
action $\Gamma_{B,\Lambda}$ was first derived by
T.~Morris\cite{Morris:1993qb}.  We reproduce his results in the
following.

Our first task is to derive the relation of the generating functional
$W_{B,\Lambda}$ and the Wilson action $S_\Lambda$.  Using
(\ref{derivation-SBL}) and (\ref{derivation-WBLdef}), we obtain
\begin{equation}
\exp \left[ W_{B,\Lambda} [J]\right] = \int [d\phi] \exp \left[ -
    \frac{1}{2} \int_p 
    \frac{p^2 + m^2}{K_0-K} \phi (-p) \phi (p) + S_{I,B} [\phi] +
    \int_p J(-p) \phi (p) \right]\,.
\end{equation}
From the definition (\ref{derivation-defSILJ}) of $S_{I,\Lambda}
[\phi; J]$, we find
\begin{eqnarray}
W_{B,\Lambda} [J] &=& S_{I,\Lambda} [0; J]\nn\\
&=&  S_{I,\Lambda} \left[
        \frac{K_0-K}{p^2 + m^2} J \right] + \frac{1}{2} \int_p
    \frac{K_0-K}{p^2 + m^2} J(-p) J(p) \,.
\label{derivation-WBLvsSIL}
\end{eqnarray}
where we have used the intermediate result (\ref{derivation-Jshift}).
Using the definition (\ref{derivation-Wilson}) of the full Wilson
action $S_\Lambda$, this can be rewritten as
\begin{equation}
    W_{B,\Lambda} [J] = S_\Lambda \left[ \frac{K_0-K}{p^2 + m^2} J \right]
    + \frac{1}{2} \int_p \frac{K_0-K}{p^2 + m^2} \frac{K_0}{K} J(-p)
    J(p) \,. 
\label{derivation-WBLvsSL}
\end{equation}
Thus, we find that the Wilson action $S_\Lambda$ is basically the
generating functional for the action $S_{B,\Lambda}$.  Especially
in the limit $\Lambda \to 0+$, where $K$ vanishes, and
$W_{B,\Lambda}$ becomes $W_B$, we obtain
\begin{equation}
\lim_{\Lambda \to 0+} S_{I,\Lambda} \left[\frac{K_0}{p^2 + m^2} J \right]
= W_B [J] - \frac{1}{2} \int_p \frac{K_0}{p^2 + m^2} J(-p) J(p)\,.
\label{derivation-SIzero}
\end{equation}
Namely, in the $\Lambda \to 0+$ limit, the interaction part of the
Wilson action becomes that of the generating functional $W_B$.

We now use the above result to express $\Gamma_{B,\Lambda}$ in terms
of $S_\Lambda$.  Substituting (\ref{derivation-WBLvsSL}) into
(\ref{derivation-GammaBLdef}), we obtain
\begin{equation}
\Gamma_{B,\Lambda} [\Phi] = \frac{1}{2} \int_p \frac{K_0-K}{p^2+m^2}
\frac{K_0}{K} J J + S_\Lambda \left[ \frac{K_0-K}{p^2 + m^2} J \right]
- \int_p J \Phi\,,
\end{equation}
where
\begin{equation}
\Phi (p) = \frac{K_0-K}{p^2 + m^2} \left[
\frac{K_0}{K} J(p) + \frac{\delta S_\Lambda [\phi]}{\delta \phi (-p)}
\Big|_{\phi = \frac{K_0-K}{p^2+m^2} J} \right]\,.
\end{equation}
It is more convenient to use 
\begin{equation}
\phi (p) \equiv \frac{\Kz{p} - \K{p}}{p^2 + m^2}\,J(p)
\end{equation}
as field variables instead of $J$. We then obtain\cite{Morris:1993qb}
\begin{equation}
\Gamma_{B,\Lambda} [\Phi] = S_\Lambda [\phi] + \int_p
\frac{p^2 + m^2}{K_0-K} \left( \frac{1}{2} \frac{K_0}{K} \phi (p) \phi
    (-p) - \Phi (p) \phi (-p) \right)\,,\label{derivation-GammaBSL}
\end{equation}
where
\begin{equation}
    \Phi (p) = \frac{K_0}{K} \phi (p) + \frac{K_0-K}{p^2+m^2} \frac{\delta
      S_\Lambda [\phi]}{\delta \phi (-p)} \,.\label{derivation-Phiphi}
\end{equation}
Conversely, by regarding $\SL [\phi]$ as the Legendre transform of
$\Gamma_{B,\Lambda} [\Phi]$, we obtain
\begin{equation}
\phi (p) = - \frac{K_0-K}{p^2 + m^2} \frac{\delta \Gamma_{B,\Lambda}
  [\Phi]}{\delta \Phi (-p)}\,.\label{derivation-phiPhi}
\end{equation}
Let us now rewrite the above three equations using $\SIL [\phi]$ and
the interaction part of $\Gamma_{B,\Lambda}$, defined by
\begin{equation}
\Gamma_{I,B,\Lambda} [\Phi] \equiv \Gamma_{B,\Lambda} [\Phi] +
\frac{1}{2} \int_p \frac{p^2 + m^2}{K_0-K} \Phi (-p) \Phi (p)\,.
\label{derivation-GammaIBLdef}
\end{equation}
First, we can rewrite (\ref{derivation-GammaBSL}) as
\begin{equation}
\Gamma_{I,B,\Lambda} [\Phi] = S_{I,\Lambda} [\phi] + \frac{1}{2} \int_p
  \frac{p^2+m^2}{K_0-K} \left( \Phi - \phi \right) (-p)
\left( \Phi - \phi \right) (p)\,.\label{derivation-GammaISI}
\end{equation}
Then, (\ref{derivation-Phiphi}) gives
\begin{equation}
\Phi (p) = \phi (p) + \frac{K_0-K}{p^2 + m^2} \frac{\delta
  S_{I,\Lambda} [\phi]}{\delta \phi (-p)}\,,\label{derivation-Phibyphi}
\end{equation}
and (\ref{derivation-phiPhi}) gives
\begin{equation}
\phi (p) = \Phi (p) - \frac{K_0-K}{p^2 + m^2} \frac{\delta
  \Gamma_{I,B,\Lambda} [\Phi]}{\delta \Phi (-p)}\,.\label{derivation-phibyPhi}
\end{equation}
Thus, we obtain
\begin{equation}
\frac{\delta
  S_{I,\Lambda} [\phi]}{\delta \phi (-p)} = \frac{\delta
  \Gamma_{I,B,\Lambda} [\Phi]}{\delta \Phi (-p)}\,.
\end{equation}

Differentiating (\ref{derivation-Phibyphi}) with respect to $\phi
(q)$, we obtain
\begin{equation}
\frac{\delta \Phi (p)}{\delta \phi (q)}  = (2
\pi)^D \delta^{(D)} (p-q) + \frac{K_0-K}{p^2+m^2} \frac{\delta^2
  S_{I,\Lambda} [\phi]}{\delta 
  \phi (-p) \delta \phi (q)}\,, \label{derivation-dPhidphi}
\end{equation}
and differentiating (\ref{derivation-phibyPhi}) with respect to $\Phi
(q)$, we obtain
\begin{eqnarray}
\frac{\delta \phi (p)}{\delta \Phi (q)}
&=& (2 \pi)^D \delta^{(D)} (p-q) 
    - \frac{K_0-K}{p^2 + m^2} \frac{\delta^2 \Gamma_{I,B,\Lambda}
      [\Phi]}{\delta \Phi (-p) \delta \Phi (q)}\nn\\
&=& - \frac{K_0-K}{p^2 + m^2} \Gamma_{B,\Lambda}^{(2)} (p,-q)\,,
\end{eqnarray}
where we define
\begin{equation}
\Gamma_{B,\Lambda}^{(2)} (p,-q) \equiv
\frac{\delta^2 \Gamma_{B,\Lambda}
      [\Phi]}{\delta \Phi (-p) \delta \Phi (q)} \,.
\end{equation}
Since
\begin{equation}
\int_q \frac{\delta \Phi (p)}{\delta \phi (q)} \frac{\delta \phi
  (q)}{\delta \Phi (r)} = (2 \pi)^D \delta^{(D)} (p-r)\,,
\end{equation}
we obtain
\begin{equation}
\int_q \frac{\delta \Phi (p)}{\delta \phi (q)}
(-) \frac{K_0-K}{q^2 + m^2} \Gamma_{B,\Lambda}^{(2)} (q,-r)
= (2 \pi)^D \delta^{(D)} (p-r)\,.
\end{equation}
Hence, defining
\begin{eqnarray}
\left(\Gamma_{B,\Lambda}^{(2)}\right)^{-1} (p,-q) 
&\equiv& - \frac{\delta
  \Phi (p)}{\delta \phi (q)} \frac{K_0-K}{q^2 + m^2}\nn\\
&=& - (2 \pi)^D \delta^{(D)} (p-q) \frac{K_0-K}{p^2 + m^2}\nn\\
&&\qquad - \frac{K_0-K}{p^2+m^2} \frac{\delta^2 S_{I,\Lambda} [\phi]}{\delta
  \phi (-p) \delta \phi (q)} \frac{K_0-K}{q^2 + m^2}\,,
\label{derivation-Gamma2inverse}
\end{eqnarray}
where we have used \bref{derivation-dPhidphi}, we obtain
\begin{equation}
\int_q \left(\Gamma_{B,\Lambda}^{(2)}\right)^{-1} (p,-q) 
\Gamma_{B,\Lambda}^{(2)} (q,-r)
= (2 \pi)^D \delta^{(D)} (p-r) \,.
\end{equation}
This justifies our notation \bref{derivation-Gamma2inverse}.

Before deriving the Wetterich equation, let us consider the two limits of
$\Gamma_{I,B,\Lambda}$ as $\Lambda \to \Lambda_0-0$ or $\Lambda \to 0+$.
\begin{itemize}
\item[(i)] $\Lambda \to \Lambda_0-0$ limit --- From
    (\ref{derivation-Phibyphi}), we obtain $\Phi=\phi$, and combining
    this with (\ref{derivation-GammaISI}) gives
\begin{equation}
\lim_{\Lambda \to \Lambda_0} \Gamma_{I,B,\Lambda}  = \lim_{\Lambda
  \to \Lambda_0} S_{I,\Lambda} = S_{I,B}  \,.
\end{equation}
\item[(ii)] $\Lambda \to 0+$ limit --- (\ref{derivation-SBzero})
    implies
\begin{equation}
\lim_{\Lambda \to 0+} \Gamma_{B,\Lambda} =
 \Gamma_B\,,\label{derivation-Lambdazerolimit} 
\end{equation}
where $\Gamma_B$ is the effective action for the bare
	   action.\footnote{Strictly speaking,
	   \bref{derivation-Lambdazerolimit} is valid only if $\SIL$ has
	   no tadpole, i.e., no interaction term proportional to $\phi
	   (0)$ at zero momentum.}
Defining the interaction part of $\Gamma_B$ such that
\begin{equation}
\Gamma_{B} [\Phi] \equiv  - \frac{1}{2} \int_p \frac{p^2 + m^2}{K_0}
\Phi (-p) \Phi (p) + \Gamma_{I,B} [\Phi]\,,
\end{equation}
we obtain
\begin{equation}
\lim_{\Lambda \to 0+} \Gamma_{I,B,\Lambda} = \Gamma_{I,B}\,.
\end{equation}
\end{itemize}
Thus, $\Gamma_{I,B,\Lambda}\,(\Lambda_0 > \Lambda > 0)$ interpolates
between the bare action $S_{I,B}$ and its effective action
$\Gamma_{I,B}$.  In contrast, the Wilson action $S_{I,\Lambda}$
interpolates between the bare action $S_{I,B}$ and the interaction part
of the generating functional $W_B$, up to rescaling of the source as
given in (\ref{derivation-SIzero}).

\subsubsection{Deriving the Wetterich equation}

We are now ready to derive the Wetterich equation.  Using the property
(d) of the Legendre transform, the $\Lambda$ derivative of
$\Gamma_{I,B,\Lambda}$ is obtained as
\begin{equation}
    - \Lambda \frac{\partial}{\partial \Lambda} \Gamma_{I,B,\Lambda} [\Phi]
    = - \Lambda \frac{\partial}{\partial \Lambda} S_{I,\Lambda} [\phi]
    - \frac{1}{2} \int_p \frac{p^2+m^2}{\left(K_0-K\right)^2} \Delta
    (p/\Lambda) \left( \Phi - \phi \right) (-p) \left( \Phi - \phi \right)
    (p)\,.
\end{equation}
Using (\ref{derivation-polchinskieq}) and (\ref{derivation-Phibyphi}), we
obtain
\begin{equation}
    - \Lambda \frac{\partial}{\partial \Lambda} \Gamma_{I,B,\Lambda} [\Phi]
    = \frac{1}{2} \int_p \frac{\Delta (p/\Lambda)}{p^2 + m^2}
    \frac{\delta^2 S_{I,\Lambda}[\phi]}{\delta \phi (-p) \delta \phi (p)}\,.
\end{equation}
We rewrite this as
\begin{equation}
 - \Lambda \frac{\partial}{\partial \Lambda} \Gamma_{I,B,\Lambda} [\Phi]
    = \frac{1}{2} \int_p \left(p^2 + m^2\right) \frac{\Delta
      (p/\Lambda)}{(K_0-K)^2} \cdot
\frac{K_0-K}{p^2+m^2} \frac{\delta^2 S_{I,\Lambda}[\phi]}{\delta \phi
  (-p) \delta \phi (p)} \frac{K_0-K}{p^2+m^2}\,.
\end{equation}
We now note
\begin{equation}
- \Lambda \frac{\partial}{\partial \Lambda} R_\Lambda (p)
= - (p^2 + m^2) \frac{\Delta (p/\Lambda)}{(K_0-K)^2}\,,
\label{derivation-dRdL}
\end{equation}
and (\ref{derivation-Gamma2inverse})
\begin{eqnarray}
&&\frac{K_0-K}{p^2 + m^2} \frac{\delta^2 S_{I,\Lambda} [\phi]}{\delta
  \phi (-p) \delta \phi (p)} \frac{K_0-K}{p^2 + m^2}\nn\\
&=& - \left(\Gamma^{(2)}_{B,\Lambda} \right)^{-1} (p,-p)
- (2\pi)^D \delta^{(D)} (0) \cdot \frac{K_0-K}{p^2 + m^2}\,.
\end{eqnarray}
Thus, ignoring a field independent constant (proportional to the space
volume), we obtain the Wetterich
equation\cite{Nicoll:1977hi}\cite{Wetterich:1992yh}\cite{Wetterich:1993ne}
\cite{Morris:1993qb}\cite{Bonini:1992vh} 
\begin{equation}
- \Lambda \frac{\partial}{\partial \Lambda} \Gamma_{I,B,\Lambda} [\Phi]
    = \frac{1}{2} \int_p \left( - \Lambda \frac{\partial}{\partial
          \Lambda} R_\Lambda (p) \right)
    \left(\Gamma^{(2)}_{B,\Lambda} \right)^{-1} (p,-p)
    \,.\label{derivation-Wettericheq} 
\end{equation}
There is no formal integral formula for the solution of the Wetterich
equation such as (\ref{derivation-WilsonI}) for the solution of the
Polchinski equation.

\subsection{Recapitulation}

For the reader's convenience, we would like to tabulate the functionals
and their relationship, introduced thus far:
\begin{enumerate}
\item the bare and Wilson actions (Table \ref{derivation-TableI}) ---
      Given a bare action $S_B$ with cutoff $\Lambda_0$, we construct
      its Wilson action $\SL$ by integrating the field over momenta
      between $\Lambda$ and $\Lambda_0$. The generating functionals of
      the connected correlation functions for $S_B$ and $\SL$ are
      basically the same.
\begin{table}[tbh]
\begin{center}
\begin{tabular}{ll}
\hline
symbol& description\\
\hline
$S_B [\phi]$& the bare action with a UV cutoff $\Lambda_0$\\
$S_{I,B}[\phi]$& the interaction part of $S_B$\\
$W_B [J]$& the generating functional of $S_B$\\
$\Gamma_B [\Phi]$& the effective action of $S_B$\\
\multicolumn{2}{c}{
$\displaystyle
\begin{array}{c@{~=~}l@{\quad}l}
S_B [\phi] & - \frac{1}{2} \int \frac{p^2+m^2}{K_0}
\phi \phi + S_{I,B} [\phi]& (\ref{derivation-SB})\\
\exp \left[W_B [J]\right] & \int [d\phi] \exp \left[ S_B [\phi] +
    \int J \phi \right]& (\ref{derivation-WB})\\
\Gamma_B [\Phi] & W_B [J] - \int J \Phi\,,\,
\textrm{where} \,
\lb\begin{array}{c@{~=~}l}
\frac{\delta W_B [J]}{\delta J(-p)} & \Phi (p)\\
\frac{\delta \Gamma_B [\Phi]}{\delta \Phi (p)}& - J(-p)
\end{array}\right.&\end{array}
$}\\
\hline
$\SL [\phi]$& the Wilson action with a UV cutoff $\Lambda$\\
$S_{I,\Lambda} [\phi]$& the interaction part of $\SL$\\
$W_{\Lambda} [J]$& the generating functional of $\SL$\\
\multicolumn{2}{c}{
$\displaystyle 
\begin{array}{c@{~=~}l@{\quad}l}
\SL [\phi] & - \frac{1}{2} \int \frac{p^2 + m^2}{K}
  \phi \phi + S_{I,\Lambda} [\phi]& (\ref{derivation-Wilson})\\
\exp \left[ S_{I,\Lambda} [\phi] \right] & \int
  [d\phi'] \exp \left[ - 
      \frac{1}{2} \int \frac{p^2+ m^2}{K_0-K} \phi' \phi' + S_{I,B}
      [\phi + \phi']\right]& (\ref{derivation-WilsonI})\\
\exp \left[ W_\Lambda [J] \right] & \int [d\phi] \exp
  \left[ \SL [\phi] + \int J \phi \right]& (\ref{derivation-WL})
\end{array}$
}\\
\hline
\multicolumn{2}{l}{
  $W_B$ and $W_\Lambda$ are related as}\\
\multicolumn{2}{c}{
  $\displaystyle W_B [J] = W_\Lambda \left[ \frac{K_0}{K} J\right] -
  \frac{1}{2} \int \frac{K_0-K}{p^2 + m^2} \frac{K_0}{K} J J$
  (\ref{derivation-WBWL})}\\ 
\hline
\end{tabular}
\caption{The bare and Wilson actions --- \bref{derivation-WBWL} shows
 that the generating functionals $W_B$ and $W_\Lambda$ are basically
 the same.}  \label{derivation-TableI}
\end{center}
\end{table}
\item the bare action with both UV and IR cutoffs (Table
      \ref{derivation-TableII}) --- Instead of the bare action $S_B$
      with an UV cutoff $\Lambda_0$, we introduce $S_{B,\Lambda}$ with
      both UV and IR cutoffs $\Lambda_0$ \& $\Lambda$.  We can define
      its generating functional $W_{B,\Lambda}$ of the connected
      correlation functions, and the Legendre transform
      $\Gamma_{B,\Lambda}$.  The Wilson action $\SL$ is equal to
      $W_{B,\Lambda}$.
\begin{table}[tbh]
\begin{center}
\begin{tabular}{ll}
\hline
symbol& description\\
\hline
$S_{B,\Lambda} [\phi]$& the bare action with a UV cutoff $\Lambda_0$
and IR cutoff $\Lambda$\\
$W_{B,\Lambda} [J]$& the generating functional of $S_{B,\Lambda}$\\
$\Gamma_{B,\Lambda} [\Phi]$& the effective average action; the
Legendre transform of $W_{B,\Lambda}$\\
$\Gamma_{I,B,\Lambda} [\Phi]$& the interaction part of
$\Gamma_{B,\Lambda}$\\
\multicolumn{2}{c}{
$\displaystyle 
\begin{array}{c@{~=~}l@{\quad}l}
S_{B,\Lambda} [\phi] & - \frac{1}{2} \int
\frac{p^2+m^2}{K_0-K} \phi \phi + S_{I,B} [\phi]& (\ref{derivation-SBL})\\
\exp \left[ W_{B,\Lambda} [J]\right] & \int [d\phi] \exp \left[
    S_{B,\Lambda} [\phi] + \int J \phi \right]& (\ref{derivation-WBLdef})\\
\Gamma_{B,\Lambda} [\Phi] & W_{B,\Lambda} [J] - \int J \Phi\,,
\textrm{where}\,
\lb\begin{array}{c@{~=~}l}
\frac{\delta W_{B,\Lambda} [J]}{\delta J(-p)} & \Phi (p)\\
\frac{\delta \Gamma_{B,\Lambda} [\Phi]}{\delta \Phi (p)} & - J (-p)
\end{array}\right.& (\ref{derivation-GammaBLdef})\\
\Gamma_{B,\Lambda} [\Phi] & - \frac{1}{2} \int
\frac{p^2+m^2}{K_0-K} \Phi \Phi + \Gamma_{I,B,\Lambda} [\Phi]&
 (\ref{derivation-GammaIBLdef})
\end{array}$}\\ 
\hline
\multicolumn{2}{l}{the relation between $S_{I,\Lambda}$ and $W_B$}\\
\multicolumn{2}{c}{$\displaystyle W_{B,\Lambda} [J] = S_{I,\Lambda} \left[
      \frac{K_0-K}{p^2 + m^2} J\right] + \frac{1}{2} \int
  \frac{K_0-K}{p^2 + m^2} J J\quad(\ref{derivation-WBLvsSIL})$}\\
\multicolumn{2}{l}{
the relation between $\Gamma_{I,B,\Lambda}$ and $S_{I,\Lambda}$}\\
\multicolumn{2}{c}{
$\displaystyle \Gamma_{I,B,\Lambda} [\Phi] = S_{I,\Lambda} [\phi] +
\frac{1}{2} \int \frac{p^2+m^2}{K_0-K} (\Phi-\phi)(\Phi-\phi)$
 (\ref{derivation-GammaISI})}\\ 
\multicolumn{2}{c}{where
$\displaystyle \lb\begin{array}{c@{~=~}l@{\quad}c}
\Phi (p) & \phi (p) + \frac{K_0-K}{p^2 + m^2} \frac{\delta
  S_{I,\Lambda} [\phi]}{\delta \phi (-p)}&\bref{derivation-Phibyphi}\\
\phi (p) & \Phi (p) - \frac{K_0-K}{p^2 + m^2} \frac{\delta
  \Gamma_{I,B,\Lambda} [\phi]}{\delta \phi (-p)}&
\bref{derivation-phibyPhi}\end{array}\right.$}\\
\hline
\end{tabular}
\caption{We construct $S_{B,\Lambda}$ from $S_B$ by cutting off the IR
 modes with momenta below $\Lambda$.  The Wilson action $\SL$ is the
 generating functional of $S_{B,\Lambda}$.}
\label{derivation-TableII}
\end{center}
\end{table}
\item the $\Lambda \to 0+$ and $\Lambda \to \Lambda_0-$ limits (Table
      \ref{derivation-TableIII}) --- As we take the IR cutoff $\Lambda$
      to zero, $S_{B,\Lambda}$ becomes $S_B$.  As we take $\Lambda$ to
      $\Lambda_0$, the interaction parts of $\SL$ and
      $\Gamma_{B,\Lambda}$ become the interaction part of $S_B$.
\begin{table}[tbh]
\begin{center}
\begin{tabular}{c}
\hline
$\Lambda \to 0+$ limit\\
\hline
$\displaystyle
\begin{array}{r@{~=~}l}
\lim_{\Lambda \to 0} S_{B,\Lambda} & S_B\\
\lim_{\Lambda \to 0} \Gamma_{B,\Lambda} & \Gamma_B\\
\lim_{\Lambda \to 0} W_{B,\Lambda} & W_B \end{array}$\\
\hline
$\Lambda \to \Lambda_0-0$ limit\\
\hline
$\displaystyle
\lim_{\Lambda \to \Lambda_0-0} S_{I,\Lambda} [\phi] =
\lim_{\Lambda \to \Lambda_0-0} \Gamma_{I,B,\Lambda} [\phi] 
= S_{I,B} [\phi]$\\
\hline
\end{tabular}
\caption{The two limits: $\Lambda \to 0+$ and $\Lambda \to \Lambda_0-0$}
\label{derivation-TableIII}
\end{center}
\end{table}
\end{enumerate}

\subsection{Diagrammatic interpretation}

Although ERG is not limited to perturbation theory, it is
straightforward to derive all the results of the previous subsections
using Feynman diagrams.

\subsubsection{$S_{I,\Lambda}$ and the Polchinski equation}

\begin{figure}[t]
\begin{center}
\epsfig{file=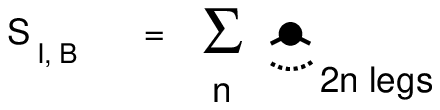, width=5cm}
\caption{$S_{I,B}$ gives elementary vertices.}
\label{derivation-vertex}
\end{center}
\end{figure}

\begin{figure}[t]
\begin{center}
\epsfig{file=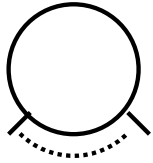, width=2cm}
\caption{$\V_{2n}$ consists of all possible connected diagrams with
  $2n$ legs}
\label{derivation-vertexfunction}
\end{center}
\end{figure}

Let us first examine $S_{I,\Lambda}$, the interaction part of the
Wilson action.  It is defined by (\ref{derivation-WilsonI}):
\begin{equation}
\exp \left[S_{I,\Lambda} [\phi]\right]
\equiv \int [d\phi'] \exp \left[ - \frac{1}{2} \int_p
    \frac{p^2 + m^2}{K_0-K} \phi' (p) \phi' (-p) 
+ S_{I,B} [\phi + \phi'] \right] \,.
\end{equation}
We can compute $S_{I,\Lambda}$ using the high momentum propagator
\begin{equation}
\frac{K_0-K}{p^2 + m^2}
\end{equation}
and the elementary vertices given by $S_{I,B}$. (Fig.~\ref{derivation-vertex})
If we expand
\begin{eqnarray}
    S_{I,\Lambda} [\phi] &=& \sum_{n=1}^\infty \frac{1}{(2n)!}
    \int_{p_1,\cdots,p_{2n}} 
    \phi (p_1) \cdots \phi (p_{2n}) \cdot (2 \pi)^D \delta^{(D)} (p_1 +
    \cdots + p_{2n})\nn\\
    && \quad \times \V_{2n} (\Lambda; p_1,\cdots,p_{2n})\,,
\end{eqnarray}
in powers of fields, the vertex function $\V_{2n}$ is obtained as the
sum of all connected Feynman diagrams with $2n$ external legs.  No
propagator is assigned for the external
leg. (Fig.~\ref{derivation-vertexfunction})

\begin{figure}[t]
\begin{center}
\epsfig{file=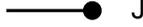, width=2cm}
\caption{The linear coupling to the external field is interpreted as an
  interaction vertex.}
\label{derivation-Jphi}
\end{center}
\end{figure}

\begin{figure}[t]
\begin{center}
\epsfig{file=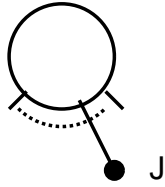, width=2.2cm}
\caption{A $J$ vertex attaches to $S_{I,\Lambda} [\phi]$ through a
  high momentum propagator.}
\label{derivation-attachJ}
\end{center}
\end{figure}

\begin{figure}[t]
\begin{center}
\epsfig{file=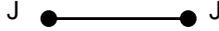, width=3cm}
\caption{Two $J$ vertices connected by a high momentum propagator.}
\label{derivation-JJgraph}
\end{center}
\end{figure}

Now, let us consider $S_{I,\Lambda} [\phi; J]$ for the bare action
$S_{I,B} [\phi; J]$ with a source $J$ coupled linearly to $\phi$.  It
is defined by
\begin{equation}
\exp \left[ S_{I,\Lambda} [\phi; J] \right]
\equiv \int [d\phi'] \exp \left[ - \frac{1}{2} \int \frac{p^2 +
      m^2}{K_0-K} \phi' \phi' + S_{I,B} [\phi+\phi'] + \int J \cdot
    (\phi + \phi') \right]\,.
\end{equation}
The source gives an extra vertex given in Fig.~\ref{derivation-Jphi}.
This extra vertex can do three things:
\begin{itemize}
\item[(i)] It attaches to one of the external lines of $\V_{2n}$
via a high momentum propagator as in Fig.~\ref{derivation-attachJ}.
This gives a shift of $\phi$:
\begin{equation}
S_{I,\Lambda} \left[ \phi + \frac{K_0-K}{p^2 + m^2} J\right]\,.
\end{equation}
\item[(ii)] Two of them get connected by a single high momentum
    propagator as in Fig.~\ref{derivation-JJgraph}.  This gives
\begin{equation}
\frac{1}{2} \int_p J(-p) \frac{K_0-K}{p^2 + m^2} J(p)\,.
\end{equation}
\item[(iii)] It remains as itself, giving
\begin{equation}
\int_p J(-p) \phi (p)\,.
\end{equation}
\end{itemize}
Hence, altogether we obtain (\ref{derivation-Jshift}), which we have
called an intermediate result:
\begin{equation}
S_{I,\Lambda} [\phi; J] = S_{I,\Lambda} \left[ \phi + \frac{K_0-K}{p^2
      + m^2} J \right] + \frac{1}{2} \int
J \frac{K_0-K}{p^2 +  m^2} J  + \int J \cdot \phi \,.
\end{equation}

\begin{figure}[t]
\begin{center}
\epsfig{file=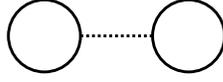, width=3cm}
\caption{$\Lambda$ derivative acts on a type 1 propagator.}
\label{derivation-type1}
\end{center}
\end{figure}

Now, let us consider the $\Lambda$ derivative of the Wilson action
$S_{I,\Lambda}$.  The propagators in each Feynman diagram can be
classified into two types:
\begin{enumerate}
\item \textbf{type 1} --- if cut, the diagram breaks into two separate
    pieces. 
\item \textbf{type 2} --- if cut, the diagram remains a
    single piece.
\end{enumerate}
\vspace{0.2cm}
When we differentiate $\V_{2n}$ with respect to $\Lambda$, only the
propagators are acted on, since the vertices $S_{I,B}$ are independent
of $\Lambda$.  When differentiation acts on a type 1 propagator, we
obtain two Feynman graphs connected to each other by the
differentiated propagator:
\begin{equation}
- \Lambda \frac{\partial}{\partial \Lambda} \frac{K_0-K}{p^2 + m^2}
= \frac{\Delta (p/\Lambda)}{p^2 + m^2}\,.
\end{equation}
Denoting this by a broken line, we obtain two vertex functions
connected by a broken line as in Fig.~\ref{derivation-type1}.
Summing over all possibilities, we obtain
\begin{equation}
\frac{1}{2} \int_p \frac{\Delta (p/\Lambda)}{p^2 + m^2} \frac{\delta
  S_{I,\Lambda}}{\delta \phi (-p)} \frac{\delta
  S_{I,\Lambda}}{\delta \phi (p)}\,,
\end{equation}
where the factor $\frac{1}{2}$ avoids overcounting.

\begin{figure}[t]
\begin{center}
\epsfig{file=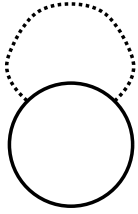, width=2cm}
\caption{$\Lambda$ derivative acts on a type 2 propagator.}
\label{derivation-type2}
\end{center}
\end{figure}

When differentiation acts on a type 2 propagator, we get a single
vertex function with a loop of the differentiated propagator as in
Fig.~\ref{derivation-type2}.
Summing over all possibilities, we obtain
\begin{equation}
\frac{1}{2} \int_p \frac{\Delta (p/\Lambda)}{p^2 + m^2} \frac{\delta^2
  S_{I,\Lambda}}{\delta \phi (p) \delta \phi (-p)}\,.
\end{equation}
Thus, altogether we obtain the Polchinski equation
\begin{equation}
- \Lambda \frac{\partial}{\partial \Lambda} S_{I,\Lambda} [\phi]
= \int_p \frac{\Delta (p/\Lambda)}{p^2 + m^2} \lb
 \frac{\delta S_{I,\Lambda}}{\delta \phi (p)}
 \frac{\delta S_{I,\Lambda}}{\delta \phi (-p)}
+ \frac{\delta^2
  S_{I,\Lambda}}{\delta \phi (p) \delta \phi (-p)} \rb\,.
\end{equation}

\subsubsection{$\Gamma_{I,B,\Lambda}$ and the Wetterich equation}

\begin{figure}[t]
\begin{center}
\epsfig{file=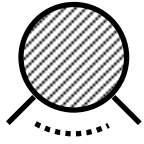, width=2cm}
\caption{$\Gamma_{2n}$ consists of 1PI diagrams with $2n$ legs}
\label{derivation-1PI}
\end{center}
\end{figure}

The average action $\Gamma_{B,\Lambda}$ is the Legendre transform of
the generating functional $W_{B,\Lambda}$.  Expanding the interaction
part $\Gamma_{I,B,\Lambda}$ of the average action as
\begin{eqnarray}
\Gamma_{I,B,\Lambda} [\Phi] &=& \sum_{n=1}^\infty \frac{1}{(2n)!}
\int_{p_1,\cdots,p_{2n}} \Phi (p_1) \cdots \Phi (p_{2n}) \cdot
(2\pi)^D \delta^{(D)} (p_1 + \cdots p_{2n})\nn\\
&& \quad \times \Gamma_{2n} (\Lambda; p_1, \cdots, p_{2n})\,.
\end{eqnarray}
we find that $\Gamma_{2n}$ consists of all 1PI (one particle
irreducible) diagrams with $2n$ external legs.  Again, no propagator
is assigned to the external legs.  All the internal lines are of type
2. We denote $\Gamma_{2n}$ by a shaded blob as in
Fig.~\ref{derivation-1PI}.

\begin{figure}[t]
\begin{center}
\epsfig{file=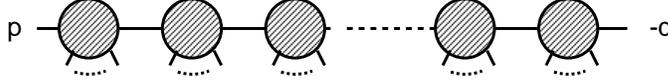, width=9cm}
\caption{A series of 1PI diagrams giving $G(p,-q)$}
\label{derivation-cut1PI}
\end{center}
\end{figure}

Differentiating $\Gamma_{2n} (\Lambda)$ with respect to $\Lambda$, we
do not get graphs as in Fig.~\ref{derivation-type1}, but only graphs
as in Fig.~\ref{derivation-type2}.  The graph after the cutting is
not necessarily 1PI; it consists of multiple 1PI graphs connected in
series as in Fig.~\ref{derivation-cut1PI}.
Hence, we obtain the graphical equation
\begin{center}
\parbox{1.2cm}{$\displaystyle - \Lambda \frac{\partial}{\partial
    \Lambda}$}
\parbox{1.8cm}{\epsfig{file=derivation-1PI, width=1.5cm}}
\parbox{0.5cm}{$=$}
\parbox{4cm}{\epsfig{file=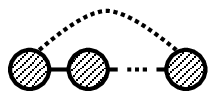, width=4cm}}
\end{center}
where the broken line on the top denotes the $\Lambda$ derivative of
the propagator.

Consider the sum over graphs in Fig.~\ref{derivation-cut1PI},
where we assign a field $\Phi$ for each external leg except for those
with momentum $p$ or $-q$.  Let us call the sum $G (p,-q)$ so that
\begin{equation}
- \Lambda \frac{\partial}{\partial \Lambda} \Gamma_{I,B,\Lambda} [\Phi]
= \int_p \frac{\Delta (p/\Lambda)}{p^2+m^2} G(p,-p)\,.
\end{equation}
Each blob of Fig.~\ref{derivation-cut1PI} gives 
\begin{equation}
\Gamma_{I,B,\Lambda}^{(2)} (q,-r) \equiv \frac{\delta^2
  \Gamma_{I,B,\Lambda} [\Phi]}{\delta \Phi (-q) \delta \Phi (r)}\,.
\end{equation}
Hence, $G(p,-q)$ is a geometric series given by
\begin{equation}
G(p,-q) =  \Gamma^{(2)}_{I,B,\Lambda} (p,-q) + \int_r
\Gamma^{(2)}_{I,B,\Lambda} (p,-r) \frac{K_0-K}{r^2+m^2} 
\Gamma^{(2)}_{I,B,\Lambda} (r,-q) + \cdots\,.
\end{equation}
Hence, we obtain
\begin{eqnarray}
&&\int_r \left( (2\pi)^D \delta^{(D)} (p-r) + \frac{K_0-K}{p^2+m^2}
    G(p,-r) \right) \frac{K_0-K}{r^2+m^2}\\
&&\quad \times \left(
- \frac{r^2+m^2}{K_0-K} (2\pi)^D \delta^{(D)} (r-q) +
\Gamma_{I,B,\Lambda}^{(2)} (r,-q)\right)
 = - (2\pi)^D\delta^{(D)} (p-q)\,.\nn
\end{eqnarray}
Therefore, from (\ref{derivation-Gamma2inverse}), we find
\begin{equation}
\left((2\pi)^D \delta^{(D)} (p-q) + \frac{K_0-K}{p^2+m^2}
    G(p,-q)\right) 
\frac{K_0-K}{q^2+m^2} = - \left(\Gamma^{(2)}_{B,\Lambda} \right)^{-1}
(p,-q)\,.
\end{equation}
Thus, ignoring an additive constant proportional to $\delta^{(D)}
(0)$, we obtain
\begin{eqnarray}
- \Lambda \frac{\partial}{\partial \Lambda} \Gamma_{I,B,\Lambda} [\Phi]
&=& \int_p (p^2+m^2) \frac{\Delta (p/\Lambda)}{(K_0-K)^2}
\left(\frac{K_0-K}{p^2+m^2} \right)^2 G(p,-p)\nn\\
&=& - \int_p (p^2+m^2) \frac{\Delta (p/\Lambda)}{(K_0-K)^2}
\left(\Gamma^{(2)}_{B,\Lambda}\right)^{-1} (p,-p)\,,
\end{eqnarray}
which is the Wetterich equation.

\subsubsection{$\Lambda$ dependence of the generating functional}

Let us understand (\ref{derivation-WBWL})
\begin{equation}
W_B [J] = W_\Lambda \left[ \frac{K_0}{K} J \right] - \frac{1}{2}
\int_p \frac{K_0 - K}{p^2 + m^2} \frac{K_0}{K} J(-p) J(p)\,,
\end{equation}
using Feynman diagrams.  Recall that this is an important equation
implying we miss no physics by the ERG transformation.

Let us first consider $W_B [J]$.  This can be calculated
perturbatively using the vertices provided by $S_{I,B}$ and the source
term $\int J \phi$.  The propagator is given by
\begin{equation}
\frac{K_0}{p^2 + m^2}\,.
\end{equation}
Note that there is a single contribution that does not involve the
interaction vertex $S_{I,B}$, which is given graphically by
Fig.~\ref{derivation-JJgraph} with the above propagator:
\begin{equation}
\frac{1}{2} \int_p J(-p) \frac{K_0}{p^2 + m^2} J(p)\,.\label{derivation-JKzJ}
\end{equation}

\begin{figure}[t]
\begin{center}
\epsfig{file=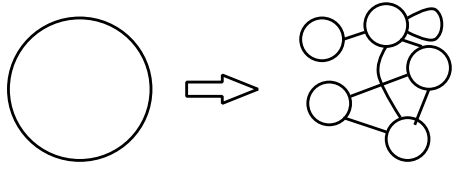, width=6cm}
\caption{On the right, the blobs contain only high momentum
  propagators.  These blobs are connected to each other by low
  momentum propagators.}
\label{derivation-decomposition}
\end{center}
\end{figure}

Now, let us consider the Feynman diagrams contributing to
\begin{equation}
W_B [J] - \frac{1}{2} \int_p J (-p) \frac{K_0}{p^2 + m^2} J(p)\,.
\end{equation}
Each internal propagator can be decomposed as the sum of high and low
momentum propagators:
\begin{equation}
 \frac{K_0}{p^2 + m^2} = \frac{K_0-K}{p^2 + m^2} +
\frac{K}{p^2 + m^2} \,.\label{derivation-highpluslow}
\end{equation}
We substitute the above decomposition to all the internal propagators.
Those parts involving only the high momentum propagators give the
vertices of the Wilson action $S_{I,\Lambda}$.  The low momentum
propagator is to be used with the vertices $S_{I,\Lambda}$.  (See
Fig.~\ref{derivation-decomposition}.)
For the external propagators ending with sources, we write
\begin{equation}
\frac{K_0}{p^2 + m^2} J(p) = \frac{K}{p^2 + m^2} \cdot
\frac{K_0}{K} J (p)\,,
\end{equation}
using the low momentum propagator.  This amounts to replacing $J$ by
$K_0/K \cdot J$.  Hence, the decomposition
(\ref{derivation-highpluslow}) gives
\begin{equation}
W_B [J] - \frac{1}{2} \int_p J \frac{K_0}{p^2 + m^2} J
= W_\Lambda \left[ \frac{K_0}{K} J \right] - \frac{1}{2} \int_p
\frac{K_0}{K} J \frac{K}{p^2+ m^2} \frac{K_0}{K} J\,,
\end{equation}
where we have subtracted the contributions independent of interaction
vertices.  From this, we obtain (\ref{derivation-WBWL}).

\subsection{Fermions}

It is straightforward to extend ERG to include spin $\frac{1}{2}$
fields.  We enumerate the main results:
\begin{enumerate}
\item Given a bare action
\begin{equation}
S_B [\psi, \bar{\psi}] \equiv - \int_p \bar{\psi} (-p)
\frac{\fmslash{p} + i m}{K_0} \psi (p) + S_{I,B} [\psi,\bar{\psi}]\,,
\end{equation}
the Wilson action is defined by
\begin{eqnarray}
&&\exp \left[ S_{I,\Lambda} [\psi,\bar{\psi}] \right] 
\equiv
\int [d\psi' d\bar{\psi}'] \exp \Big[\nn\\
&&\quad\left. - \int_p \bar{\psi}' (-p) \frac{
        \fmslash{p} + i m}{K_0-K} \psi' (p) + S_{I,B} [\psi + \psi',
    \bar{\psi} + \bar{\psi}' ] \right]\,.
\end{eqnarray}
\item The Polchinski equation is obtained as
\begin{eqnarray}
&& - \Lambda \frac{\partial}{\partial \Lambda} S_{I,\Lambda}
    = (-) \int_p \frac{\Delta (p/\Lambda)}{p^2 + m^2} \cdot
\Tr \Big[ (\fmslash{p} - i m) \nn\\
&&\quad\left. \times \lb\Ld{\bar{\psi} (-p)}
    S_{I,\Lambda} \cdot S_{I,\Lambda} \Rd{\psi (p)}
    + \Ld{\bar{\psi} (-p)}
    S_{I,\Lambda} \Rd{\psi (p)} \rb \right]\,,
\end{eqnarray}
where the minus sign $(-)$ is due to the Fermi statistics.
\item By introducing anticommuting sources to the bare action
\begin{equation}
+ \int_p \left( \bar{\xi} (-p) \psi (p) + \bar{\psi} (-p) \eta (p)
\right)\,,
\end{equation}
the Wilson action obtains the following source dependence:
\begin{eqnarray}
S_{I,\Lambda} [\psi, \bar{\psi}; \bar{\xi}, \eta]
&=& S_{I,\Lambda} \left[ \psi (p)+ \frac{K_0-K}{\fmslash{p} + i m}
    \eta (p),
    \bar{\psi} (-p) + \bar{\xi} (-p) \frac{K_0-K}{\fmslash{p} + i m} \right]
\nn\\
&&\quad +
\int \left( \bar{\xi} \psi + \bar{\psi} \eta \right)
+ \int \bar{\xi} (-p) \frac{K_0-K}{\fmslash{p} + i m} \eta (p) \,.
\end{eqnarray}
\item The generating functionals of the bare and Wilson actions are
    related by
\begin{eqnarray}
&&W_B [\bar{\xi},\eta] - \int \bar{\xi} \frac{K_0}{\fmslash{p} + i m}
\eta\nn\\
&& = W_\Lambda \left[\frac{K_0}{K} \bar{\xi}, \frac{K_0}{K}
    \eta\right] - \int \frac{K_0}{K} \bar{\xi} (-p) \frac{K}{\fmslash{p} +
  i m} \frac{K_0}{K} \eta (p)\,.
\end{eqnarray}
\item  For the action with both UV and IR cutoffs
\begin{equation}
S_{B,\Lambda} [\psi, \bar{\psi}] \equiv 
- \int \bar{\psi} (-p) R_\Lambda (p) \psi (p) + S_B [\psi,\bar{\psi}]\,,
\end{equation}
where
\begin{equation}
R_\Lambda (p) \equiv \left( \fmslash{p} + i m \right) \left(
\frac{1}{K_0-K} - \frac{1}{K_0} \right)\,,
\end{equation}
we define the generating functional by
\begin{equation}
\exp \left[ W_{B,\Lambda} [\bar{\xi}, \eta] \right]
\equiv \exp \left[ S_{B,\Lambda} [\psi, \bar{\psi}]
+ \int_p \left( \bar{\xi} \psi +
    \bar{\psi} \eta \right) \right]\,.
\end{equation}
This is given in terms of the Wilson action by
\begin{equation}
W_{B,\Lambda} [\bar{\xi},\eta] 
=
S_{I,\Lambda} \left[ \bar{\xi} \frac{K_0-K}{\fmslash{p} + i m},
    \frac{K_0-K}{\fmslash{p} + i m} \eta \right] + \int \bar{\xi} (-p)
\frac{K_0-K}{\fmslash{p} + i m} \eta (p) \,.
\end{equation}
\item The effective average action $\Gamma_{B,\Lambda}
    [\Psi,\bar{\Psi}]$ is defined as its Legendre transform:
\begin{equation}
\Gamma_{B,\Lambda} [\Psi,\bar{\Psi}] \equiv W_{B,\Lambda}
[\bar{\xi},\eta] - \int_p \left( \bar{\xi} \Psi + \bar{\Psi} \eta
\right)\,,
\end{equation}
where $\bar{\xi}, \eta$ are determined by
\begin{equation}
\lb\begin{array}{c@{~=~}l}
\Psi (p) & \Ld{\bar{\xi} (-p)} W_{B,\Lambda} [\bar{\xi},\eta]\\
\bar{\Psi} (-p) &  W_{B,\Lambda} [\bar{\xi},\eta] \Rd{\eta (p)}
\end{array}\right.
\end{equation}
as functionals of $\Psi, \bar{\Psi}$.  
\item We define
$\Gamma^{(2)}_{B,\Lambda}$ by
\begin{equation}
 \Gamma_{B,\Lambda}^{(2)} (p,-q) \equiv \Ld{\bar{\Psi} (-p)}
 \Gamma_{B,\Lambda} [\Psi,\bar{\Psi}] \Rd{\Psi (q)}\,,
\end{equation}
and its inverse by
\begin{equation}
\int_q \left( \Gamma_{B,\Lambda}^{(2)} \right)^{-1} (p,-q)
\Gamma_{B,\Lambda}^{(2)} (q,-r) = (2 \pi)^D \delta^{(D)} (p-r)\,.
\end{equation}
\item The Wetterich equation for the effective average action is
given as
\begin{equation}
- \Lambda \frac{\partial}{\partial \Lambda} \Gamma_{I,B,\Lambda}
[\Psi,\bar{\Psi}]
= (-) \int_p \Tr \left( - \Lambda \frac{\partial}{\partial \Lambda}
    R_\Lambda (p) \right) 
\left(\Gamma_{B,\Lambda}^{(2)} \right)^{-1} (p,-p)\,,
\end{equation}
where
\begin{equation}
- \Lambda \frac{\partial}{\partial \Lambda} R_\Lambda (p) =
- \left( \fmslash{p} + i m \right) \frac{\Delta
  (p/\Lambda)}{(K_0-K)^2} \,.
\end{equation}
\end{enumerate}

\newpage

\section{Continuum limits\label{cl}}

In \S\ref{derivation} we have kept the UV cutoff $\Lambda_0$ finite.  In
this section we consider taking the continuum limit $\Lambda_0 \to
\infty$.  We call the theory given by a bare action $S_B$
\textbf{renormalizable} if we can obtain a limit of the Wilson action
$\SL$ as $\Lambda_0 \to \infty$, by giving appropriate $\Lambda_0$
dependence to the parameters of $S_B$.  Since the Wilson action
$S_\Lambda$ contains the same physics as the bare action $S_B$, we
expect that the physics of the continuum limit is fully contained in the
$\Lambda_0 \to \infty$ limit of $S_\Lambda$.  In this section we
restrict our discussion only to perturbative renormalization, even
though ERG is applicable non-perturbatively.

\subsection{Perturbative renormalizability}

\begin{figure}[t]
\begin{center}
\epsfig{file=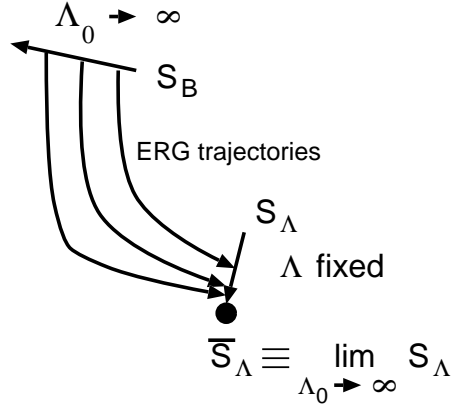, width=6cm}
\caption{$S_\Lambda$ is the solution of the Polchinski equation with
  the initial condition $S_{\Lambda_0} = S_B$. Renormalizability
  amounts to the existence of a limit of $\SL$ as we take $\Lambda_0$
  to infinity.}
\label{cl-renormalizability}
\end{center}
\end{figure}

For concreteness, let us consider the $\phi^4$ theory in $D=4$.  We
give the bare action in the following form:
\begin{equation}
S_B = - \frac{1}{2} \int_p \frac{p^2 + m^2}{\Kz{p}} \phi (-p) \phi (p)
+ S_{I,B}\,,
\end{equation}
where the interaction part has three terms:
\begin{equation}
S_{I,B} = - \int d^4 x \left[ \Delta m^2 \,\frac{\phi^2}{2} 
+ \Delta z \,\frac{1}{2} \partial_\mu \phi \partial_\mu \phi + 
\lambda_0 \,\frac{\phi^4}{4!} \right]\,.
\end{equation}
We determine the cutoff dependence of the coefficients $\Delta m^2,
\Delta z, \lambda_0$ so that the Wilson action $S_\Lambda$ has a
finite limit as $\Lambda_0 \to \infty$:
\begin{equation}
\bar{S}_\Lambda \equiv \lim_{\Lambda_0 \to \infty} S_\Lambda\,,
\end{equation}
where $S_\Lambda$ is the solution of the Polchinski equation
satisfying the initial condition $S_{\Lambda_0} = S_B$.\footnote{Here,
we denote the continuum limit of the Wilson action putting a bar above
$\SL$.  This notation is used only in \S \ref{cl}, and the continuum limit
is simply written as $\SL$ in the later sections.  The bar notation is
adopted again in \S \ref{AF} to denote a Wilson action in the presence
of antifields.}  The
$\Lambda_0$ dependence of $S_\Lambda$ must go away in the
limit. (Fig.~\ref{cl-renormalizability})
In fact it was for this proof of renormalizability that Polchinski's
differential equation was first
introduced.\cite{Polchinski:1983gv}\footnote{The first differential
equation for ERG, introduced by K.~Wilson\cite{Wilson:1973jj}, is
somewhat different from Polchinski's, which suits perturbation theory
better.  In Appendix \ref{diffusion} we derive the original ERG
differential equation by Wilson.}  The idea is not to look at the
continuum limit of the correlation functions, but to look instead at the
limit of the Wilson action.  It is beyond the scope of this review,
however, to explain how to use the Polchinski equation to prove
perturbative renormalizability.  We refer the reader to the original
paper\cite{Polchinski:1983gv} and the references cited in \S\ref{intro}.

We now recall an important result (\ref{derivation-WBWL}) from \S
\ref{derivation}.  Taking the limit $\Lambda_0 \to \infty$, we obtain
\begin{equation}
W_\infty [J]  =  W_\Lambda \left[ \frac{J}{K}
    \right] + \frac{1}{2} \int_p \frac{1 - 1/K}{p^2 + m^2} J(-p)
    J(p)\,,
\label{cl-WinfWL}
\end{equation}
where $W_\infty [J]$ is the generating functional of connected correlation
functions in the continuum limit.  Equivalently, by differentiating
the above with respect to $J$'s (or taking the limit $\Lambda_0 \to
\infty$ of (\ref{derivation-vevSBSL})), we obtain
\begin{equation}
\lb\begin{array}{r@{~=~}l}
\vev{\phi (p) \phi (-p)}^\infty & \frac{1}{K^2} \vev{\phi (p) \phi
  (-p)}_{\bar{S}_\Lambda} + \frac{1 - 1/K}{p^2 + m^2}\,,\\
\vev{\phi (p_1) \cdots \phi (p_n)}^\infty & \prod_{i=1}^n
\frac{1}{\K{p_i}} \cdot \vev{\phi (p_1) \cdots \phi (p_n)}_{\bar{S}_\Lambda}\,,
\end{array}\right.\label{cl-vevinfSL}
\end{equation}
where we denote the continuum limit by a superscript ${}^\infty$.
These imply that the correlation functions of the continuum limit are
fully obtained from the Wilson action $\bar{S}_\Lambda$ which has a
finite momentum cutoff.  Hence, if we expect some symmetry in the
continuum limit, it should be a symmetry of the Wilson action.  The
above two relations (\ref{cl-WinfWL}) \& (\ref{cl-vevinfSL}) play a
key role justifying our approach to realization of symmetry.

\begin{figure}[t]
\begin{center}
\epsfig{file=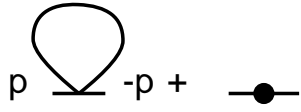, width=4cm}
\caption{Two-point vertex at 1-loop is calculated with a high momentum
  propagator.  The second graph denotes a counterterm $- (\Delta
  m^2)^{(1)}$.}
\label{cl-2pt1loop}
\end{center}
\end{figure}

In the remaining part of this subsection, let us compute the two-point
vertex function of $S_{I,\Lambda}$ at 1-loop level. 
This is only for
an illustration; it is no substitute for a general proof of
renormalizability using the Polchinski equation.  
Let us first recall the expansion of the Wilson action in powers of
fields:
\begin{equation}
\SL [\phi] = \sum_{n=1}^\infty \frac{1}{(2n)!} \int_{p_1 + \cdots + p_{2n} = 0}
\V_{2n} (\Lambda; p_1, \cdots, p_{2n}) \,\phi (p_1) \cdots \phi (p_{2n})\,.
\end{equation}
We wish to compute $\V_2$ at 1-loop.  Denoting the 1-loop contribution
by a superscript ${}^{(1)}$, we obtain
\begin{equation}
\V_2^{(1)} (\Lambda; p, -p) = - \frac{\lambda}{2} \int_q \frac{\Kz{q}
  - \K{q}}{q^2 + m^2} - \left(\Delta m^2\right)^{(1)}\,,
\end{equation}
where $\lambda$ is the tree level part of
$\lambda_0$. (Fig.~\ref{cl-2pt1loop})  This is independent of $p$.
We evaluate the integral as follows:
\begin{eqnarray}
&&\int_q \frac{\Kz{q}  - \K{q}}{q^2 + m^2}
= \int_q \left(\Kz{q}  -
    \K{q}\right) \left( \frac{1}{q^2} - \frac{m^2}{q^4} +
    \frac{m^4}{q^4 (q^2 + m^2)} \right)\nn\\
&&\qquad = \int_q \frac{\Kz{q}  -
    \K{q}}{q^2} - m^2 \int_q \frac{\Kz{q}  -
    \K{q}}{q^4} \nn\\
&&\qquad\qquad + \int_q \left(\Kz{q}  -
    \K{q}\right)\frac{m^4}{q^4 (q^2 + m^2)} \,.
\end{eqnarray}
The last integral has a finite limit:
\begin{equation}
\lim_{\Lambda_0 \to \infty} \int_q \left(\Kz{q}  -
    \K{q}\right)\frac{m^4}{q^4 (q^2 + m^2)} = \int_q \left(1 -
    \K{q}\right)\frac{m^4}{q^4 (q^2 + m^2)}\,.
\end{equation}
The first integral gives\footnote{Please note that we are not using the
short-hand notation (\ref{derivation-shorthand}) of \S\ref{derivation}
here; $K(q)$ on the right does not stand for $K(q/\Lambda)$.  The
dimensionless argument $q$ is obtained from $q/\Lambda$ or $q/\Lambda_0$
by change of variables.  In this section, we do not use the short-hand
notation.}
\begin{equation}
    \int_q \frac{\Kz{q}  - \K{q}}{q^2} = \left( \Lambda_0^2 -
        \Lambda^2 \right) \int_q \frac{K(q)}{q^2}\,,
\end{equation}
where the integral on the right-hand side is a finite constant,
dependent on the particular cutoff function chosen.  It needs a little
more work to calculate the second integral.  Let
\begin{equation}
F(\Lambda_0,\Lambda) \equiv \int_q \frac{\Kz{q}-\K{q}}{q^4}
\end{equation}
so that
\begin{equation}
F(\Lambda,\Lambda) = 0\,.
\end{equation}
Differentiating $F$ with respect to $\Lambda_0$, we obtain
\begin{equation}
\Lambda_0 \frac{\partial}{\partial \Lambda_0} F(\Lambda_0, \Lambda)
= \int_q \frac{\Delta (q/\Lambda_0)}{q^4} = \int_q \frac{\Delta
  (q)}{q^4}
\end{equation}
which is a constant.  This constant is independent of the choice of
$K$ due to the following general formula:
\begin{equation}
\int_q \frac{\Delta (q) \left(1 - K(q)\right)^n}{q^4} = \frac{2}{(4
  \pi)^2} \frac{1}{n+1}\,.\label{cl-integral}
\end{equation}
(Proof) Using the spherical coordinates and the definition $\Delta (q)
\equiv - 2 q^2 dK(q)/dq^2$, we calculate
\begin{eqnarray}
\int_q \frac{\Delta (q) \left(1 - K(q)\right)^n}{q^4}
&=& \frac{2 \pi^2}{(2\pi)^4} \int_0^\infty \frac{q^2 dq^2}{2} (-2 q^2)
\frac{d K}{d q^2} (1-K)^n \frac{1}{q^4} \nn\\
&=& \frac{2}{(4 \pi)^2} \int_0^\infty dq^2 \frac{1}{n+1}
\frac{d}{dq^2} \left(1 - K\right)^{n+1}\nn\\
&=& \frac{2}{(4 \pi)^2} \frac{1}{n+1}\,,
\end{eqnarray}
using $K(0)=1$ and $K(\infty) = 0$. (End of proof)\\
Hence, we obtain
\begin{equation}
\Lambda_0 \frac{\partial}{\partial \Lambda_0} F (\Lambda_0,\Lambda) =
\frac{2}{(4 \pi)^2}\,.
\end{equation}
Thus,
\begin{equation}
F(\Lambda_0, \Lambda) = \frac{2}{(4 \pi)^2} \ln
\frac{\Lambda_0}{\Lambda}\,.
\end{equation}

Altogether, we obtain
\begin{eqnarray}
\V_2^{(1)} (\Lambda) &=& - \frac{\lambda}{2} \left[
\left( \Lambda_0^2 - \Lambda^2 \right) \int_q \frac{K(q)}{q^2}
- m^2 \frac{2}{(4\pi)^2} \ln \frac{\Lambda_0}{\Lambda} +
\frac{m^4}{\Lambda^2} \int_q 
\frac{1 - K(q)}{q^4 \left(q^2 + m^2/\Lambda^2\right)} \right]\nn\\
&& - \left(\Delta m^2\right)^{(1)}\,.
\end{eqnarray}
For this to have a limit as $\Lambda_0 \to \infty$, we take
\begin{equation}
\left(\Delta m^2\right)^{(1)} = \frac{\lambda}{2} \left( - \Lambda_0^2
    \int_q \frac{K(q)}{q^2} + m^2 \frac{2}{(4\pi)^2}\ln
    \frac{\Lambda_0}{\mu} \right) \,,
\end{equation}
where we have introduced an an arbitrary momentum scale $\mu$ to make
the argument of the logarithm dimensionless.  This gives the continuum
limit
\begin{equation}
\bar{\V}_2^{(1)} (\Lambda) = - \frac{\lambda}{2} \left[
- \Lambda^2  \int_q \frac{K(q)}{q^2} + m^2 \frac{2}{(4\pi)^2}\ln
\frac{\Lambda}{\mu} 
 + \frac{m^4}{\Lambda^2} \int_q
\frac{1 - K(q)}{q^4 \left(q^2 + m^2/\Lambda^2\right)} \right]
\label{cl-v2oneloop}
\end{equation}
as the 1-loop self-energy correction from the high-momentum modes.

In addition, the contribution of the low-momentum modes is given by
\begin{equation}
- \frac{\lambda}{2} \int_q \frac{\K{q}}{q^2 + m^2} \,.
\end{equation}
Hence, the 1-loop self-energy correction is
\begin{equation}
- \Sigma^{(1)} \equiv \bar{\V}_2^{(1)} (\Lambda) - \frac{\lambda}{2}
 \int_q \frac{\K{q}}{q^2 + m^2} \,,
\end{equation}
which is independent not only of $\Lambda$, but also of the external
momentum.  Thus, up to 1-loop, the inverse propagator is given by
\begin{equation}
1/\vev{\phi (p) \phi (-p)}^\infty = p^2 + m^2 + \Sigma^{(1)}
= p^2 + m_{\mathrm{ph}}^2\,,
\end{equation}
where we have defined the physical squared mass $m_{\mathrm{ph}}^2$ by
\begin{equation}
 m_{\mathrm{ph}}^2 \equiv m^2 + \Sigma^{(1)} =
m^2  - \bar{\V}_2^{(1)} (\Lambda) + \frac{\lambda}{2}
 \int_p \frac{\K{p}}{p^2 + m^2}\,.
\end{equation}
This is independent of $\Lambda$, but depends on $\mu$.  

To all orders in perturbation theory, we expect the following
$\Lambda_0$ dependence of the counterterms:
\begin{eqnarray}
\Delta m^2 &=& \Lambda_0^2 \,A_2 (\ln \Lambda_0/\mu) + m^2 B_2 (\ln
\Lambda_0/\mu) = \mathrm{O} (\lambda)\\
\Delta z &=& C_2 (\ln \Lambda_0/\mu) = \mathrm{O} (\lambda^2)\\
\lambda_0 &=& A_4 (\ln \Lambda_0/\mu) = \lambda + \mathrm{O} (\lambda^2)
\end{eqnarray}
The $\Lambda_0$ dependence is determined so that $S_\Lambda$, for a
fixed $\Lambda$, has a limit as $\Lambda_0 \to \infty$.  To fix the
$\Lambda_0$ independent additive constants in $B_2, C_2, A_4$, we must
provide three renormalization conditions.  For example, we can adopt
the BPHZ condition at zero momentum so that
\begin{eqnarray}
1/\vev{\phi (p) \phi (-p)}^\infty &=& m^2 + p^2 + p^2 \cdot \mathrm{O}
\left(\frac{p^2}{m^2}\right)\,,\\ 
\vev{\phi (0) \phi (0) \phi (0) \phi (0)}^\infty &=& - \lambda\,.
\end{eqnarray}
At 1-loop, the BPHZ condition amounts to choosing $\mu$ so that
\begin{equation}
\Sigma^{(1)} = 0\,.
\end{equation}

\subsection{Asymptotic behaviors}

\begin{figure}[t]
\begin{center}
\epsfig{file=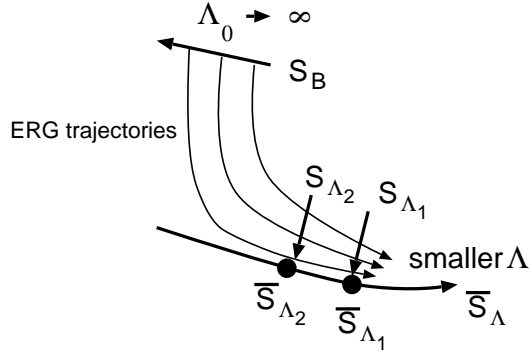, width=7cm}
\caption{The Wilson action $\bar{S}_\Lambda$ of the continuum limit
  obeys its own Polchinski differential equation and asymptotic conditions.}
\label{cl-renormalizedtrajectory}
\end{center}
\end{figure}
For realization of symmetry, it is more convenient to parameterize the
theory by directly specifying the ``asymptotic'' behavior of the
continuum limit $\bar{S}_\Lambda$ as $\Lambda \to
\infty$.\cite{Becchi:1996an, Pernici:1998tp, Sonoda:2002pb} (See
Fig.~\ref{cl-renormalizedtrajectory}.)  To explain this parameterization,
let us expand the continuum limit $\bar{S}_\Lambda$ of the Wilson action
in powers of fields:
\begin{equation}
\bar{S}_{I,\Lambda} [\phi] = \sum_{n=1}^\infty \frac{1}{(2n)!}
 \int_{p_1+\cdots+p_{2n} = 0} \bar{\V}_{2n} (\Lambda; p_1, \cdots,
 p_{2n}) \,\phi (p_1) \cdots \phi (p_{2n})\,.
\end{equation}
The vertex functions $\bar{\V}_{2n}$ are local, meaning that we can
expand them in powers of $m^2$ and the external momenta.  Since
$\bar{\V}_{2n}$ has mass dimension $4-2n$, we expect the following
expansions:
\begin{equation}
\begin{array}{r@{~=~}l}
\bar{\V}_2 (\Lambda; p,-p) & \Lambda^2 a_2 (\ln\Lambda/\mu) + m^2 b_2
 (\ln \Lambda/\mu) + p^2 c_2 (\ln \Lambda/\mu) + \cdots\,,\\
\bar{\V}_4 (\Lambda; p_1, \cdots, p_4) & a_4 (\ln\Lambda/\mu) + \cdots\,,\\
\bar{\V}_{2n} (\Lambda; p_1, \cdots, p_{2n}) &
 \frac{1}{\Lambda^{2n-4}} a_{2n} (\ln \Lambda/\mu) + \cdots\,.\qquad (2n \ge
 6)
\end{array}
\end{equation}
At each order of loop expansions, the coefficients $a_2$, $b_2$, etc.,
are finite polynomials of the logarithm of $\Lambda$.  The scale
parameter $\mu$ is introduced to make the argument of the logarithm
dimensionless, as has been done in the previous subsection.  Since the
Polchinski differential equation \bref{derivation-polchinskieq}
determines only the $\Lambda$ dependence, it leaves $\Lambda$
independent parts undetermined.  More explicitly, the undetermined
$\Lambda$ independent parts are additive constants in
\begin{equation}
b_2 (\ln \Lambda/\mu)\,,\quad c_2 (\ln \Lambda/\mu)\,,\quad
a_4 (\ln \Lambda/\mu)\,.
\end{equation}
We can fix these additive constants by specifying the values of $b_2,
c_2, a_4$ at $\Lambda = \mu$.  Once the additive constants are fixed,
the vertex functions, including $a_2$ and $a_{2n\ge 6}$, are determined
unambiguously by the Polchinski differential equation.  Thus, we have
argued that the Wilson action can be determined uniquely by imposing
a convention on the three additive constants
\begin{equation}
b_2 (0)\,,\quad c_2 (0)\,,\quad a_4 (0)\,.
\end{equation}
The simplest choice would be
\begin{equation}
b_2 (0) = c_2 (0) = 0\,,\quad
a_4 (0) = - \lambda\,,\label{cl-phi4MS}
\end{equation}
which is an analog of the MS scheme for dimensional
regularization.\cite{Hooft:1973mm}\footnote{In Appendix
\ref{appcomp-beta} we derive a mass independent RG equation using the
choice \bref{cl-phi4MS}.}

In lieu of expanding the vertex functions in powers of $m^2$ and
external momenta, we often refer to the ``asymptotic behavior'' of the
Wilson action:
\begin{eqnarray}
\bar{S}_{I,\Lambda} &\asym& \int d^4 x \, \left[ \left(\Lambda^2 a_2 (\ln
    \Lambda/\mu) + m^2 b_2 (\ln \Lambda/\mu) \right)
\frac{\phi^2}{2} \right.\nn\\
&&\quad \left. + c_2 (\ln \Lambda/\mu) \frac{1}{2} \partial_\mu
\phi \partial_\mu \phi + a_4 (\ln \Lambda/\mu) \frac{1}{4!} \phi^4
\right]\,.\label{cl-phi4asymp}
\end{eqnarray}
This is a convenient way of summarizing the leading terms of the
expansions of the vertex functions in powers of $m^2$ and external
momenta.  The above ``asymptotic'' expansion is valid in two ways:
\begin{enumerate}
\item For fixed $m^2$ and external momenta, we take $\Lambda$
      asymptotically large.
\item For fixed $\Lambda$, we take $m^2$ and external momenta small.
\end{enumerate}
Hence, calling \bref{cl-phi4asymp} an asymptotic expansion is a bit
imprecise.  When we refer to asymptotic expansions, we often consider
the second meaning.  Please note that we only keep the part of the
vertex functions multiplied by non-negative powers of $\Lambda$ in the
asymptotic behavior; we do not write a $\phi^6$ term in the asymptotic
behavior, since the leading term of $\bar{\V}_6$ is multiplied by
$1/\Lambda^2$.

For concreteness, let us consider the examples of two- and four-point
vertex functions at 1-loop.  First, let us consider the differential
equation for the two-point vertex:
\begin{equation}
- \Lambda \frac{\partial}{\partial \Lambda} \bar{\V}_2^{(1)} (\Lambda)
= - \frac{\lambda}{2} \int_q \frac{\Delta (q/\Lambda)}{q^2 + m^2}
\label{cl-ERG2pt1loop}
\end{equation}
Expanding the right-hand side
in $\frac{m^2}{\Lambda^2}$, we obtain
\begin{equation}
- \Lambda \frac{\partial}{\partial \Lambda} \bar{\V}_2^{(1)} (\Lambda)
= - \frac{\lambda}{2} \Lambda^2 \int_q \frac{\Delta (q)}{q^2} \left(
1 - \frac{m^2}{\Lambda^2} \frac{1}{q^2} + \mathrm{O} \left(
    \frac{m^4}{\Lambda^4} \right) \right)\,.
\end{equation}
Using
\begin{equation}
\int_q \frac{\Delta (q)}{q^2} = 2 \int_q \frac{K(q)}{q^2},\quad
\int_q \frac{\Delta (q)}{q^4} = \frac{2}{(4 \pi)^2}\,,
\end{equation}
we obtain the asymptotic behavior
\begin{equation}
\bar{\V}_2^{(1)} (\Lambda) \asym - \frac{\lambda}{2} \Lambda^2 \left[
- \frac{1}{2} \int_q \frac{\Delta (q)}{q^2}  + \frac{2}{(4
  \pi)^2} \frac{m^2}{\Lambda^2} \ln \frac{\Lambda}{\mu}  \right]
+ \mathrm{const}\,,
\label{cl-v21loopasymp}
\end{equation}
where we have introduced $\ln \mu$ as part of the additive constant.  
Hence, we obtain
\begin{equation}
\lb\begin{array}{c@{~=~}l}
a_2^{(1)} & \frac{\lambda}{4} \int_q \frac{\Delta (q)}{q^2}
= \frac{\lambda}{2} \int_q \frac{K(q)}{q^2}\,,\\
b_2^{(1)} (\ln\Lambda/\mu) & - \frac{\lambda}{(4 \pi)^2} \ln
\frac{\Lambda}{\mu} + \mathrm{const}\,.
\end{array}\right.
\end{equation}
Then, by imposing the convention
\begin{equation}
b_2^{(1)} (0) = 0\,,
\end{equation}
we obtain the asymptotic behavior of the solution given by
\bref{cl-v2oneloop}.

\begin{figure}[t]
\begin{center}
\epsfig{file=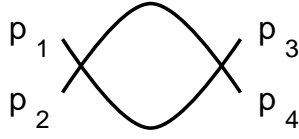, width=4cm}
\caption{The s-channel contribution to $\bar{\V}_4^{(1)}$}
\label{cl-4pt1loop}
\end{center}
\end{figure}

\begin{figure}[t]
\begin{center}
\epsfig{file=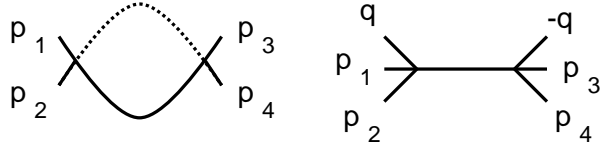, width=8cm}
\caption{The left represents (\ref{cl-4pt1loopdiff}), where the broken
  line gives the propagator proportional to $\Delta$.  The right
  represents $v_6 (\Lambda; p_1+p_2+q,p_3+p_4-q)$.}
\label{cl-4pt1loopdiffgraph}
\end{center}
\end{figure}

Let us next consider a 1-loop contribution to the four-point vertex
function corresponding to the s-channel Feynman diagram in
Fig.~\ref{cl-4pt1loop}.  In the rest of this section, we consider
only the continuum limit, and for simplicity we omit a bar to denote the
limit.
Calling the s-channel vertex by $v_4 (\Lambda; p_1+p_2)$, we obtain
the differential equation
\begin{equation}
- \Lambda \frac{\partial}{\partial \Lambda} v_4 (\Lambda;
p_1+p_2) = \int_p \frac{\Delta (p/\Lambda)}{p^2 + m^2} \, v_6 (\Lambda;
p_1+p_2+q, p_3 + p_4 - q)\,,\label{cl-4pt1loopdiff}
\end{equation}
where $v_6$ is part of the tree-level six-point vertex function
$\V_6^{(0)} (\Lambda; p_1,\cdots,p_4, q, -q)$ corresponding to the
right diagram in Fig.~\ref{cl-4pt1loopdiffgraph}.
Now, $v_6$ satisfies
\begin{equation}
- \Lambda \frac{\partial}{\partial \Lambda} v_6 (\Lambda; p,-p) =
\lambda^2 \frac{\Delta (p/\Lambda)}{p^2 + m^2}\,,
\end{equation}
and the asymptotic condition
\begin{equation}
v_6 (\Lambda; p,-p) \asym 0\,,
\end{equation}
since no six-point vertex survives the $\Lambda \to \infty$
limit.\footnote{Recall our terminology.  Without referring to
``asymptotic'' behaviors, we can say the same thing as follows.
Expanding $\V_6$ in powers of $m^2$ and momenta, the leading term is
proportional to a negative power of $\Lambda$.}  Hence, the solution
is
\begin{equation}
v_6 (\Lambda; p,-p) = \lambda^2 \frac{1 - \K{p}}{p^2 + m^2}\,.
\end{equation}
Substituting this into (\ref{cl-4pt1loopdiff}), we obtain
\begin{equation}
- \Lambda \frac{\partial}{\partial \Lambda} v_4 (\Lambda;
p) = \lambda^2 \int_p \frac{\Delta (p/\Lambda)}{p^2 + m^2}
\frac{1 - \K{(q+p)}}{(q+p)^2 + m^2}\,.
\end{equation}
Now, $v_4 (\ln \Lambda/\mu;p)$ is determined by this differential
equation and an additive constant in its asymptotic behavior.  To find
the asymptotic behavior, we consider the integral
\begin{eqnarray}
\int_q \frac{\Delta (q/\Lambda)}{q^2 + m^2} \frac{1 -
  \K{(q+p)}}{(q+p)^2 + m^2} &=& \int_q
\frac{\Delta (q)}{q^2 + m^2/\Lambda^2} \frac{1 -
  K(q+p/\Lambda)}{\left( q + p/\Lambda \right)^2 + m^2/\Lambda^2}\nn\\
&\asym& \int_q \frac{\Delta (q) \left( 1 - K(q) \right)}{q^4}
= \frac{1}{(4 \pi)^2}\,.
\end{eqnarray}
Hence, the asymptotic behavior of $v_4$ is given by
\begin{equation}
v_4 (\ln \Lambda/\mu; p) \asym - \frac{\lambda^2}{(4 \pi)^2} \ln
\frac{\Lambda}{\mu} + \mathrm{const}\,.
\end{equation}
This implies, after summing over three channels,
\begin{equation}
a_4^{(1)} (\ln \Lambda/\mu) = - \frac{3 \lambda^2}{(4 \pi)^2} \ln
\frac{\Lambda}{\mu} + \mathrm{const}\,.
\end{equation}

Now, we obtain
\begin{eqnarray}
&&- \Lambda \frac{\partial}{\partial \Lambda} \left( v_4 (\ln \Lambda/\mu;
    p) + \frac{\lambda^2}{(4 \pi)^2} \ln \frac{\Lambda}{\mu}
    \right)\nn\\
&=& \lambda^2 \int_q \left[ \frac{\Delta (q/\Lambda)}{q^2 +
      m^2} \frac{1 - K\left((q+p)/\Lambda\right)}{(q +
          p)^2 + m^2} 
- \frac{\Delta (q/\Lambda) \left( 1 - \K{q} \right)}{q^4} \right]\,.
\end{eqnarray}
This is integrated to give
\begin{eqnarray}
&&v_4 (\ln\Lambda/\mu; p) = - \frac{\lambda^2}{(4 \pi)^2} \ln
\frac{\Lambda}{\mu} + \mathrm{const}\nn\\
&&\quad + \frac{\lambda^2}{2} \int_q \left[ \frac{1 - \K{q}}{q^2 +
      m^2} \frac{1 - K\left((q+p)/\Lambda\right)}{(q +
          p)^2 + m^2} 
- \frac{\left( 1 - \K{q} \right)^2}{q^4} \right]\,.
\end{eqnarray}
The integral on the right-hand side vanishes asymptotically, i.e., as we
take $m^2$ and $p$ to zero.  The additive constant must be zero if we
adopt the convention $a_4 (0) = - \lambda$.

\subsection{Asymptotic behaviors of the Wilson action and the
  effective average action}

In \S \ref{derivation-Wetterich} we have introduced the effective
average action $\Gamma_{B,\Lambda} [\Phi]$ as the Legendre transform
of the Wilson action $\SL [\phi]$.  Their precise relation is given by
(\ref{derivation-GammaBSL}) and (\ref{derivation-Phiphi}).  In the
continuum limit $\Lambda_0 \to \infty$, let us write $\Gamma_\Lambda$
for $\Gamma_{B,\Lambda}$.  (We continue omitting a bar to denote the
continuum limit.) Then, we can write (\ref{derivation-GammaBSL}) and
(\ref{derivation-Phiphi}) as
\begin{equation}
\Gamma_\Lambda [\Phi] = \SL [\phi] + \int_p \frac{p^2 + m^2}{1 - \K{p}}
\left( \frac{1}{\K{p}} \frac{1}{2} \phi (p) \phi (-p) - \Phi (p) \phi (-p)
\right)\,
\label{cl-GammaLSL}
\end{equation}
and
\begin{equation}
\Phi (p) = \frac{1}{\K{p}} \phi (p) + \frac{1 - \K{p}}{p^2 + m^2} \frac{\delta
  \SL [\phi]}{\delta \phi (-p)}\,.
\end{equation}
For the interaction parts defined by
\begin{equation}
\lb\begin{array}{c@{~\equiv~}l}
\SIL [\phi] & \frac{1}{2} \int_p \frac{p^2 + m^2}{\K{p}} \phi (-p)
\phi (p) + \SL [\phi]\,,\\
\Gamma_{I,\Lambda} [\Phi] & \frac{1}{2} \int_p \frac{p^2 + m^2}{1 -
  \K{p}} \Phi (-p) \Phi (p) + \Gamma_\Lambda [\Phi]\,,
\end{array}\right.
\end{equation}
we can write (\ref{cl-GammaLSL}) as
\begin{equation}
\Gamma_{I,\Lambda} [\Phi] = \SIL [\phi] +
\frac{1}{2} \int_p \frac{p^2+m^2}{1 - \K{p}} \left( \Phi - \phi
\right) (-p) \left( \Phi - \phi \right) (p)\,.
\label{cl-GammaILSIL}
\end{equation}
Since $\Phi$ and $\phi$ become equal as $\Lambda \to \infty$,
(\ref{cl-GammaILSIL}) implies that $\Gamma_{I,\Lambda} [\Phi]$ and $\SIL
[\phi]$ have the same asymptotic behaviors.  In the case of the $\phi^4$
theory in $D=4$, the asymptotic behavior (\ref{cl-phi4asymp}) of the
Wilson action implies\footnote{This is the parameterization adopted in
ref.~\citen{Pernici:1998tp}, where $\mu$ is denoted as $\Lambda_R$.}
\begin{eqnarray}
\Gamma_{I,\Lambda} [\Phi] &\asym& \int d^4 x\, \left[ \left(\Lambda^2
        a_2 (\ln \Lambda/\mu) +
        m^2 b_2 (\ln \Lambda/\mu) \right) \frac{\Phi^2}{2} \right.\nn\\
&& \quad \left.  + c_2 (\ln \Lambda/\mu) \frac{1}{2} \left( \partial_\mu \Phi
    \right)^2 + a_4 (\ln \Lambda/\mu) \frac{\Phi^4}{4!}\right]\,.
\end{eqnarray}

\subsection{Alternative parameterization}

We have shown how to specify the Wilson action of the $\phi^4$ theory
in $D=4$ by specifying $b_2 (0), c_2 (0), a_4 (0)$ by hand.
Alternatively, we can specify the behavior of the two- and four-point
vertex functions at zero external momentum as an analog of the BPHZ
scheme.  More concretely, we can impose
\begin{eqnarray}
\V_2 (\Lambda = \mu; p=0) &=& 0\,,\\
\frac{\partial}{\partial p^2} \V_2 (\Lambda=\mu; p,-p)\Big|_{p=0} &=& 0\,,\\
\V_4 (\Lambda = \mu; p_i = 0) &=& - \lambda\,,
\end{eqnarray}
for a given choice of the renormalization scale $\mu$.

Let us see how the above conditions change the choice of $b_2 (0), a_4
(0)$ at 1-loop.  The two-point vertex function becomes
\begin{equation}
\V_2^{(1)} (\Lambda) = - \frac{\lambda}{2} \left[
\left( - \Lambda^2 + \mu^2\right) \int_q \frac{K(q)}{q^2} +
m^2 \frac{2}{(4 \pi)^2} \ln \frac{\Lambda}{\mu} 
+ m^4 \int_q \frac{K(q/\mu) - K(q/\Lambda)}{q^4 (q^2 + m^2)}
\right]\,.
\end{equation}
Hence,
\begin{equation}
b_2 (0) = - \frac{\lambda}{2} \left[ \frac{\mu^2}{m^2} \int_q
    \frac{K(q)}{q^2} + m^2 \int_q \frac{K(q/\mu)-1}{q^4 (q^2 + m^2)}
\right]\,. 
\end{equation}
Note the non-trivial dependence on $m^2$.

The four-point vertex function, for a single channel, becomes
\begin{equation}
v_4 (\ln \Lambda/\mu; p) = 
\frac{\lambda^2}{2} \int_q \left[
\frac{1 - \K{q}}{q^2 + m^2} \frac{1 - \K{(q+p)}}{(q+p)^2 + m^2} -
\left( \frac{1 - K(q/\mu)}{q^2 + m^2} \right)^2 \right]\,,
\end{equation}
where we have used
\begin{equation}
\int_q \left[ \left(\frac{1 -
          K(q/\mu)}{q^2} \right)^2 - \left(\frac{1 -
          \K{q}}{q^2}\right)^2 \right] = \frac{2}{(4 \pi)^2} \ln
\frac{\Lambda}{\mu}\,. 
\end{equation}
Hence,
\begin{equation}
a_4 (0) = - \lambda + 3 \times \frac{\lambda^2}{2} \int_q \left(1 -
    K(q/\mu)\right)^2 \left[\frac{1}{q^4} - \frac{1}{(q^2+m^2)^2}
\right]\,. 
\end{equation}
where the factor $3$ is for three channels (s,t,u). This also has
non-trivial dependence on $m^2$.\footnote{The non-trivial dependence of
$b_2 (0), a_4 (0)$ on $\frac{m^2}{\mu^2}$ implies that the beta function
and anomalous dimension of the squared mass, explained in Appendix
\ref{appcomp-beta}, are mass dependent.}

In general, a renormalization scheme corresponds to a convention for
the coefficients $b_2 (0), c_2 (0), a_4 (0)$.  In Appendix
\ref{appcomp-univ}, we will show that changing the three coefficients
amounts to changing $m^2, \lambda$, and the normalization of $\phi$
under the same renormalization scheme.

\newpage

\section{Composite operators\label{comp}}

In this section we introduce \textbf{composite
operators},\cite{Becchi:1996an} which provide 
the most important tool for our program of symmetry realization.  By the
name we do not mean any random functionals of field variables.  Only
when a cutoff dependent functional satisfies a particular property, we
call it a composite operator.

\subsection{Definition\label{comp-def}}

Let $\Delta S_\Lambda$ be an infinitesimal deformation of a Wilson
action $S_\Lambda$ so that both $S_\Lambda$ and $S_\Lambda + \Delta
S_\Lambda$ satisfy the same Polchinski equation.  Then, the difference
satisfies the following linear equation:
\begin{eqnarray}
- \Lambda \frac{\partial}{\partial \Lambda} \Delta S_\Lambda &=&
\int_p \frac{\Delta (p/\Lambda)}{p^2+ m^2} \left[
\frac{p^2+m^2}{\K{p}} \phi (p) \frac{\delta}{\delta \phi
  (p)}\right.\nn\\
&&\left. + \frac{\delta
      S_{\Lambda}}{\delta \phi (-p)} \frac{\delta}{\delta \phi (p)}
+ \frac{1}{2} \frac{\delta^2}{\delta \phi (-p)\delta \phi (p)} \right]
\Delta S_\Lambda\,.
\end{eqnarray}
This can be rewritten using the interaction part $S_{I,\Lambda}$ as
\begin{equation}
- \Lambda \frac{\partial}{\partial \Lambda} \Delta S_\Lambda =
\int_p \frac{\Delta (p/\Lambda)}{p^2+ m^2} \left[ \frac{\delta
      S_{I,\Lambda}}{\delta \phi (-p)} \frac{\delta}{\delta \phi (p)}
+ \frac{1}{2} \frac{\delta^2}{\delta \phi (-p)\delta \phi (p)} \right]
\Delta  S_\Lambda\,.
\end{equation}

In general we call a $\Lambda$ dependent functional $\Op_\Lambda
[\phi]$ a composite operator if it satisfies the same differential
equation as above:
\begin{eqnarray}
- \Lambda \frac{\partial}{\partial \Lambda} \Op_\Lambda &=& 
\int_p \frac{\Delta (p/\Lambda)}{p^2+ m^2} \left[
\frac{p^2+m^2}{\K{p}} \phi (p) \frac{\delta}{\delta \phi
  (p)}\right.\nn\\
&&\left. \qquad+ \frac{\delta
      S_{\Lambda}}{\delta \phi (-p)} \frac{\delta}{\delta \phi (p)}
+ \frac{1}{2} \frac{\delta^2}{\delta \phi (-p)\delta \phi (p)} \right]
\Op_\Lambda\nn\\
&=&
\int_p \frac{\Delta (p/\Lambda)}{p^2+ m^2} \left[ \frac{\delta
      S_{I,\Lambda}}{\delta \phi (-p)} \frac{\delta}{\delta \phi (p)}
+ \frac{1}{2} \frac{\delta^2}{\delta \phi (-p)\delta \phi (p)} \right]
\Op_\Lambda\,.\label{comp-compdiffeq}
\end{eqnarray}
In other words, any $\Lambda$ dependent functional $\Op_\Lambda
[\phi]$  is a
composite operator, if
\begin{equation}
S_{I,\Lambda} + \ep \Op_\Lambda
\end{equation}
satisfies the Polchinski equation up to first order in $\ep$.  Note
that the right-hand side of (\ref{comp-compdiffeq}) is a linear
differential operator acting on $\Op_\Lambda$.  Hence, denoting the
differential operator by $\mathcal{D}$, we may express
\bref{comp-compdiffeq} symbolically as
\begin{equation}
- \Lambda \frac{\partial}{\partial \Lambda} \Op_\Lambda = \mathcal{D}
\cdot \Op_\Lambda\,.\label{comp-ERGdiffO}
\end{equation}

To determine $\Op_\Lambda$ unambiguously, we must supply either an
initial condition at an UV scale $\Lambda_0$, or an asymptotic
condition in the case of a renormalizable theory.  In the former case,
given $\Op_{\Lambda_0} = \Op_B$, we obtain an integral formula for the
solution of the differential equation (\ref{comp-compdiffeq}) as
follows:
\begin{equation}
\Op_\Lambda [\phi] \exp \left[ S_{I,\Lambda} [\phi] \right]
= \int [d\phi'] \Op_B [\phi+\phi']
\exp \left[ - \frac{1}{2} \int_p \frac{p^2+m^2}{K_0-K} \phi' \phi'
+ S_{I,B} [\phi + \phi'] \right]\,.\label{comp-compLcompB}
\end{equation}
This is obtained by taking an infinitesimal variation of the
definition (\ref{derivation-WilsonI}) of the Wilson action:
\begin{equation}
\exp \left[ S_{I,\Lambda} [\phi] \right] \equiv
\int [d\phi'] \exp \left[ - \frac{1}{2} \int \frac{p^2 + m^2}{K_0-K}
    \phi' \phi' + S_{I,B} [\phi' + \phi] \right]\,.
\label{comp-WilsonI}
\end{equation}

If the theory is renormalizable, we can choose the $\Lambda_0$
dependence of $\Op_B$ such that
\begin{equation}
\lim_{\Lambda_0 \to \infty} \Op_\Lambda [\phi]
\end{equation}
exists for any fixed $\Lambda$.  Alternatively, we can impose an
asymptotic condition such as
\begin{equation}
\Op_\Lambda [\phi] \asym \sum_i \Lambda^{x-x_i} c_i (\ln \Lambda/\mu)
\,\Op_i \,,
\end{equation}
where $x$ is the scale dimension of the composite operator
$\Op_\Lambda$, and $x_i$ is the scale dimension of $\Op_i$, a product of
derivatives of $\phi$.  At a given order in perturbation theory, the
coefficient $c_i$ is a polynomial of $\ln \Lambda/\mu$.  If $x_i < x$,
the $\Lambda$ dependence of the coefficient of $\Op_i$ is completely
determined by the differential equation (\ref{comp-compdiffeq}).  For
$x_i = x$, though, $c_i$ is left ambiguous by an additive constant
independent of $\ln \Lambda/\mu$.  Hence, the additive constants $c_i
(0)$, satisfying $x_i = x$, parameterize the composite operator.  We will
give three examples from the $\phi^4$ theory in $D=4$ in
subsect.~\ref{comp-asymp}.  It is important to note that if
\begin{equation}
 c_i (0) = 0
\end{equation}
for all $i$ such that $x_i = x$, the operator itself vanishes
\begin{equation}
\Op_\Lambda [\phi] = 0\,,
\end{equation}
due to the linearity of the differential equation
(\ref{comp-compdiffeq}) and uniqueness of its solution.

To understand the correlation functions of composite operators, let us
introduce a source $J$.  We recall the relation
(\ref{derivation-WBWL}) between $W_B$ and $W_\Lambda$, which can be
written in the form of functional integrals as
\begin{eqnarray}
&&\int [d\phi] \exp \left[ S_B [\phi] + \int J \phi \right]\nn\\
&&= \int [d\phi] \exp \left[ S_\Lambda [\phi] + \int \frac{K_0}{K} J
    \phi - \frac{1}{2} \int_p \frac{K_0-K}{p^2 + m^2}
    \frac{K_0}{K} J J \right] \,.\label{comp-WBWL}
\end{eqnarray}
Changing the action infinitesimally by $\Delta S_B = \ep \Op_B$, the
Wilson action changes by $\Delta S_\Lambda = \ep \Op_\Lambda$.  Hence,
we obtain
\begin{eqnarray}
    &&\int [d\phi] \Op_B [\phi] \exp \left[ S_B [\phi] +
        \int J \phi \right]\nn\\ 
    &&= \int [d\phi] \Op_\Lambda [\phi] \exp \left[ S_\Lambda [\phi]
        + \int \frac{K_0}{K} J  \phi - \frac{1}{2} \int_p
        \frac{K_0-K}{p^2 + m^2} \frac{K_0}{K} J J \right] \,.
\label{comp-compBJcompLJ}
\end{eqnarray}
Differentiating this with respect to $J$'s, we obtain\footnote{As
  usual, we only consider connected correlations.  In the presence of
  $J (p)$ for $p \ne 0$, we do not factor out the delta function for
  momentum conservation from the correlation functions.}
\begin{equation}
\vev{\Op_B \,\phi (p_1) \cdots \phi (p_n)}_{S_B}
= \prod_{i=1}^n \frac{\Kz{p_i}}{\K{p_i}} \cdot \vev{\Op_\Lambda \,\phi
  (p_1) \cdots \phi (p_n)}_{S_\Lambda}\,. \label{comp-compvev}
\end{equation}
This is independent of $\Lambda$.  If the theory is renormalizable, we
can take $\Lambda_0 \to \infty$ to obtain the continuum limit
\begin{equation}
\vev{\Op \,\phi (p_1) \cdots \phi (p_n)}^\infty
= \prod_{i=1}^n \frac{1}{\K{p_i}} \cdot \vev{\Op_\Lambda \,\phi
  (p_1) \cdots \phi (p_n)}_{S_\Lambda}\,,
\end{equation}
where $\Op$ denotes the continuum limit of $\Op_B$.

We now wish to show that (\ref{comp-WBWL}) is equivalent to
(\ref{comp-WilsonI}) as we have promised in \S\ref{derivation-source}.
In other words, the relation (\ref{comp-WBWL}), between the generating
functionals $W_B$ and $W_\Lambda$, is equivalent to the definition
(\ref{comp-WilsonI}) of the Wilson action; taking an infinitesimal
change of $S_B$ and $S_\Lambda$, we obtain (\ref{comp-compBJcompLJ})
from (\ref{comp-WBWL}), and (\ref{comp-compLcompB}) from
(\ref{comp-WilsonI}).  Since we have already derived (\ref{comp-WBWL})
from (\ref{comp-WilsonI}), we only need to derive (\ref{comp-WilsonI})
from (\ref{comp-WBWL}).  For this purpose we use the functional Fourier
inverse:
\begin{equation}
\exp \left[ S_\Lambda [\phi] \right] = \int [dJ] [d\phi'] \exp \left[
    S_\Lambda [\phi'] + \int_p \frac{K_0}{K} J(-p) \left( \phi' - \phi
    \right) (p) \right]\,,
\end{equation}
where $J (-p)$ (or its Fourier transform in real space) is integrated
along the imaginary axis.  Using (\ref{comp-WBWL}) to rewrite
the right-hand side, we obtain
\begin{equation}
\exp \left[ S_\Lambda [\phi] \right] = \int [dJ] [d\phi'] \exp \left[
    S_B [\phi'] + \int J \cdot \left( \phi' - \frac{K_0}{K} \phi \right) +
    \frac{1}{2} \int \frac{K_0-K}{p^2 + m^2} \frac{K_0}{K} J J
\right]\,.
\end{equation}
The integral over $J$ is gaussian, and we obtain
\begin{eqnarray}
&&\exp \left[ S_\Lambda [\phi] \right] = \int [d\phi'] \exp \Big[ S_B
    [\phi']\nn\\
&&\qquad\qquad\left. - \frac{1}{2} \int (p^2 + m^2) \frac{K}{K_0 (K_0-K)}
    \left( \phi' - \frac{K_0}{K} \phi\right)
    \left( \phi' - \frac{K_0}{K} \phi\right)\right]\nn\\
&&= \int [d\phi'] \exp \left[ S_{I,B} [\phi'] -
    \int \frac{p^2 + m^2}{K_0-K} (\phi'-\phi)(\phi'-\phi)  
- \frac{1}{2} \int \frac{p^2 + m^2}{K} \phi \phi \right]\nn\\
&&= \exp \left[ - \frac{1}{2} \int \frac{p^2 + m^2}{K} \phi \phi
\right] \int [d\phi'] \exp \left[ S_{I,B} [\phi'+\phi] -
    \int \frac{p^2 + m^2}{K_0-K} \phi' \phi' \right]\,.
\end{eqnarray}
This gives (\ref{comp-WilsonI}).

To recapitulate, we have introduced three equivalent criteria for a
functional $\Op_\Lambda [\phi]$ to be a composite operator:
\begin{enumerate}
\item $\Op_\Lambda$ satisfies the differential equation
    (\ref{comp-compdiffeq}). 
\item $\Op_\Lambda$ satisfies the integral formula
    (\ref{comp-compLcompB}) for some bare functional $\Op_B$.
\item $\Op_\Lambda$ satisfies the integral formula
    (\ref{comp-compBJcompLJ}) for some bare functional $\Op_B$.
    (\ref{comp-compBJcompLJ}) is further equivalent to
    (\ref{comp-compvev}). 
\end{enumerate}

\subsection{General examples\label{comp-examples}}

We introduce some general examples of composite operators that play
important roles in realization of symmetry.

\subsubsection{$[\phi]_\Lambda (p)$}

First of all, note that the elementary field $\phi (p)$ is
\textbf{not} a composite operator.  But
\begin{eqnarray} [\phi]_\Lambda (p) &\equiv& \frac{K_0}{K} \phi
    (p) + \frac{K_0 -
      K}{p^2 + m^2} \frac{\delta S_\Lambda}{\delta \phi (-p)}\nn\\
    &=& \phi (p) + \frac{K_0 - K}{p^2 + m^2} \frac{\delta
      S_{I,\Lambda}}{\delta \phi (-p)} \label{comp-phi}
\end{eqnarray}
is.  We can check this directly by differentiating this with respect
to $\Lambda$ and using the Polchinski equation.  Alternatively, we
differentiate (\ref{comp-WBWL})  with respect to $J(-p)$ to
obtain
\begin{eqnarray}
&&\int [d\phi] \phi (p) \exp \left[ S_B [\phi] + \int J \phi
\right] = \int [d\phi] \left( \frac{K_0}{K} \phi (p) - \frac{K_0-K}{p^2 +
      m^2} \frac{K_0}{K} J (p) \right) \nn\\
&&\qquad\times \exp \left[ S_\Lambda [\phi] + \int \frac{K_0}{K} J \phi - 
\frac{1}{2} \int \frac{K_0-K}{p^2 + m^2} \frac{K_0}{K} J J \right]\,.
\end{eqnarray}
Since
\begin{eqnarray}
&&\frac{K_0}{K} J(p) \int [d\phi] \exp \left[ S_\Lambda [\phi] + \int
    \frac{K_0}{K} J \phi \right]\nn\\
&&\quad= \int [d\phi] \exp \left[S_\Lambda [\phi]\right]
\frac{\delta}{\delta \phi (-p)} \exp \left[ \int
    \frac{K_0}{K} J \phi \right] \nn\\
&&\quad= - \int [d\phi] \frac{\delta S_\Lambda [\phi]}{\delta \phi (-p)}
\exp \left[ S_\Lambda [\phi] + \int
    \frac{K_0}{K} J \phi \right]\,,\label{comp-JWL}
\end{eqnarray}
we obtain
\begin{eqnarray}
&&\int [d\phi] \phi (p) \exp \left[ S_B [\phi] + \int J \phi
\right] \nn\\
&&= \int [d\phi] \left( \frac{K_0}{K} \phi (p) + \frac{K_0-K}{p^2 +
      m^2} \frac{\delta S_\Lambda}{\delta \phi (-p)} \right)\nn\\
&&\quad \times \exp \left[ S_\Lambda [\phi] + \int
    \frac{K_0}{K} J \phi  - \frac{1}{2} \int \frac{K_0-K}{p^2 + m^2}
    \frac{K_0}{K} J J \right]\,.
\end{eqnarray}
Thus, $[\phi]_\Lambda (p)$ is a composite operator according to
(\ref{comp-compBJcompLJ}). Taking the limit $\Lambda_0 \to \infty$, we
obtain $[\phi]_\Lambda (p)$ that has the asymptotic behavior
\begin{equation}
[\phi]_\Lambda (p) \asym \phi (p)\,.\label{comp-Phiasymp}
\end{equation}
For any $n = 1,2,\cdots$, we find
\begin{equation}
\prod_{i=1}^n \frac{1}{\K{p_i}} \cdot \vev{[\phi]_\Lambda (p) \phi
  (p_1) \cdots \phi (p_n)}_{S_\Lambda} 
\end{equation}
is independent of $\Lambda$.

\subsubsection{$\K{p} \frac{\delta \SL}{\delta \phi (-p)}$}

Another example is
\begin{equation}
\K{p} \frac{\delta S_\Lambda}{\delta \phi (-p)}\,.\label{comp-EofM}
\end{equation}
To see this is a composite operator, we multiply (\ref{comp-WBWL}) by
$J(p)$.  An analog of (\ref{comp-JWL}) gives, for $S_B$,
\begin{equation}
J(p) \int [d\phi] \exp \left[ S_B [\phi] + \int J \phi \right]
= - \int [d\phi] \frac{\delta S_B}{\delta \phi (-p)}  \exp \left[ S_B
    [\phi] + \int J \phi \right] \,.\label{comp-JWB}
\end{equation}
Combining this with (\ref{comp-JWL}), we obtain
\begin{eqnarray}
&&\int [d\phi] \frac{\delta S_B}{\delta \phi (-p)}  \exp \left[ S_B
    [\phi] + \int J \phi \right] \nn\\
&& = \frac{K}{K_0} \int [d\phi] \frac{\delta S_\Lambda
    [\phi]}{\delta \phi (-p)} \exp \left[ S_\Lambda [\phi] + \int
    \frac{K_0}{K} J - \frac{1}{2} \int \frac{K_0-K}{p^2 + m^2}
    \frac{K_0}{K} J J \right]\,.
\end{eqnarray}
Hence, according to (\ref{comp-compBJcompLJ}), (\ref{comp-EofM}) is a
composite operator.  This composite operator may be called the
equation of motion operator, since
\begin{eqnarray}
&&\prod_{i=1}^n \frac{1}{\K{p_i}} \cdot \vev{
\K{p} \frac{\delta \SL}{\delta \phi (-p)} \,
\phi (p_1) \cdots \phi (p_n)}_{\SL}\nn\\
&& = - \sum_{i=1}^n (2 \pi)^D
\delta^{(D)} (p+p_i) \vev{ \phi(p_1) \cdots \widehat{\phi (p_i)}
\cdots \phi (p_n)}^\infty\,,
\end{eqnarray}
where $\phi (p_i)$ is omitted on the right-hand side.

\subsubsection{$\frac{\delta \SIL}{\delta \phi (-p)}$}

We find
\begin{equation}
\Kz{p} \frac{\delta \SIL}{\delta \phi (-p)} = (p^2 + m^2) 
\left[\phi\right]_\Lambda (p) + \K{p} \frac{\delta \SL}{\delta \phi
  (-p)}\,.
\end{equation}
Hence,
\begin{equation}
\frac{\delta \SIL}{\delta \phi (-p)} \label{comp-dSIdphi}
\end{equation}
is a composite operator, since it is a sum of two composite operators.

\subsubsection{$ \K{p} \left( \frac{\delta \Op_\Lambda}{\delta \phi (-p)} +
    \Op_\Lambda \frac{\delta S_\Lambda}{\delta \phi (-p)} \right)$}

We next consider an arbitrary composite operator $\Op_\Lambda$ and the
corresponding bare operator $\Op_B$ so that the two are related by
(\ref{comp-compBJcompLJ}).  Multiplying this by $J(-p)$, and
integrating by parts as in (\ref{comp-JWL}) and (\ref{comp-JWB}), we
obtain
\begin{eqnarray}
&&\int [d\phi] \left( \frac{\delta \Op_B}{\delta \phi (-p)} + \Op_B
    \frac{\delta S_B}{\delta \phi (-p)} \right) \exp \left[ S_B
    [\phi] + \int J \phi \right] \nn\\
&&= \frac{K}{K_0} \int [d\phi] \left( \frac{\delta \Op_\Lambda}{\delta
      \phi (-p)} + \Op_\Lambda  \frac{\delta S_\Lambda}{\delta \phi
      (-p)} \right) \nn\\
&&\quad \times \exp \left[ S_\Lambda [\phi] + \int
    \frac{K_0}{K} J  - \frac{1}{2} \int \frac{K_0-K}{p^2 + m^2}
    \frac{K_0}{K} J J \right]\,.
\end{eqnarray}
Hence, according to (\ref{comp-compBJcompLJ}),
\begin{equation}
\Op'_\Lambda (p) 
\equiv \K{p} \left( \frac{\delta \Op_\Lambda}{\delta \phi (-p)} +
    \Op_\Lambda \frac{\delta S_\Lambda}{\delta \phi (-p)} \right)
\label{comp-dOdphi}
\end{equation}
is a composite operator.

The correlation functions of $\Op'_\Lambda$ is given in terms of those
of $\Op_\Lambda$:
\begin{eqnarray}
&&\vev{\Op'_\Lambda (p)\,\phi (p_1)
  \cdots \phi (p_n)}_{S_\Lambda} \nn\\
&&= \K{p} \int [d\phi] \phi (p_1) \cdots \phi (p_n) \frac{\delta}{\delta
  \phi (-p)} \left( \Op_\Lambda [\phi] \exp \left[ S_\Lambda [\phi]
    \right] \right)\nn\\
&&= - \sum_{i=1}^n (2 \pi)^D \delta^{(D)} (p+p_i) \K{p_i}
\vev{\Op_\Lambda (p)\,\phi
  (p_1) \cdots \widehat{\phi (p_i)} \cdots \phi (p_n)}_{S_\Lambda}\,,
\label{comp-dOdphivev}
\end{eqnarray}
where $\phi (p_i)$ is absent on the right-hand side.  Divided by
$\prod_{i=1}^n \K{p_i}$, this becomes independent of $\Lambda$.

We may call $\Op'_\Lambda$ a generalized equation of motion operator,
since $\Op'_\Lambda = 1$ gives the equation of motion operator $\K{p}
\frac{\delta \SL}{\delta \phi (-p)}$.

\subsubsection{$\frac{\delta S_\Lambda [\phi; J]}{\delta J(-p)}$}

Suppose that the Wilson action depends on a classical external field
$J$, which is not necessarily coupled linearly to the field $\phi$, so
that we may not have an explicit formula such as
(\ref{derivation-Jshift}).  We find that the derivative with respect
to the external field
\begin{equation}
\frac{\delta S_\Lambda [\phi; J]}{\delta J(-p)} = \frac{\delta S_{I,\Lambda}
  [\phi; J]}{\delta J(-p)}\label{comp-dSdJ}
\end{equation}
is a composite operator, since the derivative can be regarded as an
infinitesimal deformation of the Wilson action.  We indeed find that
\begin{eqnarray}
&&\prod_{i=1}^n \frac{1}{\K{p_i}} \cdot \vev{\frac{\delta S_\Lambda
    [\phi; J]}{\delta J(-p)} \phi (p_1) \cdots \phi (p_n)}_{S_\Lambda}\nn\\
&&\quad = \frac{\delta}{\delta J(-p)} \prod_{i=1}^n \frac{1}{\K{p_i}} \cdot
\vev{\phi (p_1) \cdots \phi (p_n)}_{S_\Lambda} 
\end{eqnarray}
is independent of $\Lambda$.  Hence, substituting $\Op_\Lambda =
\frac{\delta S_\Lambda [\phi; J]}{\delta J(-p)}$ into
(\ref{comp-dOdphi}), we obtain
\begin{equation}
\K{q} \lb \frac{\delta S_\Lambda [\phi;J]}{\delta J(-p)} \frac{\delta
  S_\Lambda [\phi; J]}{\delta \phi (q)}  + \frac{\delta^2 S_\Lambda [\phi;
  J]}{\delta J(-p) \delta \phi (q)}\rb
\label{comp-Sigmatype}
\end{equation}
is a composite operator.  The composite operators $\Sigma$ and
$\bar{\Sigma}$ which play important roles for realization of symmetry
are given in this form.  (See sects. \ref{WT} \& \ref{AF}.)

\subsubsection{$ \frac{\delta \Op_\Lambda [\phi;J]}{\delta J(-p)} + \Op_\Lambda
    [\phi; J] \frac{\delta S_{I,\Lambda} [\phi; J]}{\delta J(-p)}$}

Let us give one more example from a Wilson action in the presence of
$J$.  Given a composite operator satisfying
\begin{eqnarray}
&&\Op_\Lambda [\phi; J] \exp \left[ S_{I,\Lambda} [\phi; J] \right]
= \int [d\phi'] \Op_B [\phi+\phi'; J] \nn\\
&&\quad \times \exp \left[ - \frac{1}{2} \int_p \frac{p^2 +
      m^2}{K_0-K} + S_{I,B} [\phi +\phi'; J] \right]\,,
\end{eqnarray}
we differentiate this with respect to $J(-p)$ to find that
\begin{equation}
    \frac{\delta \Op_\Lambda [\phi;J]}{\delta J(-p)} + \Op_\Lambda
    [\phi; J] \frac{\delta S_{I,\Lambda} [\phi; J]}{\delta J(-p)}
\label{comp-dOdJ}
\end{equation}
is also a composite operator according to (\ref{comp-compLcompB}). By
choosing $\Op_\Lambda = \K{q} \frac{\delta \SL [\phi; J]}{\delta \phi
(q)}$ as the equation of motion operator, we obtain
(\ref{comp-Sigmatype}) again.

\subsubsection{Summary}

To summarize, we have introduced the composite operators of the following
types:
\begin{table}[h]
\begin{center}
\begin{tabular}{l}
composite operator\\
\hline
$[\phi]_\Lambda (p)$ defined by (\ref{comp-phi})\\
$\K{p} \delta \SL/\delta \phi (-p)$, the equation of motion\\
$\delta \SIL/\delta \phi (-p)$, a linear combination of the above two\\
$\K{p} \left( \delta \Op_\Lambda/\delta \phi (-p) + \Op_\Lambda \delta
 \SL/\delta \phi (-p)\right)$, generalized equation of motion\\
$\delta \SL/\delta J(-p)$, for an arbitrary external source $J$\\
$\delta \Op_\Lambda/\delta J (-p) + \Op_\Lambda \delta \SIL/\delta 
 J(-p)$\\
\hline
\end{tabular}
\caption{Various types of composite operators}
\label{comp-table}
\end{center}
\end{table}

\subsection{Composite operators defined by their asymptotic behaviors
  \label{comp-asymp}} 

In this subsection, we give three concrete examples specifically from
the $\phi^4$ theory in $D=4$.  Since the theory is renormalizable, we
can introduce composite operators by specifying their asymptotic
behaviors.
\begin{itemize}
\item[(i)] $[\phi^2/2]_\Lambda$ is defined by the differential equation
	   (\ref{comp-compdiffeq}) and the asymptotic condition
\begin{equation}
\left[\frac{\phi^2}{2}\right]_\Lambda (p) \asym b (\ln \Lambda/\mu)
\frac{1}{2} \int_q \phi (q) \phi (p-q)\,.
\end{equation}
The normalization is determined by the value $b (0)$.  For example, we
can take 
\begin{equation}
b(0) = 1\,.
\end{equation}
If we choose $b(0)=0$, then the operator vanishes identically.
\item[(ii)] The dimension $4$ operator $[\phi^4/4!]_\Lambda$ with zero
    momentum has the following asymptotic behavior
\begin{eqnarray}
&&\left[\frac{\phi^4}{4!}\right]_\Lambda (0) \asym d (\ln \Lambda/\mu)
\frac{1}{4!} \int_{p_1,p_2,p_3} \phi (p_1) \phi (p_2) \phi (p_3) \phi
(-p_1-p_2-p_3) \nn\\
&&\qquad + \frac{1}{2} \int_p \phi (p) \phi (-p)
\lb \Lambda^2 a (\ln \Lambda/\mu) + m^2 b (\ln \Lambda/\mu) + p^2 c
(\ln \Lambda/\mu) \rb \,,
\end{eqnarray}
where $d(0), b(0), c(0)$ parameterize the operator.
As a simplest choice, we can take
\begin{equation}
d(0) = 1,\quad b(0) = c(0) = 0\,.
\end{equation}
\item[(iii)] A related composite operator $[\frac{1}{2} \partial_\mu
\phi \partial_\mu \phi] (0)$ can be defined by
\begin{equation}
c(0) = 1,\quad b(0) = d(0) = 0\,.
\end{equation}
If we choose $b(0)=c(0)=d(0)=0$, the operator vanishes.
\end{itemize}
Any scalar composite operator $\Op_\Lambda$ of dimension $4$ with zero
momentum is given as a linear combination
\begin{equation}
\Op_\Lambda  = c_1 \,
m^2 \left[\frac{\phi^2}{2}\right]_\Lambda (0)
+ c_2\, \left[\frac{\phi^4}{4!}\right]_\Lambda (0)
+ c_3\, \left[\frac{1}{2} \partial_\mu
\phi \partial_\mu \phi\right] (0)\,,
\end{equation}
where the coefficients $c_{1,2,3}$ are independent of $\Lambda$.

\subsection{1PI composite operators\label{comp-1PI}}

We recall from \S \ref{comp-def} that a composite operator
$\Op_\Lambda [\phi]$ is an infinitesimal change of the interaction
action $\SIL$.  Let $\Op^{1PI}_\Lambda [\Phi]$ be the corresponding
infinitesimal change in the effective average action
$\Gamma_{B,\Lambda} [\Phi]$.  Using the general relations
(\ref{derivation-deltaGammadeltaW}) and (\ref{derivation-WBLvsSL}), we
obtain
\begin{equation}
\Op^{1PI}_\Lambda [\Phi] = \Op_\Lambda [\phi]\,,
\label{comp-OO1PI}
\end{equation}
where $\Phi$ is given by (\ref{derivation-Phiphi})
\begin{equation}
\Phi (p) =  \frac{K_0}{K}  \phi (p) +
\frac{K_0 - K}{p^2 + m^2} \frac{\delta \SL}{\delta \phi (-p)}\,.
\label{comp-Phiphi}
\end{equation}
We immediately notice that $\Phi$, which has been introduced for the
Legendre transformation, coincides with the composite operator
$[\phi]_\Lambda$ defined by (\ref{comp-phi}):
\begin{equation}
\Phi (p) = [\phi]_\Lambda (p)\,.
\end{equation}
Since $\Phi$ and $\phi$ are identical asymptotically, we find that in
renormalized theories $\Op_\Lambda [\phi]$ and $\Op^{1PI}_\Lambda
[\Phi]$ have the same asymptotic behaviors:
\begin{equation}
\lb\begin{array}{c@{~\asym~}l}
\Op_\Lambda [\phi] & \sum_i \Lambda^{x-x_i} c_i (\ln \Lambda/\mu)\,
\Op_i [\phi]\,,\\
\Op^{1PI}_\Lambda [\Phi] & \sum_i \Lambda^{x-x_i} c_i (\ln \Lambda/\mu)\,
\Op_i [\Phi]\,,
\end{array}\right.
\end{equation}
where $\phi$ is fixed for $\Op_\Lambda$, and $\Phi$ for
$\Op^{1PI}_\Lambda$. 

The $\Lambda$ dependence of
$\Op^{1PI}_\Lambda$ for fixed $\Phi$ can be obtained easily from the
Wetterich equation (\ref{derivation-Wettericheq}).  Varying the
effective average action infinitesimally by $\Op_\Lambda^{1PI}
[\Phi]$, we obtain
\begin{eqnarray}
 - \Lambda \frac{\partial}{\partial \Lambda} \Op^{1PI}_\Lambda [\Phi]
    &=& - \frac{1}{2} \int_{p,q,r} \left( - \Lambda
        \frac{\partial}{\partial \Lambda} R_\Lambda (p) \right)\nn\\
&&\quad \times \left(\Gamma_{B,\Lambda}^{(2)}\right)^{-1} (p,-q)
     \frac{\delta^2 \Op^{1PI}_\Lambda [\Phi]}{\delta \Phi (-q) \delta \Phi (r)}
\left(\Gamma_{B,\Lambda}^{(2)}\right)^{-1} (r,-p)\,,
\label{comp-dOdL}
\end{eqnarray}
where we have used the following formula for an infinitesimal variation:
\begin{equation}
\delta \left(\Gamma_{B,\Lambda}^{(2)}\right)^{-1} (p, -p) 
= - \int_{q,r} \left(\Gamma_{B,\Lambda}^{(2)}\right)^{-1} (p,-q) \cdot
\delta \Gamma_{B,\Lambda}^{(2)} (q,-r) \cdot
\left(\Gamma_{B,\Lambda}^{(2)}\right)^{-1} (r, -p)\,.
\end{equation}
The simplest example satisfying the above differential equation is 
\begin{equation}
\Op_\Lambda^{1PI} [\Phi] = \Phi (p)\,,
\end{equation}
for which the right-hand side of \bref{comp-dOdL} vanishes.  $\Phi (p)$
has no $\Lambda$ dependence as a functional of $\Phi$; regarding this as
a functional of $\phi$, however, this has non-trivial $\Lambda$
dependence as given by \bref{comp-Phiphi}.

\newpage

\section{Realization of symmetry by WT identity\label{WT}}

\subsection{WT identity in the continuum limit}

In the continuum limit, the Ward-Takahashi (WT) identity for
continuous symmetry is given in the following form:
\begin{equation}
\sum_{i=1}^n \vev{\phi (p_1) \cdots \mathcal{O} (p_i) \cdots \phi
  (p_n)}^\infty = 0\,,
\end{equation}
where $\mathcal{O} (p)$ is a composite operator, giving an infinitesimal
symmetry transformation of the elementary field $\phi (p)$.

We wish to realize the above WT identity in the ERG formalism as an
invariance of the Wilson action.  Denoting by $\Op_\Lambda (p)$ a
composite operator that gives $\Op (p)$ in the continuum limit, we
consider an infinitesimal transformation of the elementary field as
\begin{equation}
\delta \phi (p) = \K{p}\,\Op_\Lambda (p)\,.\label{WT-deltaphi}
\end{equation}
As we will explain shortly, the change of the Wilson action under the
above transformation is given by the \textbf{WT composite operator}:
\begin{equation}
\Sigma_\Lambda \equiv \int_p \K{p} \left( \Op_\Lambda (p) \frac{\delta
      S_\Lambda}{\delta \phi (p)} + \frac{\delta \Op_\Lambda
      (p)}{\delta \phi (p)} \right)\,.\label{WT-Sigma}
\end{equation}
This is a composite operator of type (\ref{comp-dOdphi}), integrated
over momentum $p$.  Hence, its correlation functions are given by
\begin{equation}
\vev{\Sigma_\Lambda \,\phi (p_1) \cdots \phi (p_n)}^\infty
= \sum_{i=1}^n \vev{\phi (p_1) \cdots \Op_\Lambda (p_i) \cdots
  \phi (p_n)}^\infty\,.
\end{equation}
Therefore, the WT identity is equivalent to the vanishing of the WT
composite operator $\Sigma_\Lambda$:
\begin{equation}
    \Sigma_\Lambda = 0\,.\label{WT-vanishingSigma}
\end{equation}
The WT operator \bref{WT-Sigma} is easy to understand.  The first part
of $\Sigma_\Lambda$ 
\begin{equation}
\int_p \K{p} \Op_\Lambda (p) \frac{\delta \SL}{\delta \phi (p)}
\end{equation}
denotes the change of the Wilson action under the infinitesimal
transformation (\ref{WT-deltaphi}).  The second part
\begin{equation}
\int_p \K{p} \frac{\delta \Op_\Lambda (p)}{\delta \phi (p)}
\end{equation}
denotes the contribution of the transformation (\ref{WT-deltaphi}) to
the jacobian.  Hence, (\ref{WT-Sigma}) is the ``quantum'' change of the
Wilson action, and (\ref{WT-vanishingSigma}) amounts to the
\textbf{quantum invariance}.

Let us consider $\Sigma_\Lambda$ further.  This is determined by $\SL$
and $\Op_\Lambda$.  Recall how $\SL$ and $\Op_\Lambda$ are determined in
renormalizable theories.  The interaction part $\SIL$ of the Wilson
action is determined by the Polchinski equation
\begin{equation}
- \Lambda \frac{\partial}{\partial \Lambda} \SIL = \int_p
\frac{\Delta (p/\Lambda)}{p^2 + m^2} \frac{1}{2} \lb \frac{\delta
  S_{I,\Lambda}}{\delta \phi (-p)} \frac{\delta
  S_{I,\Lambda}}{\delta \phi (p)} + \frac{\delta^2
  S_{I,\Lambda}}{\delta \phi (-p)\delta \phi (p)}  \rb\,,
\end{equation}
and its asymptotic behavior.  For example, for the $\phi^4$ theory in
$D=4$, the Wilson action $S_{I,\Lambda}$ is specified uniquely by the
additive constants $b_2 (0), c_2 (0), a_4 (0)$ in the asymptotic
behavior:
\begin{eqnarray}
S_{I,\Lambda} &\asym& \int d^4 x\, \left[ \left( \Lambda^2 a_2 (\ln
    \Lambda/\mu) + m^2 b_2 (\ln \Lambda/\mu) \right) \frac{1}{2}
\phi^2 \right.\nn\\
&&\left. \quad + c_2 (\ln \Lambda/\mu) \frac{1}{2} \partial_\mu
\phi \partial_\mu \phi + a_4 (\ln \Lambda/\mu) \frac{1}{4!} \phi^4
\right]\,.
\end{eqnarray}
Similarly, the composite operator $\Op_\Lambda$ is determined by the ERG
linear differential equation:
\begin{equation}
- \Lambda \frac{\partial}{\partial \Lambda} \Op_\Lambda = \int_p
\frac{\Delta (p/\Lambda)}{p^2 + m^2} \lb 
 \frac{\delta S_{I,\Lambda}}{\delta \phi (-p)} 
 \frac{\delta}{\delta \phi (p)} + \frac{1}{2}  \frac{\delta^2}{\delta
   \phi (-p)\delta \phi (p)}  \rb \Op_\Lambda\,,
\end{equation}
and its asymptotic behavior:
\begin{equation}
\Op_\Lambda \asym \sum_i \Lambda^{x-x_i} d_i (\ln \Lambda/\mu) \Op_i \,,
\end{equation}
where $x \ge x_i$, and $\Op_i$ is a local polynomial of $\phi$ and its
derivatives.\footnote{$\Op_i$ may also contain a product of mass
  parameters such as $m^2$ of the $\phi^4$ theory.}  $\Op_\Lambda$ is
completely determined by $d_i (0)$ for those $i$ satisfying $x_i = x$.

Let $c_i (0)\,(i=1,\cdots,N)$ be the generic parameters of the Wilson
action, and $d_i (0)\,(i=1,\cdots,K)$ be those of the composite operator
$\Op_\Lambda$, so that the theory has altogether $N+K$ parameters that
we can adjust.  We call the theory symmetric if we can adjust these
parameters so that
\begin{equation}
\Sigma_\Lambda = 0\,.\label{WT-WTid}
\end{equation}
Note that $\Sigma_\Lambda$ itself is a composite operator that
satisfies the linear ERG differential equation.  So, if we show
\begin{equation}
\lim_{\Lambda \to \infty} \Sigma_\Lambda = 0\,,
\end{equation}
then it vanishes for any finite $\Lambda$.

Let us go over perturbative construction.  We expand the parameters in
the number of loops:
\begin{equation}
c_i (0) = \sum_{l=0}^\infty c_i^{(l)}\,,\quad
d_i (0) = \sum_{l=0}^\infty d_i^{(l)}\,.
\end{equation}
Let us also expand $\SIL$ and $\Sigma_\Lambda$ in loops:
\begin{equation}
\SIL = \sum_{l=0}^\infty \SIL^{(l)}\,,\quad
\Sigma_\Lambda = \sum_{l=0}^\infty \Sigma_\Lambda^{(l)}\,.
\end{equation}
Suppose $\Sigma_\Lambda = 0$ is made to vanish up to $l$-loop
\begin{equation}
\Sigma_\Lambda^{(0)} = \cdots = \Sigma_\Lambda^{(l)} = 0
\end{equation}
by adjusting $c_i (0), d_i (0)$ also up to $l$-loop.  Under this
assumption, we consider if we can make $\Sigma_\Lambda^{(l+1)}$ vanish
by choosing $c_i^{(l+1)}, d_i^{(l+1)}$ appropriately.

Due to the induction hypothesis, $\Sigma_\Lambda^{(l+1)}$ satisfies
the ERG differential equation
\begin{equation}
- \Lambda \frac{\partial}{\partial \Lambda} \Sigma_\Lambda^{(l+1)}
= \int_p \frac{\Delta (p/\Lambda)}{p^2 + m^2} \frac{\delta
  \SIL^{(0)}}{\delta \phi (-p)} \frac{\delta \Sigma_\Lambda^{(l+1)}
  }{\delta \phi (p)} 
\,.
\end{equation}
Since for fixed momentum $p$
\begin{equation}
\Delta (p/\Lambda) \asym 0\,,
\end{equation}
we find that the right-hand side vanishes and that
$\Sigma_\Lambda^{(l+1)}$ is independent of $\Lambda$ for large
$\Lambda$:
\begin{equation}
\Sigma_\Lambda^{(l+1)} \asym \sum_i s^{(l+1)}_i \,\Op_i\,,\label{WT-si}
\end{equation}
where $s_i^{(l+1)}$ is independent of $\Lambda$, and $\Op_i$ is a local
polynomial of $\phi$ and its derivatives.\footnote{$\Op_i$ may contain
  a product of mass parameters such as $m^2$ of the $\phi^4$ theory
 which has scale dimension $2$.}
The scale dimension of $\Op_i$ must be the same as that of
$\Sigma_\Lambda$.

On the other hand, from the definition (\ref{WT-Sigma}), we obtain
\begin{eqnarray}
\Sigma_\Lambda^{(l+1)} &=& \int_p \K{p} \left( \Op_\Lambda^{(l+1)}
    \frac{\delta \SIL^{(0)}}{\delta \phi (p)} + \Op_\Lambda^{(0)}
    \frac{\delta \SIL^{(l+1)}}{\delta \phi (p)} \right)\nn\\
&& + \int_p \K{p} \left( \sum_{k=1}^l \Op_\Lambda^{(l+1-k)}
\frac{\delta \SIL^{(k)}}{\delta \phi (p)} + \frac{\delta
  \Op_\Lambda^{(l)}}{\delta \phi (p)} \right)\,.\label{WT-Sigmal+1}
\end{eqnarray}
Only the first integral depends on the $(l+1)$-loop parameters
$c_i^{(l+1)}, d_i^{(l+1)}$.  The dependence is linear.
Asymptotically, neither $\SIL^{(0)}$ nor $\Op_\Lambda^{(0)}$ has
$\Lambda$ dependence.  Hence, the part of $\Sigma_\Lambda^{(l+1)}$
that depends on $c_i^{(l+1)}, d_i^{(l+1)}$ are independent of
$\Lambda$.  For large $\Lambda$, the $\Lambda$ dependence of the first
integral cancels that of the second integral, and we find that 
$s_i$ of (\ref{WT-si}) is written as the sum
\begin{equation}
s^{(l+1)}_i = t^{(l+1)}_i + u^{(l+1)}_i\,.
\end{equation}
Here, $t_i$ is linear in $c^{(l+1)}$'s and $d^{(l+1)}$'s with
$l$-independent constant coefficients:
\begin{equation}
t_i^{(l+1)} =  \sum_{j=1}^N A_{ij}\, c_j^{(l+1)} + \sum_{j=1}^K B_{ij}\,
d_j^{(l+1)}\,.
\end{equation}
$u_i^{(l+1)}$ denotes the contribution from the second integral of
(\ref{WT-Sigmal+1}), and is determined only by the lower loop
coefficients.

Whether or not we can choose $c^{(l+1)}$'s and $d^{(l+1)}$'s so that
\begin{equation}
s_i^{(l+1)} = 0\label{WT-zerosi}
\end{equation}
must be examined case by case, and cannot be answered in general.  In \S
\ref{WTexamples}, we discuss four concrete examples.  In the cases of
QED, the Wess-Zumino model, and the two dimensional O(N) non-linear
sigma model, we can prove the possibility of realizing
(\ref{WT-zerosi}).  In the case of YM theories we can also satisfy
(\ref{WT-zerosi}), but the possibility cannot be proven within the
present formalism; we must resort to the antifield formalism of \S
\ref{AF}.  In general, if a symmetry present at the tree level cannot be
extended to arbitrary loop levels, we call the symmetry
\textbf{anomalous}.  The anomaly is expressed by the non-vanishing
$\Sigma_\Lambda$.  We discuss how the axial and chiral anomalies arise
in the ERG formalism in \S \ref{anomaly}.

\subsection{WT identity for the bare action}

In the previous subsection we have discussed the Wilson action in the
continuum limit, and shown how to satisfy the Ward-Takahashi identity by
adjusting renormalized parameters.  The symmetry is checked with the
Wilson action, but not with the correlation functions, which correspond
to the Wilson action in the limit of $\Lambda \to 0$.  The advantage of
our method is twofold:
\begin{enumerate}
\item No problem with massless theories --- Low momentum modes are
      crucially important for the manifestation of the underlying
      symmetry of the theory.  But they are totally irrelevant to the
      question of the existence of the underlying symmetry.  By keeping
      the cutoff $\Lambda$ non-vanishing, we stay away safely from the
      low momentum modes.  Massless theories cause no problem, since we
      never look at the correlation functions.
\item Only a few parameters to deal with --- We examine the vanishing of
      the composite operator $\Sigma_\Lambda$ for large $\Lambda$.
      The existence of symmetry is a matter of adjusting a few
      parameters, if it is possible.
\end{enumerate}

In the following we ask if it is possible to realize symmetry using the
bare action $S_B$ with a large cutoff $\Lambda_0$.  Under the symmetry
transformation
\begin{equation}
\delta \phi (p) = \Kz{p} \Op_B [\phi] (p)
\end{equation}
the bare action changes by
\begin{equation}
\Sigma_B \equiv \int_p \Kz{p} \left( \frac{\delta S_B}{\delta \phi
      (p)} \Op_B [\phi] (p) + \frac{\delta}{\delta \phi (p)} \Op_B
    [\phi] (p) \right)\,.
\end{equation}
If the symmetry exists, the above must vanish.  Since we are only
interested in the continuum limit, we only demand $\Sigma_B$ to vanish
in the continuum limit $\Lambda_0 \to \infty$.  But what does it mean
that $\Sigma_B$ vanishes in the limit?

Unless it vanishes identically, it is very difficult to see if
$\Sigma_B$ vanishes or not.  To see if it vanishes, we must lower the
cutoff from $\Lambda_0$ to a finite $\Lambda$, and take the limit
$\Lambda_0 \to \infty$ while fixing $\Lambda$.  In other words, we must
look at $\Sigma_\Lambda$, defined by
\begin{equation}
\Sigma_\Lambda \exp \left[ S_{I,\Lambda} [\phi] \right]
= \int [d\phi'] \Sigma_B [\phi+\phi'] \exp \left[ - \frac{1}{2} \int_p
    \frac{p^2 + m^2}{K_0-K} \phi' \phi' + S_{I,B} [\phi+\phi'] \right]\,,
\end{equation}
and check if we can make it vanish in the limit $\Lambda_0 \to \infty$.

In realizing symmetry in the continuum limit, it is difficult to use the
bare action $S_B$ and bare WT operator $\Sigma_B$.  To appreciate this
difficulty, let us consider a bare operator
\begin{equation}
\Op_B \equiv \frac{m^2}{\Lambda_0^2}  \frac{\phi^4}{4!}
\end{equation}
in the $\phi^4$ theory in $D=4$.  Naively, this operator appears to
vanish in the limit $\Lambda_0 \to \infty$.  But that is false.
Lowering the cutoff from $\Lambda_0$ to $\Lambda$, we obtain the
following 1-loop correction to $\Op_\Lambda$:
\begin{equation}
\Op_\Lambda^{(1)} = m^2 c (\ln \Lambda_0/\mu) \frac{\phi^2}{2} +
\cdots\,,
\end{equation}
where
\begin{equation}
c (\ln \Lambda_0/\mu) = \frac{1}{\Lambda_0^2} \int_p
\frac{K_0-K}{p^2 + m^2} \simeq \int_p \frac{K (p)}{p^2} = \mathrm{const}\,.
\end{equation}
This does not vanish even in the limit $\Lambda_0 \to \infty$.
Hence, the small coefficient proportional to $\frac{1}{\Lambda_0^2}$ in
$\Op_B$ is misleading. 

Thus, in general it is difficult to judge the existence of symmetry from
the bare action and its transformation property, unless the bare action
is exactly invariant, i.e., the symmetry is manifest.  We must lower the
cutoff from $\Lambda_0$ to a finite fixed value $\Lambda$.  We have
already pointed out the potential problem with choosing $\Lambda = 0$.
Thus, we advocate the examination of symmetry using a Wilson action at
non-vanishing $\Lambda$.

\newpage

\section{Examples\label{WTexamples}}

In this section we give a few examples to apply the formalism introduced
in the previous section and realize symmetry as the WT identity
\bref{WT-vanishingSigma}.

\subsection{QED\label{WTexamples-QED}}

The WT identities of QED have been discussed using the ERG formalism in
various references such as \citen{Bonini:1993kt, Pernici:1997ie,
Sonoda:2007dj} to name only a few.  Our presentation here follows closely
that of \citen{Sonoda:2007dj}.\footnote{For scalar QED, the ERG
formalism was first applied in \citen{Reuter:1994sg}.}

The classical action of QED is given by
\begin{eqnarray}
S_{cl} &\equiv& \int d^4 x\, \left[ - \frac{1}{2} \partial_\mu
    A_\nu \partial_\mu A_\nu + \frac{1}{2} (1 - 1/\xi)
    \left(\partial_\mu A_\mu\right)^2 - \partial_\mu
    \bar{c} \partial_\mu c \right.\nn\\
&&\qquad \left.- \bar{\psi} \left( \frac{1}{i}
        \fmslash{\partial} + i m \right) \psi + e \bar{\psi}
    \fmslash{A} \psi \right]\,,\label{WTexamples-QED-Scl}
\end{eqnarray}
where $\xi$ is a gauge fixing parameter, and $c,\bar{c}$ are the free
Faddeev-Popov ghost, antighost fields.  The action is invariant under
the classical BRST transformation:
\begin{equation}
\lb\begin{array}{c@{~=~}l}
\de A_\mu & \ep \frac{1}{i} \partial_\mu c\,,\\
\de c & 0\,,\\
\de \bar{c} & \ep \frac{1}{\xi} \frac{1}{i} \partial_\mu A_\mu\,,\\
\de \psi & e \ep c \psi\,,\\
\de \bar{\psi} & - e \ep c \bar{\psi}\,,
\end{array}\right. \label{WTexamples-QED-classicalBRST}
\end{equation}
where $\ep$ is an arbitrary anticommuting constant.  In the following
we quantize this system using the ERG formalism.

The Wilson action of QED is given as the sum of free and interacting
parts:
\begin{equation}
S_\Lambda = S_{F,\Lambda} + S_{I,\Lambda}\,,
\end{equation}
where the free action is given by
\begin{eqnarray}
S_{F,\Lambda} &=& - \frac{1}{2} \int_k \frac{1}{\K{k}} \left( k^2
    \delta_{\mu\nu} - 
    \left(1 - 1/\xi\right) k_\mu k_\nu \right) A_\mu (-k)
A_\nu (k) \nn\\
&&\quad - \int_k \frac{k^2}{\K{k}} \bar{c} (-k) c(k) 
 - \int_p \frac{1}{\K{p}} \bar{\psi} (-p) \left( \fmslash{p} +
    i m \right) \psi (p)\,.\label{WTexamples-QED-SFL}
\end{eqnarray}
Here, we have chosen the same cutoff function for all the fields; it
is for the sake of simplicity, and by no means necessary.  The cutoff
propagators are given by
\begin{equation}
\lb\begin{array}{c@{~=~}l}
\vev{A_\mu (k) A_\nu (-k)}_{S_{F,\Lambda}} & \frac{\K{k}}{k^2} \left(
    \delta_{\mu\nu} - (1-\xi) \frac{k_\mu k_\nu}{k^2}\right)\,,\\
\vev{c (k) \bar{c} (-k)}_{S_{F,\Lambda}} & \frac{\K{k}}{k^2} \,,\\
\vev{\psi (p) \bar{\psi} (-p)}_{S_{F,\Lambda}} &
\frac{\K{p}}{\fmslash{p} + i m}\,.
\end{array}\right.
\end{equation}

The ghost and antighost fields are free, and the interaction action
depends only on $A_\mu, \psi, \bar{\psi}$.  The Polchinski equation is
given by
\begin{eqnarray}
&&- \Lambda \frac{\partial}{\partial \Lambda} S_{I,\Lambda} \nn\\
&& = \int_k \frac{\Delta (k/\Lambda)}{k^2} \left(
    \delta_{\mu\nu} - (1 - \xi) \frac{k_\mu k_\nu}{k^2} \right) 
\cdot \frac{1}{2} \lb \frac{\delta S_{I,\Lambda}}{\delta A_\mu
  (-k)} \frac{\delta S_{I,\Lambda}}{\delta A_\nu (k)}
+  \frac{\delta^2 S_{I,\Lambda}}{\delta A_\mu
  (-k)\delta A_\nu (k)} \rb\nn\\
&&\quad - \int_p \frac{\Delta (p/\Lambda)}{p^2 + m^2} \Tr (\fmslash{p} - i
m) \left[ \Ld{\bar{\psi} (-p)} S_{I,\Lambda} \cdot S_{I,\Lambda}
    \Rd{\psi (p)} + \Ld{\bar{\psi} (-p)} S_{I,\Lambda} \Rd{\psi (p)}
\right]\,,
\end{eqnarray}
where the minus sign before the second integral is due to the Fermi
statistics.  The asymptotic behavior of $S_{I,\Lambda}$ is given in the
following form:
\begin{eqnarray}
S_{I,\Lambda} &\asym& \int d^4 x \,\left[ \left(\Lambda^2 a_2 (\ln
        \Lambda/\mu) + m^2 b_2 (\ln \Lambda/\mu) \right) \frac{1}{2}
    A_\mu^2\right.\nn\\
&&\quad + c_2 (\ln \Lambda/\mu) \frac{1}{2} \partial_\mu
A_\nu \partial_\mu A_\nu + d_2 (\ln \Lambda/\mu)
\frac{1}{2} \left(\partial_\mu A_\mu\right)^2\nn\\
&&\quad + a_4 (\ln \Lambda/\mu) \frac{1}{4!} \left( A_\mu^2 \right)^2\nn\\
&&\quad + a_f (\ln \Lambda/\mu) \bar{\psi} \frac{1}{i}
\fmslash{\partial} \psi + b_f (\ln \Lambda/\mu) i m \bar{\psi}
\psi\nn\\
&&\quad + a_3 (\ln \Lambda/\mu) \bar{\psi} \fmslash{A} \psi
\Big]\,.\label{WTexamples-QED-SIasymp}
\end{eqnarray}
Hence, the theory has 7 parameters:
\begin{equation}
b_2 (0), c_2 (0), d_2 (0), a_f (0), b_f (0), a_3 (0), a_4 (0)\,.
\end{equation}
Out of these, three are unphysical, since they only change
normalization of the fields and the mass parameter $m$.  So, we can
arbitrarily impose the normalization condition
\begin{equation}
c_2 (0) = a_f (0) = b_f (0) = 0\,.
\end{equation}
This leaves us four arbitrary parameters:
\begin{equation}
b_2 (0), d_2 (0), a_3 (0), a_4 (0)\,.\label{WTexamples-QED-parameters}
\end{equation}
We will show how to fix these for the theory to become QED.

The quantum BRST transformation is defined by
\begin{equation}
\lb\begin{array}{c@{~=~}l}
\delta A_\mu (k) & k_\mu \ep c (k)\,,\\
\delta c (k) & 0\,,\\
\delta \bar{c} (-k) & \ep \frac{1}{\xi} k_\mu [A_\mu]_\Lambda (-k)\,,\\
\delta \psi (p) & e \int_k \ep c(k) [\psi]_\Lambda (p-k)\,,\\
\delta \bar{\psi} (-p) & - e \int_k \ep c (k) [\bar{\psi}]_\Lambda
(-p-k)\,,
\end{array}\right.\label{WTexamples-QED-BRST}
\end{equation}
where $e$ is the gauge coupling.  We have introduced an anticommuting
constant $\ep$ so that the transformation preserves the statistics of
the fields.  The composite operators $[A_\mu]_\Lambda$,
$[\psi]_\Lambda$, $[\bar{\psi}]_\Lambda$ are defined by
\begin{equation}
\lb\begin{array}{c@{~\equiv~}l}
\left[A_\mu\right]_\Lambda (-k) & \frac{1}{\K{k}} A_\mu (-k) + \frac{1
  - \K{k}}{k^2} \left( \delta_{\mu\nu} - (1-\xi) \frac{k_\mu
      k_\nu}{k^2} \right) \frac{\delta \SL}{\delta A_\nu (k)}\,,\\
\left[\psi\right]_\Lambda (p) & \frac{1}{\K{p}} \psi (p) + \frac{1 -
  \K{p}}{\fmslash{p} + 
i m} \Ld{\bar{\psi} (-p)} S_\Lambda\,,\\
\left[\bar{\psi}\right]_\Lambda (-p) & \frac{1}{\K{p}} \bar{\psi} (-p) +
S_\Lambda \Rd{\psi (p)} \frac{1 - \K{p}}{\fmslash{p} + i m}\,.
\end{array}\right.
\end{equation}
The corresponding WT composite operator is
\begin{eqnarray}
\Sigma_\Lambda &\equiv& \int_k \K{k} \left[ \frac{\delta
      S_\Lambda}{\delta A_\mu (k)} \delta A_\mu (k) + \delta \bar{c}
    (-k) \Ld{\bar{c} (-k)} S_\Lambda \right]\nn\\ 
&& + \int_p \K{p} \left[ S_\Lambda \Rd{\psi (p)} \delta \psi (p) - \Tr
    \delta \psi (p) \Rd{\psi (p)} \right.\nn\\
&&\left.\quad + \delta \bar{\psi} (-p) \Ld{\bar{\psi} (-p)} S_\Lambda - \Tr 
\Ld{\bar{\psi} (-p)} \delta \bar{\psi} (-p) \right]\,.
\end{eqnarray}
For the Wilson action to be invariant under the BRST transformation
(\ref{WTexamples-QED-BRST}), this must vanish.

Substituting the transformation (\ref{WTexamples-QED-BRST}) into the
above, we obtain\footnote{It is straightforward and not tedious.}
\begin{equation}
\Sigma_\Lambda = \ep \int_k c(k) \lb k_\mu \frac{\delta
  S_{I,\Lambda}}{\delta A_\mu (k)} - \Phi (-k) \rb\,,
\end{equation}
where
\begin{equation}
    J_\mu (-k) \equiv \frac{\delta
      S_{I,\Lambda}}{\delta A_\mu (k)}
\end{equation}
is a composite operator of type (\ref{comp-dSIdphi}), and denotes an
electric current.  $\Phi$ is a composite operator of type
(\ref{comp-dOdphi}) defined by
\begin{eqnarray}
    &&\Phi (-k) \equiv e \int_p \K{p} \Tr \left[ [\psi]_\Lambda (p-k) \cdot
        S_\Lambda \Rd{\psi (p)} + [\psi]_\Lambda (p-k) \Rd{\psi (p)}
    \right]\nn\\
    && \quad - e \int_p \K{p} \Tr \left[ \Ld{\bar{\psi} (-p)} S_\Lambda
        \cdot [\bar{\psi}]_\Lambda (-p-k) + \Ld{\bar{\psi} (-p)}
        [\bar{\psi}]_\Lambda (-p-k) \right] .
\end{eqnarray}
Thus, the WT identity is equivalent to the operator identity:
\begin{equation}
k_\mu J_\mu (-k) = \Phi (-k)\,
\end{equation}
for arbitrary $k$.  This equation is valid if and only if the
asymptotic behaviors of the both hand sides match.  

The asymptotic behavior of the left-hand side is obtained from
(\ref{WTexamples-QED-SIasymp}) as
\begin{eqnarray}
k_\mu J_\mu (-k)
&\asym& \left( \Lambda^2 a_2 (\ln \Lambda/\mu) + m^2 b_2 (\ln
    \Lambda/\mu) + k^2 (c_2+d_2) (\ln \Lambda/\mu) \right) k_\mu A_\mu
(-k)\nn\\
&& + a_3 (\ln \Lambda/\mu) \int_p \bar{\psi} (-p-k) \fmslash{k} \psi
(p)\nn\\
&& + a_4 (\ln \Lambda/\mu) \frac{1}{2} \int_{k_1,k_2} A_\nu (k_1)
A_\nu (k_2) k_\mu A_\mu (-k-k_1-k_2)\,.
\end{eqnarray}

To obtain the asymptotic behavior of $\Phi (-k)$, we first note that
$\Phi (-k)$ vanishes at $k=0$ due to the fermion number conservation.
For $k=0$, substituting the definition of the composite operators, we
obtain
\begin{equation}
\Phi (0) = e \int_p \left[ - \SL \Rd{\psi (p)} \psi (p) + \bar{\psi}
    (-p) \Ld{\bar{\psi} (-p)} \SL \right]\,.
\end{equation}
Since $\SL$ contains the same number of $\psi$'s and $\bar{\psi}$'s,
this vanishes:
\begin{equation}
\Phi (0) = 0\,.
\end{equation}

Therefore, $\Phi (-k)$ is a total derivative.  Hence, its asymptotic
behavior must be proportional to $k$.  Since $\Phi$ has dimension $4$,
the most general form is given by
\begin{eqnarray}
\Phi (-k) &\asym& \left( \Lambda^2 \bar{a}_2 (\ln \Lambda/\mu) + m^2
    \bar{b}_2 (\ln 
    \Lambda/\mu) + k^2 \bar{d}_2 (\ln \Lambda/\mu) \right) k_\mu A_\mu
(-k)\nn\\
&& + \bar{a}_3 (\ln \Lambda/\mu) \int_p \bar{\psi} (-p-k) \fmslash{k} \psi
(p)\nn\\
&& + \bar{a}_4 (\ln \Lambda/\mu) \frac{1}{2} \int_{k_1,k_2} A_\nu (k_1)
A_\nu (k_2) k_\mu A_\mu (-k-k_1-k_2)\,.
\end{eqnarray}
It is crucial that this has the same form as the asymptotic behavior
of $k_\mu J_\mu (-k)$.

Thus, the WT identity is equivalent to the following four equations:
\begin{equation}
\lb
\begin{array}{r@{~=~}l}
b_2 (0) & \bar{b}_2 (0)\,,\\
d_2 (0) & \bar{d}_2 (0)\,,\\
a_3 (0) & \bar{a}_3 (0)\,,\\
a_4 (0) & \bar{a}_4 (0)\,,
\end{array}\right.\label{WTexamples-QED-matching}
\end{equation}
where we have used the normalization condition $c_2 (0) = 0$. If the
right-hand sides do not depend on $b_2 (0), d_2 (0), a_3 (0), a_4
(0)$, we can use the above equations to determine these parameters.

To show that the above equations can be solved uniquely, we must
examine the asymptotic behavior of $\Phi (-k)$ further.  Splitting the
action into the free and interaction parts, we can rewrite
\begin{eqnarray}
\Phi (-k) &=& e \int_p \K{p} \left[ - \SL \Rd{\psi (p)} [\psi]_\Lambda
    (p-k) + [\bar{\psi}]_\Lambda (-p-k) \Ld{\bar{\psi} (-p)} \SL
\right]\nn\\
&& \quad + e \int_p \Tr U(-p-k,p) \Ld{\bar{\psi} (-p)} \SIL \Rd{\psi
  (p+k)}\,,\label{WTexamples-QED-PhiU}
\end{eqnarray}
where
\begin{equation}
U (-p-k,p) \equiv \K{p+k} \frac{1 - \K{p}}{\fmslash{p} + i m} -
\frac{1 - \K{(p+k)}}{\fmslash{p} + \fmslash{k} + i m} \K{p}\,.
\label{WTexamples-QED-U}
\end{equation}
It is straightforward to obtain the asymptotic behavior of the first
integral on the right-hand side:
\begin{eqnarray}
&&e \int_p \K{p} \left[ - \SL \Rd{\psi (p)} [\psi]_\Lambda
    (p-k) + [\bar{\psi}]_\Lambda (-p-k) \Ld{\bar{\psi} (-p)} \SL
\right]\nn\\
&& \asym e \left(1 - a_f (\ln \Lambda/\mu) \right) \int_p
\bar{\psi} (-p-k) \fmslash{k} \psi (p)\,.
\end{eqnarray}
Since $a_f (0) = 0$ by the normalization condition, we find that the
above contributes $e$ to $\bar{a}_3 (0)$, and nothing at all to
$\bar{b}_2 (0), \bar{d}_2 (0), \bar{a}_4 (0)$.

We now introduce the loop expansions of the coefficients:
\begin{equation}
\lb\begin{array}{c@{~=~}l}
b_2 (0) & \sum_{l=1}^\infty b_2^{(l)}\,,\\
d_2 (0) & \sum_{l=1}^\infty d_2^{(l)}\,,\\
a_3 (0) & e + \sum_{l=1}^\infty a_3^{(l)}\,,\\
a_4 (0) & \sum_{l=1}^\infty a_4^{(l)}\,,
\end{array}\right.\quad
\lb\begin{array}{c@{~=~}l}
\bar{b}_2 (0) & \sum_{l=1}^\infty \bar{b}_2^{(l)}\,,\\
\bar{d}_2 (0) & \sum_{l=1}^\infty \bar{d}_2^{(l)}\,,\\
\bar{a}_3 (0) & e + \sum_{l=1}^\infty \bar{a}_3^{(l)}\,,\\
\bar{a}_4 (0) & \sum_{l=1}^\infty \bar{a}_4^{(l)}\,,
\end{array}\right.
\end{equation}
where each term of the series is a function of $e$ and $\xi$.  The sum
is over 1- and higher-loop contributions; we have fixed the tree level
of $a_3 (0)$ to satisfy the WT identity
(\ref{WTexamples-QED-matching}) at tree level.

Now, let us suppose we have fixed $b_2 (0), d_2 (0), a_3 (0), a_4 (0)$
up to $l$-loop level by the WT identity
(\ref{WTexamples-QED-matching}).  This fixes $\SIL$ up to l-loop
level.  Using this, we can compute (\ref{WTexamples-QED-PhiU}) up to
(l+1)-loop level, where the extra loop is provided by the integral
over $p$ for the second integral.  Then, using
(\ref{WTexamples-QED-matching}), we can determine $b_2^{(l+1)},
d_2^{(l+1)}, a_3^{(l+1)}, a_4^{(l+1)}$ unambiguously.  We can repeat
this procedure to all orders.  This is how we can construct QED
perturbatively.

We end this subsection by giving concrete 1-loop calculations.  We
need to evaluate the second integral on the right-hand side of
(\ref{WTexamples-QED-PhiU}). We only need the tree level vertex $e
\gamma_\mu$ and the definition (\ref{WTexamples-QED-U}) of $U$ for the
calculations.
\begin{itemize}
\item[(i)] $b_2^{(1)},d_2^{(1)}$ --- The corresponding Feynman graph
    is given by Fig.~\ref{WTexamples-QED-b2d2}.
\begin{figure}[t]
\begin{center}
\epsfig{file=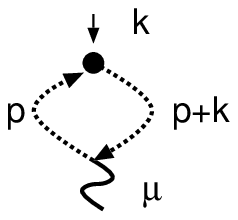, width=3cm}
\caption{The broken curve gives $e U$ (\ref{WTexamples-QED-U}).}
\label{WTexamples-QED-b2d2}
\end{center}
\end{figure}
\begin{eqnarray}
&&\int_p \Tr e \gamma_\mu e U(-p-k,p) \nn\\
&\asym& e^2 k_\mu \left[ - 2 \Lambda^2 \int_p \frac{1}{q^2} \Delta (q)
    \left( 1 - K(q)\right) + \left( m^2 + \frac{k^2}{3} \right) \int_p
    \frac{\Delta (p)}{p^4} \right]\nn\\
&=& \left( \Lambda^2 a_2^{(1)}+ m^2 b^{(1)}_2  + k^2
    d_2^{(1)}  \right) k_\nu\,.
\end{eqnarray}
Using the integral formula (\ref{cl-integral}), we obtain
\begin{equation}
b^{(1)}_2 = \frac{2 e^2}{(4 \pi)^2},\quad
d_2^{(1)} = \frac{2 e^2}{3 (4 \pi)^2}\,.
\end{equation}
We also obtain
\begin{equation}
a_2^{(1)} = - 2 e^2 \int_p \frac{\Delta (p) \left( 1 -
      K(p)\right)}{p^2}\,,
\end{equation}
which is a constant dependent on the choice of $K$.
\item[(ii)] $a_3^{(1)}$ --- The corresponding Feynman graph is given
    by Fig.~\ref{WTexamples-QED-a3}.
\begin{figure}[t]
\begin{center}
\epsfig{file=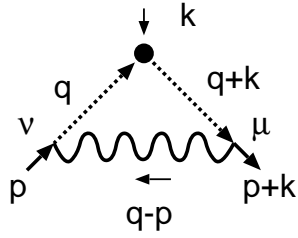, width=4cm}
\caption{The solid curve gives a high-momentum photon propagator.}
\label{WTexamples-QED-a3}
\end{center}
\end{figure}
\begin{eqnarray}
&&(-) \int_q e \gamma_\mu e U (-q-k,q) e \gamma_\nu\nn\\
&&\qquad \times \frac{1 - \K{(q-p)}}{(q-p)^2} \left( \delta_{\mu\nu} -
    (1-\xi) \frac{(q-p)_\mu (q-p)_\nu}{(q-p)^2} \right)\\
&\asym& - e^3 \fmslash{k} \int_q \frac{1}{q^4} \lb
\xi K(q) (1-K(q))^2 + \frac{3 - \xi}{4} (1-K(q)) \Delta (q) \rb
= a_3^{(1)} \fmslash{k}\,,\nn
\end{eqnarray}
where the first minus sign is due to the interchange of $\psi (p)$ and
$\bar{\psi} (-p-k)$.  Hence, we obtain
\begin{equation}
a_3^{(1)} = - e^3 \left( \xi \int_q \frac{K(1-K)^2}{q^4} +
    \frac{3 - \xi}{4 (4\pi)^2} \right)\,.
\end{equation}
\item[(iii)] $a_4^{(1)}$ --- The corresponding Feynman graph is given
    by Fig.~\ref{WTexamples-QED-a4}.
\begin{figure}[t]
\begin{center}
\epsfig{file=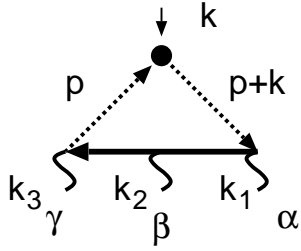, width=4cm}
\caption{The solid line gives a high-momentum electron propagator.}
\label{WTexamples-QED-a4}
\end{center}
\end{figure}
Writing $k_3 = - k - k_1 - k_2$, we obtain
\begin{eqnarray}
&&\int_p \Tr e \gamma_\alpha e U (-p-k,p) e \gamma_\gamma 
 \frac{1 -
  \K{(q-k_3)}}{\fmslash{q} - \fmslash{k}_3 + i m}
e \gamma_\beta \frac{1 - \K{(q - k_2 - k_3)}}{\fmslash{q} -
  \fmslash{k}_2 - \fmslash{k}_3}\nn\\
&& \qquad + \,(\textrm{5 permutations})\nn\\
&&\asym \left( k_\alpha \delta_{\beta\gamma} + k_\beta
    \delta_{\gamma\alpha} + k_\gamma \delta_{\alpha\beta} \right) 2
e^4 \int_p \frac{\Delta (q)}{q^4} \left( 1 - K(q)\right)^2\nn\\
&&\quad = \left( k_\alpha \delta_{\beta\gamma} + k_\beta
    \delta_{\gamma\alpha} + k_\gamma \delta_{\alpha\beta} \right)
a_4^{(1)}\,. 
\end{eqnarray}
Hence, we obtain
\begin{equation}
a_4^{(1)} = \frac{4}{3} \frac{e^4}{(4 \pi)^2}\,.
\end{equation}
\end{itemize}

It is interesting to note the presence of $a_2, b_2, a_4$.  The first
and second are photon mass terms, and the third is a four-photon
coupling.  These break the invariance of the action under the classical
BRST transformation \bref{WTexamples-QED-classicalBRST}.  As has been
emphasized in \S \ref{WT}, the theory has full gauge (or BRST)
invariance, but it requires order-by-order fine tuning, and is not
realized manifestly.

\subsection{Yang-Mills theories\label{WTexamples-YM}}

In the ERG framework YM theories have been discussed usually in the
antifield formalism explained in \S \ref{AF}.  We give relevant
references at the beginning of \S \ref{AFexamples-YM}.

The classical action of a generic YM theory is given by
\begin{equation}
S_{cl} = \int d^4 x\, \left[ - \frac{1}{4} \left(F_{\mu\nu}^a
    \right)^2 - \frac{1}{2 \xi} \left( \partial_\mu A_\mu^a \right)^2
    - \partial_\mu \bar{c}^a \left(D_\mu c\right)^a \right]\,,
\end{equation}
where the field strength is defined by
\begin{equation}
F_{\mu\nu}^a \equiv \partial_\mu A_\nu^a - \partial_\nu A_\mu^a - g
f^{abc} A_\mu^b A_\nu^c\,,
\end{equation}
and the covariant derivative is defined by
\begin{equation}
\left( D_\mu c \right)^a \equiv \partial_\mu c^a - g f^{abc} A_\mu^b
c^c\,.
\end{equation}
$f^{abc}$ is a totally antisymmetric structure constant of whatever
gauge group under consideration, satisfying the Jacobi identity
\begin{equation}
f^{abd} f^{dce} + f^{bcd} f^{dae} + f^{cad} f^{dbe} = 0\,.
\end{equation}
The classical action is invariant
under the following classical BRST transformation:
\begin{equation}
\lb\begin{array}{c@{~=~}l}
\de A_\mu^a & \ep \frac{1}{i} \left( D_\mu c \right)^a\,,\\
\de c^a & \ep \frac{g}{2 i} f^{abc} c^b c^c\,,\\
\de \bar{c}^a & \ep \frac{1}{\xi} \frac{1}{i} \partial_\mu
A_\mu^a\,,
\end{array}\right.
\end{equation}
where $\ep$ is an arbitrary anticommuting constant.  Transforming
twice, we obtain
\begin{equation}
\lb\begin{array}{c@{~=~}l}
\delta_\eta \de A_\mu^a & 0\,,\\
\delta_\eta \de c^a & 0\,,\\
\delta_\eta \de \bar{c}^a & - \ep \eta \frac{1}{\xi} \partial_\mu
\left( D_\mu c\right)^a\,.
\end{array}\right.
\end{equation}
The last one vanishes by the ghost equation of motion
\begin{equation}
\partial_\mu \left( D_\mu c\right)^a = 0\,.
\end{equation}
In the following we quantize the system using the ERG
formalism.\footnote{We can make the classical BRST transformation
nilpotent by introducing an auxiliary field $B^a$, equal to
$\frac{1}{\xi} \frac{1}{i} \partial_\mu A_\mu^a$ by the equation of
motion.  We do not do it here because $B^a$ does not help the
perturbative construction described here and in \S\ref{AFexamples-YM}.
In \S\ref{BVexamples}, however, we introduce $B^a$ in order to keep the
bare action $\bar{S}_B$ at most linear in the antifields.}

The Wilson action is given as the sum of free and interacting parts:
\begin{equation}
\SL = \SFL + \SIL\,,
\end{equation}
where the free action is given by
\begin{eqnarray}
\SFL &=& - \frac{1}{2} \int_k \frac{1}{\K{k}} \left( k^2
    \delta_{\mu\nu} - \left( 1 - 1/\xi \right) k_\mu k_\nu \right)
A_\mu^a (-k) A_\nu^a (k)\nn\\
&&\quad - \int_k \frac{k^2}{\K{k}} \bar{c}^a (-k) c^a (k)\,.
\label{WTexamples-YM-SFL}
\end{eqnarray}
Hence, the cutoff propagators are given by
\begin{equation}
\lb\begin{array}{c@{~=~}l}
\vev{A_\mu^a (k) A_\nu^b (-k)}_{\SFL} & \delta^{ab} \frac{\K{k}}{k^2} \left(
    \delta_{\mu\nu} - (1-\xi) \frac{k_\mu k_\nu}{k^2} \right)\,,\\
\vev{c^a (k) \bar{c}^b (-k)}_{\SFL} & \delta^{ab} \frac{\K{k}}{k^2}\,.
\end{array}\right.
\end{equation}
The Polchinski equation for the interaction action is given by
\begin{eqnarray}
&&- \Lambda \frac{\partial}{\partial \Lambda} S_{I,\Lambda} \nn\\
&& = \int_k \frac{\Delta (k/\Lambda)}{k^2} \left(
    \delta_{\mu\nu} - (1 - \xi) \frac{k_\mu k_\nu}{k^2} \right) 
\cdot \frac{1}{2} \lb \frac{\delta S_{I,\Lambda}}{\delta A_\mu^a
  (-k)} \frac{\delta S_{I,\Lambda}}{\delta A_\nu^a (k)}
+  \frac{\delta^2 S_{I,\Lambda}}{\delta A_\mu^a
  (-k)\delta A_\nu^a (k)} \rb\nn\\
&&\quad + \int_k \frac{\Delta (k/\Lambda)}{k^2} \left(\SIL \Rd{c^a
      (k)} \cdot \Ld{\bar{c}^a (-k)} \SIL -  \Ld{\bar{c}^a (-k)} \SIL
\Rd{c^a (k)} \right)\,.
\end{eqnarray}
The possible asymptotic behavior depends on the gauge group.  For
simplicity we consider only the group SU(2).  For SU(2), the structure
constant $f^{abc}$ becomes the familiar antisymmetric symbol
$\ep^{abc}$, where the indices run from $1$ to $3$. It satisfies
\begin{equation}
\ep^{abc} \ep^{cde} = \delta^{ad} \delta^{ce} - \delta^{ae}
\delta^{cd}\,.
\end{equation}
The asymptotic behavior of the interaction action is now given in the
following form:
\begin{eqnarray}
\SIL &\asym& \int d^4 x\, \left[ \Lambda^2 a_0 (\ln \Lambda/\mu)
    \frac{1}{2} \left(A_\mu^a\right)^2\right.\nn\\
&&\quad + a_1  (\ln \Lambda/\mu) \frac{1}{2} \left(
\partial_\mu A_\nu^a \right)^2 + a_2  (\ln \Lambda/\mu) \frac{1}{2}
\left( \partial_\mu A_\mu^a \right)^2\nn\\
&&\quad + a_3 (\ln \Lambda/\mu) g
\ep^{abc} \partial_\mu A_\nu^a \cdot A_\mu^b A_\nu^c\nn\\
&&\quad + a_4 (\ln \Lambda/\mu) \frac{g^2}{4}
\left( A_\mu^a A_\mu^a \right)^2 + a_5 (\ln \Lambda/\mu)
\frac{g^2}{4} \left( A_\mu^a A_\nu^a  \right)^2\nn\\
&&\left.\quad - a_6 (\ln \Lambda/\mu) \partial_\mu
\bar{c}^a \partial_\mu c^a - a_7 (\ln \Lambda/\mu)
g \ep^{abc} \partial_\mu \bar{c}^a A_\mu^b c^c\right]\,.
\end{eqnarray}
Note that the action depends only on the derivative of the antighost
field $\bar{c}^a$; this property certainly holds for the tree level
action, and it is further inherited by the Wilson action, since the
property is preserved by the Polchinski equation.  Hence, besides the
gauge coupling $g$, the theory has 7 parameters:
\begin{equation}
a_1 (0),\, a_2 (0),\, a_3 (0),\, a_4 (0),\, a_5 (0),\,
a_6 (0),\, a_7 (0)\,.\label{WTexamples-YM-parameters}
\end{equation}
At tree level, we obtain
\begin{equation}
a_3 (0) = 1,\,
a_4 (0) = -1,\,
a_5 (0) = 1,\,
a_7 (0) = -1,
\end{equation}
and the rest vanishing.

Let us now consider the quantum BRST transformation, given by
\begin{equation}
\lb\begin{array}{c@{~=~}l}
\delta A_\mu^a (k) & \ep \left( k_\mu \left[c^a\right] (k) + \A_\mu^a
    (k) \right)\,,\\
\delta c^a (k) & \ep \,\C^a (k)\,,\\
\delta \bar{c}^a (-k) & \ep \frac{1}{\xi} (- k_\mu)
\left[A_\mu^a\right] (-k)\,,
\end{array}\right.\label{WTexamples-YM-BRST}
\end{equation}
where the composite operators $[c^a], [A^a_\mu]$ are defined as
usual:
\begin{equation}
\lb\begin{array}{c@{~=~}l}
\left[c^a\right] (k) & \frac{1}{\K{k}} c^a (k) + \frac{1 - \K{k}}{k^2}
\Ld{\bar{c}^a (-k)} \SL\,,\\
\left[A_\mu^a\right] (-k) & \frac{1}{\K{k}} A_\mu^a (-k) + \frac{1 -
  \K{k}}{k^2} \left( \delta_{\mu\nu} - (1-\xi) \frac{k_\mu k_\nu}{k^2}
\right) \frac{\delta \SL}{\delta A_\nu (k)}\,.
\end{array}\right.
\end{equation}
We define the composite operator $\A_\mu^a (k)$ so that $k_\mu \delta
A_\mu^a (k)$ gives the composite operator for the ghost equation of
motion:
\begin{equation}
k_\mu \left( k_\mu [c^a] (k) + \A_\mu^a (k) \right)
= - \K{k} \Ld{\bar{c}^a (-k)} \SL\,.
\label{WTexamples-YM-ghostEOM}
\end{equation}
This is equivalent to
\begin{equation}
k_\mu \A_\mu^a (k) = - \Ld{\bar{c}^a (-k)} \SIL\,.
\end{equation}
This implies that the asymptotic behavior of $\A_\mu^a (k)$ is
determined by that of the action:
\begin{equation}
\A_\mu^a (k) \asym a_6 (\ln \Lambda/\mu) k_\mu c^a (k) + a_7 (\ln
\Lambda/\mu) \frac{1}{i} g \ep^{abc} \int_l A_\mu^b (k-l) c^c (l)\,.
\end{equation}
On the other hand, the asymptotic behavior of the composite operator
$\C^a (k)$ is not determined by the action, and it is given by
\begin{equation}
\C^a (k) \asym a_8 (\ln \Lambda/\mu) \frac{g}{2i}
\ep^{abc} \int_l c^b (k-l) c^c (l)\,.
\end{equation}
The parameter $a_8 (0)$, which is $1$ at tree level, is the
normalization constant of $\C^a$.  This makes the eighth parameter of
the theory, besides the seven parameters
(\ref{WTexamples-YM-parameters}).  We wish to fix these eight by
demanding the invariance of the Wilson action under the BRST
transformation (\ref{WTexamples-YM-BRST}).  Here, we note that the gauge
invariance is expected to leave three parameters free to choose: $a_2
(0), a_6 (0)$ are the normalization constants of the gauge and ghost
fields, respectively, and $a_3 (0)$ normalizes the coupling constant
$g$.  Whatever constraints we find among the eight parameters must leave
these three parameters free.

Before going further, let us take a moment to understand the
requirement (\ref{WTexamples-YM-ghostEOM}).  The WT identity for the
two-point function is
\begin{equation}
\vev{\delta A_\mu^a (k) \bar{c}^b (-k)}^\infty + \vev{A_\mu^a (k)
  \delta \bar{c}^b (-k)}^\infty = 0\,,
\end{equation}
which gives
\begin{equation}
\vev{\left( k_\mu c^a (k) + \A_\mu^a (k) \right) \bar{c}^b
  (-k)}^\infty - \frac{1}{\xi} \vev{A_\mu^a (k) k_\nu A_\nu^b
  (-k)}^\infty = 0\,.
\end{equation}
Thus, (\ref{WTexamples-YM-ghostEOM}) gives
\begin{equation}
\frac{1}{\xi} \vev{k_\mu A_\mu^a (k) k_\nu A_\nu^b (-k)}^\infty =
\delta^{ab} \,,
\end{equation}
implying that the longitudinal part of the gauge boson propagator does
not receive any radiative correction.

The WT composite operator\cite{Igarashi:2008bb} is now defined by
\begin{eqnarray}
\Sigma_\Lambda &\equiv& \int_k \K{k} \left[ \frac{\delta \SL}{\delta
      A_\mu^a (k)} \delta A_\mu^a (k) + \frac{\delta}{\delta
      A_\mu^a (k)} \delta A_\mu^a (k) \right.\nn\\
&&\quad + \SL \Rd{c^a (k)} \delta c^a (k) - \delta c^a (k) \Rd{c^a
  (k)}\nn\\
&&\quad \left. + \delta \bar{c}^a (-k) \Ld{\bar{c}^a (-k)} \SL -
    \Ld{\bar{c}^a (-k)} \delta \bar{c}^a (-k) \right]\,.
\label{WTexamples-YM-Sigma}
\end{eqnarray}
This is a dimension $5$ scalar composite operator with ghost number
$1$.  Hence, for SU(2), its asymptotic behavior must have the
following form:
\begin{eqnarray}
\Sigma_\Lambda &\asym& \ep \int d^4 x \left[
\Lambda^2 s_0 \,c^a \partial_\mu A_\mu^a 
+ s_1 \,\frac{1}{i}
c^a \partial^2 \partial_\mu A_\mu^a\right.\nn\\
&&\quad + s_2 \,\frac{1}{i} g \ep^{abc} c^a \partial^2 A_\mu^b \cdot
A_\mu^c + s_3 \,\frac{1}{i} g \ep^{abc} c^a \partial_\mu
\partial_\nu A_\nu^b \cdot A_\mu^c \nn\\
&&\quad + s_4 \,i g^2  c^a \partial_\mu A_\mu^a \cdot A_\nu^b A_\nu^b + s_5\,
i g^2 c^a \partial_\mu A_\nu^a \cdot A_\mu^b A_\nu^b + s_6\, i g^2 c^a
A_\mu^a \partial_\nu A_\nu^b A_\mu^b \nn\\
&&\quad + s_7 \,i g^2 c^a A_\mu^a \partial_\nu
A_\mu^b A_\nu^b + s_8 \,i g^2 c^a A_\mu^a \partial_\mu A_\nu^b A_\nu^b\nn\\
&&\left.\quad + s_{9} \, g \ep^{abc} \partial_\mu c^a c^b
\frac{1}{i} \partial_\mu \bar{c}^c + s_{10} \,
g^2 c^a c^b \frac{1}{i} \partial_\mu \bar{c}^a  A_\mu^a \right]\,,
\label{WTexamples-YM-asymptoticSigma}
\end{eqnarray}
where the coefficients $s_i\,(i=0, 1, \cdots,10)$ depend on $\ln
\frac{\Lambda}{\mu}$ and the gauge coupling constant $g$.  As a
composite operator, $\Sigma_\Lambda$ has ten parameters:
\begin{equation}
s_i (\ln \Lambda/\mu = 0) = s_i (0)\quad (i=1,\cdots,10)\,.
\end{equation}
We must make these vanish by arranging the eight parameters. 

Substituting the BRST transformation (\ref{WTexamples-YM-BRST}) into
(\ref{WTexamples-YM-Sigma}), we can simplify the expression of
$\Sigma_\Lambda$ somewhat:
\begin{eqnarray}
\Sigma_\Lambda &=& \ep \int_k \K{k} \left( \A_\mu^a (k) \frac{\delta
      \SL}{\delta A_\mu^a (k)} + \frac{\delta}{\delta A_\mu^a (k)}
    \A_\mu^a (k) \right)\nn\\
&& - \ep \int_k \K{k} \left( \SL \Rd{c^a (k)} \C^a (k) - \C^a (k)
    \Rd{c^a (k)} \right)\nn\\
&& + \ep \int_k \left[ k_\mu \frac{\delta \SIL}{\delta A_\mu^a (k)}
    c^a (k) - \frac{1}{\xi} k_\mu A_\mu^a (-k) \Ld{\bar{c}^a (-k)}
    \SIL \right]\,.
\end{eqnarray}
Each of the three integrals is a composite operator by itself.

Let us now introduce loop expansions of $\SL$ and $\Sigma_\Lambda$:
\begin{equation}
\SL = \sum_{l=0}^\infty \SL^{(l)}\,,\quad
\Sigma_\Lambda = \sum_{l=0}^\infty \Sigma_\Lambda^{(l)}\,.
\end{equation}
Correspondingly, let us expand the parameters:
\begin{equation}
a_i (0) = \sum_{l=0}^\infty a_i^{(l)}\,.\quad(i=1,\cdots,8)
\end{equation}
Suppose we have made
\begin{equation}
\Sigma_\Lambda^{(0)} = \cdots = \Sigma_\Lambda^{(l)} = 0
\end{equation}
by arranging $a_i^{(0)}, \cdots, a_i^{(l)}$.  Then, as $s_i$'s at
$(l+1)$-loop are all constants, independent of $\ln \Lambda/\mu$, as a
consequence of the ERG differential equation for $\Sigma_\Lambda$.  We
ask if we can choose the $(l+1)$-loop coefficients $a_i^{(l+1)}$ to
make all $s_i$ vanish at $(l+1)$-loop, which is equivalent to
\begin{equation}
\Sigma_\Lambda^{(l+1)} = 0\,.
\end{equation}

Examining the structure of $\Sigma_\Lambda$
above, we find that at $(l+1)$-loop the coefficients $s_i$
are given as the sum:
\begin{equation}
s_i = t_i + u_i\quad (i=1,\cdots,10)\,,
\end{equation}
where $t$'s are linear in $a^{(l+1)}$'s, and $u$'s are determined by the
action only up to $l$-loop.  Not all $t$'s are independent; only
the following five are independent:
\begin{subequations}
\label{WTexamples-YM-tbya}
\begin{eqnarray}
t_1 &=& a^{(l+1)}_1 + a^{(l+1)}_2 \,,\\
t_2 &=& a^{(l+1)}_1 + a^{(l+1)}_3 + a^{(l+1)}_6 + a^{(l+1)}_7 \,,\\
t_4 &=& a^{(l+1)}_3 + a^{(l+1)}_4 - a^{(l+1)}_6 - a^{(l+1)}_7 \,,\\
t_5 &=& - a^{(l+1)}_3 + a^{(l+1)}_5 + a^{(l+1)}_6 + a^{(l+1)}_7 \,,\\
t_9 &=& a^{(l+1)}_7 + a^{(l+1)}_8 \,.
\end{eqnarray}
\end{subequations}
The remaining $t$'s are given by
\begin{subequations}
\label{WTexamples-YM-tbyt}
\begin{eqnarray}
t_3 &=& t_1 - t_2\,,\\
t_6 &=& t_5\,,\\
t_7 &=& 2 \,t_4\,,\\
t_8 &=& t_5\,,\\
t_{10} &=& - t_9\,.
\end{eqnarray}
\end{subequations}
It is possible to make all $s_i = 0$ by adjusting $a^{(l+1)}$'s if and
only if $u$'s satisfy the same relations as $t$'s. If $t$'s and $u$'s
satisfy the above relations, so do $s$'s:
\begin{subequations}
\label{WTexamples-YM-srelations}
\begin{eqnarray}
s_3 &=& s_1 - s_2\,,\\
s_6 &=& s_5\,,\\
s_7 &=& 2 \,s_4\,,\\
s_8 &=& s_5\,,\\
s_{10} &=& - s_9\,.
\end{eqnarray}
\end{subequations}
If these relations hold, we have five linearly independent conditions
on eight parameters, leaving three of them arbitrary.  This is as
expected, since the YM gauge symmetry does not constrain the
normalization of fields and the coupling constant $g$. For example, we
can adopt the normalization condition:
\begin{equation}
a_1 (0) = a_6 (0) = 0\,,\quad a_3 (0) = 1\,.
\end{equation}
It is impossible to derive the five relations
\bref{WTexamples-YM-srelations} within the formalism of the WT identity.
We will resort to the antifield formalism of \S \ref{AF} for derivation.

\subsection{Wess-Zumino model\label{WTexamples-WZ}}

The ERG formalism has been applied to the Wess-Zumino model, for
example, in \citen{Bonini:1998ec}, \citen{Pernici:1998ex}, and
\citen{Rosten:2008ih}, where the supersymmetry transformation is
linearized via auxiliary fields.  Without the auxiliary fields, the
supersymmetry transformation is non-linear, and the model was
first formulated with ERG in \citen{Sonoda:2008dz}.  

The Wess-Zumino model\cite{Wess:1973kz} is the simplest supersymmetric
model defined in 4 dimensions.  Without auxiliary fields, the classical
action of the model is given as
\begin{eqnarray}
&&S_{cl} \equiv - \int d^4 x\, \Big[ \partial_\mu
    \bar{\phi} \partial_\mu \phi + \bar{\chi}_L
    \sigma_\mu \partial_\mu \chi_R \nn\\
&& \quad + \left( m + g \phi \right) \frac{1}{2} \bar{\chi}_R \chi_R +
\left( \bar{m} + \bar{g} \bar{\phi} \right) \frac{1}{2} \bar{\chi}_L
\chi_L \nn\\
&& \quad + \left( m \phi + g \frac{\phi^2}{2} \right) \left(
        \bar{m} \bar{\phi} + \bar{g} \frac{\bar{\phi}^2}{2} \right)
\Big]\,,
\end{eqnarray}
where $\chi_{R,L}$ are 2-component spinors, and we define
\begin{equation}
\sigma_\mu \equiv (\vec{\sigma}, -i)\,\quad
\bar{\sigma}_\mu \equiv (\vec{\sigma}, i)\,,\quad
\bar{\chi}_R \equiv \chi_R^T \sigma_y\,,\quad
\bar{\chi}_L \equiv \chi_L^T \sigma_y\,.
\end{equation}
This is invariant under the following supersymmetry transformation:
\begin{equation}
\lb\begin{array}{c@{~\equiv~}l}
\delta_{cl} \phi & \bar{\xi}_R  \chi_R\,, 
\\ \delta_{cl} \bar{\phi} & \bar{\xi}_L \chi_L \,,
\\ \delta_{cl} \chi_R & \bar{\sigma}_\mu \xi_L \partial_\mu \phi -
\left( \bar{m} \bar{\phi} + \bar{g} \,
\frac{{\bar{\phi}}^2}{2} \right) \xi_R \,,
\\ \delta_{cl} \chi_L & \sigma_\mu \xi_R \partial_\mu \bar{\phi}
 - \left(  m \phi + g \, \frac{\phi^2}{2} \right) \xi_L \,,
\end{array}\right.\label{WTexamples-WZ-classicalsusy}
\end{equation}
where $\xi_{R,L}$ are anticommuting constant spinors.

The above classical action has a spontaneously broken $\mathbf{Z_2}$
invariance.  To make the symmetry manifest, we shift the scalar fields
as follows:
\begin{equation}
\lb\begin{array}{c@{~\longrightarrow~}l}
\phi + \frac{m}{g} & \phi\,,\\
\bar{\phi} + \frac{\bar{m}}{\bar{g}} & \bar{\phi}\,.
\end{array}\right.
\end{equation}
Denoting
\begin{equation}
v \equiv \frac{m}{g}\,\quad \bar{v} \equiv \frac{\bar{m}}{\bar{g}}\,,
\end{equation}
we can rewrite the action with manifest $\mathbf{Z_2}$ invariance:
\begin{eqnarray}
S_{cl} &=& - \int \left[ \partial_\mu \bar{\phi} \partial_\mu \phi +
\bar{\chi}_L \sigma_\mu \partial_\mu \chi_R + \frac{g}{2} \phi
\bar{\chi}_R \chi_R + \frac{\bar{g}}{2} \bar{\phi} \bar{\chi}_L \chi_L
\right.\nn\\
&&\left. \quad + \frac{g}{2} \left( \phi^2 - v^2 \right) 
\frac{\bar{g}}{2}\left( \bar{\phi}^2 - \bar{v}^2\right) \right]
\end{eqnarray}
Under the $\mathbf{Z_2}$ transformation, the fields transform as
follows:
\begin{equation}
\begin{array}{c@{~\longrightarrow~}l@{\quad}c@{~\longrightarrow~}l}
\phi& - \phi\,,& \bar{\phi} & - \bar{\phi}\\
\chi_R & i \chi_R\,,& \chi_L & \frac{1}{i} \chi_L
\end{array}
\end{equation}
In terms of the shifted scalar fields, the supersymmetry transformation
is given by
\begin{equation}
\lb\begin{array}{c@{~\equiv~}l}
\delta_{cl} \phi & \bar{\xi}_R  \chi_R\,, 
\\ \delta_{cl} \bar{\phi} & \bar{\xi}_L \chi_L \,,
\\ \delta_{cl} \chi_R & \bar{\sigma}_\mu \xi_L \partial_\mu \phi -
\frac{\bar{g}}{2} \left( \bar{\phi}^2 - \bar{v}^2\right) \xi_R\,,\\ 
\delta_{cl} \chi_L & \sigma_\mu \xi_R \partial_\mu \bar{\phi}
 - \frac{g}{2} \left( \phi^2 - v^2 \right) \xi_L \,.
\end{array}\right.\label{WTexamples-WZ-classicalsusyshifted}
\end{equation}

The classical action is parameterized by two complex parameters $g$ and
$v$.  If we ask if their phases have any physical significance, we
immediately see that their phase changes can be canceled by the
corresponding phase changes of the fields:
\begin{enumerate}
\item $g \to g \e^{i \alpha}$, $\bar{g} \to \bar{g} \e^{- i \alpha}$ are
      canceled by
\begin{equation}
\phi \to \e^{- i \alpha} \phi\,,\quad
\bar{\phi} \to \e^{i \alpha} \bar{\phi}\,.
\label{WTexamples-WZ-alpha}
\end{equation}
\item $v \to v \e^{i \beta}$, $\bar{v} \to \bar{v} \e^{- i \beta}$ are
      canceled by
\begin{equation}
\phi \to \e^{i \beta} \phi\,,\quad
\bar{\phi} \to \e^{- i \beta} \bar{\phi}\,,\quad
\chi_R \to \e^{- i \frac{\beta}{2}} \chi_R\,,\quad
\chi_L \to \e^{i \frac{\beta}{2}} \chi_L\,.
\label{WTexamples-WZ-beta}
\end{equation}
\end{enumerate}
Hence, the physics of the model depends only on $|g|$ and $|v|$, and we
make sure that this property is carried over upon
quantization.\footnote{The following results in this section are taken
from a recent unpublished work of H.~Sonoda and K.~\"Ulker.  The use of
the $\mathbf{Z_2}$ symmetry makes the construction much simpler than
that given in the earlier work of the same authors
\cite{Sonoda:2008dz}.}

We now consider the Wilson action:
\begin{equation}
\SL = \SFL + \SIL\,,
\end{equation}
where the free action is
\begin{equation}
\SFL \equiv - \int_p \left[
\frac{1}{\Kb{p}} p^2 \bar{\phi} (-p) \phi (p) + \frac{1}{\Kf{p}}
\bar{\chi}_L (-p) i p \cdot \sigma \chi_R (p) \right]\,.
\end{equation}
For generality we take the cutoff functions of the bosons and fermions
independent, since their equality is not demanded by supersymmetry.

The interaction action is determined by the ERG differential equation
\begin{eqnarray}
&&- \Lambda \frac{\partial}{\partial \Lambda} \SIL
= \int_p \frac{\Delta_b (p/\Lambda)}{p^2} \lb
\frac{\delta \Si}{\delta \bar{\phi} (-p)} \frac{\delta \Si}{\delta
  \phi (p)} + \frac{\delta^2 \Si}{\delta \bar{\phi} (-p) \delta
  \phi (p)} \rb\\
&&\, - \int_p \frac{\Delta_f (p/\Lambda)}{p^2} \Tr (- i p) \cdot
\bar{\sigma} \lb \Ld{\bar{\chi}_L (-p)} \Si \cdot \Si \Rd{\chi_R (p)}
+  \Ld{\bar{\chi}_L (-p)} \Si \Rd{\chi_R (p)} \rb\,,\nn
\end{eqnarray}
and the asymptotic behavior
\begin{eqnarray}
\SIL &\stackrel{\Lambda \to \infty}{\longrightarrow}&
\int d^4 x \, \Big[ \Lambda^2 a (\ln \Lambda/\mu) |\phi|^2
+ c_1 (\ln \Lambda/\mu) \partial_\mu \bar{\phi} \partial_\mu \phi +
c_2 (\ln \Lambda/\mu) \bar{\chi}_L \sigma \cdot \partial
\chi_R\nn\\
&& \quad + c_3 (\ln \Lambda/\mu) \left( g \phi \frac{1}{2}
    \bar{\chi}_R \chi_R + \bar{g} \bar{\phi} \frac{1}{2} \bar{\chi}_L
    \chi_L \right) + c_4 (\ln \Lambda/\mu) |g|^2 \frac{1}{4} |\phi|^4
\nn\\
&&\quad + c_5 (\ln \Lambda/\mu) |g|^2 \frac{1}{4} \left(
\bar{v}^2 \phi^2 + v^2 \bar{\phi}^2 \right) 
 + c_6 (\ln \Lambda/\mu) \frac{1}{4} |g|^2 |v|^4\,
\Big]\,.\label{WTexamples-WZ-SIlargeL}
\end{eqnarray}
Hence, the action has six parameters:
\begin{equation}
c_i (0)\quad (i=1,\cdots, 6)\,.
\end{equation}
We have included $c_6 (0)$ as a parameter, even though it corresponds to
an additive constant to the action.  With the constant term, the
derivative of the action with respect to $v^2$ gives an integral of the
composite operator
\begin{equation}
\left[ \frac{\phi^2}{2} \right]_\Lambda (x)
\end{equation}
over space:
\begin{equation}
\frac{2}{\bar{g}} \frac{\partial}{\partial \bar{v}^2} \SIL = \int d^4 x
 \, g \left[\frac{\phi^2}{2}\right]_\Lambda (x)\,.
\end{equation}
Similarly, the derivative with respect to $v^2$ gives
\begin{equation}
\frac{2}{g} \frac{\partial}{\partial v^2} \SIL = \int d^4 x \, \bar{g}
 \left[\frac{\bar{\phi}^2}{2} \right]_\Lambda (x)\,.
\end{equation}
The asymptotic behavior (\ref{WTexamples-WZ-SIlargeL}) implies the
following asymptotic behavior of the two composite operators in
coordinate space:
\begin{equation}
\lb\begin{array}{c@{\stackrel{\Lambda\to\infty}{\longrightarrow}}l}
\left[ \frac{\phi^2}{2} \right]_\Lambda & c_5 (\ln\Lambda/\mu)
\frac{\phi^2}{2} + c_6 (\ln \Lambda/\mu) \frac{v^2}{2}\,,\\
\left[ \frac{\bar{\phi}^2}{2} \right]_\Lambda &  c_5 (\ln\Lambda/\mu)
\frac{\bar{\phi}^2}{2} + c_6 (\ln \Lambda/\mu) \frac{\bar{v}^2}{2}\,.
\end{array}\right.
\end{equation}

We now define the WT identity for supersymmetry by
\begin{eqnarray*}
\Sigma_\Lambda &\equiv& \int_p \Kb{p} \left[ \delta \phi (p)
      \frac{\delta}{\delta \phi (p)} \SL + \frac{\delta}{\delta \phi
        (p)} \delta \phi (p) + \delta \bar{\phi} (p)
      \frac{\delta}{\delta \bar{\phi} (p)} \SL + \frac{\delta}{\delta
        \bar{\phi} (p)} \delta \bar{\phi} (p) \right]\\
&&\, + \int_p \Kf{p} \left[ \SL \Rd{\chi_R (p)} \delta \chi_R (p) - \Tr
    \delta \chi_R (p) \Rd{\chi_R (p)} \right.\\
&&\qquad\qquad\qquad \left.+ \SL \Rd{\chi_L (p)} \delta
    \chi_L (p) - \Tr \delta \chi_L (p) \Rd{\chi_L (p)} \right]\,,
\end{eqnarray*}
where the composite operators $\delta \phi (p)$, etc., are defined by
\begin{equation}
\lb\begin{array}{c@{~=~}l}
\delta \phi (p) & \bar{\xi}_R [\chi_R] (p)\,,\\
\delta \chi_R (p) & [\phi] (p) i p \cdot \bar{\sigma} \xi_L - \bar{g}
\left[ \frac{\bar{\phi}^2}{2} \right] (p) \xi_R\,,\\
\delta \bar{\phi} (p) & \bar{\xi}_L [\chi_L] (p)\,,\\
\delta \chi_L (p) & [\bar{\phi}] (p) i p \cdot \sigma \xi_R - g \left[
    \frac{\phi^2}{2} \right] (p) \xi_L\,.
\end{array}
\right.
\end{equation}
Under the phase changes of $g, v$, the composite operator
$\Sigma_\Lambda$ remains invariant, 
if we change the phases of the fields according to
(\ref{WTexamples-WZ-alpha}) \& (\ref{WTexamples-WZ-beta}), and at the
same time change the phases of $\xi_{R,L}$ according to
\begin{equation}
\xi_R \to \e^{- i \alpha + i \frac{3}{2} \beta} \xi_R,\quad
\xi_L \to \e^{i \alpha - i \frac{3}{2} \beta} \xi_L\,.
\label{WTexamples-WZ-xialphabeta}
\end{equation}

In the remainder of this subsection, we wish to show that we can satisfy
the WT identity
\begin{equation}
\Sigma_\Lambda = 0
\end{equation}
by tuning four of the six parameters, leaving $c_1 (0)$ and $c_3 (0)$
arbitrary.  The physical meaning of the arbitrary parameters is
clear:
\begin{enumerate}
\item $c_1 (0)$ --- overall normalization of the fields; the relative
    normalization of $\phi$ and $\chi_R$ is fixed by supersymmetry.
\item $c_3 (0)$ --- normalization of $g$.
\end{enumerate}
Note that the relative normalization of $\phi$ and $\bar{g}
[\bar{\phi}^2]$ is fixed by supersymmetry.  Hence, the relative
normalization of $\phi$ and $g v^2$ is also fixed, since $\bar{g}
[\bar{\phi}^2]$ is the derivative of the action with respect to $g v^2$.

We first enumerate all possible terms in the asymptotic behavior of the
composite operator $\Sigma_\Lambda$:
\begin{eqnarray*}
\Sigma_\Lambda &\stackrel{\Lambda \to \infty}{\longrightarrow}&
\int d^4 x \left[ \Lambda^2 b (\ln \Lambda/\mu) \left( \bar{\chi_R}
    \xi_R \bar{\phi} + \bar{\chi_L} \xi_L \phi \right)\right.\\
&& \, + s_1 (\ln \Lambda/\mu) \left( \partial^2 \bar{\phi}
    \bar{\chi}_R \xi_R + \partial^2 \phi \bar{\chi}_L \xi_L
  \right)\\
&&\, + s_2 (\ln \Lambda/\mu) \left( \bar{g} \bar{\chi}_L \sigma_\mu
  \xi_R \partial_\mu \frac{\bar{\phi}^2}{2} + g \bar{\chi}_R
  \bar{\sigma}_\mu \xi_L \partial_\mu \frac{\phi^2}{2} \right)\\
&&\, + s_3 (\ln \Lambda/\mu) |g|^2 \left( \bar{\chi}_R \xi_R \phi
  \frac{\bar{\phi}^2}{2} + \bar{\chi}_L \xi_L \bar{\phi}
  \frac{\phi^2}{2} \right)\\
&&\left. \, + s_4 (\ln \Lambda/\mu) |g|^2 \left( \frac{\bar{v}^2}{2}
    \phi \bar{\chi}_R \xi_R + \frac{v^2}{2} \bar{\phi} \bar{\chi}_L
    \xi_L \right) \right]\,.
\end{eqnarray*}
We have enumerated all terms of dimension up to $4$, which are invariant
under the simultaneous phase changes of the parameters and fields given
by (\ref{WTexamples-WZ-alpha}), (\ref{WTexamples-WZ-beta}),
(\ref{WTexamples-WZ-xialphabeta}).

At tree level, we find that the choice
\begin{equation}
a = c_1 = c_2 = 0,\quad
c_3 = c_4 = -1,\quad c_5 = 1,\quad c_6 = -1
\end{equation}
satisfies the tree level WT identity
\begin{equation}
\Sigma_\Lambda^{(0)} = 0\,.
\end{equation}
We now assume the vanishing of $\Sigma_\Lambda$ up to $l$-loop level:
\begin{equation}
\Sigma_\Lambda^{(1)} = \cdots = \Sigma_\Lambda^{(l)} = 0\,.
\end{equation}
Then, the ERG differential equation for $\Sigma_\Lambda^{(l+1)}$ becomes
\begin{eqnarray}
&&- \Lambda \frac{\partial}{\partial \Lambda} \Sigma_\Lambda^{(l+1)}
= \int_p \frac{\Delta_b (p/\Lambda)}{p^2} \left( \frac{\delta
      \SIL^{(0)}}{\delta \phi (p)} \frac{\delta \Sigma_\Lambda^{(l+1)}}{\delta
      \bar{\phi} (-p)} + \frac{\delta
      \SIL^{(0)}}{\delta \bar{\phi} (p)} \frac{\delta
      \Sigma_\Lambda^{(l+1)}}{\delta \phi (-p)} \right)\nn\\
&& \quad + \int_p \frac{\Delta_f (p/\Lambda)}{p^2} \left(
\SIL^{(0)} \Rd{\chi_R (p)} (- i p) \cdot \bar{\sigma} \Ld{\bar{\chi}_L
  (-p)} \Sigma_\Lambda^{(l+1)}\right.\nn\\
&&\qquad\qquad\qquad \left. + \SIL^{(0)} \Rd{\chi_L (p)} (- i p) \cdot
\sigma \Ld{\bar{\chi}_R  (-p)} \Sigma_\Lambda^{(l+1)} \right)\,.
\end{eqnarray}
This vanishes as $\Lambda \to \infty$.
Hence, we obtain that
\begin{equation}
b^{(l+1)} (\ln \Lambda/\mu) = 0\,,
\end{equation}
and that $s_i^{(l+1)}$'s are independent of $\Lambda$.  In the following
we wish to show that we can choose $c_i^{(l+1)} (0)\,(i=1,\cdots,6)$ so that
\begin{equation}
s_i^{(l+1)} = 0\quad (i=1,\cdots,4)\,.
\end{equation}

We split $s^{(l+1)}$'s into $t^{(l+1)}$'s linear in $c^{(l+1)}$'s and
$u^{(l+1)}$'s independent of them:
\begin{equation}
s_i^{(l+1)} = t_i^{(l+1)} + u_i^{(l+1)}\,.
\end{equation}
The coefficients $t^{(l+1)}$'s can be calculated from
\begin{eqnarray}
\Sigma_\Lambda^{(l+1),t} &\equiv& \int_p \Kb{p} \left( \delta \phi^{(0)} (p)
  \frac{\delta \SL^{(l+1)}}{\delta \phi (p)} + \delta \bar{\phi}^{(0)}
  (p) \frac{\delta \SL^{(l+1)}}{\delta \bar{\phi} (p)} \right)\nn\\
&& \, + \int_p \Kf{p} \left( \SL^{(l+1)} \Rd{\chi_R (p)} \delta
  \chi_R^{(0)} (p) + \SL^{(l+1)} \Rd{\chi_L (p)} \delta \chi_L^{(0)} (p)
\right.\nn\\
&&\quad\left.  + \SL^{(0)} \Rd{\chi_R (p)} \delta \chi_R^{(l+1)} (p) + \SL^{(0)}
  \Rd{\chi_L (p)} \delta \chi_L^{(l+1)} (p) \right)\,.
\end{eqnarray}
We then obtain, by simple substitution of the asymptotic behavior of
$\SL^{(l+1)}$ \& $\delta \chi_{R,L}^{(l+1)}$ into the above, the following
results:
\begin{equation}
\lb\begin{array}{c@{~=~}l}
t^{(l+1)}_1 & - c^{(l+1)}_1 + c^{(l+1)}_2\,,\\
t^{(l+1)}_2 & c^{(l+1)}_5 - c^{(l+1)}_2 + c^{(l+1)}_3\,,\\
t^{(l+1)}_3 & c^{(l+1)}_5 - c^{(l+1)}_3 + c^{(l+1)}_4\,,\\
t^{(l+1)}_4 & c^{(l+1)}_6 + c^{(l+1)}_3 + c^{(l+1)}_5\,.
\end{array}\right.
\end{equation}
Thus, we must choose $c^{(l+1)}$'s so that
\begin{equation}
\lb\begin{array}{c@{~=~}l}
 - c^{(l+1)}_1 + c^{(l+1)}_2 & - u_1^{(l+1)}\,,\\
 c^{(l+1)}_5 - c^{(l+1)}_2 + c^{(l+1)}_3& - u_2^{(l+1)}\,,\\
 c^{(l+1)}_5 - c^{(l+1)}_3 + c^{(l+1)}_4& - u_3^{(l+1)}\,,\\
 c^{(l+1)}_6 + c^{(l+1)}_3 + c^{(l+1)}_5& - u_4^{(l+1)}\,.
\end{array}\right.
\end{equation}
For arbitrary $u^{(l+1)}$'s, this is solved by
\begin{equation}
\lb\begin{array}{c@{~=~}l}
c_2^{(l+1)} & c_1^{(l+1)} - u_1^{(l+1)}\,,\\
c_4^{(l+1)} & - c_1^{(l+1)} + 2 c_3^{(l+1)} + u_1^{(l+1)} +
u_2^{(l+1)} - u_3^{(l+1)}\,,\\
c_5^{(l+1)} & c_1^{(l+1)} - c_3^{(l+1)} - u_1^{(l+1)} - u_2^{(l+1)}\,,\\
c_6^{(l+1)} & - c_1^{(l+1)} + u_1^{(l+1)} + u_2^{(l+1)} - u_4^{(l+1)}\,,
\end{array}\right.
\end{equation}
where $c_1^{(l+1)}$ and $c_3^{(l+1)}$ remain arbitrary.  With the above choice
we can make
\begin{equation}
\Sigma_\Lambda^{(l+1)} = 0\,.
\end{equation}

\begin{figure}[t]
\begin{center}
\epsfig{file=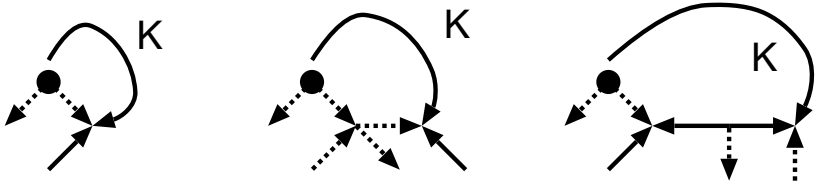, width=11cm} \caption{Feynman graphs
for $u_i^{(1)}$'s --- The solid curve marked $K$ indicates the
multiplication of the cutoff function $K$.  The dot denotes the
$\bar{\phi}^2$ vertex. We use a broken line for the scalar, a solid line
for the spinor.}  \label{WTexamples-WZ-phi2}
\end{center}
\end{figure}

We end this subsection by giving the results of 1-loop calculations.
For simplicity, we take
\begin{equation}
K_b = K_f = K\,.
\end{equation}
For the interaction action, we obtain
\begin{equation}
\lb\begin{array}{c@{~=~}l}
a^{(1)} & |g|^2 \int_q \frac{\Delta (q)}{q^2} \left( - \frac{1}{2} + K
					       (q) \right)\,,\\
c_1^{(1)} (\ln \Lambda/\mu) & \frac{|g|^2}{(4 \pi)^2} \ln
 \frac{\Lambda}{\mu} + c_1^{(1)} (0)\,,\\
c_2^{(1)} (\ln \Lambda/\mu) & \frac{|g|^2}{(4 \pi)^2} \ln
 \frac{\Lambda}{\mu} + c_2^{(1)} (0)\,,\\
c_3^{(1)} (\ln \Lambda/\mu) & c_3^{(1)} (0)\,,\\
c_4^{(1)} (\ln \Lambda/\mu) & - \frac{|g|^2}{(4 \pi)^2} \ln
 \frac{\Lambda}{\mu} + c_4^{(1)} (0)\,,\\
c_5^{(1)} (\ln \Lambda/\mu) & \frac{|g|^2}{(4 \pi)^2} \ln
 \frac{\Lambda}{\mu} + c_5^{(1)} (0)\,,\\
c_6^{(1)} (\ln \Lambda/\mu) & - \frac{|g|^2}{(4 \pi)^2} \ln
 \frac{\Lambda}{\mu} + c_6^{(1)} (0)\,.\\
\end{array}\right.
\end{equation}
The coefficients $u_i^{(1)}$'s can be calculated from
\begin{equation}
\Sigma^{(1), u}_\Lambda = - \int_p \K{p} \left[
\bar{g} \left[\frac{\bar{\phi}^2}{2}\right]^{(0)}_\Lambda (p) \Rd{\chi_R
(p)} \xi_R + g \left[ \frac{\phi^2}{2}\right]_\Lambda^{(0)} \Rd{\chi_L
(p)} \xi_L \right]
\end{equation}
as follows (Fig.~\ref{WTexamples-WZ-phi2}):
\begin{equation}
\lb\begin{array}{c@{~=~}l}
u_1^{(1)} & |g|^2 \int_p \frac{1}{p^4} \Delta (p) \left( \frac{1}{4} K
  (p) - \frac{1}{8}  \Delta (p) \right)\,,\\
u_2^{(1)} & 0\,,\\
u_3^{(1)} & - 2 |g|^2 \int_p \frac{1}{p^4} K(1-K)^2 (2 - K)\,,\\
u_4^{(1)} & 0\,.
\end{array}\right.
\end{equation}
If we choose
\begin{equation}
c_2^{(1)} (0) = c_3^{(1)} (0) = 0\,,
\end{equation}
then we obtain
\begin{equation}
\lb\begin{array}{c@{~=~}l}
c_1^{(1)} (0) & |g|^2 \int_p \frac{1}{p^4} \Delta \left( \frac{1}{4} K -
		  \frac{1}{8}  \Delta\right)\,,\\
c_4^{(1)} (0) & 2 |g|^2 \int_p \frac{1}{p^4} K(1-K)^2 (2 - K)\,,\\
c_5^{(1)} (0) & c_6^{(1)} (0) = 0\,.
\end{array}\right.
\end{equation}

\subsection{O(N) non-linear sigma
  model\label{WTexamples-ON}}

The first application of the ERG formalism to the two-dimensional O(N)
non-linear sigma model was done by Becchi \cite{Becchi:1996an}.  The
discussion here follows a recent unpublished work by L\"utf\"uo\u{g}lu
and Sonoda.\footnote{The ERG formalism has been applied to the
supersymmetric non-linear sigma models extensively in
\citen{Higashijima:2002mh,Higashijima:2003rp,Higashijima:2005qk,
Higashi:2007ie}.}  

The classical action of the two-dimensional O(N) non-linear sigma
model is given by
\begin{eqnarray}
S_{cl} &=& - \frac{1}{2 g} \int d^2 x\, \left[ g \partial_\mu
    \phi_i \partial_\mu \phi_i + \partial_\mu \sqrt{ 1 - g \phi^2
    } \,\partial_\mu \sqrt{1 - g \phi^2}\right]\nn\\
&=& \int d^2 x\, \left[ \frac{1}{2} \phi_i \partial^2 \phi_i 
- \partial^2 \left(\frac{\phi^2}{2}\right) \cdot \frac{1}{4} \ln
\left( 1 - g \phi^2 \right) \right]\,,
\end{eqnarray}
where the index $i$ runs from $1$ to $N-1$, the repeated indices
are summed, and
\begin{equation}
\phi^2 \equiv \phi_i \phi_i = \sum_{i=1}^{N-1} \phi_i \phi_i \,.
\end{equation}
The classical action is invariant under the following infinitesimal
rotation:
\begin{equation}
\de \phi_i = \ep_i \,\Phi_N\,,\quad (i=1,\cdots,N-1)
\end{equation}
where $\ep_i$ are arbitrary infinitesimal constants, and
\begin{equation}
\Phi_N \equiv \sqrt{1 - g \phi^2}
\end{equation}
is the N-th component of an N-dimensional unit vector whose $i$-th
component is $\sqrt{g}~\phi_i$.

Note that the usual IR problem with the massless scalar does not arise
here, since the Wilson action is constructed by integration of high
momentum modes $p^2 > \Lambda^2$.  Hence, we do not need to introduce a
mass term (or an external field) to break the symmetry explicitly to
O(N$-$1).

The Wilson action is given by
\begin{equation}
\SL = \SFL + \SIL\,,
\end{equation}
where the free action is given by
\begin{equation}
\SFL = - \frac{1}{2} \int_p \frac{p^2}{\K{p}} \phi_i (-p) \phi_i
(p)\,.
\end{equation}
Hence, the cutoff propagator is
\begin{equation}
\vev{\phi_i (p) \phi_j (-p)}_{\SFL} = \delta_{ij} \frac{\K{k}}{k^2}
\,.
\end{equation}
The interaction action is defined by the Polchinski equation
\begin{equation}
- \Lambda \frac{\partial}{\partial \Lambda} \SIL
= \int_k \frac{\Delta (k/\Lambda)}{k^2} \frac{1}{2}
\lb \frac{\delta \SIL}{\delta \phi_i (-p)} 
\frac{\delta \SIL}{\delta \phi_i (p)} + 
\frac{\delta^2 \SIL}{\delta \phi_i (-p) \delta \phi_i (p)} \rb\,,
\end{equation}
and the asymptotic behavior
\begin{eqnarray}
\SIL &\asym& \int d^2 x \left[ \Lambda^2 a \left(\ln \Lambda/\mu;
        \phi^2/2 \right)\right.\\
&&\left.\quad + A \left(\ln \Lambda/\mu; \phi^2/2\right) 
\left( - \partial^2 \right) \frac{\phi^2}{2}
+ B \left(\ln \Lambda/\mu; \phi^2/2\right)
    \phi_i \left( - \partial^2 \right) \phi_i
\right]\,,\nn
\end{eqnarray}
where the coefficient functions can be expanded in powers of $\phi^2/2$ as
\begin{equation}
\lb\begin{array}{c@{~=~}l}
a (\ln \Lambda/\mu; x) & \sum_{n=1}^\infty \frac{x^n}{n!} a_n (\ln
\Lambda/\mu)\,,\\
A (\ln \Lambda/\mu; x) & \sum_{n=1}^\infty \frac{x^n}{n!} A_n (\ln
\Lambda/\mu)\,,\\
B (\ln \Lambda/\mu; x) & \sum_{n=0}^\infty \frac{x^n}{n!} B_n (\ln
\Lambda/\mu)\,.
\end{array}\right.
\end{equation}
The series for $A$ starts from $n=1$, since the $n=0$ would correspond
to a total derivative in the action.  At tree level, we find
\begin{equation}
a (\ln \Lambda/\mu; x) = 0\,,\quad
A (\ln \Lambda/\mu; x) = \frac{1}{4} \ln \left( 1 - 2 g x
\right)\,,\quad
B (\ln \Lambda/\mu; x) = 0\,.
\end{equation}

The infinitesimal transformation is given by
\begin{equation}
\delta \phi_i (p) = \ep_i \PN (p)\,,
\end{equation}
where the composite operator $\PN$ is the composite operator
corresponding to the classical $\Phi_N$.  $\PN$ satisfies an ERG
differential equation, and is specified by the asymptotic behavior
\begin{equation}
\int_p \e^{i p x} \PN (p) \asym  P \left( \ln
    \Lambda/\mu; \phi (x)^2/2 \right)\,,
\end{equation}
where the function $P$ can be expanded as
\begin{equation}
P (\ln \Lambda/\mu; x) = \sum_{n=0}^\infty \frac{x^n}{n!} P_n (\ln
\Lambda/\mu)\,.
\end{equation}
At tree level, we find
\begin{equation}
P (\ln \Lambda/\mu; x) = \sqrt{1 - 2 g x}\,.
\end{equation}

Thus, the action is parameterized by two functions:
\begin{equation}
A (0; x),\quad B (0; x)\,,
\end{equation}
while the transformation is parameterized by a function
\begin{equation}
P(0;x)\,.
\end{equation}
Each Taylor coefficient of $A(0;x), B(0;x)$ is a parameter of the
Wilson action.  Similarly, $P(0;x)$ parameterizes the transformation.
Hence, there are an infinite number of parameters.  We wish to show
how the WT identity determines the three functions.

The WT composite operator is defined by
\begin{equation}
\Sigma_\Lambda \equiv \int_p \K{p} \left[ \frac{\delta \SL}{\delta
      \phi_i (p)} \delta \phi_i (p) + \frac{\delta}{\delta \phi_i (p)}
    \delta \phi_i (p) \right]\,.
\end{equation}
This has the following asymptotic behavior:
\begin{eqnarray}
    \Sigma_\Lambda &\asym& \int d^2 x\, \ep_i \phi_i
    \Big[ \partial_\mu
    \phi_j \partial_\mu \phi_j \cdot s_1 (\ln \Lambda/\mu; \phi^2/2) \\
    && \, + \phi_j \partial^2 \phi_j \cdot s_2 (\ln \Lambda/\mu; \phi^2/2)
    + (\phi_j \partial_\mu \phi_j)^2 \cdot s_3 (\ln \Lambda/\mu;
    \phi^2/2) \Big]\,. 
\end{eqnarray}
We would like to fine-tune the three functions $A(0; x), B(0; x), P(0;
x)$ so that
\begin{equation}
s_1 (0; x) = s_2 (0; x) = s_3 (0; x) = 0\,.
\end{equation}
We prove this possibility by mathematical induction on the number of
loops.

Let us introduce loop expansions using the by now familiar notation
\begin{equation}
\SL = \sum_{l=0}^\infty \SL^{(l)}\,,\,\,
\PN = \sum_{l=0}^\infty \PN^{(l)}\,,\,\,
\Sigma_\Lambda = \sum_{l=0}^\infty \Sigma_\Lambda^{(l)}\,,\,\,
A (0;x) = \sum_{l=0}^\infty A^{(l)} (x)\,,\,\cdots
\end{equation}
Suppose we have determined $\SL^{(0)},\cdots, \SL^{(l)}$ (or
equivalently $A^{(0,\cdots,l)} (x)$, $B^{(0,\cdots,l)} (x)$) and
$\PN^{(0)}, \cdots, \PN^{(l)}$ (equivalently $P^{(0,\cdots,l)} (x)$)
so that
\begin{equation}
\Sigma_\Lambda^{(0)} = \cdots = \Sigma_\Lambda^{(l)} = 0\,.
\end{equation}
Then, the asymptotic behavior of $\Sigma_\Lambda^{(l+1)}$ is independent
of $\Lambda$, and we can write it as
\begin{eqnarray}
\Sigma_\Lambda^{(l+1)} &\asym& 
\int d^2 x\, \ep_i \phi_i
\Big[ \partial_\mu
    \phi_j \partial_\mu \phi_j \cdot s_1 (\phi^2/2) \\
&& \, + \phi_j \partial^2 \phi_j \cdot s_2 (\phi^2/2)
+ (\phi_j \partial_\mu \phi_j)^2 \cdot s_3 (\phi^2/2) \Big]\,,
\end{eqnarray}
where $s_i (x)\,(i=1,2,3)$ are functions of $x$, independent of $\ln
\Lambda/\mu$.  

The definition of $\Sigma_\Lambda$ gives
\begin{equation}
\Sigma_\Lambda^{(l+1)} = \Sigma_\Lambda^{(l+1),t} + \Sigma_\Lambda^{(l+1),u}\,,
\end{equation}
where
\begin{eqnarray}
\Sigma_\Lambda^{(l+1),t} &=& \ep_i \int_p \K{p} \left[ \frac{\delta
  \SL^{(l+1)}}{\delta \phi_i (p)} \PN^{(0)} (p) +
\frac{\delta \SL^{(0)}}{\delta \phi_i (p)} \PN^{(l+1)} (p)\right]\,,\\
\Sigma_\Lambda^{(l+1),u} &=& 
\ep_i \int_p \K{p} \left[ \sum_{k=1}^{l} \frac{\delta
  \SL^{(k)}}{\delta \phi_i (p)} \PN^{(l+1-k)} (p)
+ \frac{\delta \PN^{(l)} (p)}{\delta \phi_i (p)} \right]\,.
\end{eqnarray}
Only $\Sigma_\Lambda^{(l+1),t}$ depends on 
$A^{(l+1)} (x), B^{(l+1)} (x), P^{(l+1)} (x)$, and
$\Sigma_\Lambda^{(l+1),u}$ are determined by $\SL$ and $\PN$ up to
$l$-loop. 

Therefore, the functions $s_i (x)$ are given as the sum
\begin{equation}
s_i (x) = t_i (x) + u_i (x)\,,\quad (i=1,2,3)
\end{equation}
where $t_i (x)$ are linear in $A^{(l+1)} (x), B^{(l+1)} (x), P^{(l+1)} (x)$,
and $u_i (x)$ are determined by the lower loop functions.  We obtain
explicitly
\begin{eqnarray}
    t_1 (x) &=& P^{(l+1)'} - (2 A^{(l+1)'}+B^{(l+1)'}) P^{(0)}
    \nn\\
    &&\quad - 2 A^{(0)'}
    P^{(l+1)} - 2 P^{(0)'} B^{(l+1)}\,,\\
    t_2 (x) &=& P^{(l+1)'} - 2 (A^{(l+1)'}+B^{(l+1)'})
    P^{(0)}\nn\\
    &&\quad  -  2 A^{(0)'}
    P^{(l+1)} - 2 P^{(0)'} B^{(l+1)} \,,\\
    t_3 (x) &=& P^{(l+1)''} - (A^{(l+1)''} + B^{(l+1)''})
    P^{(0)}\nn\\ &&\quad -
    A^{(0)''} P^{(l+1)} - 2 P^{(0)''} B^{(l+1)}
    - 2 B^{(l+1)'} P^{(0)'}\,,
\end{eqnarray}
where the primes denote derivatives with respect to $x$.  There is no
relation among the $t (x)$'s.  Thus, whatever $u (x)$'s are, we can
solve the equations
\begin{equation}
s_i (x) = t_i (x) + u_i (x) = 0\,.\quad (i=1,2,3)
\end{equation}
Using
\begin{equation}
A^{(0)} (x) = \frac{1}{4} \ln (1 - 2 g x)\,,\quad
B^{(0)} (x) = 0\,,\quad
P^{(0)} (x) = \sqrt{1 - 2 g x}\,.
\end{equation}
the solution is obtained as follows:
\begin{eqnarray}
B^{(l+1)} (x) &=& B^{(l+1)} (0) + \int_0^x dy \frac{- u_1 (y) + u_2
  (y)}{\sqrt{1 - 2 g y}}\,,\\ 
\frac{d}{dx} A^{(l+1)} (x) &=& \frac{1}{(1 - 2 g x)^2} \Big[
A^{(l+1)'} (0) \nn\\
&&\, + \int_0^x dy \lb - 2 g^2 B^{(l)} (y) + (1-2 g y) B^{(l)''}
    (y)\right.\nn\\
&&\,\,\, + g \sqrt{1 - 2 g y} \left( - 2 u_1 (y) + u_2 (y) \right)
+ (1-2 g y)^{\frac{3}{2}} \left( 2 u'_1 (y) - u'_2 (y) \right)\nn\\
&&\left.\,\,\,- (1 - 2 g y)^{\frac{3}{2}} u_3 
(y) \rb\Big]\,,\\
P^{(l+1)} (x) &=& \sqrt{1 - 2 g x} 
\left[ P^{(l+1)} (0) + \int_0^x dy\, \lb
2 A^{(l+1)'} (y) \right.\right.\nn\\
&&\quad \left.\left.- \frac{2 g}{1 - 2 g y} B^{(l+1)}
    (y) + \frac{- 2 u_1 (y) + u_2 (y)}{\sqrt{1 - 2 g y}} \rb \right] \,.
\end{eqnarray}
Note that
\begin{equation}
A^{(l+1)'} (0)\,,\quad
B^{(l+1)} (0) \,,\quad
P^{(l+1)} (0)
\end{equation}
are left undetermined as constants of integration.  This is expected,
since $A^{(l+1)'} (0)$ normalizes the coupling $g$, $B^{(l+1)} (0)$
normalizes the field $\phi^i$, and $P^{(l+1)} (0)$ normalizes the
composite operator $\PN$.  For example, we can adopt the convention
\begin{equation}
\frac{\partial}{\partial x} A (0;x)\Big|_{x=0} = - \frac{g}{2}\,,\quad
B(0;0) = 0\,,\quad
P(0;0) = 1\,,
\end{equation}
or equivalently
\begin{equation}
\lb\begin{array}{l}
A^{(0)'} (0) = - \frac{g}{2}\,,\quad A^{(l)'} (0) = 0\,,\quad (l \ge 1)\\
B^{(l)} (0) = 0\,,\quad (l \ge 0)\\
P^{(0)} (0) = 1\,,\quad P^{(l)} (0) = 0\,.\quad (l \ge 0)
\end{array}\right.
\end{equation}
This concludes our perturbative proof of the existence of the O(N)
non-linear sigma model.

\begin{figure}[t]
\begin{center}
\epsfig{file=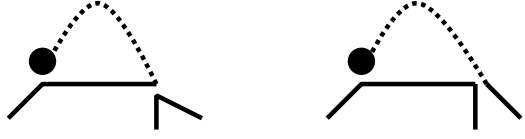, width=7cm}
\caption{The broken curve denotes the cutoff function $K$, and the
  solid the high-momentum propagator. Each curve follows the same O(N)
  index $i = 1,\cdots,N-1$. The left vertex is from $\PN$, and the
  right from $\SIL$.}
\label{WTexamples-ON-jacobian1}
\end{center}
\end{figure}

\begin{figure}[t]
\begin{center}
\epsfig{file=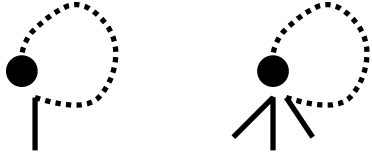, width=5cm}
\caption{These graphs give no contribution to the $u$ coefficients.}
\label{WTexamples-ON-jacobian2}
\end{center}
\end{figure}

Before concluding this example, let us compute $B^{(1)'} (0)$ and
$P^{(1)'} (0)$.  (Note $A^{(1)'} (0) = 0$ by the above
convention.)  For this, we compute the asymptotic behavior of
\begin{equation}
\Sigma_\Lambda^{(1),u} = \int_p \K{p} \frac{\delta \PN^{(0)}
  (p)}{\delta \phi_i (p)} 
\end{equation}
up to third order in $\phi$'s.  The two relevant Feynman
diagrams given in Fig.~\ref{WTexamples-ON-jacobian1}.
Note that the two graphs in Fig.~\ref{WTexamples-ON-jacobian2} give
only the $\Lambda^2$ terms, but no contribution to the $u$ coefficients.
Ignoring the $\Lambda^2$ terms, we obtain
\begin{eqnarray}
\Sigma_\Lambda^{(1),u} &\asym& - \int d^2 x \,\ep_i \phi_i \left[
\partial_\mu \phi_j \partial_\mu \phi_j \, g^2 \left( \frac{1}{4 \pi}
    + \int_q \frac{K(q)(1-K(q))}{q^2} \right)\right.\nn\\
&&\left.\quad + \phi_j \partial^2 \phi_j\, g^2 \int_q \frac{1}{q^2} \left(
    \frac{1}{4} \Delta (q)^2 + 2 K(q)(1-K(q))\right)\right]
\end{eqnarray}
up to third order in $\phi$'s.  Hence,
\begin{equation}
\lb\begin{array}{c@{~=~}l}
u_1 (0) & - g^2 \left( \frac{1}{4 \pi}
    + \int_q \frac{1}{q^2} K(q)(1-K(q))\right)\,,\\
u_2 (0) & - g^2  \int_q \frac{1}{q^2} \left(
    \frac{1}{4} \Delta (q)^2 + 2 K(q)(1-K(q))\right)\,.\\
\end{array}\right.
\end{equation}
This gives
\begin{equation}
\lb\begin{array}{c@{~=~}l}
{B^{(1)}}'(0) & u_2 (0) - u_1 (0) = g^2 \left( \frac{1}{4\pi} - \int_q
    \frac{1}{q^2} \left( K(1-K) + \frac{1}{4} \Delta^2 \right) \right)\,,\\
{P^{(1)}}'(0) & u_2 (0) - 2 u_1 (0) = g^2 \left( \frac{1}{2\pi} -
   \int_q \frac{1}{q^2}  \frac{1}{4} \Delta^2 \right)\,.
\end{array}\right.
\end{equation}

\subsection{The axial symmetry with a massive
  propagator\label{WTexamples-axial}}

We consider a Wilson action for the spin $\frac{1}{2}$ field:
\begin{equation}
\SL = - \int_p \frac{1}{\K{p}} \bar{\psi} (-p) \left( \fmslash{p} + i
    m \right) \psi (p) + \SIL\,,
\end{equation}
where the mass $m$ is non-vanishing.  The free part of the action breaks
the axial symmetry explicitly.  Nevertheless, the breaking can be
compensated by $\SIL$, and the theory can still possess the axial
symmetry.

In the presence of the axial symmetry, the correlation functions in
the continuum limit satisfy the following WT identity:
\begin{eqnarray}
&&\sum_{i=1}^N \left[ \vev{\psi (p_1) \cdots \gamma_5 \psi (p_i) \cdots
      \psi (p_N) \bar{\psi} (-q_1) \cdots \bar{\psi} (-q_N)}^\infty
\right.\nn\\
&&\left.\qquad + \vev{\psi (p_1) \psi (p_i) \cdots
      \psi (p_N) \bar{\psi} (-q_1) \cdots \bar{\psi} (-q_i) \gamma_5
      \bar{\psi} (-q_N)}^\infty \right] = 0\,.
\end{eqnarray}
The WT operator for the U(1) axial symmetry is given by
\begin{eqnarray}
\Sigma_\Lambda &\equiv& \int_p \K{p} \left[ S_\Lambda \Rd{\psi (p)}
    \gamma_5 [\psi] (p) + [\bar{\psi}] (-p) \gamma_5
    \Ld{\bar{\psi} (-p)} S_\Lambda \right.\nn\\
&& \left.- \Tr \gamma_5 \lb [\psi] (p) \Rd{\psi (p)} +
\Ld{\bar{\psi} (-p)} [\bar{\psi}] (-p) \rb \right] = 0\,,
\end{eqnarray}
where 
\begin{equation}
    \lb\begin{array}{c@{~\equiv~}l}
        \left[ \psi \right] (p) & \frac{1}{\K{p}} \psi (p) + \frac{1 -
          \K{p}}{\fmslash{p} + i 
          m} \Ld{\bar{\psi} (-p)} S_{\Lambda} \\
        \left[ \bar{\psi} \right] (-p) & \frac{1}{\K{p}} \bar{\psi}
        (-p) + S_{\Lambda} 
        \Rd{\psi (p)} \frac{1 - \K{p}}{\fmslash{p} + i m}
\end{array}\right.
\end{equation}
are the composite operators corresponding to $\psi, \bar{\psi}$.  
We can further rewrite the above as
\begin{eqnarray}
\Sigma_\Lambda 
&=& \int_p \left[ S_\Lambda \Rd {\psi (p)} \gamma_5 \psi (p) +
    \bar{\psi} (-p) \gamma_5 \Ld{\bar{\psi} (-p)}
    S_\Lambda - \frac{K(1-K)}{p^2 + m^2} (- 2 i
    m) \right.\nn\\ 
&&\quad \left.\times \Tr \gamma_5 \lb
\Ld{\bar{\psi} (-p)} S_\Lambda \cdot S_\Lambda \Rd{\psi (p)} 
+ \Ld{\bar{\psi} (-p)} S_\Lambda  \Rd{\psi (p)} \rb \right]\,.
\label{Sigma for axial sym}
\end{eqnarray}
This must vanish for the axial symmetry.

In the simplest case, the theory is free, and the Wilson action is
quadratic:
\begin{equation}
S_\Lambda = - \int_p \bar{\psi} (-p) D_\Lambda (p) \psi (p)
\end{equation}
Substituting this into $\Sigma_\Lambda = 0$, we obtain the
Ginsparg-Wilson relation\cite{Ginsparg:1981bj}:
\begin{equation}
\lb D_\Lambda (p)~,~ \gamma_5\rb = \frac{K(1-K)}{p^2 + m^2} (- 2 i m)
D_\Lambda (p) \gamma_5 D_\Lambda (p)\,.
\label{GW relation}
\end{equation}
The right-hand side is non-vanishing only for $p^2 > \Lambda^2$ if we
choose $\K{p} = 1$ for $p^2 < \Lambda^2$.
Expanding
\begin{equation}
D_\Lambda (p) = \fmslash{p} A_\Lambda (p) + i m B_\Lambda (p)
\end{equation}
with scalar coefficients, the above relation gives
\begin{equation}
B_\Lambda (p)  = \frac{K(1-K)}{p^2 + m^2} \left( p^2 A_\Lambda (p)^2
    + m^2 B_\Lambda (p)^2 \right)\,.
\label{WTexamples-axial-ABGW}
\end{equation}
Moreover, the Polchinski equation for $S_\Lambda$ gives
\begin{equation}
\Lambda \frac{\partial}{\partial \Lambda} D_\Lambda (p)
= - 2 \frac{\Delta}{K} D + \frac{\Delta}{p^2 + m^2} D (\fmslash{p} - i
m) D\,,
\end{equation}
which implies
\begin{subequations}
\label{WTexamples-axial-ABERG}
\begin{eqnarray}
\Lambda \frac{\partial}{\partial \Lambda} A_\Lambda  &=& - 2
\frac{\Delta}{K} A_\Lambda + \frac{\Delta}{p^2 + m^2} \left( p^2 A_\Lambda^2 -
    m^2 B_\Lambda^2 + 2 A_\Lambda B_\Lambda m^2 \right)\,,\\ 
\Lambda \frac{\partial}{\partial \Lambda} B_\Lambda &=& - 2
\frac{\Delta}{K} B_\Lambda + \frac{\Delta}{p^2 + m^2} \left( - p^2 A_\Lambda^2
    + m^2 B_\Lambda^2 + 2 A_\Lambda B_\Lambda p^2 \right)\,.
\end{eqnarray}
\end{subequations}

To obtain a general solution, we cheat a little.  In this case we know
the general solution for the continuum limit:
\begin{equation}
    \vev{\psi (p) \bar{\psi} (-p)}^\infty = \frac{c(p)}{\fmslash{p}}\,,
\end{equation}
where $c$ is a scalar function, independent of $\Lambda$.\footnote{We
assume parity.}  Using
\begin{equation}
\vev{\psi (p) \bar{\psi} (-p)}^\infty = \frac{1}{\K{p}^2} \vev{\psi
  (p) \bar{\psi} (-p)}_{S_\Lambda} + \frac{1 - 1/\K{p}}{\fmslash{p} + i
  m}\,,
\end{equation}
we obtain
\begin{equation}
\frac{c (p)}{\fmslash{p}} = \frac{1}{\K{p}^2}
\frac{1}{D_\Lambda (p)} + \frac{1 - 1/\K{p}}{\fmslash{p} + i m}\,.
\end{equation}
Hence,
\begin{eqnarray}
\frac{1}{D_\Lambda (p)} &=& K(p)^2 \left( \frac{c (p)}{\fmslash{p}} -
    \frac{1 - 1/\K{p}}{\fmslash{p} + i m} \right)\nn\\
&=& K^2 \lb c \left( \fmslash{p} + i m \right) -
    \left( 1 - 1/K \right) \fmslash{p} \rb \frac{1}{\fmslash{p}
  (\fmslash{p} + i m)}\nn\\
&=& K^2 \lb (c - 1 + 1/K) \fmslash{p} + i m c \rb \frac{1}{\fmslash{p}
  (\fmslash{p} + i m)}\,.
\end{eqnarray}
Inverting this, we obtain
\begin{eqnarray}
D_\Lambda (p) &=& \frac{1}{K^2} \frac{p^2 + i m
  \fmslash{p}}{(c-1+1/K)^2 p^2 + m^2 c^2} \lb (c-1+1/K) \fmslash{p} -
i m c \rb 
\nn\\
&=& \frac{1}{K^2} \frac{ \left( p^2 (c-1+1/K) + m^2 c \right)
  \fmslash{p} - i m c p^2 + i m p^2 (c-1+1/K)}{(c-1+1/K)^2 p^2 + m^2
  c^2} \nn\\
&=& \frac{1}{K} \frac{\left( p^2 \left( K c + 1 - K \right) + K c m^2
  \right) \fmslash{p} + i m p^2 (1 - K)}{\left(K c + 1 - K\right)^2 p^2 +
      K^2 m^2 c^2}\,.
\end{eqnarray}
This gives
\begin{subequations}
\begin{eqnarray}
A_\Lambda (p) &=& \frac{1}{K} \frac{p^2 (K c + 1 - K) + m^2 K c}{ p^2 (K c
  + 1 - K)^2 + m^2 K^2 c^2}\\
B_\Lambda (p) &=& \frac{1}{K} \frac{p^2 (1-K)}{ p^2 (K c
  + 1 - K)^2 + m^2 K^2 c^2}
\end{eqnarray}
\end{subequations}
which satisfy both the Polchinski equation \bref{WTexamples-axial-ABERG}
and the Ginsparg-Wilson relation \bref{WTexamples-axial-ABGW}.  For
small and large momenta, $D_\Lambda (p)$ behaves as
\begin{equation}
D_\Lambda (p) \lb \begin{array}{l@{\quad}l}
= \frac{\fmslash{p}}{c(p)}\,,& (p^2 < \Lambda^2)\\
\simeq \frac{1}{\K{p}} \left( \fmslash{p} + i m \right)\,.& (p^2 \gg
\Lambda^2)
\end{array}\right.
\end{equation}
For the simplest case
\begin{equation}
c (p) = 1\,,
\end{equation}
we obtain
\begin{eqnarray}
D_\Lambda (p) &=& \frac{1}{K} \frac{\left( p^2 + K m^2 \right)
  \fmslash{p} + i m p^2 (1-K)}{p^2 + K^2 m^2}\nn\\
&=& \frac{\fmslash{p} + i m}{K} + \frac{(1-K) m^2 \fmslash{p} - i m (p^2 + m^2
  K)}{p^2 + m^2 K^2}\,.
\end{eqnarray}

In the above, we have seen that the Wilson action of a fully interacting
theory must satisfy \bref{Sigma for axial sym} (or its analog), which
reduces to the GW relation \bref{GW relation} only for free theories.
However, it is possible to build up a fully interacting theory starting
from \bref{GW relation}.\cite{Ichinose:1999ke, Igarashi:2001cv}
Particular solutions of \bref{Sigma for axial sym} that correspond to
interacting theories have been constructed explicitly in
\citen{Igarashi:2002bs,Igarashi:2002ba}.

\newpage

\section{Realization of symmetry in the antifield formalism\label{AF}}

The purpose of this section is to introduce a method of symmetry
realization more powerful than the WT identity discussed in the previous
two sections.  The method is an adaptation of the antifield formalism of
Batalin and
Vilkovisky\cite{Batalin:1981jr,BVreview-Gomis,Henneaux-Teitelboim} to
the ERG formulation of field
theory.\cite{Igarashi:1999rm,Igarashi:2000vf,Igarashi:2001jq,Igarashi:2001mf}
We introduce only what is needed for perturbative construction of
theories, postponing more general discussions till \S \ref{BV}.  Most
results of this section were first obtained by
Becchi.\cite{Becchi:1996an}.  Our presentation here follows \S\S 6,7 of
\citen{Becchi:1996an}.\footnote{For a formulation with the effective
average action, see \citen{Pernici:1997ie}, for example.}

\subsection{Quantum master equation\label{AF-def}}

We introduce an external source $\phi^*$ in order to generate the
symmetry transformation\footnote{The symmetry can be any continuous
symmetry, either local or global.  For the application of the BV
formalism to global symmetries, see
Refs.~\citen{Brandt:1996uv,Brandt:1997cz} for example.}  of $\phi$ as
the functional derivative of the Wilson action with respect to $\phi^*$.
Following the general method by Batalin and Vilkovisky, we give $\phi^*$
the statistics opposite to that of $\phi$.  In what follows, we treat
$\phi$ as a generic bosonic field (scalars and vectors) and $\phi^*$ as
the corresponding fermionic antifield.  Extension to a fermionic field
and its bosonic antifield is straightforward.

In the antifield formalism, the transformation of $\phi$ is called the
\textbf{BRST transformation},\footnote{For YM theories, this BRST
  transformation coincides with the ordinary BRST transformation in the
  limit of vanishing antifields.} and it is given as the derivative of
the Wilson action with respect to the corresponding antifield:
\begin{equation}
\delta \phi (p) = \Ld{\phi^* (-p)} \bSL\,.
\end{equation}
To show the presence of antifields explicitly, we use a bar above the
Wilson action and other related quantities.  At the vanishing
antifields, the above transformation reduces to the infinitesimal
transformation for the WT identity.  The WT composite operator is
generalized to the \textbf{quantum master operator}
\begin{equation}
\bar{\Sigma}_\Lambda \equiv \int_p \K{p} \left[ \frac{\delta
      \bSL}{\delta \phi (p)} \cdot \Ld{\phi^* (-p)} \bSL
+ \Ld{\phi^* (-p)} \frac{\delta \bSL}{\delta \phi (p)} \right]\,,
\label{AF-Sigmabar}
\end{equation}
which reduces to the WT composite operator at $\phi^*=0$.  Note that
$\bar{\Sigma}_\Lambda$ is a composite operator.  (Quick proof)
$\Ld{\phi^* (-p)} \bSL$ is a composite operator of type
(\ref{comp-dSdJ}).  Hence, $\bar{\Sigma}_\Lambda$ is a composite
operator of type (\ref{comp-dOdphi}).

We now replace the WT identity by the \textbf{quantum master
  equation}:
\begin{equation}
\bar{\Sigma}_\Lambda = 0\,,\label{AF-qme}
\end{equation}
which reduces to the WT identity at the vanishing antifields.
The quantum master equation implies, for the correlation functions,
the following identities:
\begin{eqnarray}
&&\sum_{i=1}^n \vev{\phi (p_1) \cdots \delta \phi (p_i)
     \cdots \phi (p_n)}^\infty \nn\\
&&= \sum_{i=1}^n \Ld{\phi^* (-p_i)} \vev{\phi (p_1) \cdots \widehat{\phi
    (p_i)} \cdots \phi (p_n)}^\infty = 0\,.
\end{eqnarray}

Given a functional $\Op_\Lambda [\phi, \phi^*]$ in general, we define
its \textbf{quantum BRST transformation} by
\begin{eqnarray}
\delta_Q \Op_\Lambda &\equiv& \int_p \K{p} \left[ \frac{\delta
      \bSL}{\delta \phi (p)} 
    \cdot \Ld{\phi^* (-p)} \Op_\Lambda + \Ld{\phi^* (-p)} \bSL \cdot
    \frac{\delta}{\delta \phi (p)} \Op_\Lambda \right.\nn\\
&&\qquad\qquad\qquad \left.+
\Ld{\phi^* (-p)} \frac{\delta}{\delta \phi (p)} \Op_\Lambda
\right]\,.\label{AF-deltaQ}
\end{eqnarray}
This definition is obtained from $\bar{\Sigma}_\Lambda$ by changing
$\bSL$ infinitesimally by $\Op_\Lambda$.  For example, we obtain
\begin{equation}
\delta_Q \phi (p) = \K{p} \Ld{\phi^* (-p)} \bSL\,,
\end{equation}
so that we can write
\begin{equation}
\bar{\Sigma}_\Lambda = \int_p \left[ \frac{\delta \bSL}{\delta \phi
      (p)} \delta_Q \phi (p) + \Ld{\phi (p)} \delta_Q \phi (p)
\right]\,.
\end{equation}
The first term is the change of $\bSL$ under the infinitesimal field
transformation $\delta_Q \phi$, and the second term is the jacobian.

Suppose $\Op_\Lambda$ is a composite operator.  Then,
\begin{equation}
\Op'_\Lambda \equiv \K{p} \left( \frac{\delta \bSL}{\delta \phi (p)}
    \Op_\Lambda + \frac{\delta \Op_\Lambda}{\delta \phi (p)} \right)
\end{equation}
is a composite operator of type (\ref{comp-dOdphi}).  We then note
that
\begin{equation}
\Op''_\Lambda \equiv \Ld{\phi^* (-p)} \bSL \cdot \Op'_\Lambda +
\Ld{\phi^* (-p)} \Op'_\Lambda
\end{equation}
is a composite operator of type (\ref{comp-dOdJ}). Since
\begin{equation}
\delta_Q \Op_\Lambda = \int_p \Op''_\Lambda - \bar{\Sigma}_\Lambda
\cdot \Op_\Lambda\,,
\end{equation}
$\delta_Q \Op_\Lambda$ is also a composite operator, if
$\bar{\Sigma}_\Lambda = 0$.  If $\bar{\Sigma}_\Lambda \ne 0$, $\delta_Q
\Op_\Lambda$ is not a composite operator in general.

Note also that if $\bar{\Sigma}_\Lambda = 0$, we can rewrite
\begin{equation}
\delta_Q \Op_\Lambda = \exp \left[ - \bSL \right] \int_p \K{p}
\frac{\delta}{\delta \phi 
  (p)} \Ld{\phi^* (-p)} \left( \exp \left[ \bSL \right] \Op_\Lambda \right)
\end{equation}
for an arbitrary functional $\Op_\Lambda$, not necessarily a composite
operator.  Since
\begin{equation}
\int_p \K{p} \frac{\delta}{\delta \phi (p)} \Ld{\phi^* (-p)} 
\end{equation}
is a fermionic differential operator acting on the functionals of $\phi$
and $\phi^*$, its square vanishes:
\begin{equation}
\delta_Q \delta_Q \Op_\Lambda = \exp \left[ - \bSL \right]
\left( \int_p \K{p} \frac{\delta}{\delta \phi 
  (p)} \Ld{\phi^* (-p)} \right)^2 \left( \exp \left[ \bSL \right]
\Op_\Lambda \right) = 0\,.
\end{equation}
Thus, we obtain the nilpotency:
\begin{equation}
\delta_Q^2 = 0\,,
\end{equation}
if $\bar{\Sigma}_\Lambda = 0$.

To summarize, the quantum master equation (\ref{AF-qme}), i.e.,
$\bar{\Sigma}_\Lambda = 0$, implies the following:
\begin{itemize}
\item[(i)] $\delta_Q \Op_\Lambda$ is a composite operator, if
    $\Op_\Lambda$ is.
\item[(ii)] $\delta_Q$ is nilpotent, $\delta_Q^2 = 0$.
\end{itemize}
As an example of (i), we find, under the assumption
$\bar{\Sigma}_\Lambda = 0$,
\begin{equation}
\delta_Q [\phi]_\Lambda (p) =  \Ld{\phi^* (-p)} \bSL\,,
\end{equation}
which is a composite operator.

Finally, we come to an important algebraic identity
\begin{equation}
\delta_Q \bar{\Sigma}_\Lambda = 0\,.\label{AF-identity}
\end{equation}
This simply follows from the definitions (\ref{AF-Sigmabar}) of
$\bar{\Sigma}_\Lambda$ and (\ref{AF-deltaQ}) of $\delta_Q$.  The
identity is valid no matter what $\bSL$ is.  Eq.~(\ref{AF-identity}) is
the algebraic structure we have been missing in our discussion of the WT
identity in \S \ref{WT}.  We will see, in \S\ref{AF-perturbation}, how
useful this identity is for the realization of continuous symmetry in
the ERG approach.

\subsection{Outline of perturbative construction\label{AF-perturbation}}

We now outline a perturbative construction of the theory that
satisfies the quantum master equation (\ref{AF-qme}).  Let us first
introduce loop expansions:
\begin{equation}
\bSL = \sum_{l=0}^\infty \bSL^{(l)},\quad
\bar{\Sigma}_\Lambda = \sum_{l=0}^\infty
\bar{\Sigma}_\Lambda^{(l)}\,.
\end{equation}
If the theory is renormalizable, the asymptotic behavior of the Wilson
action is given in the form
\begin{equation}
\bSIL \asym \int d^D x \,\sum_i \Lambda^{D-x_i} c_i (\ln \Lambda/\mu)
\,\Op_i \,,
\end{equation}
where $\Op_i$ is a local polynomial of $\phi, \phi^*$, and their
derivatives, and $x_i \le D$ is the scale dimension of $\Op_i$.  The
Wilson action is parameterized only by $c_i (0)$ that has $x_i = D$. We
introduce the loop expansion of $c_i (0)$ as
\begin{equation}
c_i (0) = \sum_{l=0}^\infty c_i^{(l)}\,.
\end{equation}

Our starting point is the classical action $\bScl$ which is
independent of the cutoff $\Lambda$.  Since the interaction part of
the tree level action $\bSL^{(0)}$ is the sum of tree graphs
consisting of the vertices of $\bScl$ and high-momentum
propagators, we obtain
\begin{equation}
\bScl = \lim_{\Lambda \to \infty} \bSL^{(0)}\,.
\label{AF-limit}
\end{equation}
We assume that this satisfies the classical master equation:
\begin{equation}
\bar{\Sigma}_{cl} \equiv \int_p \Ld{\phi^* (-p)} \bScl \cdot
\frac{\delta \bScl}{\delta \phi (p)} = 0\,.\label{AF-clme}
\end{equation}
Since $\bar{\Sigma}_{cl}$ is the asymptotic part of
$\bar{\Sigma}_\Lambda^{(0)}$, we obtain
\begin{equation}
\bar{\Sigma}_\Lambda^{(0)} = 0\,.
\end{equation}

Now, our induction hypothesis is
\begin{equation}
\bar{\Sigma}_\Lambda^{(0)} = \cdots = \bar{\Sigma}_\Lambda^{(l)} =
0\,.\label{AF-indhypo}
\end{equation}
For this we have adjusted the parameters of the theory up to $l$-loop
order.  In other words we have adjusted 
\begin{equation}
c_i^{(0)}, \cdots, c_i^{(l)}\,.\quad (x_i = D)
\end{equation}
Since $\bar{\Sigma}_\Lambda$ is a composite operator satisfying the
ERG differential equation, the induction hypothesis implies that
$\bar{\Sigma}_\Lambda^{(l+1)}$ is independent of $\Lambda$ for large
$\Lambda$.  In the following we wish to show that we can fine-tune
$c_i^{(l+1)}$ so that
\begin{equation}
    \lim_{\Lambda \to \infty} \bar{\Sigma}_\Lambda^{(l+1)} =
     0\,,\label{AF-toshow} 
\end{equation}
which is equivalent to
\begin{equation}
 \bar{\Sigma}_\Lambda^{(l+1)} = 0\,.
\end{equation}
Hence, (\ref{AF-toshow}) completes the perturbative construction by
mathematical induction.

To show the possibility of such fine-tuning, we must examine the
structure of $\bar{\Sigma}_\Lambda^{(l+1)}$ further:
\begin{eqnarray}
\bar{\Sigma}_\Lambda^{(l+1)}
&=& \int_p \K{p} \left[
\sum_{k=0}^{l+1} \Ld{\phi^* (-p)} \bSL^{(k)} \cdot
\frac{\delta \bSL^{(l+1-k)}}{\delta \phi (p)} + \Ld{\phi^* (-p)}
\frac{\delta \bSL^{(l)}}{\delta \phi (p)} \right]\nn\\
&=& \bar{\Sigma}_\Lambda^{(l+1),t} + \bar{\Sigma}_\Lambda^{(l+1),u}\,,
\end{eqnarray}
where
\begin{eqnarray}
\bar{\Sigma}_\Lambda^{(l+1),t} &=&
\int_p \K{p} \left[\Ld{\phi^* (-p)} \bScl \cdot
\frac{\delta \bSL^{(l+1)}}{\delta \phi (p)}
+ \Ld{\phi^* (-p)} \bSL^{(l+1)} \cdot
\frac{\delta \bScl}{\delta \phi (p)} \right]\,,\label{AF-Sigmabar-t}\\
\bar{\Sigma}_\Lambda^{(l+1),u}  &=& \int_p \K{p}
\left[ \sum_{k=1}^{l} \Ld{\phi^* (-p)} \bSL^{(k)} \cdot
\frac{\delta \bSL^{(l+1-k)}}{\delta \phi (p)} 
+ \Ld{\phi^* (-p)}
\frac{\delta \bSL^{(l)}}{\delta \phi (p)}\right]\,.\label{AF-Sigmabar-u}
\end{eqnarray}
For large $\Lambda$, $\bar{\Sigma}_\Lambda^{(l+1)}$ is independent of
$\Lambda$.  Hence, the $\Lambda$ dependence of
$\bar{\Sigma}_\Lambda^{(l+1),t}$ and $\bar{\Sigma}_\Lambda^{(l+1),u}$
must cancel for large $\Lambda$.  Note that only
$\bar{\Sigma}_\Lambda^{(l+1),t}$ depends on $c_i^{(l+1)}$, and
$\bar{\Sigma}_\Lambda^{(l+1),u}$ is completely determined by the lower
loop parameters.

For later convenience, we introduce the following notation.  We first
denote
\begin{equation}
\bar{\Sigma}_\infty^{(l+1)} \equiv \lim_{\Lambda \to \infty}
\bar{\Sigma}_\Lambda^{(l+1)} \,.
\end{equation}
We then denote the $\Lambda$ independent part (namely the part that does
not vanish at $\ln \Lambda/\mu = 0$) of the asymptotic behavior of
$\bSL^{(l+1)}$ by $\bSinf^{(l+1)}$ so that
\begin{equation}
\bSinf^{(l+1)} = \int d^D x \, \sum_i c_i^{(l+1)}\, \Op_i\,.
\label{AF-bSinf}
\end{equation}
Similarly, we denote the $\Lambda$ independent parts of
$\bar{\Sigma}_\Lambda^{(l+1),t}$ and $\bar{\Sigma}_\Lambda^{(l+1),u}$
by $\bar{\Sigma}_\infty^{(l+1),t}$, $\bar{\Sigma}_\infty^{(l+1),u}$,
respectively.  In defining these, we ignore the part multiplied by
positive powers of either $\Lambda$ or $\ln \Lambda/\mu$. By
definition, we obtain
\begin{equation}
\bar{\Sigma}_\infty^{(l+1)} ~=~ \bar{\Sigma}_\infty^{(l+1), t} ~+~
\bar{\Sigma}_\infty^{(l+1), u} \,.\label{AF-Sigmabar-stu}
\end{equation}

Now, for large $\Lambda$, (\ref{AF-Sigmabar-t}) gives
\begin{equation}
\bar{\Sigma}_\infty^{(l+1),t} = \int d^D x \left[
\Ld{\phi^*} \bScl \cdot \frac{\delta \bSinf^{(l+1)}}{\delta \phi} +
\Ld{\phi^*} \bSinf^{(l+1)} \cdot \frac{\delta \bScl}{\delta \phi}
\right]\,.
\end{equation}
Introducing the classical BRST transformation
\begin{equation}
\delta_{cl} \equiv \int d^D x \left[ \Ld{\phi^*} \bScl \cdot
    \frac{\delta}{\delta \phi} + \frac{\delta \bScl}{\delta \phi}
    \Ld{\phi^*}  \right]\,,\label{AF-deltacl}
\end{equation}
we can rewrite the above as
\begin{equation}
\bar{\Sigma}_\infty^{(l+1), t} = \delta_{cl} \bSinf^{(l+1)}\,.
\end{equation}
The classical master equation (\ref{AF-clme}) implies the nilpotency
of the classical BRST transformation:
\begin{equation}
\delta_{cl}^2 = 0\,.\label{AF-classicalnilpotency}
\end{equation}
Hence, we obtain
\begin{equation}
\delta_{cl} \bar{\Sigma}_\infty^{(l+1), t} = 0\,.\label{AF-deltaSigma-t}
\end{equation}

We now consider the algebraic identity (\ref{AF-identity}).
At $(l+1)$-loop, this gives
\begin{equation}
\int_p \K{p} \left[
\Ld{\phi^* (-p)} \bSL^{(0)} \cdot
    \frac{\delta}{\delta \phi (p)} \bar{\Sigma}_\Lambda^{(l+1)} 
+ \frac{\delta}{\delta \phi (p)} \bSL^{(0)} \cdot
 \Ld{\phi^* (-p)}    \bar{\Sigma}_\Lambda^{(l+1)} \right] = 0
\end{equation}
due to the induction hypothesis (\ref{AF-indhypo}).
Considering large $\Lambda$, we obtain
\begin{equation}
\delta_{cl} \bar{\Sigma}_\infty^{(l+1)} = 0\,.
\end{equation}
This gives
\begin{equation}
\delta_{cl} \bar{\Sigma}_\infty^{(l+1), u} =
0\,, \label{AF-Sigmabar-u-closed} 
\end{equation}
using (\ref{AF-Sigmabar-stu}) and (\ref{AF-deltaSigma-t}).

To summarize so far, our goal is to show that we can fine-tune the
$(l+1)$-loop parameters $c_i^{(l+1)}$ so that
$\bar{\Sigma}_\infty^{(l+1)} = 0$.  This condition is equivalent to
solving
\begin{equation}
\delta_{cl} \bSinf^{(l+1)} = - \bar{\Sigma}_\infty^{(l+1), u}\,,
\label{AF-tosolve}
\end{equation}
where 
\begin{enumerate}
\item $\delta_{cl}$, defined by (\ref{AF-deltacl}), satisfies nilpotency
      (\ref{AF-classicalnilpotency}),
\item $\bar{\Sigma}_\infty^{(l+1), u}$ satisfies
      (\ref{AF-Sigmabar-u-closed}), i.e., closed under $\delta_{cl}$.
\item $\bSinf^{(l+1)}$ is given by (\ref{AF-bSinf}), the most general
local functional of fields with dimension $D$.  
\end{enumerate}
Thus, we have converted the problem of perturbative construction into a
question of classical algebra.\footnote{This is called the BRST
cohomology.}  For \bref{AF-tosolve} to have a solution, it is sufficient
that $\delta_{cl} \Op = 0$ implies $\Op = \delta_{cl} \Op'$; using a
more proper mathematical language, the BRST cohomology defined by
$\delta_{cl}$ must be trivial.

The final algebraic question \bref{AF-tosolve} does not have any general
solution.  The answer depends on the the set of available fields and
antifields and the choice of the classical action $\bScl$.  But it is
clear what we must do.  We first consider the most general
$\bar{\Sigma}_\infty^{(l+1), u}$ that satisfies
(\ref{AF-Sigmabar-u-closed}), using the available set of fields and
antifields.  If it can be given as $\delta_{cl} \bSinf^{(l+1)}$, the
theory can be constructed.  If not, the symmetry is anomalous, and we
cannot construct a symmetric quantum theory based on the classical
action.

For YM theories, the algebraic problem has a well-known solution.
We refer the reader to
\citen{Barnich:1994ve,Barnich:1994mt,Barnich:1994db,Barnich:2000zw}.
In \S\ref{AFexamples-YM}, we show a concrete solution of \bref{AF-tosolve}
for the SU(2) gauge theory.

\newpage

\section{Examples\label{AFexamples}}

\subsection{QED\label{AFexamples-QED}}

In \S \ref{WTexamples-QED} we have already shown how to construct QED
with the help of a WT composite operator.  Though we do not need the
antifield formalism for assurance of the theory's existence, it is
instructive to see how the formalism applies to QED.

Our starting point is the classical action:
\begin{equation}
\bScl = S_{cl} + \int d^4 x\, \left[ A_\mu^*
    \frac{1}{i} \partial_\mu c + \frac{1}{\xi} \bar{c}^*
    \frac{1}{i} \partial_\mu A_\mu 
+ \frac{1}{2 \xi} \left(\bar{c}^*\right)^2 +
e \bar{\psi}^* c \psi + e \bar{\psi} c \psi^* \right]\,,
\end{equation}
where $S_{cl}$ is given by (\ref{WTexamples-QED-Scl}).  Here, the
fields with the superscript ${}^*$ are antifields that have the
opposite statistics to the respective conjugate fields.  Let us
tabulate them:
\begin{center}
\begin{tabular}{c|ccc}
field& antifield& statistics& dimension\\
\hline
$A_\mu$& $A_\mu^*$& Fermi& $2$\\
$\bar{c}$& $\bar{c}^*$& Bose& $2$\\
$\psi$& $\bar{\psi}^*$& Bose& $\frac{3}{2}$\\
$\bar{\psi}$& $\psi^*$& Bose& $\frac{3}{2}$\\
\hline
\end{tabular}
\end{center}
\vspace{0.2cm}
The classical action satisfies the following classical master
equation:
\begin{eqnarray}
    &&\int d^4 x \left[ \frac{\delta \bScl}{\delta A_\mu}
        \Ld{A_\mu^*} \bScl + \frac{\delta \bScl}{\delta
          \bar{c}^*} \Ld{\bar{c}} \bScl \right.\nn\\
    &&\left.\quad + \bScl \Rd{\psi} \cdot \Ld{\bar{\psi}^*} \bScl
        + \bScl \Rd{\psi^*} \cdot \Ld{\bar{\psi}} \bScl \right]
    = 0\,.
\end{eqnarray}
Except for the term quadratic in $\bar{c}^*$, the antifield dependence
of the action is determined so that the derivative of the action with
respect to the antifield gives the BRST transformation of the
corresponding field:
\begin{equation}
\lb\begin{array}{c@{~=~}l}
\delta A_\mu & \Ld{A_\mu^*} \bScl = \frac{1}{i} \partial_\mu
c\,,\\
\delta \bar{c}& \frac{\delta \bScl}{\delta \bar{c}^*} =
\frac{1}{\xi} \left( \frac{1}{i} \partial_\mu A_\mu + \bar{c}^*
\right)\,,\\
\delta \psi& \Ld{\bar{\psi}^*} \bScl = e c \psi\,,\\
\delta \bar{\psi}& \bScl \Rd{\psi^*} = - e c\bar{\psi}\,.
\end{array}\right.
\end{equation}
We note that all the antifields are coupled linearly to the dynamical
fields.  The $\psi^*, \bar{\psi}^*$ are coupled to the quadratic
fields $\bar{\psi}c, c \psi$, respectively, but these are basically
linear since the free field $c$ acts almost like an external field.

Let us consider the dependence of the Wilson action on each antifield.
\begin{itemize}
\item[(i)] $A_\mu^*$ --- This is coupled linearly to $A_\mu^*$,
    contributing a term
\begin{equation}
\int_k A_\mu^* (-k) k_\mu c (k)
\end{equation}
to the Wilson action.  Since $c$ is free, no further $A_\mu^*$
dependence is generated.
\item[(ii)] $\bar{c}^*$ --- The linear coupling
\begin{equation}
\int_k \bar{c}^* (-k) \frac{1}{\xi} k_\mu A_\mu (k)
\end{equation}
amounts to an external source
\begin{equation}
\frac{1}{\xi} k_\mu \bar{c}^* (-k)
\end{equation}
coupled linearly to the gauge field $A_\mu (k)$.  From the general
result (\ref{derivation-Jshift}), the linear coupling generates the
quadratic term:
\begin{eqnarray}
&&\frac{1}{2} \int_k \frac{1}{\xi} k_\mu \bar{c}^* (-k) \frac{1 -
  \K{k}}{k^2} \left( \delta_{\mu\nu} - (1-\xi) \frac{k_\mu k_\nu}{k^2}
\right) \frac{1}{\xi} (-k_\nu) \bar{c}^* (k)\nn\\
&& = - \frac{1}{2 \xi} \int_k \left( 1 - \K{k}\right) \bar{c}^* (-k)
\bar{c}^* (k)\,.
\end{eqnarray}
Furthermore, the shift 
\begin{eqnarray}
A_\mu (k) &\longrightarrow& A_\mu^{sh} (k) \equiv A_\mu (k) + \frac{1
  - \K{k}}{k^2} \left( 
    \delta_{\mu\nu} - (1-\xi) \frac{k_\mu k_\nu}{k^2} \right)
\frac{1}{\xi} (- k_\nu) \bar{c}^* (k)\nn\\
&&= A_\mu (k) - \frac{1 - \K{k}}{k^2} k_\mu \bar{c}^* (k)
\label{AFexamples-QED-Ashift}
\end{eqnarray}
is generated in the interaction part $\SIL$ of the Wilson action.
\item[(iii)] $\psi^*, \bar{\psi}^*$ --- We can regard the free field
    $c$ as part of the external source coupled to $\psi, \bar{\psi}$:
\begin{equation}
\int_p \left( e \int_k \bar{\psi}^* (-p-k) c (k) \cdot \psi (p)
+ \bar{\psi} (-p) \cdot e \int_k c (k) \psi^* (p-k) \right)\,.
\end{equation}
Hence, from the general result (\ref{derivation-Jshift}), we find
first that the quadratic term
\begin{equation}
\int_p  e \int_k \bar{\psi}^* (-p-k) c (k) \cdot
\frac{1 - \K{p}}{\fmslash{p} + i m} \cdot
\int_q e \int_l c(l) \psi^* (p-l)
\end{equation}
is generated, and second that the fields are shifted as
\begin{equation}
\lb\begin{array}{c@{~\longrightarrow~}l}
\psi (p)& \psi^{sh} (p) \equiv \psi (p) + \frac{1-\K{p}}{\fmslash{p} +
  i m} e \int_k c(k) \psi^* (p-k)\,,\\
\bar{\psi} (-p)& \bar{\psi}^{sh} (-p) \equiv \bar{\psi} (-p) + e
\int_k \bar{\psi}^* (-p-k) c(k) \frac{1 - \K{p}}{\fmslash{p} + i m}\,,
\end{array}\right.\label{AFexamples-QED-psishift}
\end{equation}
in $\SIL$.
\end{itemize}

Altogether, we obtain the following action in the presence of
antifields:\cite{Igarashi:2007fw}\cite{Higashi:2007ax}
\begin{eqnarray}
    \bSL &=& S_{F,\Lambda} \left[ A_\mu, c, \bar{c}, \psi, \bar{\psi}
    \right]\nn\\
    && + \int_k \left( A_\mu^* (-k) k_\mu c (k) +
    \bar{c}^* (-k) \frac{1}{\xi} k_\mu A_\mu (k) \right)
    + \frac{1}{2 \xi} \int_k \K{k} \bar{c}^* (-k) \bar{c}^* (k)\nn\\
    && + \int_{p,k} \left( e \bar{\psi}^* (-p-k) c (k) \cdot \psi (p)
+ \bar{\psi} (-p) \cdot e c (k) \psi^* (p-k) \right)\nn\\
&& + \int_{p,k,l}  e \bar{\psi}^* (-p-k) c (k) \cdot
\frac{1 - \K{p}}{\fmslash{p} + i m} \cdot
e c(l) \psi^* (p-l)\nn\\
&& + \SIL \left[ A^{sh}_\mu (k)\,,\, 
\psi^{sh} (p)\,,\,
\bar{\psi}^{sh} (-p) \right]\,,\label{AFexamples-QED-bSL}
\end{eqnarray}
where $S_{F,\Lambda}$ is the free part defined by
(\ref{WTexamples-QED-SFL}), and $\SIL$ is the interaction part
constructed in \S \ref{WTexamples-QED}. The fields are shifted in the
interaction part.  If $\SL = S_{F,\Lambda} + \SIL$, without shifts,
satisfies the Polchinski equation for QED, so does $\bSL$ given
above.\footnote{In \S\ref{BVexamples-QED}, we rederive this result using
a functional method.}

We now wish to show that $\bSL$ thus constructed indeed satisfies the
quantum master equation.  The quantum master operator is defined by
\begin{eqnarray}
\bar{\Sigma}_\Lambda &\equiv& \int_k \K{k} \left[ \Ld{A_\mu^* (-k)} \bSL \cdot
        \frac{\delta \bSL}{\delta A_\mu (k)} + \frac{\delta
          \bSL}{\delta \bar{c}^* (k)}\cdot \Ld{\bar{c}(-k)} \bSL
    \right]\nn\\
    && - \int_p \K{p} \Tr \left[ \Ld{\bar{\psi}^* (-p)} \bSL \cdot
        \bSL \Rd{\psi (p)}
        + \Ld{\bar{\psi}^* (-p)} \bSL \Rd{\psi (p)}\right]\nn\\
&& + \int_p \K{p} \Tr \left[ \Ld{\bar{\psi}(-p)} \bSL \cdot \bSL
    \Rd{\psi^* (p)} + \Ld{\bar{\psi}(-p)} \bSL
    \Rd{\psi^* (p)} \right]\,.
\end{eqnarray}
We now substitute (\ref{AFexamples-QED-bSL}) into the above to express
it in terms of $\SIL$.  We skip the intermediate steps which are long
but straightforward.\footnote{The details of a similar calculation are
given for the O(N) linear sigma model in Appendix \ref{AFexamples-ON}.}
Toward the end, we use the anticommuting nature of $c$:
\begin{equation}
\int_{k,l} c(k) c(l) F(k,l) = 0
\end{equation}
for any symmetric integration kernel $F(k,l) = F(l,k)$.  We finally
obtain the result
\begin{equation}
\bar{\Sigma}_\Lambda = \Sigma_\Lambda \left[ A^{sh}_\mu, c, \psi^{sh},
    \bar{\psi}^{sh} \right]\,,
\label{AFexamples-QED-SigmabarSigma}
\end{equation}
where $\Sigma_\Lambda$ is the WT composite operator of QED defined in
\S \ref{WTexamples-QED}, and the fields are shifted as
(\ref{AFexamples-QED-Ashift}) and
(\ref{AFexamples-QED-psishift}). Since $\Sigma_\Lambda = 0$ for
arbitrary field variables, we obtain
\begin{equation}
\bar{\Sigma}_\Lambda = 0\,.
\end{equation}

\subsection{YM theories\label{AFexamples-YM}}

For YM theories, the antifield formalism is essential for proving the
possibility of their construction.  We recall, from
\S\ref{WTexamples-YM}, that we still have to derive the algebraic
relations (\ref{WTexamples-YM-srelations}), necessary in order to show
the possibility of realizing $\Sigma_\Lambda = 0$ by fine-tuning the
parameters.  The first application of the ERG formulation and antifield
formalism was done by Becchi\cite{Becchi:1996an}.  This work was
extended subsequently in
\citen{Bonini:1993sj,Bonini:1994dz,Bonini:1994kp,Bonini:1995tx,Bonini:2000wr}
among others.  Our presentation here follows closely \S 7 of
Ref.~\citen{Becchi:1996an} with quite a few additions for
pedagogy.\cite{Sonoda:2007av}

In the presence of the antifields, the classical action is given by
\begin{eqnarray}
\bScl &=& \int d^4x \, \left[ - \frac{1}{4} \left(
        F_{\mu\nu}^a\right)^2 + \frac{1}{2 \xi} \left( \bar{c}^{a*} +
        \frac{1}{i} \partial_\mu
        A_\mu^a \right)^2 \nn\right.\\
&&\left.\quad + \left( A_\mu^{a*} +
        \frac{1}{i} \partial_\mu \bar{c}^a \right) \left( D_\mu
        c\right)^a + c^{a*} \frac{g}{2 i} f^{abc} c^b c^c\right]\,.
\end{eqnarray}
For convenience, we tabulate the antifields below:
\begin{center}
\begin{tabular}{c|cccc}
field& antifield& statistics& dimension& ghost number\\
\hline
$A_\mu^a$& $A_\mu^{a*}$& Fermi& $2$& $-1$\\
$c^a$& $c^{a*}$& Bose& $2$& $-2$\\
$\bar{c}^a$& $\bar{c}^{a*}$& Bose& $2$& $0$\\
\hline
\end{tabular}
\end{center}
\vspace{0.2cm} The ghost numbers are assigned so that the action has
zero ghost number.  The classical action satisfies the classical
master equation:
\begin{equation}
\int d^4 x \left[ \frac{\delta \bScl}{\delta A_\mu^a}
        \Ld{A_\mu^{a*}} \bScl + \frac{\delta \bScl}{\delta
          \bar{c}^{a*}} \Ld{\bar{c}^a} \bScl 
- \bScl \Rd{c^a} \frac{\delta \bScl}{\delta c^{a*}}
\right] = 0\,.
\end{equation}
As in the case of QED, the BRST transformation of a field is given by
the functional derivative of the classical action with respect to the
conjugate field:
\begin{equation}
\lb\begin{array}{c@{~=~}l}
\delta A_\mu^a & \Ld{A_\mu^{a*}} \bScl = \frac{1}{i} \left(
    D_\mu c \right)^a\,,\\
\delta \bar{c}^a & \frac{\delta \bScl}{\delta \bar{c}^{a*}} =
\frac{1}{\xi} \left( \frac{1}{i} \partial_\mu A_\mu^a + \bar{c}^{a*}
\right)\,,\\
\delta c^a & \frac{\delta \bScl}{\delta c^{a*}} = \frac{g}{2 i}
f^{abc} c^b c^c\,. 
\end{array}\right.
\end{equation}

We now consider the Wilson action $\bSL$ in the presence of
antifields.  As in QED, we can determine the dependence of the Wilson
action on $\bar{c}^{a*}$ and $A_\mu^{a*}$.  Let us consider them one
by one:
\begin{itemize}
\item[(i)] $\bar{c}^{a*}$ --- This antifield is coupled linearly to
    the gauge field as
\begin{equation}
\int_k \frac{1}{\xi} k_\mu \bar{c}^{a*} (k)  A_\mu^a (-k)\,.
\end{equation}
This generates the quadratic term
\begin{equation}
- \frac{1}{2} \int_k \left( 1 - \K{k} \right) \bar{c}^{a*} (-k)
\bar{c}^{a*} (k)
\end{equation}
and the shift
\begin{equation}
A_\mu^{a} (k) \longrightarrow A_\mu^{a} (k) - \frac{1 - \K{k}}{k^2}
k_\mu \bar{c}^{a*} (k)
\end{equation}
in the interaction part of the Wilson action.  This dependence on
$\bar{c}^{a*}$ is the same as in QED.
\item[(ii)] $A_\mu^{a*}$ --- The ghost equation of motion
	   (\ref{WTexamples-YM-ghostEOM}) is now
    generalized to
\begin{equation}
k_\mu \Ld{A_\mu^{a*} (-k)} \bSL = - \K{k} \Ld{\bar{c}^a (-k)} \bSL\,,
\label{AFexamples-YM-ghostEOM}
\end{equation}
since the BRST transformation of $A_\mu^a$ is now given by
\begin{equation}
\delta A_\mu^a (k) = \Ld{A_\mu^{a*} (-k)} \bSL\,.
\end{equation}
(\ref{AFexamples-YM-ghostEOM}) implies that the dependence of $\bSL$
on $A_\mu^{a*}$ and $\bar{c}^a$ is through a linear
combination\footnote{We recall that the Wilson action $\SL$ of a YM
  theory depends on $\bar{c}^a (-k)$ only through the derivative
  $k_\mu \bar{c}^a (-k)$.}
\begin{equation}
A_\mu^{a*} (-k) - \frac{1}{\K{k}} k_\mu \bar{c}^a (-k)\,.
\end{equation}
\end{itemize}

Now that we have determined the dependence on $\bar{c}^{a*}$ and
$A_\mu^{a*}$, the Wilson action has the following form:
\begin{equation}
\bSL = S_{F,\Lambda} + \bSIL\,,\label{AFexamples-YM-bSL}
\end{equation}
where $S_{F,\Lambda}$ is given by (\ref{WTexamples-YM-SFL}), and
\begin{eqnarray}
    \bSIL &=&
 \frac{1}{\xi} \int_k \bar{c}^{a*} (-k) k_\mu A_\mu^a (k) +
\frac{1}{2 \xi} \int_k \K{k} \bar{c}^{a*} (-k) \bar{c}^{a*} (k)
 + \int_k A_\mu^{a*} (-k) k_\mu c^a (k)\nn\\
        &&\, + \,\SIL' \left[ A_\mu^a (k) - \frac{1 - \K{k}}{k^2} k_\mu
        \bar{c}^{a*} (k)\,,\,
        A_\mu^{a*} (-k) - \frac{1}{\K{k}} k_\mu \bar{c}^a (-k)\,,\right.\nn\\
    &&\qquad\qquad c^a (k)\,,\, c^{a*} (-k) \Big]\,.\label{AFexamples-YM-bSIL}
\end{eqnarray}
$\bSL$ reduces to the Wilson action $\SL$ of \S \ref{WTexamples-YM} for
the vanishing antifields.  Especially, we find
\begin{equation}
\SIL' \left[ A_\mu^a (k)\,,\, - \frac{1}{\K{k}} k_\mu \bar{c}^a
    (-k)\,,\, c^a (k)\,,\, 0 \right]
= \SIL \left[ A_\mu^a (k), \bar{c}^a (-k), c^a (k)\right]\,.
\end{equation}

The interaction action $\bSIL$ is determined by the Polchinski
equation
\begin{eqnarray}
- \Lambda \frac{\partial}{\partial \Lambda} \bSIL &=&
\int_k \frac{\Delta (k/\Lambda)}{k^2} \left[\,
\left( \delta_{\mu\nu} - (1-\xi)
    \frac{k_\mu k_\nu}{k^2} \right)\right.\nn\\
&&\qquad \times \frac{1}{2} \lb
\frac{\delta \bSIL}{\delta A_\mu^a (-k)} 
\frac{\delta \bSIL}{\delta A_\nu^a (k)} + \frac{\delta^2 \bSIL}{\delta
  A_\mu^a (-k) \delta A_\nu^a (k)} \rb\nn\\
&&\left. + \bSIL \Rd{c^a (k)} \cdot \Ld{\bar{c}^a (-k)} \bSIL -
\Ld{\bar{c}^a (-k)} \bSIL \Rd{c^a (k)} \,\right]\,,
\end{eqnarray}
and an asymptotic behavior.  For SU(2), the latter is given in the
following form:
\begin{eqnarray}
\bSIL &\asym& \int d^4 x\, \left[ 
\Lambda^2 a_0 (\ln \Lambda/\mu)
    \frac{1}{2} \left(A_\mu^a\right)^2\nn\right.\\
&&\quad + a_1  (\ln \Lambda/\mu) \frac{1}{2} \left(
\partial_\mu A_\nu^a \right)^2 + a_2  (\ln \Lambda/\mu) \frac{1}{2}
\left( \partial_\mu A_\mu^a \right)^2\nn\\
&&\quad + a_3 (\ln \Lambda/\mu) g
\ep^{abc} \partial_\mu A_\nu^a \cdot A_\mu^b A_\nu^c\nn\\
&&\quad + a_4 (\ln \Lambda/\mu) \frac{g^2}{4}
\left( A_\mu^a A_\mu^a \right)^2 + a_5 (\ln \Lambda/\mu)
\frac{g^2}{4} \left( A_\mu^a A_\nu^a  \right)^2\nn\\
&&\quad + a_6 (\ln \Lambda/\mu) 
\left( A_\mu^{a*} + \frac{1}{i} \partial_\mu
\bar{c}^a \right) \frac{1}{i} \partial_\mu c^a \nn\\
&&\quad + a_7 (\ln \Lambda/\mu)
g \ep^{abc} \left( A_\mu^{a*} + \frac{1}{i} \partial_\mu \bar{c}^a
\right)\frac{1}{i} A_\mu^b c^c\nn\\
&&\quad+ a_8 (\ln \Lambda/\mu) c^{a*} \frac{g}{2i} \ep^{abc} c^b c^c
\bigg]\,.
\end{eqnarray}
We note that all the coefficients are the same as in \S
\ref{WTexamples-YM}; as for $a_0, \cdots, a_7$, they survive the limit
of zero antifields, and as for $a_8$, we must find
\begin{equation}
\frac{\delta \bSIL}{\delta c^{a*} (-k)}\Big|_{A^*=c^*=\bar{c}^*=0} =
\C^a (k)\,.
\end{equation}
Hence, the introduction of the antifields does not increase the number
of parameters of the theory.  We still have the same eight parameters
\begin{equation}
a_1 (0),\, \cdots\,,\, a_8 (0)
\label{AFexamples-YM-parameters}
\end{equation}
at our disposal.

Now, the quantum master operator is defined by
\begin{eqnarray}
\bar{\Sigma}_\Lambda &\equiv& \int_k \K{k} \Bigg[ \Ld{A_\mu^{a *} (-k)} \bSL
\cdot \frac{\delta \bSL}{\delta A_\mu^a (k)} +
\Ld{A_\mu^{a *} (-k)} \frac{\delta \bSL}{\delta A_\mu^a
  (k)}\nn\\
&&\quad + \frac{\delta \bSL}{\delta \bar{c}^{a *} (k)} \Ld{\bar{c}^a
  (-k)} \bSL + \frac{\delta}{\delta \bar{c}^{a *} (k)}\Ld{\bar{c}^a
  (-k)} \bSL\nn\\
&&\quad - \frac{\delta \bSL}{\delta c^{a *} (-k)} \cdot \bSL \Rd{c^a (k)}
- \frac{\delta \bSL}{\delta c^{a *} (-k)} \Rd{c^a (k)} \Bigg]\,.
\label{AFexamples-YM-bSigma}
\end{eqnarray}
This reduces to the WT composite operator $\Sigma_\Lambda$
(\ref{WTexamples-YM-Sigma}) of \S \ref{WTexamples-YM} at the vanishing
antifields.  We wish to fine-tune the eight parameters
\bref{AFexamples-YM-parameters} to make $\bar{\Sigma}_\Lambda$ vanish.
The possibility of such fine-tuning can be proven if we can derive the
algebraic constraints (\ref{WTexamples-YM-srelations}).

The derivation of (\ref{WTexamples-YM-srelations}) takes some steps.  We
first rewrite the quantum master operator using all we know about the
antifield dependence of $\bSL$.  Substituting (\ref{AFexamples-YM-bSL})
and (\ref{AFexamples-YM-bSIL}) into the definition
(\ref{AFexamples-YM-bSigma}), we obtain
\begin{eqnarray}
\bar{\Sigma}_\Lambda
&=& \int_k \K{k} \left[
\lb \Ld{A_\mu^{a*} (-k)} \tSL \cdot \frac{\delta \tSL}{\delta A_\nu^a
  (k)} + \Ld{A_\mu^{a*} (-k)} \frac{\delta
  \tSL}{\delta A_\nu^a (k)} \rb P_{\mu\nu} (k/\Lambda)\right.\nn\\
&& \left.\qquad\qquad - \frac{\delta \tSL}{\delta c^{a*} (-k)} \cdot
    \tSL \Rd{c^a   (k)} - \frac{\delta \tSL}{\delta c^{a*} (-k)}
    \Rd{c^a (k)}\, \right]\,,
\end{eqnarray}
where we define
\begin{equation}
P_{\mu\nu} (k) \equiv \delta_{\mu\nu} + \frac{1 - K(k)}{K(k)}
\frac{k_\mu k_\nu}{k^2}\,,
\end{equation}
and
\begin{eqnarray}
\tSL &\equiv& \bSL\nn\\
&& + \frac{1}{2 \xi} \int_k \frac{1}{\K{k}} \lb k_\mu A_\mu^a (k) +
\K{k} \bar{c}^{a*} (k) \rb \lb k_\nu A_\nu^a (-k) - \K{k} \bar{c}^{a*}
(-k) \rb\nn\\
&=& - \frac{1}{2} \int_k \frac{1}{\K{k}} A_\mu^a (k) A_\nu^a (-k)
\left( k^2 \delta_{\mu\nu} - k_\mu k_\nu \right)\nn\\
&& \quad+ \int_k \lb A_\mu^{a*} (-k) - \frac{1}{\K{k}} k_\mu \bar{c}^a (-k)
\rb k_\mu c^a (k)\nn\\
&& \quad+ \SIL' \left[ A_\mu^a (k) - \frac{1-\K{k}}{k^2} k_\mu \bar{c}^{a*}
    (k)\,,\, A_\mu^{a*} (-k) - \frac{1}{\K{k}} k_\mu \bar{c}^a
    (-k)\,,\,\nn\right.\\
&& \quad\qquad\qquad c^a (k)\,,\, c^{a*} (-k)\,\Big]\,.
\end{eqnarray}
Note that the quadratic kinetic term of the gauge field in $\tSL$ is
purely transverse.  Hence, $\tSL$ depends only on the following
combinations of the field variables:
\begin{equation}
A_\mu^a (k) - \frac{1-\K{k}}{k^2} k_\mu \bar{c}^{a*}
    (k)\,,\, A_\mu^{a*} (-k) - \frac{1}{\K{k}} k_\mu \bar{c}^a
    (-k)\,,\,c^a (k)\,,\, c^{a*} (-k)\,.
\end{equation}
Therefore, shifting $A_\mu^a$ and its antifield by terms proportional to
$k_\mu$, we define a modified
quantum master operator by
\begin{eqnarray}
&&\hat{\Sigma}_\Lambda \left[ A_\mu^a, A_\mu^{a*}, c^a,
    c^{a*}\right]\nn\\
&& \equiv \int_k \K{k} \left[ P_{\mu\nu} (k/\Lambda) \lb
\Ld{A_\mu^{a*} (-k)} \hSL \cdot \frac{\delta \hSL}{\delta A_\nu^a (k)}
+ \Ld{A_\mu^{a*} (-k)} \frac{\delta \hSL}{\delta A_\nu^a (k)}
\rb\right.\nn\\
&&\quad - \left. \frac{\delta \hSL}{\delta c^{a*} (-k)} \cdot \hSL
    \Rd{c^a (k)} - \frac{\delta \hSL}{\delta c^{a*} (-k)} \Rd{c^a (k)}
\right]\,,\label{AFexamples-YM-hSigma}
\end{eqnarray}
where $\hSL$ is $\tSL$ for the shifted variables, and defined by
\begin{eqnarray}
&&\hSL \left[A_\mu^a (k), A_\mu^{a*} (-k), c^a (k), c^{a*} (-k)
\right]\nn\\
 &\equiv& - \frac{1}{2} \int_k A_\mu^a (k) A_\nu^a
(-k) \frac{1}{\K{k}} \left( k^2 \delta_{\mu\nu} - k_\mu k_\nu
\right)
+ \int_k A_\mu^{a*} (-k) k_\mu c^a (k)\nn\\
&&\, + \SIL' \left[ A_\mu^a (k), A_\mu^{a*} (-k), c^a (k), c^{a*} (-k)
\right]\,.\label{AFexamples-YM-hSL}
\end{eqnarray}
Then, the quantum master equation
\begin{equation}
\bar{\Sigma}_\Lambda = 0
\end{equation}
is equivalent to the modified quantum master equation
\begin{equation}
\hat{\Sigma}_\Lambda = 0\,.\label{AFexamples-YM-hSigmazero}
\end{equation}

It is the modified action $\hSL$ (\ref{AFexamples-YM-hSL}) and the
corresponding quantum master operator $\hat{\Sigma}_\Lambda$
(\ref{AFexamples-YM-hSigma}) that we apply the general formalism
developed in \S \ref{AF}.  Let us identify the classical algebraic
problem corresponding to (\ref{AF-tosolve}):
\begin{equation}
\hat{\delta}_{cl} \hat{S}_\infty^{(l+1)} = -
\hat{\Sigma}_\infty^{(l+1), u}\,.
\end{equation}

First, the classical action is given by
\begin{equation}
\hat{S}_{cl} \equiv \int d^4 x \left[ - \frac{1}{4} \left(
        F_{\mu\nu}^a \right)^2 + A_\mu^{a*} \left( D_\mu c \right)^a
+ c^{a*} \frac{g}{2i} \ep^{abc} c^b c^c \right]\,,
\end{equation}
which satisfies the classical master equation
\begin{equation}
\int d^4 x \left[ \frac{\delta
      \hat{S}_{cl}}{\delta A_\mu^a} \cdot \Ld{A_\mu^{a*}} \hat{S}_{cl}
+ \frac{\delta \hat{S}_{cl}}{\delta c^{a*}} \cdot \Ld{c^a} \hat{S}_{cl}
 \right] = 0\,.
\end{equation}
Then, the classical BRST transformation is defined by
\begin{equation}
\hat{\delta}_{cl} \equiv \int d^4 x \left[ \frac{\delta
      \hat{S}_{cl}}{\delta A_\mu^a} \Ld{A_\mu^{a*}}
+ \Ld{A_\mu^{a*}} \hat{S}_{cl} \frac{\delta}{\delta A_\mu^a}
+ \frac{\delta \hat{S}_{cl}}{\delta c^{a*}} \Ld{c^a}
- \hat{S}_{cl} \Rd{c^a} \frac{\delta}{\delta c^{a*}} \right]\,,
\end{equation}
which is nilpotent
\begin{equation}
\hat{\delta}_{cl}^2 =0
\end{equation}
due to the classical master equation.  The most general form of
$\hat{S}_\infty^{(l+1)}$, for SU(2), is given by
\begin{eqnarray}
\hat{S}_\infty^{(l+1)} &=& \int d^4 x\, \left[ a_1^{(l+1)}  \frac{1}{2}
    \left( \partial_\mu A_\nu^a \right)^2 + a_2^{(l+1)} 
    \frac{1}{2} \left( \partial_\mu A_\mu^a \right)^2\right.\nn\\
&&\quad + a_3^{(l+1)}  g \ep^{abc} \partial_\mu A_\nu^a \cdot
A_\mu^b A_\nu^c \nn\\
&&\quad + a_4^{(l+1)}  \frac{g^2}{4} \left(A_\mu^a A_\mu^a\right)^2 
+ a_5^{(l+1)}  \frac{g^2}{4} \left(A_\mu^a A_\nu^a \right)^2\nn\\
&&\quad + a_6^{(l+1)}  A_\mu^{a*} \frac{1}{i} \partial_\mu c^a
+ a_7^{(l+1)}  g \ep^{abc} A_\mu^{a*} \frac{1}{i} A_\mu^b c^c\nn\\
&&\left.\quad + a_8^{(l+1)}  c^{a*} \frac{g}{2i} \ep^{abc} c^b
    c^c\,
\right]\,,
\end{eqnarray}
where we use the familiar notation for the loop expansion
\begin{equation}
a_i (0) = \sum_{l=0}^\infty a_i^{(l)}\,.
\end{equation}

Now, $\hat{\Sigma}_\Lambda$ is a dimension $5$ fermionic scalar
composite operator of ghost number $+1$.  Hence, the most general form
of $\hat{\Sigma}_\infty^{(l+1)}$ is given by
\begin{eqnarray}
\hat{\Sigma}_\infty^{(l+1)}
&=& \int d^4 x\, 
\Bigg[\, s_1 \frac{1}{i} c^a \partial^2 \partial_\mu A_\mu^a\nn\\
&&\quad + s_2 \frac{1}{i} g \ep^{abc} c^a
\partial^2 A_\mu^b \cdot A_\mu^c + s_3 \frac{1}{i} g \ep^{abc} c^a \partial_\mu
\partial_\nu A_\nu^b \cdot A_\mu^c \nn\\
&&\quad + s_4 i g^2  c^a \partial_\mu A_\mu^a \cdot A_\nu^b A_\nu^b + s_5
i g^2 c^a \partial_\mu A_\nu^a \cdot A_\mu^b A_\nu^b + s_6 i g^2 c^a
A_\mu^a \partial_\nu A_\nu^b A_\mu^b \nn\\
&&\quad + s_7 i g^2 c^a A_\mu^a \partial_\nu
A_\mu^b A_\nu^b + s_8 i g^2 c^a A_\mu^a \partial_\mu A_\nu^b A_\nu^b\nn\\
&&\quad+ s_{9}  g \ep^{abc} \partial_\mu c^a c^b A_\mu^{c*} + s_{10} 
g^2 c^a c^b A_\mu^{a*} A_\mu^a \,\Bigg]\,,
\end{eqnarray}
where $s_1, \cdots, s_{10}$ are constants.  Replacing $A_\mu^{a*}$ by
$\frac{1}{i} \partial_\mu \bar{c}^a$, this is the same as the asymptotic
behavior of $\Sigma_\Lambda^{(l+1)}$ in the absence of the antifields.
(See \bref{WTexamples-YM-asymptoticSigma}.)  Similarly,
$\hat{\Sigma}_\infty^{(l+1), t}$ and $\hat{\Sigma}_\infty^{(l+1), u}$
are parameterized by $t_i$ and $u_i$, respectively, instead of $s_i$.
Then,
\begin{equation}
\hat{\Sigma}_\infty^{(l+1)} = \hat{\Sigma}_\infty^{(l+1), t} +
\hat{\Sigma}_\infty^{(l+1), u} 
\end{equation}
implies
\begin{equation}
s_i = t_i + u_i\,.\quad (i=1,\cdots,10)
\end{equation}
Since
\begin{equation}
\hat{\Sigma}_\infty^{(l+1), t} = \hat{\delta}_{cl}
\hat{S}_\infty^{(l+1)}\,,\label{AFexamples-YM-SigmadS}
\end{equation}
$t_i$ is determined by $a^{(l+1)}$'s.  But $u_i$ is determined only by
the lower loop parameters.

Let us consider $t_i$'s.  (\ref{AFexamples-YM-SigmadS}) gives
\begin{subequations}
\begin{eqnarray}
t_1 &=& a^{(l+1)}_1 + a^{(l+1)}_2\\
t_2 &=& a^{(l+1)}_1 + a^{(l+1)}_3 + a^{(l+1)}_6 + a^{(l+1)}_7\\
t_4 &=& a^{(l+1)}_3 + a^{(l+1)}_4 - a^{(l+1)}_6 - a^{(l+1)}_7\\
t_5 &=& - a^{(l+1)}_3 + a^{(l+1)}_5 + a^{(l+1)}_6 + a^{(l+1)}_7\\
t_9 &=& a^{(l+1)}_7 + a^{(l+1)}_8\,,
\end{eqnarray}
\end{subequations}
and
\begin{subequations}
\begin{eqnarray}
t_3 &=& t_1 - t_2\,,\\
t_6 &=& t_5\,,\\
t_7 &=& 2 t_4\,,\\
t_8 &=& t_5\,,\\
t_{10} &=& - t_9\,.
\end{eqnarray}
\end{subequations}
These are the same results as \bref{WTexamples-YM-tbya} \&
\bref{WTexamples-YM-tbyt} that we have found in \S \ref{WTexamples-YM}.

Next, we consider $u_i$'s.  The algebraic constraint
\begin{equation}
\hat{\delta}_{cl} \hat{\Sigma}_\infty^{(l+1), u} = 0\,,
\end{equation}
implies
\begin{subequations}
\begin{eqnarray}
u_3 &=& u_1 - u_2\,,\\
u_6 &=& u_5\,,\\
u_7 &=& 2 u_4\,,\\
u_8 &=& u_5\,,\\
u_{10} &=& - u_9\,.
\end{eqnarray}
\end{subequations} 
These are the relations that were missing in \S \ref{WTexamples-YM}.  

Thus, $\hat{\Sigma}_\infty^{(l+1)} = 0$ gives only five independent
equations
\begin{equation}
s_1 = s_2 = s_4 = s_5 = s_9 = 0\,.
\end{equation}
The solution is given by
\begin{subequations}
\begin{eqnarray}
a^{(l+1)}_2 &=& - u_1 - a^{(l+1)}_1\\ 
a^{(l+1)}_4 &=& - u_2 - \left(a^{(l+1)}_1+a^{(l+1)}_3+a^{(l+1)}_6\right)\\
a^{(l+1)}_5 &=& - u_2 - u_4 - a^{(l+1)}_1 - 2 a^{(l+1)}_3\\
a^{(l+1)}_7 &=& u_2 - u_5 + a^{(l+1)}_1 + 2 a^{(l+1)}_3 \\
a^{(l+1)}_8 &=& u_2 - u_9 + a^{(l+1)}_1 + a^{(l+1)}_3 + a^{(l+1)}_6\,,
\end{eqnarray}
\end{subequations}
where we have treated the normalization constants $a^{(l+1)}_{1,3,6}$
as independent parameters.  The simplest choice is to set them zero for
$l \ge 0$.

\newpage

\section{Axial and chiral anomalies\label{anomaly}}

In this section we formulate the axial and chiral anomalies in the ERG
formalism.  We restrict our discussions only to abelian anomalies, and
it suffices to use only the WT identity of \S\ref{WT}.  The anomalies
have been computed in the ERG formalism in
Refs.~\citen{Bonini:1997yv,Bonini:1994xj}, \citen{Pernici:1997ie}, but
here we follow an unpublished work by Igarashi, Itoh, and Sonoda.  For
references, we would like to mention that the chiral anomalies have been
studied using the antifield formalism (but not ERG) in
\citen{Troost:1989cu}, \citen{DeJonghe:1993zc}, and using the BRST
cohomology (but not ERG) in
\citen{Barnich:1994ve,Barnich:1994mt,Barnich:1994db,Barnich:2000zw}.

\subsection{Axial anomaly in QED}

In the following we wish to consider the axial vector current $J_{5,
\mu}$ in QED, and compute its divergence
\begin{equation}
k_\mu J_{5,\mu} (-k)\,.
\end{equation}
To define the axial vector current precisely, it is convenient to
couple an external field $B_\mu$ to the axial vector current.  In the
Wilson action, we only consider $B_\mu$ up to first order.  $B_\mu$
introduces the following extra terms into the asymptotic behavior of
the action:
\begin{equation}
\SIL \asym \int d^4 x \left[ \cdots + a'_3 (\ln \Lambda/\mu) \bar{\psi}
    \gamma_5 \fmslash{B} \psi + a_5 (\ln \Lambda/\mu)
    \ep_{\alpha\beta\gamma\delta} B_\alpha A_\beta
    \frac{1}{i} \partial_\gamma A_\delta \right]\,,
\label{anomaly-axial-SIL}
\end{equation}
where we have suppressed the terms independent of $B_\mu$.  We define
the axial vector current as the composite operator:
\begin{equation}
J_{5, \mu} (-k) \equiv \frac{\delta \SIL}{\delta B_\mu (k)}\,.
\end{equation}
This has the asymptotic behavior
\begin{eqnarray}
J_{5, \mu} (-k) &\asym& a'_3 (\ln \Lambda/\mu) \int_p \bar{\psi} (-p-k)
\gamma_5 \gamma_\mu \psi (p)\nn\\
&&\quad  + a_5 (\ln \Lambda/\mu)
\ep_{\mu\alpha\beta\gamma} \int_{k_1 + k_2 = -k} A_\alpha (k_1) k_{2
  \beta} A_\gamma (k_2)\,.
\end{eqnarray}
The parameter $a'_3 (0)$ normalizes the axial vector current, and the
ratio $a_5(0)/a'_3(0)$ is determined by the WT identity of QED as we will
see shortly.  The above asymptotic behavior implies
\begin{eqnarray}
k_\mu J_{5,\mu} (-k) &\asym& a'_3 (\ln \Lambda/\mu) \int_p \bar{\psi} (-p-k)
\gamma_5 \fmslash{k} \psi (p)\nn\\
&&\quad  + a_5 (\ln \Lambda/\mu)
\ep_{\alpha\beta\gamma\delta} \int_{k_1 + k_2 = -k} k_{1\alpha}
k_{2\beta} A_\gamma (k_1) A_\delta (k_2)\,.
\end{eqnarray}

To determine $a_5 (0)$, we consider the WT identity in the presence of
$B_\mu$.  In \S \ref{WTexamples-QED} we have found that the WT
identity is given by
\begin{equation}
k_\mu J_\mu (-k) = \Phi (-k)\,.
\end{equation}
The two composite operators are defined by
\begin{equation}
J_\mu (-k) \equiv \frac{\delta \SIL}{\delta A_\mu (k)}\,,
\end{equation}
and
\begin{eqnarray}
\Phi (-k) &\equiv& e \int_p \K{p} \left[ - \SL \Rd{\psi (p)} [\psi]_\Lambda
    (p-k) + [\bar{\psi}]_\Lambda (-p-k) \Ld{\bar{\psi} (-p)} \SL \right.\nn\\
&& \quad\left. + \Tr  [\psi]_\Lambda (p-k) \Rd{\psi (p)}
- \Tr \Ld{\bar{\psi} (-p)}  [\bar{\psi}]_\Lambda (-p-k) 
 \right]\,.
\end{eqnarray}
In the presence of $B_\mu$, the asymptotic behaviors of both $k_\mu
J_\mu (-k)$ and $\Phi (-k)$ get an extra term linear in $B_\mu$:
\begin{eqnarray}
k_\mu J_\mu (-k) &\asym& \cdots - a_5 (\ln \Lambda/\mu)
\ep_{\alpha\beta\gamma\delta} \int_{k_1+k_2=-k} k_{1\gamma}
k_{2\delta} B_\alpha (k_1) A_\beta (k_2)\,,\\
\Phi (-k) &\asym& \cdots + b (\ln \Lambda/\mu)
\ep_{\alpha\beta\gamma\delta} \int_{k_1+k_2=-k} k_{1\gamma}
k_{2\delta} B_\alpha (k_1) A_\beta (k_2)\,,
\end{eqnarray}
where the terms independent of $B_\mu$ are suppressed.  Thus, the
gauge invariance requires
\begin{equation}
a_5 (0) = - b (0)\,.\label{anomaly-axial-a5b}
\end{equation}
We leave open the convention for the normalization constant $a'_3 (0)$
of the axial vector current for later convenience.

We now consider the composite operator of type (\ref{comp-dOdphi}) at
$B_\mu = 0$:
\begin{eqnarray}
\Phi_5 (-k) &\equiv&  \int_p \K{p} \left[ \SL \Rd{\psi (p)}
\gamma_5 [\psi]_\Lambda (p-k) +
[\bar{\psi}]_\Lambda (p-k) \gamma_5 \Ld{\bar{\psi} (-p)} \SL\right.\nn\\
&&\quad \left.- \Tr \gamma_5 [\psi]_\Lambda (p-k) \Rd{\psi (p)}
- \Tr \Ld{\bar{\psi} (-p)} [\bar{\psi}]_\Lambda (p-k) \gamma_5
\right]\,.
\end{eqnarray}
This has the correlation functions:
\begin{eqnarray}
&&\vev{\Phi_5 (-k)\, A_{\mu_1} (k_1) \cdots A_{\mu_M} (k_M) \psi (p_1)
  \cdots \psi (p_N) \bar{\psi} (-q_1) \cdots \bar{\psi}
  (-q_N)}^\infty\nn\\
&& = - \sum_{i=1}^N \lb
\vev{\cdots \gamma_5 \psi (p_i-k) \cdots}^\infty
+ \vev{\cdots \bar{\psi} (-q_i-k) \gamma_5 \cdots}^\infty\rb\,.
\end{eqnarray}
The asymptotic behavior of $\Phi_5 (-k)$ has the following
form:\footnote{$\Phi_5$ is a pseudo scalar even under charge
  conjugation.  Hence, the dimension $4$ pseudo scalars $\bar{\psi}
  \gamma_5 \fmslash{A} \psi$, $\bar{\psi} \gamma_5
  \overleftrightarrow{\fmslash{\partial}} \psi$ are forbidden.}
\begin{eqnarray}
\Phi_5 (-k) &\asym& a' (\ln \Lambda/\mu) \int_p \bar{\psi} (-p-k)
\gamma_5 \fmslash{k} \psi (p)\nn\\
&&\, + b' (\ln \Lambda/\mu) \ep_{\alpha\beta\gamma\delta}
\int_{k_1+k_2=-k} k_{1\alpha} k_{2\beta} A_\gamma (k_1) A_\delta
(k_2)\nn\\
&&\, + c' (\ln \Lambda/\mu) i m \int_p \bar{\psi} (-p-k) \gamma_5
\psi (p)\,.\label{anomaly-axial-Phi5asymp}
\end{eqnarray}

To rewrite $\Phi_5$, we introduce two composite operators:
\begin{itemize}
\item[(i)] $[\frac{1}{2} F \tilde{F}]_\Lambda$ with the asymptotic behavior
\begin{eqnarray}
\left[F \tilde{F}/2\right]_\Lambda (-k) &\asym& f_1 (\ln
\Lambda/\mu) 
\ep_{\alpha\beta\gamma\delta} \int_{k_1+k_2=-k} k_{1\alpha} k_{2\beta}
A_\gamma (k_1) A_\delta (k_2)\nn\\
&& + f_2 (\ln \Lambda/\mu) \int_p \bar{\psi} (-p-k) \gamma_5 \fmslash{k} \psi
(p)\,,
\end{eqnarray}
where we choose the convention 
\begin{equation}
f_1 (0) = 1\,,\quad f_2 (0) = 0\,.
\end{equation}
\item[(ii)] $J_5$ with the asymptotic behavior
\begin{equation}
J_5 (-k) \asym j (\ln \Lambda/\mu) \int_p \bar{\psi} (-p-k) \gamma_5
\psi (p)\,.
\end{equation}
We will make a convenient choice for $j(0)$ shortly.
\end{itemize}
We can now expand
\begin{equation}
\Phi_5 (-k) = s\, k_\mu J_{5, \mu} (-k) + t \, \left[ F \tilde{F}/2
\right]_\Lambda + u\cdot i m J_5 (-k)\,,
\end{equation}
where $s, t, u$ are $\Lambda$ independent constants, dependent only on
$e^2$ and $\xi$.  The right-hand side has the asymptotic behavior
\begin{eqnarray}
\mathrm{(RHS)} &\asym& \left( s\, a'_3 (\ln \Lambda/\mu) + t\, f_2 (\ln
    \Lambda/\mu) \right) \int_p \bar{\psi} (-p-k) \gamma_5 \fmslash{k} \psi
(p)\nn\\
&&\, + \left( s\, a_5 (\ln \Lambda/\mu) + t\, f_1 (\ln \Lambda/\mu)
\right) \ep_{\alpha\beta\gamma\delta} \int_{k_1+k_2=-k} k_{1\alpha}
k_{2\beta} A_\gamma (k_1) A_\delta (k_2)\nn\\
&&\, + u\, j(\ln \Lambda/\mu) i m \int_p \bar{\psi} (-p-k)
\gamma_5 \psi (p)\,.
\end{eqnarray}
Comparing this with (\ref{anomaly-axial-Phi5asymp}), we obtain
\begin{equation}
    \lb\begin{array}{c@{~=~}l}
        a' (\ln \Lambda/\mu) & s\, a'_3 (\ln \Lambda/\mu) + t\, f_2 (\ln
        \Lambda/\mu) \,,\\
        b' (\ln \Lambda/\mu) & - s\, b (\ln \Lambda/\mu) + t\, f_1
        (\ln \Lambda/\mu) \,,\\ 
        c' (\ln \Lambda/\mu) & u\, j(\ln \Lambda/\mu)\,,
\end{array}\right.
\end{equation}
where we have used (\ref{anomaly-axial-a5b}).
By adopting the convention
\begin{eqnarray}
a'_3 (0) &=& a' (0)\,,\\
j(0) &=& c' (0)\,,
\end{eqnarray}
we obtain
\begin{eqnarray}
s &=& 1\,,\\
t &=& b(0) + b' (0)\,,\\
u &=& 1\,.
\end{eqnarray}
Hence, we obtain
\begin{equation}
\Phi_5 (-k) = k_\mu J_{5,\mu} (-k) + i m J_5 (-k)
+ \left( b(0) + b'(0) \right) \left[ F \tilde{F}/2 \right] (-k)\,.
\end{equation}
The last term gives the axial anomaly.  We will show the 1-loop
calculations of the axial anomaly in the next subsection.  From
(\ref{anomaly-chiral-b}, \ref{anomaly-chiral-bprime}), setting $e'=1$,
we obtain
\begin{equation}
b(0) = e^2 \frac{8}{3} \frac{1}{(4 \pi)^2}\,,\quad
b'(0) = e^2 \frac{4}{3} \frac{1}{(4 \pi)^2}\,,
\end{equation}
so that the coefficient of the axial anomaly is
\begin{equation}
 b(0) + b'(0) = \frac{e^2}{4 \pi^2}\,.
\end{equation}

\subsection{Chiral anomaly}

\subsubsection{ $U(1)_V \times U(1)_A$ gauge theory}

Let us try to construct a $U(1)_V \times U(1)_A$ gauge theory with
massless fermions.  Its classical action is given by
\begin{eqnarray}
S_{cl} &=& \int d^4 x\, \left[ - \frac{1}{4} F_{\mu\nu}^2 - \frac{1}{2
      \xi} (\partial_\mu A_\mu)^2 - \frac{1}{4} G_{\mu\nu}^2 -
    \frac{1}{2 \xi'} (\partial_\mu B_\mu)^2\right.\nn\\
&&\left.\, - \partial_\mu \bar{c} \partial_\mu c - \partial_\mu
\bar{c}' \partial_\mu c'
 + \bar{\psi} \left( - \frac{1}{i} \fmslash{\partial} 
+ e \fmslash{A} + e' \gamma_5 \fmslash{B} \right) \psi\,\right]\,,
\end{eqnarray}
where
\begin{equation}
F_{\mu\nu} \equiv \partial_\mu A_\nu - \partial_\nu A_\mu\,,\quad
G_{\mu\nu} \equiv \partial_\mu B_\nu - \partial_\nu B_\mu\,.
\end{equation}
This action is invariant under the following classical BRST
transformation:
\begin{equation}
\lb\begin{array}{c@{~=~}l@{\quad}c@{~=~}l}
\de A_\mu & \ep \frac{1}{i} \partial_\mu c\,,&
\de B_\mu & \ep \frac{1}{i} \partial_\mu c'\,,\\
\de c & 0\,,& \de c' & 0\,,\\
\de \bar{c} & \ep \frac{1}{\xi} \frac{1}{i} \partial_\mu A_\mu\,,
& \de \bar{c}' & \ep \frac{1}{\xi'} \frac{1}{i} \partial_\mu B_\mu\,,\\
\de \psi & \ep \left( e c - e' c' \gamma_5 \right) \psi\,,&
\de \bar{\psi}& \bar{\psi} \ep \left( - e c - e' c' \gamma_5\right)\,,
\end{array}\right.
\end{equation}
where $\ep$ is an arbitrary anticommuting constant.

The quantization follows the same line as for QED.  We omit writing
down the Polchinski equation for the interaction action $\SIL$.  Its
asymptotic behavior is given in the following form:
\begin{eqnarray}
&&\SIL \asym \int d^4 x \,\left[ \Lambda^2 a_2 (\ln
        \Lambda/\mu) \frac{1}{2}
    A_\mu^2 + \Lambda^2 a'_2 (\ln \Lambda/\mu) \frac{1}{2} B_\mu^2 \right.\nn\\
&&\qquad + c_2 (\ln \Lambda/\mu) \frac{1}{2} \partial_\mu
A_\nu \partial_\mu A_\nu + d_2 (\ln \Lambda/\mu)
\frac{1}{2} \left(\partial_\mu A_\mu\right)^2\nn\\
&&\qquad + c'_2 (\ln \Lambda/\mu) \frac{1}{2} \partial_\mu
B_\nu \partial_\mu B_\nu + d'_2 (\ln \Lambda/\mu)
\frac{1}{2} \left(\partial_\mu B_\mu\right)^2\nn\\
&&\qquad  + a_4 (\ln \Lambda/\mu) \frac{1}{8} \left( A_\mu^2
\right)^2 + a'_4 (\ln \Lambda/\mu) \frac{1}{8} \left( B_\mu^2
\right)^2 + a''_4 (\ln \Lambda/\mu) \frac{1}{4} A_\mu^2 B_\nu^2\nn\\
&&\qquad + a_5 (\ln \Lambda/\mu) \ep_{\alpha\beta\gamma\delta} B_\alpha
A_\beta \frac{1}{i} \partial_\gamma A_\delta\nn\\
&&\left.\qquad + \bar{\psi} \lb a_f (\ln \Lambda/\mu) \frac{1}{i}
\fmslash{\partial} 
+ a_3 (\ln \Lambda/\mu) \fmslash{A} 
 + a'_3 (\ln \Lambda/\mu) \gamma_5 \fmslash{B}\rb \psi\right]\,.
\label{anomaly-chiral-SIasym}
\end{eqnarray}
Here the $a''_4, a_5$ terms are new features, not present in QED.
Given $A_\mu$ is a vector, and $B_\mu$ is an axial vector, we find
that the $a''_4, a_5$ terms are allowed by parity.  (Note
$\ep_{\alpha\beta\gamma\delta} A_\alpha B_\beta \partial_\gamma
B_\delta$ and $B_\mu A_\mu A_\nu^2$ are forbidden by parity.)

We define the quantum BRST transformation by
\begin{eqnarray}
&&\begin{array}{c@{~=~}l@{\quad}c@{~=~}l}
\delta A_\mu (k) & k_\mu \ep c (k)\,,&
\delta B_\mu (k) & k_\mu \ep c' (k)\,,\\
\delta c (k) & 0\,,&
\delta c' (k) & 0\,,\\
\delta \bar{c} (-k) & \ep \frac{1}{\xi} k_\mu [A_\mu]_\Lambda (-k)\,,&
\delta \bar{c}' (-k) & \ep \frac{1}{\xi'} k_\mu [B_\mu]_\Lambda
(-k)\,,
\end{array}\nn\\
&&\begin{array}{c@{~=~}l}
\delta \psi (p) & \int_k \left( e \ep c (k) - e' \ep c' (k) \gamma_5 \right)
[\psi]_\Lambda (p-k)\,,\\
\delta \bar{\psi} (-p) & \int_k [\bar{\psi}]_\Lambda (-p-k) \left( - e
    \ep c (k) - e' \ep c' (k) \gamma_5 \right)\,,
\end{array}
\end{eqnarray}
where the composite operators in the square brackets are defined as
usual.

The corresponding WT composite operator is given by
\begin{eqnarray}
\Sigma_\Lambda &\equiv& \int_k \K{k} \left[ \frac{\delta \SL}{\delta
      A_\mu (k)} \delta A_\mu (k) + \delta \bar{c} (-k)
    \Ld{\bar{c} (-k)} \SL \right.\nn\\
&&\qquad \left.+ \frac{\delta \SL}{\delta
      B_\mu (k)} \delta B_\mu (k) + \delta \bar{c}' (-k)
    \Ld{\bar{c}' (-k)} \SL \right]\,\nn\\
&& + \int_p \K{p} \left[ \SL \Rd{\psi (p)} \delta \psi (p) - \Tr
    \delta \psi (p) \Rd{\psi (p)}\right.\nn\\
&&\left.\qquad + \delta \bar{\psi} (-p) \Ld{\bar{\psi} (-p)} \SL
- \Tr \Ld{\bar{\psi} (-p)} \delta \bar{\psi} (-p) \right]\,.
\end{eqnarray}
Substituting the above BRST transformation, we can rewrite this as
\begin{equation}
\Sigma_\Lambda = \ep \int_k c (k) \left( k_\mu J_\mu (-k) - \Phi
    (-k)\right)
 + \ep \int_k c' (k) \left( k_\mu J_{5,\mu} (-k) - \Phi_5 (-k)
\right)\,,
\end{equation}
where the composite operators are defined by
\begin{equation}
J_\mu (-k) \equiv \frac{\delta \SIL}{\delta A_\mu (k)}\,,\quad
J_{5,\mu} (-k) \equiv \frac{\delta \SIL}{\delta B_\mu (k)}\,,
\end{equation}
and
\begin{eqnarray}
\Phi (-k) &\equiv& e \int_p \K{p} \left[ - \SL \Rd{\psi (p)}
    [\psi]_\Lambda (p-k) 
+ \Tr [\psi]_\Lambda (p-k) \Rd{\psi (p)}\right]\nn\\
&&\quad + e \int_p \Tr U(-p-k,p) \Ld{\bar{\psi} (-p)} \SIL \Rd{\psi
  (p+k)}\,,\label{anomaly-chiral-Phi}\\
\Phi_5 (-k) &\equiv&  e' \int_p \K{p} \left[ \SL \Rd{\psi (p)}
\gamma_5 [\psi]_\Lambda (p-k) 
- \Tr \gamma_5 [\psi]_\Lambda (p-k) \Rd{\psi (p)}\right]\nn\\
&&\quad + e' \int_p \Tr U(-p-k,p) \gamma_5 \Ld{\bar{\psi} (-p)} \SIL 
\Rd{\psi  (p+k)}\,,\label{anomaly-chiral-Phi5}
\end{eqnarray}
where the matrix $U$ is defined by (\ref{WTexamples-QED-U}) with
$m=0$:
\begin{equation}
U (-p-k,p) \equiv \K{(p+k)} \frac{1 - \K{p}}{\fmslash{p}} - 
\frac{1 - \K{(p+k)}}{\fmslash{p} + \fmslash{k}} \K{p}\,.\label{anomaly-chiral-U}
\end{equation}

In the following we examine the asymptotic behavior of the WT
composite operator $\Sigma_\Lambda$.  We will restrict our discussion
to the part relevant to chiral anomalies, i.e., the part proportional
to the $\epsilon$ tensor.  The asymptotic behavior
(\ref{anomaly-chiral-SIasym}) implies 
\begin{equation}
\lb\begin{array}{c@{~\asym~}l}
k_\mu J_\mu (-k) & - a_5 \,\ep_{\alpha\beta\gamma\delta}
\int_{k_1+k_2=-k} k_{1\gamma} k_{2\delta} B_\alpha (k_1) A_\beta
(k_2)\,,\\
k_\mu J_{5, \mu} (-k) & + a_5\, \ep_{\alpha\beta\gamma\delta}
\int_{k_1+k_2=-k} k_{1\gamma} k_{2\delta} A_\alpha (k_1) A_\beta
(k_2)\,.
\end{array}\right.
\end{equation}
We have only one parameter $a_5 (0)$ at our disposal.

On the other hand, assuming parity, the asymptotic behaviors (only the
part proportional to the $\ep$ tensor) of $\Phi, \Phi_5$ are expected to
be of the following form:
\begin{equation}
\lb\begin{array}{c@{~\asym~}l}
\Phi (-k) & \ep_{\alpha\beta\gamma\delta} \int_{k_1+k_2=-k}
k_{1\gamma} k_{2\delta}  \,b \,B_\alpha (k_1) A_\beta (k_2)\,,\\
\Phi_5 (-k) &  \ep_{\alpha\beta\gamma\delta} \int_{k_1+k_2=-k}
k_{1\gamma} k_{2\delta} \left( b' A_\alpha (k_1) A_\beta (k_2) + b''
    B_\alpha (k_1) B_\beta (k_2) \right)\,.
\end{array}\right.
\end{equation}
Thus, for $\Sigma_\Lambda = 0$, we need to find
\begin{equation}
b'' = 0\,\textrm{, and}\quad
b+b' = 0\,.
\end{equation}
If this is the case, we can make $\Sigma_\Lambda$ vanish by choosing
\begin{equation}
a_5 = - b\,.
\end{equation}

Let us compute $b, b', b''$ at 1-loop.  Only the second integrals of
(\ref{anomaly-chiral-Phi}) and (\ref{anomaly-chiral-Phi5}) contribute.
\begin{itemize}
\item[(i)] $b$ --- The $BA$ term of $\Phi$ is given by the Feynman
    diagram in Fig.~\ref{anomaly-chiral-triangle} and the one with
    $B_\alpha$ and $A_\beta$ interchanged.
\begin{figure}[t]
\begin{center}
\epsfig{file=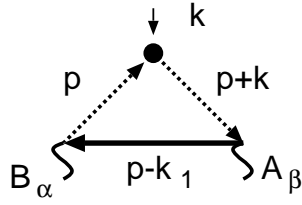, width=4cm}
\caption{A Feynman diagram contributing to $\Phi (-k)$. The broken
  line gives $e U$.}
\label{anomaly-chiral-triangle}
\end{center}
\end{figure}
The graph gives the integral
\begin{equation}
T_{\alpha\beta} (k_1,k_2) \equiv
\int_p \Tr e U(-p-k, p) e' \gamma_5 \gamma_\alpha \frac{1 -
  \K{(p-k_1)}}{\fmslash{p} - \fmslash{k}_1} e \gamma_\beta\,,
\end{equation}
where $U$ is defined by (\ref{anomaly-chiral-U}). Its $\Lambda \to
\infty$ limit can be calculated as follows:
\begin{eqnarray}
&&T_{\alpha\beta} (k_1,k_2) = e' e^2 \int_p U (-p+k_2,p+k_1) \gamma_5
\gamma_\alpha \frac{1 - \K{p}}{\fmslash{p}} \gamma_\beta \nn\\
&=& e' e^2 \int_p \Tr \left[
\gamma_5 \gamma_\alpha \frac{1 - \K{p}}{\fmslash{p}} \gamma_\beta
\cdot \lb \K{(p-k_2)} \frac{1 -
  \K{(p+k_1)}}{\fmslash{p} + \fmslash{k}_1}\right.\right.\nn\\
&&\qquad\qquad \left.\left. - \frac{1 -
  \K{(p-k_2)}}{\fmslash{p} - \fmslash{k}_2} \K{(p+k_1)} \rb
\right]\nn\\
&=& e' e^2 \int_p \frac{1 - \K{p}}{p^2} \left[ \frac{\K{(p-k_2)}
      \left( 1 - \K{(p+k_1)}\right)}{(p+k_1)^2} 
    \Tr \gamma_5 \gamma_\alpha \fmslash{p} \gamma_\beta
    \fmslash{k}_1\right.\nn\\
&&\left.\qquad - \frac{\K{(p+k_1)} \left( 1 -
            \K{(p-k_2)}\right)}{(p-k_2)^2} 
\Tr \gamma_5 \gamma_\alpha
\fmslash{p} \gamma_\beta \left( - \fmslash{k}_2\right) \right]\nn\\
&=& 4 e' e^2 \int_p \frac{1 - \K{p}}{p^2} \left[ \frac{\K{(p-k_2)}
      \left( 1 - \K{(p+k_1)}\right)}{(p+k_1)^2}
    \ep_{\alpha\gamma\beta\delta} p_\gamma
    k_{1\delta}\right.\nn\\
&&\left.\qquad - \frac{\K{(p+k_1)} \left( 1 -
            \K{(p-k_2)}\right)}{(p-k_2)^2} 
\ep_{\alpha\gamma\beta\delta} p_\gamma (- k_{2\delta})\right]\,.
\end{eqnarray}
Finally, using
\begin{eqnarray}
&&\int_p \frac{\left(1 - \K{p}\right)^2}{p^4} \K{(p-k_2)} p_\gamma\nn\\
&=& \Lambda \int_p \frac{(1-K(p))^2}{p^4} K \left( p -
    \frac{k_2}{\Lambda} \right) p_\gamma\nn\\
&\asym& \int_p \frac{(1 - K(p))^2}{p^4} \Delta (p) \frac{(p k_2)
  p_\gamma}{p^2} = \frac{1}{6 (4 \pi)^2} k_{2\gamma}\,,
\end{eqnarray}
we obtain
\begin{equation}
T_{\alpha\beta} (k_1, k_2) \asym e' e^2 \frac{4}{3} \frac{1}{(4
  \pi)^2} \ep_{\alpha\beta\gamma\delta} k_{1\gamma} k_{2\delta}\,.
\end{equation}
The other graph gives the same contribution, and altogether we obtain
\begin{equation}
b = e' e^2 \frac{8}{3} \frac{1}{(4 \pi)^2}\,.\label{anomaly-chiral-b}
\end{equation}
\item[(ii)] $b'$ --- The $AA$ term of $\Phi_5$ is given by the diagram
    in Fig.~\ref{anomaly-chiral-triangle2}, similar to the previous
    diagram.  
\begin{figure}[t]
\begin{center}
\epsfig{file=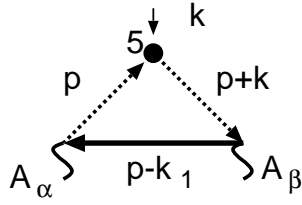, width=4cm}
\caption{The broken line gives $e' U \gamma_5$.}
\label{anomaly-chiral-triangle2}
\end{center}
\end{figure}
The difference is in the use of $U \gamma_5$ instead of $U$, and
replacement of $B$ by $A$.  There is one more graph obtained by
interchanging $A_\alpha (k_1)$ and $A_\beta (k_2)$.  The graph in
Fig.~\ref{anomaly-chiral-triangle2} gives
\begin{eqnarray}
&& \int_p \Tr e' U (-p-k,p) \gamma_5 e \gamma_\alpha \frac{1 -
  \K{(p-k_1)}}{\fmslash{p} - \fmslash{k}_1} e \gamma_\beta\nn\\
&=& T_{\alpha\beta} (k_1,k_2)
\asym e' e^2 \frac{4}{3} \frac{1}{(4 \pi)^2}
\ep_{\alpha\beta\gamma\delta} k_{1\gamma} k_{2\delta}\,.
\end{eqnarray}
The other graph gives the same.  Hence, we obtain
\begin{equation}
b' = e' e^2 \frac{4}{3} \frac{1}{(4 \pi)^2}
= \frac{b}{2}\,.\label{anomaly-chiral-bprime}
\end{equation}
\item[(iii)] $b''$ --- The $BB$ term of $\Phi_5$ can be calculated
    similarly.  We obtain
\begin{equation}
b'' = (e')^3 \frac{4}{3} \frac{1}{(4 \pi)^2}\,.
\end{equation}
\end{itemize}

Thus, we find
\begin{equation}
b'' \ne 0\,,\quad
b + b' \ne 0\,.
\end{equation}
No matter what we choose as $a_5$, we cannot satisfy $\Sigma_\Lambda =
0$.  At best, we can save the vector gauge invariance by choosing
\begin{equation}
a_5 = - b = - e' e^2 \frac{8}{3} \frac{1}{(4 \pi)^2}\,.
\end{equation}
This leaves us the following chiral anomaly:
\begin{eqnarray}
\Sigma_\Lambda &\asym& -  \frac{1}{(4
  \pi)^2} \int_{k_1,k_2} 
\ep_{\alpha\beta\gamma\delta} k_{1\gamma} k_{2\delta}\,
c'(-k_1-k_2)\nn\\
&&\quad \times \left( 4 e' e^2 A_\alpha (k_1) A_\beta (k_2) +
    \frac{4}{3} (e')^3  B_\alpha (k_1) B_\beta (k_2) \right)\,.
\end{eqnarray}

\subsubsection{$U(1)_R$ gauge theory}

We construct a quantum theory whose classical action is given by
\begin{equation}
S_{cl} = \int d^4 x \left[ - \frac{1}{4} F_{\mu\nu}^2 + \frac{1}{2
      \xi} (\partial_\mu A_\mu)^2 
- \partial_\mu \bar{c} \partial_\mu c
+ \bar{\psi}_R \left( - \frac{1}{i} \fmslash{\partial} + e
    \fmslash{A} \right) \psi_R \right]\,,
\end{equation}
where $\psi_R$ and $\bar{\psi}_R$ are chiral:
\begin{equation}
\frac{1 + \gamma_5}{2} \psi_R = \psi_R\,,\quad
\bar{\psi}_R \frac{1 - \gamma_5}{2} = \bar{\psi}_R\,.
\end{equation}

For the quantum theory, the WT composite operator is given by
\begin{equation}
\Sigma_\Lambda = \ep \int_k c(k) \left( k_\mu \frac{\delta
      \SIL}{\delta A_\mu (k)} - \Phi_R (-k) \right)\,,
\end{equation}
where
\begin{eqnarray}
\Phi_R (-k) &\equiv& e \int_p \K{p} \left[ - \SL \Rd{\psi_R (p)}
    [\psi_R]_\Lambda (p-k)
 + [\bar{\psi}_R]_\Lambda (-p-k) \Ld{\bar{\psi} (-p)} \SL \right.\nn\\
&&\left.\quad
+ \Tr \frac{1+\gamma_5}{2} U (-p-k,p) \Ld{\bar{\psi}_R (-p)} \SIL
\Rd{\psi_R (p+k)} \right]\,,
\end{eqnarray}
where $U$ is defined by (\ref{anomaly-chiral-U}).
$\Phi_R$ has the asymptotic behavior
\begin{eqnarray}
\Phi_R (-k) &\asym& \left( \Lambda^2 \bar{a}_2 (\ln \Lambda/\mu) 
    + k^2 \bar{d}_2 (\ln \Lambda/\mu) \right) k_\mu A_\mu
(-k)\nn\\
&& + \bar{a}_3 (\ln \Lambda/\mu) \int_p \bar{\psi}_R (-p-k) \fmslash{k} \psi_R
(p)\nn\\
&& + b (\ln \Lambda/\mu)
\ep_{\alpha\beta\gamma\delta} \int_{k_1+k_2=-k} k_{1\alpha} k_{2\beta} 
A_\gamma (k_1) A_\delta (k_2)\nn\\
&& + \bar{a}_4 (\ln \Lambda/\mu) \frac{1}{2} \int_{k_1,k_2} A_\nu (k_1)
A_\nu (k_2) k_\mu A_\mu (-k-k_1-k_2)\,.
\end{eqnarray}
Compared with $\Phi$ for QED, $\Phi_R$ has an extra term proportional to
the $\ep$ tensor.  On the other hand, the asymptotic behavior of $\SIL$
has the same form as for QED:
\begin{eqnarray}
S_{I,\Lambda} &\asym& \int d^4 x \,\left[ \Lambda^2 a_2 (\ln
        \Lambda/\mu) \frac{1}{2}
    A_\mu^2\right.\nn\\
&&\quad + c_2 (\ln \Lambda/\mu) \frac{1}{2} \partial_\mu
A_\nu \partial_\mu A_\nu + d_2 (\ln \Lambda/\mu)
\frac{1}{2} \left(\partial_\mu A_\mu\right)^2\nn\\
&&\quad + a_4 (\ln \Lambda/\mu) \frac{1}{4!} \left( A_\mu^2 \right)^2\nn\\
&&\quad + a_f (\ln \Lambda/\mu) \bar{\psi}_R \frac{1}{i}
\fmslash{\partial} \psi_R 
 + a_3 (\ln \Lambda/\mu) \bar{\psi}_R \fmslash{A} \psi_R
\Big]\,,
\end{eqnarray}
which implies
\begin{eqnarray}
k_\mu \frac{\delta \SIL}{\delta A_\mu (k)}
&\asym& 
\left( \Lambda^2 a_2 (\ln \Lambda/\mu) 
     + k^2 (c_2 + d_2) (\ln \Lambda/\mu) \right) k_\mu A_\mu
(-k)\nn\\
&& + a_3 (\ln \Lambda/\mu) \int_p \bar{\psi}_R (-p-k) \fmslash{k} \psi_R
(p)\nn\\
&& + a_4 (\ln \Lambda/\mu) \frac{1}{2} \int_{k_1,k_2} A_\nu (k_1)
A_\nu (k_2) k_\mu A_\mu (-k-k_1-k_2)\,.
\end{eqnarray}
Hence, unless
\begin{equation}
b (0) = 0
\end{equation}
by chance, we cannot satisfy the Ward identity by adjusting the
parameters in hands.

At 1-loop, the coefficient $b$ can be calculated from
Fig.~\ref{anomaly-chiral-triangle} where $B_\alpha$ is replaced by
$A_\alpha$, and $U$ is multiplied by the projection operator $\frac{1
  + \gamma_5}{2}$ from the left.  We obtain
\begin{eqnarray}
&&\int_p \Tr \frac{1+\gamma_5}{2} e U (-p-k,p) e \gamma_\alpha \frac{1 -
  \K{(p-k_1)}}{\fmslash{p} - \fmslash{k}_1} e \gamma_\beta\nn\\
&=& - \frac{1}{2} T_{\alpha\beta} (k_1, k_2)\Big|_{e'=e}
\asym - e^3 \frac{2}{3} \frac{1}{(4 \pi)^2}
\ep_{\alpha\beta\gamma\delta} k_{1\gamma} k_{2\delta}\,,
\end{eqnarray}
where $e'$ is replaced by $e$.  There is another graph with $A_\alpha
(k_1)$ and $A_\beta (k_2)$ interchanged.  Hence, we obtain
\begin{equation}
b (0)= -  e^3 \frac{2}{3} \frac{1}{(4 \pi)^2}\,.
\end{equation}
Therefore, we cannot satisfy the WT identity.

\newpage
\section{Functional integral approach to the antifield formalism\label{BV}}

In \S\S \ref{AF} \& \ref{AFexamples} we have adapted the antifield
formalism to the ERG framework.  Though our discussion has been limited
to perturbation theory, we have obtained an important conclusion:
realization of symmetry is reduced to solving a classical algebraic
problem.  The purpose of the present section is to give a second look at
the adaptation of the antifield formalism without relying on
perturbative expansions.

The antifield formalism of Batalin and
Vilkovisky\cite{Batalin:1981jr}\cite{BVreview-Gomis}\cite{Henneaux-Teitelboim}
can provide us with a systematic description of any symmetry present in
a theory.  In the following we apply the antifield formalism to the bare
and Wilson actions.  The quantum master equation (QME), satisfied by an
action, signals the presence of symmetry, naturally extending the WT
identity discussed in \S\ref{WT}.  As opposed to \S\S \ref{AF} \&
\ref{AFexamples}, where we have constructed the continuum limit
directly, we start from a bare action at a UV scale $\Lambda_0$.
Following the procedure of \S\ref{derivation} we construct a Wilson
action by integrating the fields over the momenta between $\Lambda_0$
and $\Lambda$ ($<\Lambda_0$).  With $\Lambda_0$ kept finite, the
resulting ERG trajectory approaches the trajectory of the continuum
limit ${\bar S}_{\Lambda}$ only as $\Lambda \to 0$.  (See
Fig.~\ref{cl-renormalizedtrajectory}.)  Nevertheless, by keeping
$\Lambda_0$ finite, we can manage to write down more explicit
expressions for the antifield dependence of the Wilson
action,\cite{Igarashi:2007fw,Higashi:2007ax,Igarashi:2009ymBV} compared
with the perturbative construction in terms of the asymptotic behavior
in \S\S \ref{AF} \& \ref{AFexamples}.

We discuss only a generic gauge theory in this section, and give
concrete examples in the next section.  A global symmetry can be treated
in a similar manner,\cite{Igarashi:2001cv}\cite{Igarashi:2006sc} though
we do not discuss it here.\footnote{The QME for chiral symmetry is an
extension of the well-known Ginsparg-Wilson
relation.\cite{Ginsparg:1981bj} In \S\ref{WTexamples-axial} we have
briefly discussed the WT identity for the axial symmetry.}  The
organization of this section is as follows.  In \S\ref{BV-BV}, we first
review the classical antifield formalism, and then explain the
modification necessary for quantized systems.  For classical systems a
symmetry is realized as a classical master equation (CME), and for
quantum systems as a quantum master equation (QME).  We then show, in
\S\ref{BV-scale}, how to adapt the formalism in the presence of a UV
cutoff, and in \S\ref{BV-WT}, how the QME relates to the WT identity.
In \S\ref{BV-AFdep} we derive explicit antifield dependence of the
Wilson action, assuming a simple form for the BRST transformation of the
bare action.  In \S\ref{BV-ST} we introduce the effective average
action, and rewrite the QME for the Wilson action as the so-called
``modified Slavnov-Taylor identity''\cite{Ellwanger:1994iz} for the
average action.  Finally, in \S\ref{BV-nil} we discuss an implication of
the nilpotency of the BRST transformation.

In this section, we adopt a notation that can handle bosonic and
 fermionic fields equally in order to respect the canonical structure of
 the antifield formalism.  We use the symbol $\phi^A$ with a suffix $A$
 for fields, and distinguish their statistics by Grassmann parity
 $\ep_A$ mod $2$:
\begin{equation}
\ep (\phi^A) = \ep_A \equiv \lb\begin{array}{c@{\quad}l}
0 & \mathrm{bosonic}\,,\\
1 & \mathrm{fermionic}\,.
\end{array}\right.\quad (\textrm{mod}\,\, 2)
\end{equation}
Though not mentioned explicitly in \S\S\ref{AF} \& \ref{AFexamples}, a
field and its corresponding antifield form a canonical pair.  In order
to keep the canonical structure, we rescale an antifield by the cutoff
function $\K{p}$.  Hence, we must replace the antifield $\phi^* (p)$ of
the previous sections by $\K{p} \phi^* (p)$ to get the corresponding
results in this section.

\subsection{ The Batalin-Vilkovisky antifield formalism\label{BV-BV}}

Let us consider a classical gauge fixed action $S_{cl}[\phi]$ that is
invariant under the nilpotent BRST transformation $\delta
\phi^A$:\footnote{The right (left) derivative is indicated by the
superscript $r$ ($l$).  }
\begin{equation}
\delta \Scl \equiv
\frac{\pa^{r} \Scl}{\pa \phi^{A}}~\delta \phi^{A} =0\,.
\label{BV-BRST inv}
\end{equation}
The fields $\phi^A$ represent collectively gauge and matter fields as
well as (anti)ghosts and B-fields.\footnote{The $B$ field is an
auxiliary field which is equal to the divergence of the gauge field
$\partial_\mu A_\mu$ by the equation of motion.  We introduce this to
avoid the quadratic term $(\bar{c}^*)^2$ in $\bScl$ (\ref{BV-bScl}).}
The index $A$ represents any index to distinguish different types of
fields, such as momentum, the Lorentz index for a vector field, and the
spinor index for a spinor field.  Hence, the Einstein convention for
repeated $A$ includes integration over momenta.

By introducing an antifield $\phi_{A}^*$ for each field $\phi^{A}$, we
define an extended action
\begin{eqnarray}
\bScl [\phi,\phi^*]\equiv \Scl [\phi] + \phi_{A}^* \delta \phi^{A}~.
\label{BV-bScl}
\end{eqnarray}
The antifield $\phi^*_A$ has the opposite Grassmann parity to that of
$\phi^{A}$:
\begin{equation} 
\ep(\phi^{*}_{A}) \equiv \ep(\phi^{A}) + 1\,.\quad(\mathrm{mod}\,\,2)
\end{equation}
We also assign ghost numbers to $\phi^A$ and $\phi^*_A$ so that they add
up to $-1$.  

In the space of $\phi^{A}$ and $\phi^{*}_{A}$, we define a canonical
structure by introducing an antibracket: for any field variables $X$ and
$Y$, we define
\begin{eqnarray}
(X,~Y) \equiv \frac{\partial^{r} X}{\partial \phi^{A}}
\frac{\partial^{l}Y}{\partial \phi_{A}^*}
- \frac{\partial^{r} X}{\partial
\phi^{*}_{A}}\frac{\partial^{l}Y}{\partial \phi^{A}}~.
\label{BV-bracket}
\end{eqnarray}
Fields $\phi^A$ and their antifields $\phi^*_{A}$ are canonical
conjugate pairs, satisfying
\begin{equation}
(\phi^{A},~\phi_{B}^{*})= \delta^{A}_{~B}\,,\quad
~(\phi^{A},~\phi^{B})=(\phi^{*}_{A},~\phi^{*}_{B})=0\,.
\end{equation}
We now define the BRST transformation of an arbitrary variable $X$ by 
\begin{eqnarray}
\delta X \equiv (X,~\bScl)\,.
\label{BV-clBRST}
\end{eqnarray}
This generalizes $\delta \phi^A$ for the fields $\phi^A$; for example,
it gives the transformation of antifields by
\begin{eqnarray}
\delta \phi^*_A = - \frac{\partial^l \bScl}{\partial \phi^A}~.
\label{BV-antifield tf}
\end{eqnarray}

Following the definitions \bref{BV-bScl}, \bref{BV-bracket},
\bref{BV-clBRST}, we obtain
\begin{equation}
(\bScl~,~\bScl) = 2 \left(\delta \Scl + \phi_A^* \delta^2 \phi^A \right)\,,
\end{equation}
where $\delta \Scl$ is defined by \bref{BV-BRST inv}.  Hence, if $\delta
S_{cl} = 0$ \bref{BV-BRST inv}, and the
original BRST transformation is nilpotent
\begin{equation}
\delta^2 \phi^A = 0\,,\label{BV-originalnilpotency}
\end{equation}
we obtain the \textbf{classical master equation} (\textbf{CME}):
\begin{equation}
(\bScl~,~\bScl) = 0\,.
\label{BV-CME}
\end{equation}
Thus, CME is equivalent to the combination of the BRST invariance of
$\Scl$ \bref{BV-BRST inv} and the nilpotency
\bref{BV-originalnilpotency}.  In fact the nilpotency of the BRST
transformation \bref{BV-clBRST} remains valid, even if $\delta$ acts on
an arbitrary variable $X$, dependent on both $\phi^A$ and $\phi^*_A$;
applying $\delta$ twice, we obtain
\begin{equation}
\delta^2 X = \left(\delta X~,~\bScl \right) = \left(
	      (X~,~\bScl)~,~\bScl\right) = (X~,~ (\bScl~,~\bScl))\,.
\end{equation}
Hence, \bref{BV-CME} implies the general nilpotency:
\begin{equation}
\delta^2 X = 0\,.\label{BV-clnilpotent}
\end{equation}

We now apply the BV formalism to quantum systems with BRST invariance.
Let $\bar{S} [\phi,\phi^*]$ be an action that defines a quantum system
via functional integration over $\phi$: $\bar{S}$ is either a bare
action with a UV cutoff $\Lambda_0$, or a Wilson action with a lower
cutoff $\Lambda$.  Under the BRST transformation of fields
\begin{equation}
{\ds\delta \phi^{A}=\frac{\pa^{l} {\bar S}}{\pa
\phi_{A}^{*}}}\,,
\label{BV-BRSTtransformation}
\end{equation}
the action changes by the \textbf{quantum master
operator}:
\begin{equation}
{\bar \Sigma}[\phi,\phi^{*}]\equiv 
\frac{\partial^{r} {\bar S}}{\partial \phi^{A}}
\frac{\partial^{l}{\bar S}}{\partial \phi_{A}^*} +
\frac{\pa^r}{\pa\phi^{A}}\delta \phi^{A}
= \frac{1}{2}({\bar S},~{\bar S}) + \Delta {\bar S}\,,
\label{BV-QMoperator}
\end{equation}
where we define
\begin{equation}
\Delta \equiv (-)^{\epsilon_A +1} \frac{\pa^{r}}{\pa \phi^{A}}
\frac{\pa^{r}}{\pa \phi_{A}^{*}}\,.
\label{BV-Delta operator}
\end{equation}
The first term is the classical change of the action under the BRST
transformation \bref{BV-BRSTtransformation}.  The second term is the
contribution from the jacobian of the BRST transformation.  The system
is BRST invariant quantum mechanically if the two contributions cancel:
\begin{equation}
\bar{\Sigma} [\phi,\phi^*] = 0\,.\label{BV-QME}
\end{equation}
We call this equation the \textbf{quantum master equation}
(\textbf{QME}).  Note that the quantum master operator
\bref{BV-QMoperator} appears somewhat different from \bref{AF-Sigmabar};
the cutoff dependent factor $\K{p}$ is missing in \bref{BV-QMoperator}.
This is due to the adoption of a different normalization of the
antifields, as mentioned in the last paragraph before \S\ref{BV-BV}.

In the antifield formalism, we define the quantum BRST transformation as
\begin{eqnarray}
\delta_Q X \equiv (X, {\bar S}) + \Delta X
\label{BV-qBRST}
\end{eqnarray}
for an arbitrary variable $X$.  Without assuming QME, we obtain two
important identities:
\begin{eqnarray}
&&\delta_Q {\bar \Sigma}[\phi, \phi^*]=0\,,
\label{BV-del sigma zero}\\
&&\delta_Q^2 X = (X, {\bar \Sigma}[\phi, \phi^*])\,.
\label{BV-qBRST squared}
\end{eqnarray}
These identities are algebraic, consequences of the definitions of the
quantum master operator \bref{BV-QMoperator} and the quantum BRST
transformation \bref{BV-qBRST}.  The identity \bref{BV-del sigma zero}
is crucial for the perturbative construction of symmetric theories, as
shown in \S\S \ref{AF} \& \ref{AFexamples}.  Eq.~\bref{BV-qBRST squared}
implies that the quantum BRST transformation \bref{BV-qBRST} is
nilpotent if and only if QME \bref{BV-QME} holds.  Note the importance
of the second term in \bref{BV-qBRST}.  If we defined the BRST
transformation without it by
\begin{equation}
\delta X \equiv (X, \bar{S})\,,
\end{equation}
we would obtain
\begin{equation}
\delta^2 X = (X, (\bar{S}, \bar{S}))\,,
\end{equation}
and hence nilpotency demands CME instead of QME.  For a generic quantum
system, however, what holds is QME, but not CME.  Therefore, it is the
quantum BRST transformation that can be nilpotent.

\subsection{ Scale change in the functional integral approach\label{BV-scale}}

The action $\bar{S}$ of the previous subsection is either a bare action
$\bar{S}_B$ or the corresponding Wilson action $\bSL$.\footnote{As in
\S\ref{derivation} we denote a bare action by $S_B$ and a Wilson action
by $\SL$.  As in \S\S\ref{AF} \& \ref{AFexamples} we also put a bar to
denote the dependence on antifields.}  We recapitulate the important
results of \S\ref{derivation} using the notation adopted in the present
section.

Let $\bar{S}_B$ be an action defined at the UV scale $\Lambda_{0}$ in the
presence of antifields:
\begin{equation}
\bar{S}_B [\phi, \phi^*] \equiv S_B [\phi] + \phi^*
 \cdot \delta \phi\,,\label{BV-bare action}
\end{equation}
where $S_B$ is the sum of the gaussian and interaction terms:
\begin{equation}
S_B [\phi] \equiv -\frac{1}{2}\phi \cdot K_{0}^{-1} D \cdot \phi + 
{S}_{I,B} [\phi]\,.
\label{micro-action}
\end{equation}
Here, we have adopted the matrix notation for momentum integrals:
\begin{eqnarray}
\phi \cdot K_0^{-1} D \cdot \phi &\equiv& \int_p \phi^A (-p)
 \frac{1}{K_0 (p)} D_{AB} (p) \phi^B
 (p)\,,\\
\phi^* \cdot \delta \phi &\equiv& \int_p \phi^*_A (-p) \phi^A (p)\,.
\end{eqnarray}
We have assumed that the antifield dependence of $\bar{S}_B$ is strictly
linear.

By introducing sources $J_A$, we define the generating functional 
\begin{eqnarray}
\bar{Z}_B [J, \phi^*] \equiv \int {\cal D} \phi 
\exp\left(\bar{S}_B [\phi, \phi^*]+ K_{0}^{-1} J \cdot \phi
       \right )~.
\label{BV-part-func1}
\end{eqnarray}
In performing the above functional integral, we decompose the fields
$\phi^A$ into the IR fields $\Phi^A$ and UV fields $\chi^A$, where the
propagators of each class of fields are given by
\begin{equation}
\begin{array}{c@{\quad\textrm{for}\quad}c}
K_{0}(p)\left(D_{AB}(p)\right)^{-1}& \phi^A\,,\\
K(p)\left(D_{AB}(p)\right)^{-1}& \Phi^A\,,\\
(K_{0}(p)-K(p))\left(D_{AB}(p)\right)^{-1}& \chi^A\,.
\end{array}
\end{equation}
Note that $\Phi^A$ carry the momenta below $\Lambda$, and $\chi^A$ those
between $\Lambda_0$ and $\Lambda$.  

We now introduce the Wilson
action\cite{Igarashi:2007fw}\cite{Igarashi:2007fw}
\begin{equation}
\bSL [\Phi, \Phi^*] \equiv -\frac{1}{2}\Phi \cdot K^{-1}D
 \cdot \Phi  
+ \bSIL [\Phi, \Phi^*]\,,
\label{BV-Wilsonian}
\end{equation}
where the interaction part is defined by the functional integral over
the UV fields $\chi^A$:
\begin{eqnarray}
\hspace{-10mm}\exp \left( \bSIL [\Phi, \Phi^*]\right)
 \equiv \int{\cal D}\chi 
\exp \left[ - \frac{1}{2} \chi \cdot (K_{0}-K)^{-1}D \cdot \chi +
\bar{S}_{I,B} [\Phi + \chi, \phi^*] \right] .
\label{BV-Wilsonian-int}
\end{eqnarray}
The two antifields $\phi^*, \Phi^*$ are related simply as 
\begin{equation}
K \,\Phi_A^* = K_0\, \phi_A^*\,.
\end{equation}
We define the generating functional of the Wilson action by
\begin{equation}
\bar{Z}_\Lambda [J, \Phi^*] = \int {\cal D} \Phi 
\exp\left(\bSL [\Phi, \Phi^*]+ K^{-1}J \cdot \Phi
       \right )\,.
\label{BV-part-func2}
\end{equation}

The two generating functionals (\ref{BV-part-func1}) and
(\ref{BV-part-func2}) are related by \bref{derivation-WBWL}, which is
rewritten as
\begin{eqnarray}
\bar{Z}_B [J, \phi^*]= N_{J} \bar{Z}_\Lambda [J, \Phi^*]\,,
\label{BV-Z-relation}
\end{eqnarray}
where the normalization factor $N_{J}$ is given by
\begin{eqnarray}
\ln N_{J} &=& -\frac{(-)^{\ep_{A}}}{2} 
J_{A}  K_{0}^{-1}K^{-1}(K_{0}-K) \left(D^{-1}\right)^{AB}  J_{B}~.
\label{BV-normalization}
\end{eqnarray}
Because of its importance, we wish to rederive it using the present
notation.
The main tool is the triviality of the gaussian integral:
\begin{eqnarray}
&&\int \mathcal{D} \theta \exp \Bigg[ -\frac{1}{2}
\Bigl(\theta -  J~(K_{0}-K)D^{-1}\Bigr) \cdot
\frac{D}{K_{0}K(K_{0}-K)} \nn\\
&& \qquad\qquad\qquad \cdot 
\Bigl(\theta - (-)^{\epsilon(J)}D^{-1}(K_{0}-K)J\Bigr) \Bigg]
= \mathrm{ const}\,.
\label{gauss}
\end{eqnarray} 
We substitute this into the partition function $\bar{Z}_B [J,~\phi^{*}]$
defined by \bref{BV-part-func1}, and perform a canonical change of
variables from $\{\phi,~\phi^{*},~\theta,~\theta^{*}\}$ to
$\{\Phi,~\Phi^{*},~\chi,~\chi^{*}\}$, where
\begin{equation}
\begin{array}{l@{\quad}l}
\phi^{A} =  \Phi^{A} + \chi^{A}\,,& \theta^{A}= (K_{0}-K)\Phi^{A}-
 K\chi^{A}~,\\
\phi^{*}_{A} = K_{0}^{-1}\left[K \Phi^{*}_{A} + 
(K_{0}-K)\chi^{*}_{A}\right]\,,&
\theta^{*}_{A}= K_{0}^{-1}(\Phi^{*}_{A} - \chi^{*}_{A})~.
\end{array}
\label{new-fields}
\end{equation}
Then, we obtain 
\begin{eqnarray}
\bar{Z}_B [J,~\phi^{*}] &=& N_{J}\int {\cal D} \Phi{\cal D}\chi 
\exp\biggl[ -\frac{1}{2}\Phi \cdot K^{-1}D \cdot \Phi + K^{-1}~J \cdot
\Phi\nn\\
&{}&  -\frac{1}{2}\chi \cdot  (K_{0}-K)^{-1}D \cdot  \chi + 
\bar{S}_{I,B} [\Phi + \chi,~\phi^{*}]\biggr]~,
\label{z-relation}
\end{eqnarray}
where $N_J$ is given by \bref{BV-normalization}.  This is independent of
the antifields $\chi^*$, and we may adopt a gauge choice
\begin{equation}
\chi^{*}_{A}=0\,.
\end{equation}
This gives
\begin{eqnarray}
K_0 \,\phi^{*}_{A}=  K\,\Phi^{*}_{A}.
\label{BV-antifield-relation}
\end{eqnarray} 
We will justify this relation shortly. With the above gauge choice, we
obtain \bref{BV-Z-relation} \& \bref{BV-normalization}, where the
Wilson action is defined by \bref{BV-Wilsonian} \& \bref{BV-Wilsonian-int}.

The partition function $\bar{Z}_B$ does not depend on the IR
cutoff $\Lambda$ so that 
\begin{equation}
\Lambda \frac{\pa}{\pa \Lambda} {\bar Z}_B = 0\,.
\end{equation}
This yields the Polchinski differential equation\cite{Polchinski:1983gv}
\begin{eqnarray}
&{}&\hspace{-7mm}-\Lambda \frac{\partial}{\partial \Lambda} \bSL =  \int_p 
\bigl(K^{-1} \Delta \bigr)(p) 
\biggl[\Phi^A(p) 
\frac{\pa^l \bSL}{\pa \Phi^A(p)}-\Phi^{*}_{A}(p) 
\frac{\pa^l \bSL}{\pa \Phi^{*}_{A}(p)} 
\biggr]
\nn\\
&{}&\hspace{-4mm}+\frac{1}{2}\int_p 
(-)^{\epsilon_A}  \bigl(D^{-1} \Delta \bigr)^{AB} (p)
\biggl[
\frac{\pa^l \bSL}{\pa \Phi^B(-p)}
\frac{\pa^r \bSL}{\pa \Phi^A(p)}
+
\frac{\pa^l \pa^r \bSL}{\pa \Phi^B(-p) \pa \Phi^A(p)}
\biggr]\,,
\label{BV-flow eq}
\end{eqnarray}
where we define
\begin{equation}
\Delta (p/\Lambda) \equiv \Lambda \frac{\partial}{\partial \Lambda}
 \K{p}
\end{equation}
as in \bref{derivation-Deltadef}.  $\bar{S}_\Lambda$ is completely
determined by the Polchinski differential equation and the initial
condition
\begin{eqnarray}
\bar{S}_{\Lambda} [\Phi,~\Phi^{*}] \Big|_{\Lambda = \Lambda_0} =
\bar{S}_B [\Phi,~\Phi^{*}] \,. 
\label{ini-con}
\end{eqnarray}

We can define a composite operator ${\bar {\cal O}}_\Lambda [\Phi,
\Phi^*]$ as a functional that satisfies the differential equation
\bref{comp-ERGdiffO}.  Using the present notation, we obtain
\begin{eqnarray}
-\Lambda \frac{\partial }{\partial \Lambda} {\bar {\mathcal{O}}}_\Lambda
 = {\bar {\cal D}} \cdot {\bar {\mathcal{O}}}_\Lambda\,,
\label{BV-RGCO-BV}
\end{eqnarray}
where
\begin{equation}
\hspace{0.5cm}\bar{\cal D} \equiv
- (K^{-1} \Delta)\Phi^*_A \frac{\partial^l}{\partial \Phi^*_A}
+ (-)^{\ep_{A}}
\left(D^{-1} \Delta\right)^{AB}\Bigl(
\frac{\pa^{l} \bSIL}{\pa \Phi^{B}}\frac{\pa^{r}}{\pa \Phi^{A}}
+\frac{1}{2}
\frac{\pa^{l}\pa^{r} }{\pa \Phi^{B}\pa \Phi^{A}} \Bigr)\,.\label{BV-Dbar}
\end{equation}
Note that the first term of \bref{BV-Dbar}, rescaling of the
antifield, is a consequence of \bref{BV-antifield-relation}.
Hence, the factor $K^{-1}$ is necessary in order to make
\begin{equation}
K^{-1}\frac{\partial \bSL}{\partial \Phi_A^*}
\end{equation}
composite operators.

For later convenience, we recall an alternative definition of composite
operators.  Given an arbitrary functional ${\bar {\cal O}}_B [\phi,
\phi^*]$ at the UV scale, the corresponding IR composite operator ${\bar
{\cal O}}_\Lambda [\Phi, \Phi^*]$ is constructed as
\begin{eqnarray}
&&{\bar {\cal O}}_\Lambda  [\Phi, \Phi^*] \exp \left(\bSIL [\Phi,
						\Phi^*; \Lambda]\right) 
\equiv \int \mathcal{D} \chi\, {\bar {\cal O}}_B  [\Phi + \chi, \phi^*]
\nn\\
&&\qquad \cdot \exp \left[ - \frac{1}{2} \chi \cdot (K_0 - K)^{-1} D
    \cdot \chi + \bar{S}_{I,B} [\Phi + \chi, \phi^*] \right]\,.
\label{defO}
\end{eqnarray}
This implies
\begin{equation}
\left\langle 
{\bar {\cal O}}_\Lambda [\Phi, \Phi^*]
\right\rangle_{\bSL, K^{-1}J}
= N_J^{-1} \left\langle 
{\bar {\cal O}}_B [\phi, \phi^*]
\right\rangle_{\bar{S}_B, K_0^{-1}J} \, \label{vevO}
\end{equation}
in the presence of arbitrary sources $J$ coupled linearly to the fields.

\subsection{ WT identity and QME\label{BV-WT}}

In this subsection we consider the cutoff dependence of the quantum
master operator.  Our starting point is the BRST transformation of
fields at the UV scale $\Lambda_0$, given in terms of an anticommuting
constant $\lambda$ as
\begin{eqnarray}
\phi^{A} \rightarrow \phi^{\prime A} =\phi^{A} +  \delta_{\lambda} \phi^{A}~,
\qquad  \delta_{\lambda} \phi^{A} = \delta \phi^{A} \lambda =
K_0 \mathcal{R}^{A}[\phi]~\lambda~.\label{BRST1}
\end{eqnarray}
 
In computing the generating functional (\ref{BV-part-func1}), we
 consider changing integration variables from $\phi$ to $\phi'$.  If the
 jacobian is properly taken into account, the integral does not depend
 on the choice of integration variables.  Hence, we obtain
\begin{eqnarray}
\int {\cal D}\phi &&\Bigl(K_0^{-1} J \cdot \delta \phi + \Sigma_B [\phi]
+ \frac{\partial^r (\phi^* \cdot \delta \phi) }{\partial \phi^A} \delta
\phi^A \Bigr) \nn\\ &&\hspace{2.5cm}\times\exp\left(\bar{S}_B
[\phi, \phi^*] +K_{0}^{-1} J \cdot \phi\right) = 0\,, \label{equality}
\end{eqnarray}
where $\Sigma_B [\phi]$ is the WT operator defined by
\begin{eqnarray}
\Sigma_B [\phi] 
 \equiv 
\frac{\pa^{r} {S}_B}{\pa \phi^{A}} \delta \phi^{A} 
+ \frac{\pa^{r} }{\pa \phi^{A}} \delta \phi^{A}~. 
\label{WTop1}
\end{eqnarray} 
$\Sigma_B [\phi]$ is the sum of the change of the original
gauge fixed action $S_B [\phi]$
\begin{eqnarray}
\delta_{\lambda} S_B = \frac{\pa^{r} S_B}{\pa \phi^{A}}
\delta_{\lambda} \phi^{A}\,,
\label{delta-cal-S}
\end{eqnarray} 
and the contribution from the jacobian
\begin{eqnarray}
\delta_{\lambda} \ln {\cal D} \phi = (-)^{\ep_{A}}
 \frac{\pa^{r} }{\pa \phi^{A}} \delta_{\lambda} \phi^{A}
=\frac{\pa^{r} }{\pa \phi^{A}} \delta \phi^{A}\lambda\,.
\label{measure}
\end{eqnarray} 

The quantum master operator defined by\footnote{Note that $\Delta$ here
is \bref{BV-Delta operator}, but not the log derivative of $K$.}
\begin{equation}
{\bar \Sigma}_B = \frac{1}{2}(\bar{S}_B, \bar{S}_B) + \Delta \bar{S}_B
\end{equation}
is indeed given as the sum of the WT operator and the third term of 
\bref{equality}:
\begin{equation}
{\bar \Sigma}_B [\phi, \phi^*]
= \frac{\pa^{r} \bar{S}_B}{\pa \phi^{A}} \delta \phi^{A} 
+ \frac{\pa^{r} }{\pa \phi^{A}} \delta \phi^{A}
= \Sigma_B [\phi] 
+ \frac{\partial^r (\phi^* \cdot \delta \phi) }{\partial \phi^A} \delta \phi^A
\,,
\label{QMEop1}
\end{equation}
where we have used $\delta \phi^A =
{\partial^l {\bar S}_B}/{\partial \phi^*_A}$.
Hence, we obtain a simple relation between the QM and WT operators:
\begin{eqnarray}
{\bar \Sigma}_B = \Sigma_B + \phi^* \cdot \delta^2 \phi~.
\label{BV-QME vs WT at UV}
\end{eqnarray}
As CME implies both the BRST invariance of the action and the nilpotency
of the BRST transformation, QME also implies both.  In contrast, the WT
identity implies only the invariance of the action.

We now rewrite \bref{equality} as
\begin{equation}
\vev{\bar{\Sigma}_B [\phi,\phi^*]}_{\bar{S}_B, K_0^{-1} J}
= - K_0^{-1} J \cdot \vev{\delta \phi}_{\bar{S}_B, K_0^{-1} J}\,.
\label{BV-SigmabarBvev}
\end{equation}
Setting $\phi^* = 0$, we obtain an analogous relation for the WT
operator:
\begin{equation}
\left\langle{\Sigma}_B [\phi]\right\rangle_{S_B, K_{0}^{-1}J}
=
- K_0^{-1} J \cdot \left<\delta \phi \right\rangle_{S_B, K_{0}^{-1}J}~. 
\label{BV-SigmaBvev}
\end{equation}
Using \bref{BRST1}, we obtain
\begin{equation}
K_0^{-1} \vev{\delta \phi}_{\bar{S}_B, K_0^{-1}J}
= \vev{\mathcal{R} [\phi]}_{\bar{S}_B, K_0^{-1} J}
= \mathcal{R} \left[ K_0 \partial^l_J\right] \bar{Z}_B [J, \phi^*]\,,
\end{equation}
where the differential operator $\mathcal{R}^A [K_0 \pa^l_J]$ is called
the Slavnov operator.  Setting $\phi^* = 0$ we obtain
\begin{equation}
K_0^{-1} \vev{\delta \phi}_{S_B, K_0^{-1}J}
= \mathcal{R} \left[ K_0 \partial^l_J\right] Z_B [J]\,.
\end{equation}
Therefore, we can rewrite \bref{BV-SigmabarBvev} and \bref{BV-SigmaBvev}
as
\begin{eqnarray}
\vev{\bar{\Sigma}_B [\phi, \phi^*]}_{\bar{S}_B, K_0^{-1}J}
&=& - J \cdot \mathcal{R}\left[K_0 \partial^l_J\right] \bar{Z}_B
[J,\phi^*] \,,\label{BV-SigmabarBvev2}\\ 
\vev{\Sigma_B [\phi]}_{S_B, K_0^{-1}J}
&=& - J \cdot \mathcal{R}\left[K_0 \pa^l_J\right] Z_B [J]\,.
\label{BV-SigmaBvev2}
\end{eqnarray}
We note that the above relations, being based upon the functional
relation \bref{equality}, are valid whether or not the bare action is
invariant under the BRST transformation (\ref{BRST1}).

We now wish to transform \bref{BV-SigmabarBvev2} into an equivalent condition
on the Wilson action ${\bar S}_\Lambda$.  We first apply (\ref{defO}) 
to define the QM operator ${\bar
\Sigma}_{\Lambda} [\Phi, \Phi^*]$ for the IR theory by
\begin{eqnarray}
&&{\bar \Sigma}_{\Lambda} [\Phi, \Phi^*] \exp \left( {\bar
  S}_{I,\Lambda} [\Phi, \Phi^*]\right) 
\equiv \int \mathcal{D} \chi\, {\bar \Sigma}_B [\Phi+\chi, \phi^*]\nn\\
&&\quad \times \exp \left[ -\frac{1}{2} \chi \cdot (K_0 - K)^{-1} D \cdot \chi 
+ 
{\bar {S}}_{I,B} [\Phi + \chi, \phi^*]
\right]
\,.\label{IRWT}
\end{eqnarray}
Then, (\ref{vevO}) implies
\begin{equation}
\left\langle {\bar \Sigma}_{\Lambda}[\Phi, \Phi^*]\right\rangle_{\bSL, K^{-1} J}
= N_J^{-1} \left\langle {\bar \Sigma}_B[\phi,
	    \phi^*]\right\rangle_{\bar{S}_B, K_0^{-1} J}\,. 
\label{twoSigmas}
\end{equation}
We also define the IR operator $\delta \Phi^A$ corresponding to
$\delta \phi^A$ by
\begin{eqnarray}
    &&K^{-1} \delta \Phi^A [\Phi, \Phi^*] \exp \left({\bar S}_{I,\Lambda}
 [\Phi, \Phi^*] \right)\\
&\equiv& \int \mathcal{D} \chi\, K_0^{-1} \delta \phi^A [\Phi + \chi]
\exp \left[ -\frac{1}{2} \chi \cdot (K_0 - K)^{-1} D \cdot \chi 
+ {\bar S}_{I,B} [\Phi+\chi, \phi^*] \right]~.  \nn
\label{deltaPhi}
\end{eqnarray}
Eq. (\ref{vevO}) implies
\begin{equation}
\left\langle K^{-1}\delta \Phi^A \right\rangle_{\bSL, K^{-1} J}
= N_J^{-1} \left\langle K_0^{-1} \delta \phi^A \right\rangle_{\bar{S}_B,
K_0^{-1} J} \,.
\label{relation of tfs}
\end{equation}
Denoting
\begin{equation}
R^A [\Phi, \Phi^*] \equiv K^{-1} \delta \Phi^A [\Phi, \Phi^*] \,,
\end{equation}
we can rewrite \bref{relation of tfs} as
\begin{equation}
R^A [K \partial^l_J, \Phi^*] \bar{Z}_\Lambda [J, \Phi^*] = 
N_J^{-1} \mathcal{R}^A
[K_0 \partial^l_J] \bar{Z}_B [J, \phi^*]\,.\label{twoRs}
\end{equation}
Hence, using (\ref{twoSigmas}) and (\ref{twoRs}), we can transform
\bref{BV-SigmabarBvev2} into
\begin{equation}
\left\langle {\bar \Sigma}_{\Lambda} [\Phi, \Phi^*]\right\rangle_{\bSL,
 K^{-1} J} 
= - J \cdot R[K \partial^l_J, \Phi^*] \bar{Z}_\Lambda [J, \Phi^*]\,,
\end{equation}
which is the desired relation.

The above relation implies
\begin{equation}
{\bar \Sigma}_{\Lambda} [\Phi, \Phi^*] = \frac{\partial^r {\bar
 S}_{\Lambda} [\Phi, \Phi^*]}{\partial
  \Phi^A} \delta \Phi^A + \frac{\partial^r}{\partial \Phi^A} \delta
\Phi^A \,.
\label{IRWTbyS}
\end{equation}
This has the form expected for the quantum master operator; here the
relevant action is the Wilson action $\bSL$ defined at an IR scale
$\Lambda$, and the BRST transformation $\delta \Phi^A$ of the IR field
is a composite operator if multiplied by $K^{-1}$.


Similarly, we can transform (\ref{BV-SigmaBvev2}) into
\begin{equation}
\vev{\Sigma_\Lambda [\Phi]}_{\SL, K^{-1}J} = - J \cdot R[K \pa^l_J]
 Z_\Lambda [J]\,, 
\end{equation}
where the WT operator and the BRST transformation are obtained by
setting $\Phi^*$ to zero:
\begin{eqnarray}
\Sigma_\Lambda [\Phi] &\equiv& \bar{\Sigma}_\Lambda \left[\Phi,
      \Phi^*\right]\big|_{\Phi^*=0}
= \frac{\pa^r \SL [\Phi]}{\pa \Phi^A} \delta
 \Phi^A [\Phi] + \frac{\pa^r}{\pa \Phi^A} \delta \Phi^A [\Phi]\,,
\label{BV-WToperator}\\
\delta \Phi^A [\Phi] &\equiv& \delta \Phi^A \left[\Phi,
	    \Phi^*\right]\big|_{\Phi^*=0} = K R^A [\Phi]\,.
\label{BV-BRSTforWT}
\end{eqnarray}


So far our discussion has been kept very general; we have only assumed
that the bare action $\bar{S}_B$ takes the form of \bref{BV-bare action}
\& \bref{micro-action}.  For symmetric theories the QME must hold:
$\bar{\Sigma}_B$ and hence $\bar{\Sigma}_\Lambda$ must vanish.  For
renormalizable theories, the QME need not hold for any finite
$\Lambda_0$, but it must hold in the limit $\Lambda_0 \to \infty$.  In
general it is a difficult problem to construct a bare action that
satisfies the QME exactly; in \S\S\ref{AF} \& \ref{AFexamples} we have
discussed this problem directly in the limit $\Lambda_0 \to \infty$, but
only perturbatively.  For anomaly-free renormalizable theories, we can
exepect that the quantum master operator behaves as
$$ 
{\bar \Sigma}_B[\phi, \phi^*] = \mathrm{O} (1/\Lambda_{0} )\,, 
$$ 
for a large but finite value of $\Lambda_{0}$.\footnote{We write
$\mathrm{O} (1/\Lambda_0)$ only to mean that it vanishes as $\Lambda_0
\to \infty$.}  To be more precise, we expect
\begin{equation}
\left\langle {\bar \Sigma}_B [\phi, \phi^*] \right\rangle_{\bar{S}_B,
 K_0^{-1} J} = \mathrm{O} (1/\Lambda_0)\,,\label{WT-Sigma-phi1}
\end{equation}
or equivalently,
\begin{equation}
J \cdot \mathcal{R} [K_0 \partial^l_J]
 \bar{Z}_{B} [J, \phi^*]
= \mathrm{O} (1/\Lambda_0)
\end{equation}
for an arbitrary source $J$.\footnote{Later in eqs.~(\ref{cal-S-QED11})
  and \bref{cal-S-YM2}, we write ${\bar S}_B[\phi, \phi^*]$ explicitly
  for gauge theories.  Thanks to renormalizability, the action has only
  a finite number of parameters.  Our assumption is that we can tune the
  parameters so that eq.~(\ref{WT-Sigma-phi1}) holds.}  This is the
  statement of the BRST invariance of the bare theory defined at the UV
  scale $\Lambda_0$.
As a consequence of \bref{twoSigmas}, the QME for the Wilson action
  $\bar{S}_{\Lambda}$ is given either as
\begin{equation}
{\bar \Sigma}_\Lambda [\Phi, \Phi^*] = \mathrm{O} (1/\Lambda_0)\,,
\end{equation}
or
\begin{equation}
J \cdot R[K \partial^l_J, \Phi^*] ~\bar{Z}_\Lambda [J, \Phi^*] = \mathrm{O}
 (1/\Lambda_0)\,. 
\end{equation}

Before concluding this subsection, we consider a
  particular class of BRST transformation
\begin{eqnarray}
\delta \phi^{A} =
K_0 \mathcal{R}^{A}[\phi] = K_0
\Bigl(\mathcal{R}^{(1)A}_{~~~B} (\Lambda_{0}) ~\phi^B  
+ \frac{1}{2} \mathcal{R}^{(2)A}_{~~~BC} (\Lambda_{0})~\phi^B \phi^C \Bigr)~,
\label{form of BRST}
\end{eqnarray}
which are at most quadratic in fields.  The BRST transformation for QED
and YM theories belongs to this class.  We wish to obtain an explicit
expression for the corresponding IR composite operator $\delta \Phi^A$.
In order to rewrite the expectation value 
\[
\langle \delta \phi^A
\rangle_{K_0^{-1}J} = K_0 \mathcal{R}^A [K_0 \partial^l_J]~{\bar
  Z}_B[J, \phi^*]
\] 
for the IR theory, we need to compute two things.

First, we compute the first order differential:
\begin{eqnarray}
    K_{0} \frac{\pa^{l}}{\pa J_{A}} \bar{Z}_B[J, \phi^*]
    &=& K_{0}\frac{\pa^{l}}{\pa J_{A}}
    N_{J}\bar{Z}_\Lambda [J, \Phi^*]\nn\\
    &=& N_{J} \left[(-)^{\ep_{A} +1}
        \left(\frac{(K_{0}-K)}{K}D^{-1}\right)^{AB}J_{B} +
        K_{0}\frac{\pa^{l}}{\pa J_{A}}\right]   \bar{Z}_\Lambda [J, \Phi^*]
    \nn\\
    &=& N_{J} \left\langle K_{0}K^{-1}\Phi^{A} +
	       (K_{0}-K)\left(D^{-1}\right)^{AB} 
        \frac{\pa^{l} {\bar S}_{\Lambda} }{\pa
	\Phi^{B}}\right\rangle_{\bSL, K^{-1}J}\nn\\
    &=& N_{J}\left\langle \Phi^{A} + (K_{0}-K)\left(D^{-1}\right)^{AB}
        \frac{\pa^{l}{\bar S}_{I,\Lambda} }{\pa
	\Phi^{B}}\right\rangle_{\bSL, K^{-1}J}\nn\\
    &=&
     N_{J}\left\langle\left[\Phi^{A}\right]^*\right\rangle_{\bSL, K^{-1}J},  
\label{CO1}
\end{eqnarray}
where we define
\begin{eqnarray}
\left[\Phi^{A}\right]^* &\equiv& K_{0}K^{-1}\Phi^{A} 
+ (K_{0}-K)\left(D^{-1}\right)^{AB}
\frac{\pa^{l} \bar{S}_{\Lambda} }{\pa \Phi^{B}}\nn\\
&=& \Phi^{A} + (K_{0}-K)\left(D^{-1}\right)^{AB}
\frac{\pa^{l} {\bar S}_{I,\Lambda}}{\pa \Phi^{B}}. 
\label{def-CO}
\end{eqnarray}
Second, we compute the second order differential:
\begin{eqnarray}
K_{0} \frac{\pa^{l}}{\pa J_{A}} K_0 \frac{\pa^{l}}{\pa J_{B}}
 \bar{Z}_B[J, \phi^*]
&=& K_{0} \frac{\pa^{l}}{\pa J_{A}} K_0
\frac{\pa^{l}}{\pa J_{B}}
N_{J}\bar{Z}_\Lambda [J, \Phi^*]\nn\\
&\equiv& N_{J} \left\langle
\left[\Phi^{A}~\Phi^{B}\right]^*\right\rangle_{\bSL, K^{-1}J}~,
\label{CO2}
\end{eqnarray}
where we define
\begin{eqnarray}
\Bigl[\Phi^{A}~\Phi^{B}\Bigr]^*
&\equiv& \left[\Phi^{A}\right]^*~
\left[\Phi^{B}\right]^*\nn\\
&{}& \hspace{-6mm} + (K_{0}-K)\left(D^{-1}\right)^{AC}
(K_{0}-K)\left(D^{-1}\right)^{BD}
\frac{\pa^{l} \pa^{l}{\bar S}_{I,\Lambda}}{\pa \Phi^{C} \pa \Phi^{D}}\,.
\label{CO3}
\end{eqnarray}
Note that nontrivial contributions arise from derivatives $\partial_J$
acting on the normalization factor $N_{J}$.

Hence, using (\ref{twoRs}), we obtain
\begin{equation}
\left\langle \delta \Phi^A\right\rangle_{\bSL, K^{-1} J} = K \left\langle
\mathcal{R}^{(1) A}_{~~~B} (\Lambda_0) \left[\Phi^B\right]^*
+ \frac{1}{2} \mathcal{R}^{(2) A}_{~~~BC} (\Lambda_0) \left[\Phi^B
    \Phi^C\right]^*\right\rangle_{\bSL, K^{-1} J}\,.
\end{equation}
Since this is valid for arbitrary $J$, we obtain the operator equality
\begin{eqnarray}
\delta \Phi^A = K \Bigl(
\mathcal{R}^{(1)A}_{~~~B} (\Lambda_0) \left[\Phi^B\right]^*
+ \frac{1}{2} \mathcal{R}^{(2)A}_{~~~BC} (\Lambda_0) \left[\Phi^B
    \Phi^C\right]^*\Bigr)~. 
\label{IR BRST tf}
\end{eqnarray} 
It is important to mention the necessity of the cutoff function $K_{0}$
to make (\ref{IR BRST tf}) UV finite.\footnote{The potential UV
divergence is hard to see in the matrix notation.  It is hidden in the
loop momentum integral contained in $\mathcal{R}^{(2)A}_{~~~BC} {[\Phi^B
\Phi^C ]}$. }

\subsection{ Antifield dependence of the Wilson action\label{BV-AFdep}}

In this subsection we derive the general structure of the Wilson action
$\bar{S}_{\Lambda}[\Phi,\Phi^{*}]$, in particular its dependence on the
antifields.  We first make some assumptions for the bare action that are
general enough to accommodate the cases of QED and Yang-Mills theories.
Then, we describe how the Wilson action depends on the antifields.  In
\S\ref{BVexamples} we will give the Wilson actions for QED and Yang-Mills
theories more explicitly.

As for QED, the Wilson action first obtained in
ref.~\citen{Higashi:2007ax} has two different types of antifield
dependence, one that can be written down explicitly and the other that
appears as a linear shift of the field variables by antifields.  These
features are seen in Eqs.~\bref{cal-J-QED}-\bref{shifted-var-QED} in the
next section.\footnote{Eqs.\bref{cal-J-QED}-\bref{shifted-var-QED}
differ from those given in ref.~\citen{Higashi:2007ax}, where no UV
cutoff is introduced.}  As for non-Abelian gauge theories, additional
non-trivial antifield dependence exists.  In this section, we will show
how to modify the method of \citen{Higashi:2007ax} to treat the
non-Abelian gauge symmetry.  Based on this, we will give explicitly the
Wilson action for Yang-Mills theories in \bref{YM-MA1}.

Without losing generality, we assume that the bare action is linear
in antifields and the BRST transformation quadratic in fields:
\begin{eqnarray}
{\bar {S}}_B[\phi,~\phi^{*}] &=&
-\frac{1}{2}\phi^{A}K_{0}^{-1}D_{{A}{B}}\phi^{B} + {\bar
{S}}_{I,B}[\phi,~\phi^{*}]~, \\
{\bar {S}}_{I,B}[\phi,~\phi^{*}] &=& {S}_{I,B}[\phi] + 
K_0 \phi^{*}_{A} {\cal R}^{A}[\phi]~,\\
{\cal R}^{A}[\phi]&=& 
{\cal R}^{(1)A}_{~~~~B}\phi^{B} + \frac{1}{2}{\cal R}^{(2)A}_{~~~~BC}
\phi^{B} \phi^{C}~. 
\label{cal-SI}
\end{eqnarray}
The bare actions of QED and Yang-Mills theories can be given in this form.

The functional integral \bref{BV-Wilsonian-int} over the UV fields
$\chi$ defines the interaction part ${\bar S}_{I,\Lambda}$ of the Wilson
action, $ \bar{S}_{\Lambda}[ \Phi, \Phi^{*}]$.  We rewrite the integrand
of \bref{BV-Wilsonian-int} as
\begin{eqnarray} 
&&- \frac{1}{2} \chi \cdot (K_{0} -K)^{-1}D \cdot \chi + 
{\bar {S}}_{I,B}[\Phi + \chi,~  K_{0}^{-1}K\Phi^{*}] \nn\\
&& = -\frac{1}{2} \chi \cdot (K_{0} -K)^{-1}D \cdot \chi
+ K \Phi^{*}_{A}{\cal R}^{A}[\Phi + \chi]
 + {S}_{I,B}[\Phi + \chi]\nn\\
&&= - \frac{1}{2} \chi \cdot (K_{0} -K)^{-1}D \cdot \chi + 
K\Phi^{*}_{{A}}{\cal R}^{A}[\Phi] + {\cal J}_{{A}}\chi^{{A}}\nn\\
&&\qquad  + \frac{1}{2}K\Phi^{*}_{{A}}\chi^{{B}} 
{\cal R}^{(2){A}}_{~~~~~{B}{C}}\chi^{C} 
 + {S}_{I,B}[\Phi + \chi]\,,
\label{Wilsonian31}
\end{eqnarray}
where the effective sources coupled to $\chi^{A}$ are given by
\begin{eqnarray}
{\cal J}_{A} \equiv K\Phi^{*}_{B}\left(
{\cal R}^{(1){B}}_{~~~~{A}} + {\cal R}^{(2){B}}_{~~~~{C}{A}}\Phi^{C} \right)\,,
\label{cal-J}
\end{eqnarray} 
and 
\begin{equation}
{S}_{I,B}[\Phi + \chi]
\equiv {\bar {S}}_{I,B}[\Phi + \chi,~
K_0^{-1}K\Phi^{*}]|_{\Phi^{*}=0}\,.
\end{equation}
For the terms quadratic in $\chi$ and linear in the antifield, 
we replace $\chi^{A}$ by the derivative $\partial / \partial {\cal J}^{A}$.
The Wilson action then takes the form 
\begin{eqnarray}
&&\exp \left[{\bar S}_{I,\Lambda}[\Phi,~\Phi^{*}] \right]
= \exp \left[ K \Phi^{*}_{A} \left({\cal R}^{A}[\Phi]
+\frac{1}{2} 
{\cal R}^{(2)C}_{~~~~AB}\frac{\pa^{l}}{\pa {\cal J}_{A}}
\frac{\pa^{l}}{\pa {\cal J}_{B}}
 \right) \right]\nn\\
&& \quad\times \int{\cal D}\chi
\exp \Bigl(-\frac{1}{2} \chi \cdot (K_{0}-K)^{-1}D \cdot \chi + 
{\cal J} \cdot \chi + {S}_{I,B}[\Phi + \chi]
\Bigr) ~. 
\label{Wilsonian4} 
\end{eqnarray} 
In order to simplify the above further, we follow
ref.~\citen{Higashi:2007ax} and introduce a change of variables.  First,
we complete the square with respect to $\chi$ in the first two terms of
the integrand:
\begin{eqnarray}
&&-\frac{1}{2} \chi \cdot (K_{0}-K)^{-1}D \cdot \chi + {\cal J}\cdot \chi \nn\\
&& =-\frac{1}{2}
\chi' \cdot (K_{0}-K)^{-1}D 
 \cdot \chi' 
 + \frac{1}{2}(-)^{\epsilon({\cal J})}{\cal J}
\cdot (K_{0}-K)D^{-1}
{\cal J}~,
\label{cal-J1}
\end{eqnarray} 
where 
\begin{eqnarray}
\chi' &\equiv& \chi-  {\cal J}\cdot (K_{0}-K) D^{-1}
= \chi - (-)^{\ep(\cal J)}D^{-1}(K_{0}-K)\cdot {\cal J}\,.
\label{shifted-var}
\end{eqnarray} 
We then introduce new variables $\Phi'^{A}$ so that 
$\Phi^{A} + \chi^{A} =\Phi'^{A} + \chi'^{A} $:
\begin{eqnarray}
\Phi'^{A} \equiv \Phi^{A} 
+ {\cal J}_{B} (K_{0}-K)(D^{-1})^{{B}{A}}\,.
\label{shifted-var2}
\end{eqnarray}
Finally we obtain
\begin{eqnarray}
&& \exp \left[\bar{S}_{I,\Lambda}[\Phi,~\Phi^{*}]\right]
= \exp \left[ K\Phi^{*}_{A}{\cal R}^{A}[\Phi]
+\frac{K}{2}\Phi^{*}_A 
{\cal R}^{(2)C}_{~~~~AB}\frac{\pa^{l}}{\pa {\cal J}_{A}}
\frac{\pa^{l}}{\pa {\cal J}_{B}}
 \right]\nn\\
&&\quad \times \exp \Bigl[\frac{1}{2}(-)^{\epsilon({\cal J})}{\cal J}
\cdot (K_{0}-K)D^{-1}\cdot
{\cal J}\Bigr]\nn\\
&& \quad\times\int{\cal D}\chi'
\exp \Bigl[-\frac{1}{2} \chi' \cdot (K_{0}-K)^{-1}D \cdot \chi' + 
{S}_{I,B}[\Phi' + \chi'] \Bigr]\,.
\label{cal-W1}
\end{eqnarray} 
Using the definition of $\SIL$ in the absence of antifields
\begin{eqnarray}
\hspace{-7mm}\exp \left[S_{I,\Lambda}[\Phi'] \right] \equiv \int{\cal D}\chi'
\exp \Bigl(-\frac{1}{2} \chi' \cdot (K_{0}-K)^{-1}D \cdot \chi' + 
{S}_{I,B}[\Phi' + \chi'] \Bigr)\,,
\label{SI11}
\end{eqnarray} 
we obtain our final expression for the Wilson action:
\begin{eqnarray}
&& \bar{S}_{\Lambda}[\Phi,~\Phi^{*}] 
= -\frac{1}{2}\Phi^{A}  K^{-1}D_{{A}{B}}\Phi^{B} 
+ K \Phi^{*}_{A} {\cal R}^{A}[\Phi]
%
\nn\\
&&\quad + \ln \Biggl[ \exp \lb\frac{K}{2} \Phi^{*}_{{A}}
{\cal R}^{(2){A}}_{~~~~{B}{C}}
\frac{\pa^{l}}{\pa {\cal J}_{B}}\frac{\pa^{l}}{\pa {\cal J}_{C}}
 \rb\nn\\
&&\quad\qquad\times
\exp \lb\frac{1}{2}(-)^{\epsilon({\cal J})}{\cal J}
\cdot (K_{0}-K)D^{-1}\cdot
{\cal J} + S_{I,\Lambda}[\Phi']\rb
\Biggr]\,.
\label{Wilsonian6}
\end{eqnarray} 

With \bref{Wilsonian6} we can construct the action ${\bar
S}_{I,\Lambda}[\Phi,~\Phi^{*}]$ from $S_{I,\Lambda}[\Phi]$.  The
antifield dependence comes partly from $\mathcal{J}$ \bref{cal-J}, and
partly from the shifted variables $\Phi'$ \bref{shifted-var2}.  The
remaining antifield dependence is solely generated by the exponentiated
differential operator $(K/2)\Phi^{*} R^{(2)}\pa_{\cal J}\pa_{\cal J}$ in
the logarithm.  This complicates $\bSL$ and the BRST transformation of
the fields $\Phi^{A}$.  Once the antifields are removed, however, we
have simple expressions~\cite{Igarashi:2008bb}:
\begin{eqnarray}
\delta \Phi^{A} \vert_{\Phi^{*}= 0} &=& \left[\frac{\partial^{l} {{\bar
 S}_{I,\Lambda}}}{\partial  \Phi^{*}_{A}}\right]_{\Phi^{*}= 0}\nn\\
&=& K \biggl({\cal R}^{(1)A}_{~~~~~B}\left[\Phi^{A}\right] 
+ \frac{1}{2}{\cal R}^{(2)A}_{~~~~~BC}\Bigl[\Phi^{A}~\Phi^{B}\Bigr]\biggr)~,
\label{W-antifield}
\end{eqnarray} 
where 
\begin{eqnarray}
\left[\Phi^{A}\right] &\equiv& 
 \Phi^{A} + (K_{0}-K)\left(D^{-1}\right)^{AB}
\frac{\pa^{l} S_{I,\Lambda}}{\pa \Phi^{B}}~,
\label{W-antifield1}
\\
\left[\Phi^{A}~\Phi^{B}\right] 
&\equiv& \left[\Phi^{A}\right]~
\left[\Phi^{B}\right]\nn\\
&&\quad + (K_{0}-K)\left(D^{-1}\right)^{AC}
(K_{0}-K)\left(D^{-1}\right)^{BD}
\frac{\pa^{l} \pa^{l}S_{I,\Lambda}}{\pa \Phi^{C} \pa \Phi^{D}}~.
\end{eqnarray}
We can obtain the same results from \bref{IR BRST tf} simply by setting
the antifields to zero.  
Before closing this subsection, we make a comment on the ghost equation
of motion: we wish to point out that the actions for gauge theories
depend on the antighost and the antifield through a particular linear
combination.

Let us consider the classical action \bref{BV-bScl} either for QED or
for Yang-Mills theories.  The gauge-fixing and the ghost parts of
$S_{cl}[\phi]$ are written as\cite{Kugo:1981hm}
\begin{equation}
\delta \Bigl({\bar c^a} F^a(\phi)\Bigr)\,,
\end{equation}
where $F^a(\phi)$ is a gauge-fixing condition.  If we choose the
covariant gauge fixing
\begin{equation}
F^a(\phi) = -i \Bigl( \partial_{\mu} a_{\mu}^a +
\frac{\xi}{2} b^a \Bigr)\,,
\end{equation}
where $a_\mu^a$ is a gauge field, and $b$ an auxiliary field, we obtain
\begin{eqnarray}
\delta \Bigl({\bar c^a} F^a(\phi)\Bigr) + \phi^*_A \delta \phi^A &=& 
i b^a F^a(\phi) + \Bigl({\bar c}^a \frac{\partial^r F^a}{\partial
\phi^A}+ \phi^*_A \Bigr)\delta \phi^A \nn\\
&=& \Bigl(-i {\bar c}^a \partial_{\mu} + a_{\mu}^{a*}\Bigr) \delta
 a_{\mu}^a + \cdots \,.
\end{eqnarray}
Therefore, in $\bar{S}_{cl} [\phi,\phi^*]$, the antifield
$a_{\mu}^{a*}(-p)$ appears as a linear combination with the
antighost, $a_{\mu}^{a*}(-p)+{\bar c}(-p)p_{\mu}$.  Hence, the action
${\bar S}_{cl}[\phi, \phi^*]$ satisfies
\begin{eqnarray}
\frac{\partial^l {\bar S}_{cl}}{\partial {\bar c}(-p)} 
-p_{\mu} \frac{\partial^l {\bar S}_{cl}}{\partial a^{a*}_{\mu}(-p)} 
=0\,.
\label{ghost EOM1}
\end{eqnarray}
We now assume that the bare action satisfies the same equation:
\begin{eqnarray}
K_0 \frac{\pa^{l} {\bar S}_B}{\partial {\bar c}(-p)} 
- 
p_{\mu} K_{0}^{-1} \frac{\pa^{l} {\bar S}_B}
{\partial a^{a*}_{\mu}(-p)} 
=0~.
\label{ghost EOM2}
\end{eqnarray}
This implies the same relation for the Wilson action
\begin{eqnarray}
K \frac{\pa^{l} \bar{S}_{\Lambda}}{\partial {\bar C}(-p)} 
- 
p_{\mu} K^{-1} \frac{\pa^{l} \bar{S}_{\Lambda}}
{\partial A^{a*}_{\mu}(-p)} 
=0~,
\label{ghost EOM3}
\end{eqnarray}
because $K \partial \bar{S}_{\Lambda}/\partial \Phi^A$ and $K^{-1}
\partial \bar{S}_{\Lambda}/\partial \Phi^*_A$ are both composite
operators.  We call Eqs.~\bref{ghost EOM1}, \bref{ghost EOM2},
\bref{ghost EOM3} the ghost equations of motion.  We encountered these
before, first as \bref{WTexamples-YM-ghostEOM} and then as
\bref{AFexamples-YM-ghostEOM}.  In fact in \S\ref{AFexamples-YM},
\bref{AFexamples-YM-ghostEOM} plays an important role in simplifying the
perturbative construction of the theory.  We will assume \bref{ghost
EOM2} and hence \bref{ghost EOM3} when we discuss examples in the next
section.


\subsection{ QME and the modified Slavnov-Taylor identity\label{BV-ST}}

In \S\ref{derivation-Wetterich} we have discussed the relation between a
Wilson action $\SL$ and the corresponding effective average action
$\Gamma_{B,\Lambda}$.  The two are related by a Legendre transformation.
A theory can be constructed equally well in terms of $\SL$ or
$\Gamma_{B,\Lambda}$.  As for realization of symmetry, however, we have
so far considered only the Wilson action.  In this subsection we make a
brief detour to write down the identity for $\bar{\Gamma}_{B,\Lambda}$,
corresponding to the QME for $\bSL$.  This is called the
\textbf{modified Slavnov-Taylor identity}\cite{Ellwanger:1994iz}.  We
also give a brief note on the same subject as Appendix \ref{appgamma}.
Compared with Appendix \ref{appgamma}, we employ a more functional
method here.

As in \S\ref{derivation-Wetterich} we introduce an action with both UV
and IR cutoffs:
\begin{equation}
{\bar {S}}_{B, \Lambda}[\phi, \phi^*] \equiv
-\frac{1}{2} \phi \cdot (K_0-K)^{-1}D \cdot \phi 
+
{S}_{I,B}[\phi] + K_0 \phi^* \cdot {\cal R}[\phi]~.
\label{BV-SBL}
\end{equation}
The gaussian term suppresses the momentum modes $p^2 > \Lambda_0^2$ or
$p^2 < \Lambda^2$.  The generating functional is defined by
\begin{equation}
\exp \left[{\bar W}_{B,\Lambda}[J, \phi^*]\right] \equiv \int
 \mathcal{D} \phi\,
\exp\Bigl({{\bar {S}}_{B, \Lambda}[\phi, \phi^*] + K_0^{-1}J \cdot
\phi}\Bigr)~.
\label{gen func}
\end{equation}
The Legendre transform of $\bar{W}_{B,\Lambda}$ is defined by
\begin{equation}
{\bar \Gamma}_{B, \Lambda}[\varphi, \phi^*] \equiv {\bar W}_{B,
 \Lambda}[J, \phi^*] - K_0^{-1} J \cdot \varphi~, 
\label{Gamma-W relation}
\end{equation}
where $\varphi$ is defined in terms of $J$ as
\begin{equation}
K_0^{-1} \varphi^A (p) \equiv \frac{\partial^l {\bar W}_{B,\Lambda}[J,
 \phi^*]}{\partial J^A(-p)}~.
\end{equation}

In the following we wish to consider the functional integral over the QM
operator for the bare action:
\begin{equation}
\bar{\Sigma}^{1PI}_{B,\Lambda} \equiv
\exp \left[ - \bar{W}_{B,\Lambda} [J,\phi^*]\right]
\int \mathcal{D} \phi\,
\bar{\Sigma}_B [\phi,\phi^*] \exp \left( \bar{S}_{B,\Lambda} [\phi,
				   \phi^*] + K_0^{-1} J \cdot \phi
				  \right)\,,
\end{equation}
and express this in terms of the effective average action
$\bar{\Gamma}_{B,\Lambda}$.  

As a preparation, we first compute the derivatives of ${\bar \Gamma}_{B,
\Lambda}[\varphi, \phi^*]$:
\begin{eqnarray}
\frac{\partial^l \bar{\Gamma}_{B,\Lambda}}{\partial \phi_A^*} 
&=& 
\frac{\partial^l \bar{W}_{B,\Lambda}}{\partial \phi_A^*} 
+
\frac{\partial^l \bar{J}_{C}}{\partial \phi_A^*}
\Bigl(
\frac{\partial^l \bar{W}_{B,\Lambda}}{\partial J^C} - K_0^{-1}\varphi^C
\Bigr) 
= \frac{\partial^l \bar{W}_{B,\Lambda}}{\partial \phi_A^*}~,
\label{Gamma by AFs}\\
\frac{\partial^r \bar{\Gamma}_{B,\Lambda}}{\partial \varphi^A}
&=&
\Bigl(
\frac{\partial^r \bar{W}_{B,\Lambda}}{\partial J^C}
-(-)^{\epsilon_B}K_0^{-1} \varphi^C
\Bigr)
\frac{\partial^r J^B}{\partial \varphi^A}
- K_0^{-1}J^A
= -K_0^{-1}J^A\,.
\label{Gamma by Fs}
\end{eqnarray}
Hence, we obtain
\begin{eqnarray}
\frac{\partial^r \bar{\Gamma}_{B,\Lambda}}{\partial \varphi^A}
\frac{\partial^l \bar{\Gamma}_{B,\Lambda}}{\partial \phi_A^*} 
%
&=& -K_0^{-1}J^A
\frac{\partial^l \bar{W}_{B,\Lambda}}{\partial \phi_A^*}
= -\e^{-\bar{W}_{B,\Lambda}} \Bigl(K_0^{-1}J^A 
\frac{\partial^l}{\partial \phi_A^*}\Bigr) \e^{\bar{W}_{B,\Lambda}}\nn\\
&=&
-\e^{-\bar{W}_{B,\Lambda}} 
\int \mathcal{D} \phi\,
\frac{\partial^r }{\partial \phi^A}
\,\e^{K_0^{-1}J \cdot \phi}\cdot 
\frac{\partial^l \bar{S}_{B, \Lambda}}{\partial \phi_A^*}
\,\e^{\bar{S}_{B, \Lambda}[\phi, \phi^*]}\nn\\
&=& \e^{-\bar{W}_{B,\Lambda}} 
\int \mathcal{D} \phi\, \bar{\Sigma}_{B,\Lambda} 
\exp \left[ \bar{S}_{B, \Lambda}[\phi, \phi^*] + K_0^{-1}J \cdot
\phi \right] ,
\label{ZJ1}
\end{eqnarray}
where $\bar{\Sigma}_{B,\Lambda}$ is the QM operator for the action
$\bar{S}_{B,\Lambda}$: 
\begin{eqnarray}
\bar{\Sigma}_{B,\Lambda} 
\equiv \frac{1}{2}
(\bar{S}_{B,\Lambda},\bar{S}_{B, \Lambda})+\Delta \bar{S}_{B, \Lambda}~.
\label{QMop for S-BL}
\end{eqnarray}

To find the difference between $\bar{\Sigma}_{B,\Lambda}$ and
$\bar{\Sigma}_B$, we note
\begin{equation}
{\bar {S}}_{B}[\phi, \phi^*]
= {\bar {S}}_{B, \Lambda}[\phi, \phi^*]
+\frac{1}{2} \phi \cdot R_{\Lambda} \cdot \phi \,,
\label{K0-K action}
\end{equation}
where we define
\begin{equation}
[R_{\Lambda}(p)]_{AB} \equiv D_{AB}(p)\Bigl(
\frac{1}{K_0-K}-\frac{1}{K_0} \Bigr) \,.
\end{equation}
Hence, we obtain
\begin{equation}
\bar{\Sigma}_B = \bar{\Sigma}_{B,\Lambda} + [R_\Lambda]_{BA} 
\phi^B \frac{\pa^l \bar{S}_{B,\Lambda}}{\partial \phi^*_A}\,.
\label{BV-SigmaBSigmaBL}
\end{equation}
The functional integral over the difference gives
\begin{eqnarray}
&&\e^{- \bar{W}_{B,\Lambda}} \int \mathcal{D} \phi\,
[R_\Lambda]_{BA} \, \phi^B \frac{\partial^l \bar{S}_{B,\Lambda}}{\pa
\phi^*_A} \, \exp \left[ \bar{S}_{B,\Lambda} [\phi,\phi^*] + K_0^{-1} J
		   \cdot \phi \right]\nn\\
&=&\e^{- \bar{W}_{B,\Lambda}}[R_\Lambda]_{BA} K_0
 \frac{\partial^l}{\partial J_B}  \frac{\partial^l}{\partial
 \phi^*_A}\e^{\bar{W}_{B,\Lambda}} \nn\\
&=& [R_\Lambda]_{BA} K_0 \Bigl( 
\frac{\partial^l \varphi^C}{\partial J_B} 
\frac{\partial^l}{\partial \varphi^C}
\frac{\partial^l {\bar \Gamma}_{B,\Lambda}}{\partial \phi^*_A}  
+
K_0^{-1} \varphi^B \frac{\partial^l {\bar \Gamma}_{B,\Lambda}}{\partial
\phi^*_A}   \Bigr)~,
\label{extra term}
\end{eqnarray}
where we have used \bref{Gamma by AFs}, \bref{Gamma by Fs}.  To
summarize so far, the functional integral over \bref{BV-SigmaBSigmaBL}
gives
\begin{eqnarray}
\bar{\Sigma}_{B,\Lambda}^{1PI} &=& \frac{\partial^r
 \bar{\Gamma}_{B,\Lambda}}{\partial \varphi^A} 
\frac{\partial^l \bar{\Gamma}_{B,\Lambda}}{\partial \phi_A^*}  \nn\\
&& \quad +  [R_\Lambda]_{BA} \Bigl( K_0
\frac{\partial^l \varphi^C}{\partial J_B} 
\frac{\partial^l}{\partial \varphi^C}
\frac{\partial^l {\bar \Gamma}_{B,\Lambda}}{\partial \phi^*_A}  
+ \varphi^B \frac{\partial^l {\bar \Gamma}_{B,\Lambda}}{\partial
\phi^*_A}   \Bigr)\,.\label{BV-SigmaBintermediate}
\end{eqnarray}

To further rewrite the second term in the above, we introduce the notation:
\begin{equation}
({\bar \Gamma}_{B,\Lambda}^{(2)})_{CA} \equiv \frac{\partial^l}{\partial
 \varphi^C} 
\frac{\partial^r {\bar \Gamma}_{B,\Lambda}}{\partial \varphi^A}
= - K_0^{-1} \frac{\partial^l J^A}{\partial \varphi^C}\,,
\end{equation}
where \bref{Gamma by Fs} is used.  Then, the inverse is written as
\begin{eqnarray}
\frac{\partial^l \varphi^A}{\partial J_C} = -K_0^{-1} ({\bar
 \Gamma}_{B,\Lambda}^{(2)})^{-1}_{CA}~. 
\label{Gamma2 inverse}
\end{eqnarray}
Hence, we can rewrite \bref{BV-SigmaBintermediate} as
\begin{eqnarray}
\bar{\Sigma}_{B,\Lambda}^{1PI}
&=& \frac{\partial^r \bar{\Gamma}_{B,\Lambda}}{\partial \varphi^A}
\frac{\partial^l \bar{\Gamma}_{B,\Lambda}}{\partial \phi_A^*} \nn\\
&&\quad + [R_{\Lambda}]_{BA}
\Bigl(
- ({\bar \Gamma}_{B,\Lambda}^{(2)})^{-1}_{BC} 
\frac{\partial^l}{\partial \varphi_C}\frac{\partial^l {\bar
\Gamma}_{B,\Lambda}}{\partial \phi^*_A}
+ \varphi^B \frac{\partial^l {\bar \Gamma}_{B,\Lambda}}{\partial \phi^*_A}
\Bigr)\,.
\label{mST}
\end{eqnarray}
Imposing this to vanish, we obtain the modified Slavnov-Taylor
identity\cite{Ellwanger:1994iz}:
\begin{equation}
\bar{\Sigma}_{B,\Lambda}^{1PI} = 0\,.
\end{equation}
In the limit of $\Lambda \rightarrow 0$, $R_{\Lambda}$ goes to zero
and ${\bar \Gamma_{B, \Lambda}}$ becomes ${\bar \Gamma}_B$.  Therefore,
the modified ST identity reduces to the Zinn-Justin equation.

\subsection{ On the nilpotency without antifields\label{BV-nil}}

As shown toward the end of \S\ref{BV-BV}, the quantum master equation
\bref{BV-QME} implies the nilpotency of the quantum BRST transformation
$\delta_Q$, defined by \bref{BV-qBRST}.  Even if $\delta_Q$ is nilpotent,
the reduced BRST transformation
\begin{equation}
\delta \equiv \delta_Q \big|_{\Phi^*=0}
\end{equation}
that appears in the WT operator \bref{BV-WToperator} is not nilpotent in
general.  In this subsection we compute $\delta^2$, and study how it
differs from zero, assuming the nilpotency $\delta_Q^2 = 0$ of the full
BRST transformation.

Take a functional ${\mathcal O}_{\Lambda}[\Phi]$ of the field $\Phi$
defined at an IR scale $\Lambda$, and consider its BRST transformation:
\begin{eqnarray}
\delta_Q {\mathcal O}_{\Lambda}[\Phi] = 
\frac{\partial^r {\mathcal O}_{\Lambda} [\Phi]}{\partial \Phi^A}
\frac{\partial^l {\bar S}_{\Lambda}}{\partial \Phi^*_A}\,.
\end{eqnarray}
Expanding the Wilson action in powers of antifields
\begin{equation}
{\bar S}_{\Lambda}[\Phi, \Phi^*] = S_{\Lambda}[\Phi]+ \Phi^*_A f^A[\Phi]+
\frac{1}{2} \Phi^*_A \Phi^*_B f^{BA}[\Phi] +
\mathrm{O}\Bigl((\Phi^*)^3\Bigr)~, 
\label{BV-Sbar-expansion}
\end{equation}
we obtain an expansion
\begin{equation}
\delta_Q \Phi^A \equiv
\frac{\partial^l {\bar S}_{\Lambda}}{\partial \Phi_A^*} = f^A[\Phi] +
\Phi^*_B f^{BA}[\Phi] + \mathrm{O} \Bigl((\Phi^*)^2 \Bigr)~.
\label{BV-deltaQ-expansion}
\end{equation}
Setting $\Phi^*=0$, we obtain
\begin{equation}
\delta \Phi^A = f^A [\Phi]\,.
\end{equation}

By the definition \bref{BV-qBRST}, the nilpotency of $\delta_Q$ implies
\begin{equation}
0 = \Bigl( \frac{\partial^r}{\partial \Phi^A} \delta_Q {\mathcal
 O}_{\Lambda}  \Bigr) 
\frac{\partial^l {\bar S}_{\Lambda}}{\partial \Phi^*_A}
 - 
\Bigl( \frac{\partial^r}{\partial \Phi^*_A} \delta_Q {\mathcal
O}_{\Lambda}  \Bigr) 
\frac{\partial^l {\bar S}_{\Lambda}}{\partial \Phi^A}
-
(-)^{\epsilon_A}\frac{\partial^r}{\partial
\Phi^A}\frac{\partial^r}{\partial \Phi^*_A} 
\Bigl(\delta_Q {\mathcal O}_{\Lambda}\Bigr)~.
\end{equation}
Setting $\Phi^*$ to zero, the first term on the right gives
$\delta^2 \Op_\Lambda$.  Therefore, we obtain
\begin{equation}
\delta^2 {\mathcal O}_{\Lambda} =
\Bigl[ \frac{\partial^r}{\partial \Phi^*_A} \delta_Q {\mathcal
O}_{\Lambda}  \Bigr]_{\Phi^*=0} 
\frac{\partial^l {S}_{\Lambda}}{\partial \Phi^A}
+
(-)^{\epsilon_A}\frac{\partial^r}{\partial \Phi^A}
\Bigl[\frac{\partial^r}{\partial \Phi^*_A}
\delta_Q {\mathcal O}_{\Lambda}\Bigr]_{\Phi^*=0}\,.
\label{delta-squared}
\end{equation}
Using the expansions \bref{BV-Sbar-expansion} \&
\bref{BV-deltaQ-expansion}, we obtain
\begin{eqnarray}
\delta^2{\mathcal O}_{\Lambda}[\Phi] &=& 
- 
%
\Bigl(\frac{\partial^r {\mathcal O}_{\Lambda}}{\partial \Phi^A} f^{AB} \Bigr)
\frac{\partial^r S_{\Lambda}}{\partial \Phi^B}
-
 \frac{\partial^r}{\partial \Phi^B}
\Bigl(
\frac{\partial^r {\mathcal O}_{\Lambda}}{\partial \Phi^A} f^{AB} 
\Bigr)
\nn\\
&=& - \e^{- \SL} \frac{\partial^r}{\partial \Phi^B} \left( 
\frac{\partial^r \Op_\Lambda}{\partial \Phi^A} f^{AB} \e^{\SL}\right) \,.
\end{eqnarray}
In particular, we obtain
\begin{eqnarray}
\delta^2 \Phi^A = 
- 
%
f^{AB} 
\frac{\partial^r S_{\Lambda}}{\partial \Phi^B}
-
 \frac{\partial^r}{\partial \Phi^B}f^{AB}
%
= - \e^{-\SL} \frac{\partial^r}{\partial \Phi^B} \left( f^{AB}
\e^{\SL}\right)\,.
\end{eqnarray}

Thus, we have found that $\delta$ fails to be nilpotent in a particular
way that $\e^{\SL} \delta^2 \Op_\Lambda$ is a total derivative, or using
the terminology introduced in \S\ref{comp-examples}, $\delta^2
\Op_\Lambda$ is a generalized equation of motion.\footnote{Note that
$f^{AB} [\Phi]$ is not necessarily a composite operator.  Hence, we are
abusing the terminology introduced for genuine composite operators.}


\newpage

\section{~Examples: the Wilson actions for QED and Yang-Mills theories
\label{BVexamples}}

The purpose of this section is to apply the general result
\bref{Wilsonian6} on the antifield dependence of the Wilson action to
QED and the Yang-Mills theories.  As for QED this has already been
done in \S\ref{AFexamples-QED}, from a consideration of the ERG
differential equation.  This method works only for QED, but not for YM
theories.  For the latter we have adopted a perturbative approach in
\S\ref{AFexamples-YM}, and have shown how to construct the continuum
limit of the Wilson action by loop expansions.  In
\S\ref{AFexamples-YM}, we have discussed the antifield dependence of
only the asymptotic behavior of the Wilson action, since that is all we
need for proving the theory's existence.  In contrast, we have adopted a
unifying approach in \S\ref{BV}, and have determined the antifield
dependence of the Wilson action explicitly by starting from a bare
action that has simple antifield dependence.\footnote{The antifield
dependence of the bare action is linear in $\phi^*$ and at most
quadratic in $\phi$ as in \bref{cal-SI}.}

%
%

\subsection{ QED\label{BVexamples-QED}}

We first consider QED and reproduce the results from
\citen{Igarashi:2007fw} and \citen{Higashi:2007ax}.  Our derivation
follows closely that of \citen{Higashi:2007ax} except that we pay more
careful attention to the necessity of a UV cutoff for the bare theory.

Denoting the UV fields by
\begin{equation}
\phi^{A} = \{ a_{\mu},~b,~c,~\cbar\,~\varphi,~\bar\varphi\}\,,
\end{equation}
and the UV antifields by
\begin{equation}
\phi^{*}_{A} = \{ a^{*}_{\mu},~b^{*},~c^{*},~\cbar^{*},~\varphi^{*},~
\bar\varphi^{*}\}\,,
\end{equation}
the bare action of QED is written as follows\footnote{We have introduced
the auxiliary field $b$ to avoid the quadratic dependence of $\bar{S}_B$
on $\bar{c}^*$.  The result \bref{AFexamples-QED-bSL} is obtained from
\bref{QED-MA1} by integrating out $B$.}:
\begin{eqnarray}
\bar{S}_B[\phi,~\phi^{*}] &=& - \frac{1}{2}\phi \cdot K_{0}^{-1} D \cdot
 \phi + K_0 \phi^{*}_{A}{\cal R}^{A}[\phi] +  {S}_{I,B}[\phi] \,,
\label{cal-S-QED}
\end{eqnarray}
where
\begin{eqnarray}
\frac{1}{2}\phi \cdot K_{0}^{-1} D \cdot
 \phi &\equiv& 
\int_p K_{0}^{-1}(p) \biggl[\frac{1}{2} a_{\mu}(-p)
(p^2 \delta_{\mu\nu}-  p_{\mu}p_{\nu})a_{\nu}(p)
+i \cbar(-p) p^{2}c(p)\nn\\
&& \qquad - b(-p) \Bigl\{ip_{\mu}a_{\mu}(p) 
+ \frac{\xi}{2}b(p)\Bigr\}+ \bar\varphi(-p)(\Slash{p}+i{m})\varphi(p)
\biggr]\,,\nn\\
K_0 ~\phi^{*}_{A}{\cal R}^{A}[\phi] &\equiv& 
\int_p K_0 (p)\biggl[a_{\mu}^{*}(-p) (-i)p_{\mu}c(p) 
+ \cbar^{*}(-p)\cdot ib(p)\nn\\
&&\qquad  + ie \int_{q} \varphi^{*}(-p)c(q)\varphi(p-q)+ie
\int_{q}\bar\varphi(-p-q)c(q)\bar\varphi^{*}(p) \biggr]\,,\nn
\end{eqnarray}
and  
\begin{eqnarray}
{S}_{I,B}[\phi]&=& \int_{p} \biggl[
\frac{a_{2}}{2}\Lambda_{0}^{2} a_{\mu}(-p)a_{\mu}(p) +
\frac{z_{2}}{2} a_{\mu}(-p) p^{2} a_{\mu}(p)+
\frac{z'_{2}}{2}a_{\nu}(-p) p_{\nu}p_{\mu} a_{\mu}(p)\nn\\ 
&&\quad + \frac{z_{4}}{8}\int_{q,k,r} \delta^{4}(p+q+k+r)
 a_{\nu}(p)a_{\nu}(q)a_{\mu}(k)a_{\mu}(r)\nn\\ 
&&\quad + \bar\varphi(-p)\Bigl( z_{f}\Slash{p} + z_{m}im\Bigr) \varphi(p)+
 z_{3}\int_{q}\bar\varphi(-p)\Slash{a}(q) \varphi(p-q) \biggr]\,.
\label{cal-S-QED11}
\end{eqnarray}
The coefficients
\begin{equation}
a_{2},~z_{2},~z'_{2},~z_{3},~z_{4},~z_{f},~z_{m}
\label{BVexamples-QEDparameters}
\end{equation}
are given such dependence on $\ln \Lambda_{0}/\mu$ that we obtain a
finite continuum limit. 

Denoting the IR fields and antifields by
\begin{equation}
\Phi^{A} = \{A_{\mu}, B, C, \Cbar, \psi, \bar\psi\}\,,\quad
\Phi^{*}_{A}=\{A^{*}_{\mu}, B^{*}, C^{*}, \Cbar^{*},
\psi^{*}, \bar\psi^{*}\}\,,
\end{equation}
the effective sources defined by \bref{cal-J} are given by
\begin{subequations}
\begin{eqnarray}
{\cal J}_{c}(-p)&=& -i 
K(p )p_{\mu}A_{\mu}^{*}(-p)\nn \\
&& -ie  \int_{q}~K(q)
\left(\psi^{*}(-q)\psi(q-p)- \bar\psi(-q-p)
\bar\psi^{*}(q)\right)~,\\
{\cal J}_{B}(-p)&=& i K(p)
\Cbar^{*}(-p)~,\\
{\cal J}_{\psi}(-p)&=& ie  \int_q~ K(q)
\psi^{*}(-q) C(q-p)~,\\
{\cal J}_{\bar\psi}(-p)&=& -ie\int_{q} K(q)
C(q-p)\bar\psi^{*}(-q)~.
\end{eqnarray}
\label{cal-J-QED}
\end{subequations}

\noindent
Since the ghost and antighost remain free, and are absent in the
interaction action ${S}_{I,B}$, the source ${\cal J}_{c}$ does not
appear in the IR action.  This simplifies the application of the general
result \bref{Wilsonian6}.  The differential operators, $(K \Phi^{*}/2)
R^{(2)}\pa_{\cal J}\pa_{\cal J}$ in \bref{Wilsonian6}, generate no
contribution.  The Wilson action is thus given as
\begin{eqnarray}
\bar{S}_{\Lambda}[\Phi,~\Phi^{*}] 
&=& - \frac{1}{2}\Phi^{A}  K^{-1}D_{{A}{B}}\Phi^{B}
 + K  \Phi^{*}_{A}{\cal R}^{A}[\Phi]\nn\\
&& + \frac{1}{2}(-)^{\epsilon({\cal J})}{\cal J}
\cdot (K_{0}-K)D^{-1}{\cal J} + S_{I,\Lambda}[\Phi']\,, 
\label{QED2}
\end{eqnarray}
where the shifted variables are defined by
\begin{subequations}
\label{shifted-var-QED}
\begin{eqnarray}
A'_{\mu}(p)&=& A_{\mu}(p) + \frac{i p_{\mu}}{p^{2}}(K_{0}(p)-K(p)){\cal
 J}_{B}(p)~,\\
\psi'(p)&=& \psi(p) - \frac{(K_{0}(p)-K(p))}{\Slash{p}+i m}{\cal
 J}_{\bar\psi}(p)~,\\
\bar\psi'(-p)&=& \bar\psi(-p) 
+ {\cal J}_{\psi}(-p) \frac{(K_{0}(p)-K(p))}{\Slash{p}+i m}~~.
\end{eqnarray}
\end{subequations}
The term quadratic in the effective sources is given by
\begin{eqnarray}
\frac{1}{2}(-)^{\epsilon({\cal J})}{\cal J}
\cdot (K_{0}-K)D^{-1}
{\cal J} &=& - {\cal J}_{\psi}(-p)
\frac{(K_{0}(p)-K(p))}{\Slash{p}+im}
{\cal J}_{\bar\psi}(p)\,.
\label{J-J-QED}
\end{eqnarray} 
Hence, the total Wilson action is
\begin{eqnarray}
\bar{S}_{\Lambda}[\Phi,~\Phi^{*}] &=& \int_p ~\Biggl[ - K(p)^{-1}
 \biggl\lbrace \frac{1}{2}  
{A}_{\mu}(-p)
(p^2 \delta_{\mu\nu}-  p_{\mu}p_{\nu})A_{\nu}(p)
+\Cbar(-p)i p^{2}C(p)\nn\\
&& \qquad\qquad\quad - B(-p)\Bigl(ip_{\mu}A_{\mu}(p) 
+ \frac{\xi}{2}B(p)\Bigr) + \bar\psi(-p)(\Slash{p}+im)\psi(p) 
\biggr\rbrace\nn\\
&& \qquad+ ~{K(p)}\biggl\lbrace~\Cbar^{*}(-p)\cdot i B(p)+
 A_{\mu}^{*}(-p)(-i)p_{\mu}C(p)
\nn\\
&& \qquad\qquad\quad + i e \int_{q}
\left( \psi^{*}(-p)C(q)\psi(p-q) + \bar\psi(-p-q)C(q)\bar\psi^{*}(p) \right)
\biggr\rbrace\nn\\
&&\quad + e^{2}\int_{q,k}{K(p+q)}{K(p-k)}
\psi^{*}(-p-q)
\frac{(K_{0}(p)-K(p))}{\Slash{p}+im}\nn\\
&& \quad\qquad \times \bar\psi^{*}(p-k)C(q)C(k)\,\Biggr]
 + S_{I,\Lambda}[A'_{\mu},~\psi',~\bar\psi']~.
\label{QED-MA1}
\end{eqnarray} 

Since QED is renormalizable, we can tune the $\ln \Lambda_0/\mu$
dependence of the bare parameters \bref{BVexamples-QEDparameters} to
obtain the continuum limit of $\SIL$.  The remaining dependence of
$\bSL$ on $\Lambda_0$ comes from $K_0$ in the $\psi^* \bar{\psi}^*$ term
and the shifted variables $A'_\mu, \psi', \bar{\psi}'$.  In the limit
$\Lambda_0 \to \infty$, $K_0$ is simply replaced by $1$, and we obtain
the continuum limit of $\bSL$ as in \citen{Higashi:2007ax}.  A canonical
transformation that eliminates the quadratic terms in the antifield
gives the action obtained in ref.~\citen{Igarashi:2007fw}.  Perturbative
construction of $\SIL$ that satisfies the WT identity, $\Sigma_\Lambda =
0$, has already been discussed in \S\ref{WTexamples-QED}.

Finally, we observe that $\bSL$ given by \bref{QED-MA1}
satisfies the ghost equation of motion \bref{ghost EOM3}.  This is a
consequence of \bref{ghost EOM2} satisfied by the bare
action \bref{cal-S-QED} as we have pointed out at the end of
\S\ref{BV-AFdep}. 

\subsection{ Yang-Mills theories}

We consider a Yang-Mills theory without matter described by the
UV action\footnote{We suppress the group index and use the notations:
$A\cdot B \equiv A^{a} B^{a},~A^{2}\equiv A^{a}A^{a}$ and $A \cdot (B
\times C) \equiv f^{abc}A^{a}B^{b}C^{c}$.}
\begin{eqnarray}
\bar{S}_B[\phi,\phi^{*}] =  
- \frac{1}{2}\phi^{A} K_{0}^{-1}D_{{A}{B}}\phi^{B} 
+ {\bar {S}}_{I,B}[\phi,\phi^{*}]\,,
\label{cal-S-YM0}
\end{eqnarray}
where fields and antifields are collectively denoted as
\begin{equation}
\phi^{A} = \{ a_{\mu},~b,~c,~\cbar\}\,,\quad
\phi^{*}_{A} = \{ a^{*}_{\mu},~b^{*},~c^{*},~\cbar^{*}\}\,.
\end{equation}
The bare action consists of the kinetic part
\begin{eqnarray}
-\frac{1}{2}\phi^{A} K_{0}^{-1}D_{{A}{B}}\phi^{B} &\equiv& 
-\int_p K_{0}^{-1}(p)\biggl[\frac{1}{2} a_{\mu}(-p)\cdot
(p^2 \delta_{\mu\nu}-  p_{\mu}p_{\nu})a_{\nu}(p)\nn\\
&&  - b(-p)\cdot\bigl\{ip_{\mu}a_{\mu}(p) +
 \frac{\xi}{2}b(p)\bigr\}+ i \cbar(-p)\cdot p^2 c(p)\biggr]
\label{cal-S-YM1}
\end{eqnarray}
and the interaction part
\begin{eqnarray}
&&\bar{S}_{I,B}[\phi,~\phi^{*}] \equiv
 S_{I,B}[\phi]
+ K_0 \phi_{A}^{*} 
\Bigl(
{\cal R}^{(1){A}}_{~~~~~{B}}\phi^{{B}} 
+ \frac{1}{2}\Phi^{{B}} 
{\cal R}^{(2){A}}_{~~~~~{B}{C}}\phi^{C}\Bigr) \nn\\
&& \quad = \int_{p} \left[\frac{a_{2}}{2}\Lambda_{0}^{2} a_{\mu}(-p)a_{\mu}(p)
+ \frac{z_{1}}{2} p^{2} a_{\mu}(p)a_{\mu}(-p)
+ \frac{z_{2}}{2} p_{\mu} p_{\nu} a_{\mu}(-p) a_{\nu}(p) \right]\nn\\
&& \qquad  +\,z_{3} \int_{p,q} p_{\nu} a_{\mu}(-p) \cdot [a_{\mu}(q)
 \times a_{\nu}(p-q)] \nn\\
&& \,\, + \int_{p_{1},\cdot\cdot,p_{4}} \hspace{-7mm}\delta (\Sigma p_{i})
\left[ \frac{z_{4}}{8} ~a_{\mu}(p_{1})\cdot
a_{\mu}(p_{2})a_{\nu}(p_{3})\cdot a_{\nu}(p_{4}) 
+ \frac{z_{5}}{8} ~a_{\mu}(p_{1})\cdot a_{\nu}(p_{2})a_{\mu}(p_{3})\cdot
a_{\nu}(p_{4})\right]\nn\\ 
&&\quad+ \int_p  ~\left[K_0(p)\Bigl\{\cbar^{*}(-p)\cdot ib(p) -
i a_{\mu}^{*}(-p) \cdot p_{\mu}c(p)
+ \frac{z_{8}}{2} c^{*}(-p) \cdot \int_{q} c(q) \times c(p-q)
\Bigr\}\right.\nn\\
&&\qquad + \left.\left(K_0^{2}(p) a_{\mu}^{*}(-p) +  p_{\mu} \cbar(-p)\right)
\cdot \Bigl\{ -iz_{6}~p_{\mu}c(p) +z_{7} 
\int_{q} a_{\mu}(p-q)\times c(q)
 \Bigr\}\right] .
\label{cal-S-YM2}
\end{eqnarray}
Here, we have assumed the gauge group SU(2) to write down $S_{I,B}
[\phi]$; more terms are available and need to be introduced for higher
symmetry. The coefficients $a_{2}, z_{1},\cdots, z_{8}$ depend on $\ln
\Lambda_0/\mu$.

The Wilson action with the IR fields \& antifields
\begin{equation} 
\Phi^{A} = \{A^{a}_{\mu}, B^{a}, C^{a}, \Cbar^{a}\}\,,\quad
\Phi^{*}_{A}=\{A^{*}_{\mu}, B^{*}, C^{*}, \Cbar^{*}\}
\end{equation}
is given by the general formula \bref{Wilsonian6}.  Let us get a
concrete expression. The effective sources read
\begin{subequations}
\label{cal-J-YM} 
\begin{eqnarray}
{\cal J}_{\mu}(-p)&=& -z_{7} \int_q~ K(q) K_0 (q) A_{\mu}^{*}(q)\times
 C(-p-q)~,\\ 
{\cal J}_{B}(-p)&=& i {K(p)} \Cbar^{*}(-p)~,\\
{\cal J}_{c}(-p)&=& -i (1+ z_{6} K_0 (p)) {K(p)} p_{\mu}A_{\mu}^{*}(-p)+ 
z_{7} \int_{q}~{K(q)} K_0 (q) A_{\mu}^{*}(q) \times A_{\mu}(-p-q)\nn\\
&& \quad+ z_{8}\int_{q}~{K(q)}C^{*}(q) \times C(-p-q)~,
\end{eqnarray}
\end{subequations}
and the differential operator w.r.t. $\cal J$ is given by
\begin{eqnarray}
&& \frac{K}{2} \Phi^{*}_{C} \frac{\pa^{l}}{\pa {\cal J}_{A}}
{R}^{(2)C}_{~~~~AB} \frac{\pa^{l}}{\pa {\cal J}_{B}}=
 z_{7} \int_{p,q}~
{K(p)} K_0 (p)
A_{\mu}^{*}(-p)\cdot \left(\frac{\pa^{l}}{\pa {\cal J}_{\mu}(-p+q)}
\times \frac{\pa^{l}}{\pa {\cal J}_{c}(-q)} \right)\nn\\
&& \qquad\qquad+ \frac{z_{8}}{2}\int_{p,q}~
{K(p)}C^{*}(-p)\cdot \left(
\frac{\pa^{l}}{\pa {\cal J}_{c}(-p+q)} \times 
\frac{\pa^{l}}{\pa {\cal J}_{c}(-q)}\right)~.
\label{J-deriv}
\end{eqnarray}
The shifted variables
defined by \bref{shifted-var2} are given by
\begin{eqnarray}
A'_{\mu}(p)&=& A_{\mu}(p) + \frac{K_{0}(p) -K(p)}{p^{2}}\left[
\delta_{\mu\nu}-(1-\xi)\frac{p_{\mu}p_{\nu}}{p^{2}}\right]
{\cal J}_{\nu}(p)\nn\\
&& \quad + \frac{i p_{\mu}}{p^{2}}(K_{0}(p)-K(p)){\cal J}_{B}(p)~,\nn\\
\Cbar'(p)&=& \Cbar(p) - \frac{i}{p^{2}}(K_{0}(p)-K(p)) {\cal J}_{c}(p)~.
\label{shifted-var-YM}
\end{eqnarray} 
It is easy to see that $S_{I,\Lambda}[\Phi']$ does not depend on $B$:
\begin{eqnarray}
S_{I,\Lambda}[\Phi']=S_{I,\Lambda}[A'_{\mu},~\Cbar',~ C]~.
\label{Wilson-Master2}
\end{eqnarray}
The quadratic term in ${\cal J}$ that appears inside the exponentials
of \bref{Wilsonian6} takes the following concrete form:
\begin{eqnarray}
&&\frac{1}{2}(-)^{\epsilon({\cal J})}{\cal J}
\cdot (K_{0}-K)D^{-1}
{\cal J} = i \frac{(K_{0}(p)-K(p))}{p^{2}}p_{\mu}{\cal J}_{\mu}(-p)
{\cal J}_{B}(p) \nn\\
&&  \quad + \frac{1}{2} {\cal J}_{\mu}(-p)\frac{(K_{0}(p)-K(p))}{p^{2}}
\left[\delta_{\mu\nu}-(1-\xi)\frac{p_{\mu}p_{\nu}}{p^{2}}\right]
{\cal J}_{\nu}(p)~.
\label{cal-W-YM}
\end{eqnarray} 

Putting together the above results, we obtain the Wilson action:
\begin{eqnarray}
\bar{S}_{\Lambda}[\Phi,~\Phi^{*}] &=& \int_p ~\Biggl[
-K^{-1}(p) \biggl\lbrace \frac{1}{2} {A}_{\mu}(-p)\cdot 
(p^2 \delta_{\mu\nu}-  p_{\mu}p_{\nu})A_{\nu}(p)
 +\Cbar(-p)\cdot i p^{2}C(p)\nn\\
&& \qquad\qquad\qquad\quad - B(-p)\cdot\Bigl(ip_{\mu}A_{\mu}(p) 
+ \frac{\xi}{2}B(p)\Bigr)\biggr\rbrace \nn\\
&&\qquad + ~{K(p)} \Cbar^{*}(-p)\cdot i B(p)\nn\\
&&\qquad + ~{K(p)} A_{\mu}^{*}(-p)\cdot
 \Big\lbrace (-i)(1+ z_{6} K_0 (p)) p_{\mu} C(p) \nn\\
&&\qquad\qquad\qquad\qquad\qquad + z_{7} K_0 (p) \int_{q}~ 
A_{\mu}(p-q) \times  C(q)\Big\rbrace\nn\\
&& \qquad + K(p)~ C^{*}(-p) \cdot z_8 \frac{1}{2} \int_{q} C(p-q) \times
 C(q)\,\,\Biggr]\nn\label{YM-MA1}\\ 
&& + \log \Bigg[ 
\exp \lb \frac{K}{2} \Phi^{*}_{{A}}\frac{\pa^{l}}{\pa
		       {\cal J}_{B}} 
{R}^{(2){A}}_{~~~~{B}{C}}\frac{\pa^{l}}{\pa {\cal J}_{C}}
 \rb\nn\\
&& \quad\qquad \times
\exp \lb\frac{1}{2}(-)^{\epsilon({\cal J})}{\cal J}
\cdot (K_{0}-K)D^{-1}\cdot
{\cal J} + S_{I,\Lambda}[\Phi']\rb
\Bigg]
\,.
\end{eqnarray} 
In writing down the bare action \bref{cal-S-YM0} we have assumed the
ghost equation of motion \bref{ghost EOM2} to hold.  Hence, the Wilson
action \bref{YM-MA1} must satisfy (\ref{ghost EOM3}), i.e., the Wilson
action must depend on the antighost and the antifield for the gauge
field only through the linear combination, $p_{\mu} {\bar C}^a(-p) +
K^2(p) A_{\mu}^{a *}(-p)$.  The way we have written \bref{YM-MA1} for
$\bSL$, this is not manifest.

\newpage

\section{Concluding remarks\label{conclusion}}

We hope we have succeeded in achieving the two goals that we have set in
\S\ref{intro}.  Formulating a field theory with a momentum cutoff is the
most natural, and it accords with Wilson's non-perturbative definition
of a field theory.  It is still surprising that we can describe the
continuum limit using a finite momentum cutoff, but it does not come for
free: the cutoff function needs to be smooth, and the Wilson action
contains an infinite number of terms.  The formalism of ERG is simple
enough that we only need several pages to summarize the essence of all
done in this review.  We have given a quick summary as Appendix
\ref{summary}.

We have introduced two ways of realizing continuous symmetry: one with
the WT identity, and another with the QME.  Among the examples we have
discussed in \S\S\ref{WTexamples} \& \ref{AFexamples}, only YM theories
need the QME for proving the possibility of construction.  Let us try to
understand this.  The WT identity is the invariance of the Wilson action
under infinitesimal transformations, and it knows nothing about the
algebra formed by the transformations.  We need the antifield formalism
to incorporate an algebraic structure.  Take the example of the two
dimensional O(N) non-linear sigma model discussed in
\S\ref{WTexamples-ON}.  Commuting two generators that move the N-th
axis, we get an element of O(N$-$1), under which the action is
manifestly invariant.  Take the Wess-Zumino model.  Commuting two
supersymmetry generators, we get a translation, under which the action
is manifestly invariant.  Hence, either in the non-linear sigma model or
the Wess-Zumino model, the algebra of infinitesimal transformations does
not constrain the theory.  Hence, the WT identity suffices.  This is not
the case with YM theories; we need the antifield formalism to
incorporate the full algebra of BRST transformations.\footnote{The
antifield formalism is necessary only to prove the possibility of
perturbative construction.  The WT identity suffices for the
order-by-order construction.}

As for perturbative applications, our list of examples is not
exhaustive.  Among those missing, supersymmetric YM theories are
particularly important.  The ERG formalism has been already
applied,\cite{Bonini:1998ec, Arnone:1998zc, Bilal:2007ne} but our
understanding is still incomplete regarding the non-renormalization
properties.\cite{Seiberg:1993vc}\cite{Weinberg:1998uv} We also think it
important to prove the non-renormalization theorem for chiral anomalies
using the ERG formalism; to our knowledge it has not been done.

As for non-perturbative applications, the order-by-order solution of the
WT identity or QME is inappropriate.  Even approximate solutions will do
as long as they are given in closed form.  Progress along this line is
hoped for.

In this review we have been primarily concerned with formulating various
symmetric theories.  The formalism has matured enough that we are ready
to address more physical questions.

\section*{Acknowledgment}

The authors would like to thank Prof.~K.~Higashijima for suggestions
and encouragement.

\newpage
\appendix


\section{Generalized diffusion equations\label{diffusion}}

In \S \ref{derivation} we have derived a particular ERG differential
equation (\ref{derivation-polchinskieq}) for the interaction part of the
Wilson action, $\SIL$, defined by (\ref{derivation-WilsonI}).  The
purpose of this appendix is twofold:
\begin{enumerate}
\item to rewrite (\ref{derivation-WilsonI}) as a formula that gives the
      entire Wilson action,
\item to generalize (\ref{derivation-WilsonI}) to incorporate the
      original Wilson's differential equation \cite{Wilson:1973jj} and
      Polchinski's \cite{Polchinski:1983gv} under the same footing.
\end{enumerate}
This is done by \bref{diffusion-St}, where different choices of $A_t, Z_t$ 
correspond to Wilson's and Polchinski's ERG differential equations.

In the following we adopt a notation slightly different from that in the
main text.  We write $S_t$ for $S_\Lambda$, and $S_0$ for $S_B$, where
$t$ is defined by
\begin{equation}
t = \ln \frac{\Lambda_0}{\Lambda} \ge 0\,.
\end{equation}
The logarithmic parameter $t$ grows, as the momentum cutoff $\Lambda$
decreases.

\subsection{Derivation of the generalized diffusion
  equations\label{diffusion-derivation}} 

\begin{figure}[t]
\begin{center}
\epsfig{file=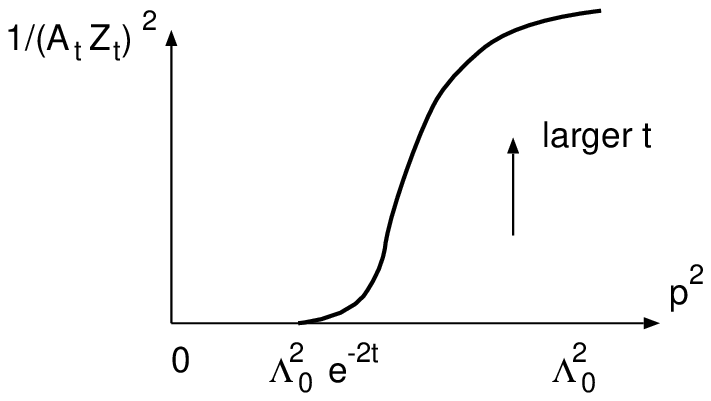, width=7cm}
\caption{More integration for higher $p^2$ and higher $t$.}
\label{diffusion-AtZt}
\end{center}
\end{figure}

We consider a one-parameter family of actions $S_t [\phi]\, (t \ge 0)$
for a real scalar field $\phi$.  We generate the $t$-dependence by a
gaussian integral transformation of the following
type:\cite{Wilson:1973jj,Wetterich:1989xg}
\begin{eqnarray}
&&\exp \left[ S_t [\phi] \right]
= \int [d\phi'] \exp \Big[\nn\\
&& \,\, - \frac{1}{2} \int_p A_t (p)^2 \left(
\phi (p) - Z_t (p) \phi' (p) \right)  \left(
\phi (-p) - Z_t (p) \phi' (-p) \right) + S_0 [\phi'] \Big]\,.
\label{diffusion-St}
\end{eqnarray}
We impose the two properties
\begin{equation}
\begin{array}{c@{\quad}l}
\mathrm{(i)}& 1/A_0 (p) = 0\,,\\
\mathrm{(ii)}& Z_0 (p) = 1\,,
\end{array}
\end{equation}
in order to assure the limit
\begin{equation}
\lim_{t \to 0+} S_t = S_0\,.
\end{equation}
The two factors $Z_t$ and $A_t Z_t$ have the following physical
 meanings:
\begin{enumerate}
\item blocking factor $Z_t (p)$ --- $Z_t (p) \phi (p)$ is the Fourier
      transform of a block spin.  Denoting the Fourier inverse of $\phi
      (p)$ and $Z_t (p)$ by the same symbols, we write
\begin{equation}
\phi (x) = \int_p \e^{i p x} \phi (p)\,,\quad
Z_t (x) = \int_p \e^{i p x} Z_t (p)\,.
\end{equation}
Then, the inverse Fourier transform of $Z_t (p) \phi (p)$ is the
block spin or average field:
\begin{equation}
\phi_t (x) \equiv \int_p \e^{i p x} Z_t (p) \phi (p)
= \int d^D y\, Z_t (x-y) \phi (y)
\end{equation}
The size of the domain of $Z_t (x)$ should grow exponentially as $\e^t$.
\item $A_t (p) Z_t(p)$ --- This has to do with incomplete functional
    integration.  The gaussian factor in (\ref{diffusion-St}) implies
      that $\phi (p)$ is equal to the block spin $Z_t (p) \phi' (p)$
    up to order $1/A_t (p)$.  Equivalently $\phi' (p)$ is integrated
      around $\phi (p)/Z_t (p)$ up to order $1/\left(A_t (p) Z_t
      (p)\right)$.  We take the squared width of integration, $1/(A_t
      Z_t (p))^2$, a growing function of both $t$ and $p^2$.
      (Fig.~\ref{diffusion-AtZt})
\end{enumerate}
We will shortly give examples of $Z_t$ and $A_t Z_t$.

We now wish to derive the differential equation for $S_t [\phi]$,
analogous to the diffusion equation.  We first consider the
$t$-derivative:
\begin{eqnarray}
&& \partial_t \exp \left[ S_t [\phi] \right] \nn\\
&=& \int [d\phi']\int_p
\Big\lbrace - \partial_t A_t (p) \cdot A_t (p) \left(\phi (p) - Z_t (p)
        \phi' (p)\right)\cdot \left( \phi (-p) - Z_t (p) \phi'
	(-p)\right)\nn\\
&&\qquad\qquad +  A_t^2 (p) \partial_t Z_t (p) \phi' (p) \left( \phi
        (-p) - Z_t (p)  \phi' (-p)\right) \Big\rbrace\nn\\
&&\qquad \times \exp \left[ - \frac{1}{2} \int_q A_t (q)^2 \left(
\phi (q) - Z_t (q) \phi' (q) \right)  \left(
\phi (-q) - Z_t (q) \phi' (-q) \right) + S_0 [\phi'] \right]\nn\\
&=& \int [d\phi'] \int_p \Big\lbrace - \left(\partial_t A_t (p) \cdot A_t
				     (p) + A_t^2(p) \partial_t \ln Z_t
				     (p) \right) \nn\\ 
&&\qquad\qquad\qquad\qquad \times \left( \phi (p) - Z_t (p) \phi' (p)\right)
\left( \phi (-p) - Z_t (p) \phi' (-p) \right)\nn\\
&&\qquad\qquad + A_t^2 (p) \partial_t \ln Z_t (p) \cdot \phi (p) \left( \phi
    (-p) - Z_t (p) \phi' (-p)\right) \Big\rbrace \exp \Big[ \cdots \Big]\,,
\end{eqnarray}
where we have suppressed the argument of the last exponential.  We
compare the above with the first and second order functional derivatives
of $\exp [S_t]$:
\begin{eqnarray}
  -  \frac{\delta}{\delta \phi (p)}\, \exp \left[S_t[\phi]\right] &=& 
    \int [d\phi']  
    A_t (p)^2 \left( \phi (-p) - Z_t (p) \phi' (-p) \right) \exp
    \Big[\cdots\Big] ,\\ 
    \frac{\delta^2}{\delta \phi (p) \delta \phi (-p)} \,\exp
    \left[S_t[\phi]\right]  &=& \int [d\phi]  
    A_t (p)^2 \left( \phi (-p) - Z_t (p) \phi' (-p) \right) \nn\\
    &&\times  A_t (p)^2 \left( \phi (-p) - Z_t (p) \phi' (-p) \right) 
    \exp \Big[\cdots\Big] ,
\end{eqnarray} 
where we have ignored an additive constant proportional to the space volume.

Thus, we obtain the generalized diffusion equation
\begin{eqnarray}
\partial_t \exp \left[ S_t [\phi] \right] &=& \int_p \left[
- \frac{1}{A_t^2 (p)} \partial_t \ln \left( A_t Z_t
\right)(p) \frac{\delta^2}{\delta \phi (p) \delta \phi (-p)}\right.\nn\\
&& \left.- \partial_t \ln Z_t (p) \cdot \phi (p) \frac{\delta}{\delta \phi (p)}
\right] \exp \left[ S_t [\phi] \right]\,.
\end{eqnarray}
This can be rewritten for $S_t$ as
\begin{eqnarray}
\partial_t S_t &=& \int_p \Bigg[ \,F_t (p)
    \cdot \phi(p) \frac{\delta S_t}{\delta \phi (p)}\nn\\
&&\quad + G_t (p) 
\cdot \frac{1}{2} \left\lbrace \frac{\delta S_t }{\delta \phi (p)}
    \frac{\delta S_t}{\delta \phi (-p)} + \frac{\delta^2 S_t}{\delta
      \phi (p) \delta \phi (-p)} \right\rbrace \Bigg]\,,
\end{eqnarray}
where
\begin{equation}
\lb
\begin{array}{c@{~\equiv~}l}
F_t (p) & - \partial_t \ln Z_t (p)\,,\\
G_t (p) & - \frac{2}{A_t (p)^2} \partial_t \ln \left( A_t Z_t \right)
(p)\,.
\end{array}\right.
\end{equation}
The equation for $S_t$ has the same form as the ERG differential equation
(\ref{derivation-polchinskieqfull}) for the full action.

We now define the generating functional of the connected correlation
functions:
\begin{equation}
\exp \left[W_t [J]\right] \equiv \int [d\phi] \exp \left[ S_t[\phi] +
     \int_p J(-p) \phi (p) \right]\,.
\end{equation}
Using the technique familiar from \S\ref{derivation-source}, we can
compute $W_t [J]$ in terms of $W_0 [J]$ as follows:
\begin{eqnarray}
\exp \left[W_t [J]\right] &=& \int [d\phi] \e^{ \int_p J(p) \phi (-p) }
\int [d\phi'] \exp \Bigg[ \nn\\
 &&  - \frac{1}{2}
\int_p A_t (p)^2 \left(\phi (p) - Z_t (p) \phi' (p)\right)
\left(\phi (-p) - Z_t (p) \phi' (-p) \right)
+ S_0 [\phi'] \Bigg]\nn\\
&=& \int [d\phi] \int [d\phi'] \exp \Bigg[ \int_p 
J(p) 
  \left( \phi (-p) - Z_t (p) \phi' (-p) \right)\nn\\
&&\,  - \frac{1}{2} \int_p A_t (p)^2 \left(\phi (p) - Z_t (p) \phi'
    (p)\right) \left(\phi (-p) - Z_t (p) \phi' (-p) \right)\nn\\
&& + S_0 [\phi'] + \int_p J(p) Z_t (p) \phi' (-p)
\Bigg]\,.
\end{eqnarray}
Now, shifting $\phi$ by $Z_t \phi'$, we obtain
\begin{eqnarray}
\exp \left[ W_t [J] \right] &=& \int [d\phi] \exp \left[ - \frac{1}{2}
    \int_p A_t (p)^2 \phi (p) \phi (-p) + 
     \int_p J(-p) \phi (p) \right]\nn\\
&& \times \int [d\phi'] \exp \left[ S_0 [\phi'] + \int_p J(p) Z_t (p)
      \phi' (-p) \right] \nn\\
&=& \exp \left[ - \frac{1}{2} \int_p \frac{1}{A_t (p)^2} J(p) J(-p) +
    W_0 \left[ Z_t (p) J(p) \right]\right]\,.
\end{eqnarray}
Hence, we obtain
\begin{equation}
W_t [J] = W_0 [Z_t J] - \frac{1}{2} \int_p \frac{1}{A_t (p)^2} J(p)
J(-p)\,.
\end{equation}
This means that nothing is lost in going from $S_0$ to $S_t$; we can
construct $W_0$ from $W_t$, and vice versa.  Differentiating the above
equation with respect to $J$, we obtain the following equation for the
connected correlation functions:
\begin{equation}
\left\lbrace\begin{array}{c@{~=~}l}
\vev{\phi (p) \phi (-p)}_{S_t} & \frac{1}{A_t (p)^2} + 
  Z_t (p)^2 \vev{\phi (p) \phi (-p)}_{S_0}\,,\\
\vev{\phi (p_1) \cdots \phi (p_n)}_{S_t} & \prod_{i=1}^n
Z_t (p_i) \cdot \vev{\phi (p_1) \cdots \phi
  (p_n)}_{S_0}\,. \quad (n > 1)\end{array}\right.
\end{equation}
This is a generalization of (\ref{derivation-vevSBSL}).  Hence, $\phi
(p)$ of $S_t$ has the same correlations as the block spin $Z_t (p) \phi
(p)$ of the original $S_0$.  The smearing factor $A_t$ affects only the
two-point correlation function.  This is a well known general
characteristic of diffusion processes.\footnote{The diffusion equation
$\partial_t P(x,t) = D \partial_x^2 P(x,t)$ preserves the connected part
of the correlation $\vev{x^n} = \int dx\, x^n P(x,t)$ for $n > 2$.}

We now choose a particular blocking function, given in terms of a cutoff
function $K$:
\begin{equation}
Z_t (p) = \frac{\Kz{p \e^t}}{\Kz{p}} \exp \left[
    \frac{1}{2} \int_0^t dt'\, \eta (t')\right]\,,
\label{diffusion-Zt}
\end{equation}
where $\eta$ is an anomalous dimension, which affects only the overall
normalization of the field.  For $\eta = 0$, we obtain
\begin{equation}
Z_t (p) \lb\begin{array}{l@{\quad}l}
= 1& (p^2 < \e^{-2t} \Lambda_0^2)\,,\\
\sim 0& (p^2 > \e^{-2t} \Lambda_0^2)\,,
\end{array}\right.
\end{equation}
if we assume $K(p/\Lambda) = 1$ for $p^2 < \Lambda^2$.
Hence, the support of the Fourier transform $Z_t (x)$ has size
$\e^t/\Lambda_0$ in space, growing exponentially as $t$ increases.
This choice of $Z_t$ corresponds to
\begin{equation}
F_t (p) \equiv - \partial_t \ln Z_t (p)
= \frac{\Delta (p\e^t/\Lambda_0)}{K(p\e^t/\Lambda_0)} - \frac{\eta
  (t)}{2}\,.
\end{equation}

With the above choice for $F_t (p)$, we obtain
\begin{eqnarray}
G_t (p) &=& - 2 \,Z_t (p)^2 \,\partial_t \ln A_t Z_t (p)\nn\\
&=& - 2 \,\e^{\int_0^t \eta}\, \frac{\Kz{p\e^t}^2}{\Kz{p}^2}
\frac{\partial_t \ln A_t Z_t (p)}{\left(A_t Z_t (p)\right)^2} \,.
\end{eqnarray}
Hence,
\begin{equation}
\partial_t \frac{1}{\left(A_t Z_t (p)\right)^2} = \e^{- \int_0^t \eta}
\frac{\Kz{p}^2}{\Kz{p \e^t}^2} G_t (p)\,,
\end{equation}
where we impose the initial condition
\begin{equation}
\lim_{t\to0+} \frac{1}{\left(A_t Z_t(p)\right)^2} = 0\,.
\label{diffusion-AtZtinitial}
\end{equation}

Let us now look at two examples of $G_t (p)$ and calculate the
corresponding $A_t Z_t (p)$.

\subsection*{Example 1 (Wilson)}

We choose
\begin{equation}
    G_t (p)  = 2 \frac{\e^{2t}}{\Lambda_0^2} \left( \frac{\Delta
          (p\e^t/\Lambda_0)}{K(p\e^t/\Lambda_0)} + 1 - 
        \frac{1}{2} \eta (t) \right)\,.
\label{diffusion-derivation-WilsonGt}
\end{equation}
This is the original choice of Wilson discussed in sect.~11 of
\citen{Wilson:1973jj}.\footnote{We have not introduced rescaling of
space.  This is discussed briefly in Appendix \ref{WilsonRG}.}
This gives
\begin{eqnarray}
\partial_t \frac{\Lambda_0^2}{(A_t Z_t (p))^2} &=& 2 \e^{2 t - \int_0^t \eta}
\frac{\Kz{p}^2}{\Kz{p\e^t}^2} \left( \frac{\Delta
      (p\e^t/\Lambda_0)}{\Kz{p\e^t}} + 1 - 
    \frac{\eta}{2} \right)\nn\\
&=& \partial_t \left( \e^{2 t - \int_0^t \eta}
    \frac{\Kz{p}^2}{\Kz{p\e^t}^2} \right) \,.
\end{eqnarray}
Hence, we obtain
\begin{equation}
\frac{\Lambda_0^2}{(A_t Z_t (p))^2} = \e^{2 t - \int_0^t \eta}
\frac{\Kz{p}^2}{\Kz{p\e^t}^2} - 1\,,
\end{equation}
using the initial condition \bref{diffusion-AtZtinitial}.  Thus, using
\bref{diffusion-Zt}, we obtain
\begin{equation}
\frac{1}{A_t (p)^2} = \frac{1}{\Lambda_0^2} \left( 1 -
\frac{\Kz{p\e^t}^2}{\Kz{p}^2}\, 
\e^{\int_0^t \eta}\right)\,.
\end{equation}
This implies
\begin{equation}
\left\lbrace\begin{array}{c@{~=~}l}
\vev{\phi (p)\phi (-p)}_{S_t} - \frac{\e^{2t}}{\Lambda_0^2} &
\e^{\int_0^t \eta}\, \frac{\Kz{p\e^t}^2}{\Kz{p}^2} \left(
\vev{\phi (p)\phi (-p)}_{S_0} - \frac{1}{\Lambda_0^2} \right)\,,\\
\vev{\phi (p_1) \cdots \phi (p_{2n})}_{S_t} &
\e^{n \int_0^t \eta}\, \prod_{i=1}^{2n} \frac{\Kz{p_i \e^t}}{\Kz{p_i}}
\cdot \vev{\phi (p_1) \cdots \phi (p_{2n})}_{S_0}\,.
\end{array}\right.
\end{equation}

Wilson's choice is characterized by the following behavior of the
two-point function as $t \to \infty$:
\begin{equation}
\vev{\phi (p) \phi (-p)}_{S_t} \stackrel{t \to
  \infty}{\longrightarrow} \frac{\e^{2 t}}{\Lambda_0^2}\,.
\end{equation}

\subsection*{Example 2 (Polchinski)}

The Polchinski equation (\ref{derivation-polchinskieqfull}) is obtained if
we choose $\eta = 0$, and
\begin{equation}
G_t (p) = \frac{\Delta (p\e^t/\Lambda_0)}{p^2 + m^2} \,.
\end{equation}
Here, we make a slightly generalized choice without assuming $\eta = 0$:
\begin{eqnarray}
G_t (p) &=& \frac{1}{p^2 + m^2 (t)} \left\lbrace \Delta
    (p\e^t/\Lambda_0) \right.\nn\\
&&\qquad \left.- \left( \eta 
        (t) + \frac{b_m (t)}{p^2 + m^2 (t)} \right) \Kz{p\e^t} \left( 1
        - \Kz{p\e^t}\right)\right\rbrace\,,
\end{eqnarray}
where $m^2 (t)$ is a $t$-dependent squared mass, and
\begin{equation}
b_m (t) \equiv \frac{d m^2 (t)}{dt}\,.
\end{equation}

Let us find the corresponding $A_t Z_t (p)$; we find
\begin{eqnarray}
&&\partial_t \frac{1}{(A_t Z_t (p))^2} = \e^{- \int_0^t \eta}
\frac{\Kz{p}^2}{\Kz{p\e^t}^2} \frac{1}{p^2 + m^2 (t)} \nn\\
&& \quad \times \left\lbrace \Delta
    (p\e^t/\Lambda_0) - \left( \eta 
        (t) + \frac{b_m (t)}{p^2 + m^2 (t)} \right) \Kz{p\e^t} \left( 1
        - \Kz{p\e^t}\right)\right\rbrace\nn\\
&&\,= \Kz{p}^2 \partial_t \lb \frac{\e^{- \int_0^t \eta}}{p^2 + m^2 (t)}
\left(\frac{1}{\Kz{p\e^t}}-1 \right)  \rb\,.
\end{eqnarray}
Hence, 
\begin{eqnarray}
&&\frac{1}{(A_t Z_t (p))^2} = \Kz{p}^2 \nn\\
&&\quad \cdot \lb \frac{\e^{- \int_0^t
        \eta}}{p^2 + m^2 (t)} \left(\frac{1}{\Kz{p\e^t}}-1
    \right)
- \frac{1}{p^2 + m^2 (0)} \left( \frac{1}{\Kz{p}}-1
    \right) \rb\,. 
\end{eqnarray}
Thus, we obtain
\begin{eqnarray}
&&\frac{1}{A_t (p)^2} = \Kz{p\e^t}^2 \nn\\
&&\quad\cdot \left\lbrace \frac{1}{p^2 + m^2 (t)}
    \left( \frac{1}{\Kz{p\e^t}} - 1 \right) - \frac{\e^{\int_0^t
        \eta}}{p^2 + m^2 (0)} \left( \frac{1}{\Kz{p}} - 1 \right)
\right\rbrace\,.
\end{eqnarray}
Hence, we obtain the following $t$ dependence of the correlation
functions: 
\begin{eqnarray}
&&\vev{\phi (p) \phi (-p)}_{S_t} - \frac{\Kz{p\e^t}
\left( 1 - \Kz{p\e^t}\right)}{p^2 + m^2 (t)} \nn\\
&& = \e^{\int_0^t \eta} \,\frac{\Kz{p\e^t}^2}{\Kz{p}^2} \left(
\vev{\phi (p)\phi (-p)}_{S_0} - \frac{\Kz{p} (1 -
\Kz{p})}{p^2 + m^2 (0)} \right)\,,
\end{eqnarray}
and
\begin{equation}
\vev{\phi (p_1) \cdots \phi (p_{2n})}_{S_t}
= \e^{n \int_0^t \eta} \prod_{i=1}^{2n} \frac{\Kz{p_i
  \e^t}}{\Kz{p_i}}\cdot
\vev{\phi (p_1) \cdots \phi (p_{2n})}_{S_0}\,.
\end{equation}
These reduce to (\ref{derivation-vevSBSL}) if we take $\eta = 0$ and
$t$-independent $m^2$.

\subsection{Extention to fermions}

In extending the generalized diffusion equations to fermions, we only
consider the spin $\frac{1}{2}$ fields in $D=4$.  We generate the
following one-parameter family of actions:
\begin{eqnarray}
&&\exp \left[S_t [\psi,\bar{\psi}]\right] \equiv \int [d\psi' d\bar{\psi}']\,
\exp \Big[ S_0 [\psi',\bar{\psi}'] \nn\\
&&\quad  - \int_p \lb \bar{\psi} (-p) - Z_t (p) \bar{\psi}'
    (-p)\rb \bar{A}_t (-p) A_t (p) \lb \psi (p) - Z_t (p) \psi' (p)
    \rb \Big]\,,
\end{eqnarray}
where $Z_t$ is a scalar, and $A_t$ \& $\bar{A}_t$ are 4-by-4 matrices.  We
choose 
\begin{equation}
Z_0 = 1\,,\quad \frac{1}{\bar{A}_0 A_0} = 0\,,
\end{equation}
so that
\begin{equation}
\lim_{t \to 0+} S_t = S_0\,.
\end{equation}

Let us define the generating functional by
\begin{equation}
\exp \left[ W_t [\bar{\xi}, \eta]\right] \equiv \int [d\psi d\bar{\psi}] \,
\exp \left[ S_t [\psi,\bar{\psi}] +   \int (\bar{\xi} \psi +
    \bar{\psi} \eta)\right] \,.
\end{equation}
We then obtain the familiar relation
\begin{equation}
W_t [\bar{\xi}, \eta] = W_0 [Z_t \bar{\xi}, Z_t \eta] + \int \bar{\xi} (-p)
\frac{1}{\bar{A}_t (-p) A_t (p)} \eta (p)\,,
\end{equation}
where $1/(\bar{A}_t A_t)$ denotes the inverse of the matrix $\bar{A}_t A_t$.

To derive the differential equation for $S_t$, it is convenient to
rewrite
\begin{eqnarray}
\exp \left[S_t [\psi,\bar{\psi}]\right] &=& \int [d\psi' d\bar{\psi}']
\exp \left[ - \int_p \bar{\psi}' (-p) \psi' (p) \right.\nn\\
&&\left.\quad + S_0 \left[ \frac{1}{Z_t} \left( \frac{1}{A_t} \psi' +
					  \psi\right), 
    \frac{1}{Z_t} \left( \bar{\psi}' \frac{1}{\bar{A}_t} +
        \bar{\psi}\right) \right] \right]\,.
\end{eqnarray}
From this, we obtain the following ERG differential equation:
\begin{eqnarray}
\partial_t S_t &=& \int_p \Bigg[ F_t (p) \lb S_t \Rd{\psi (p)} \psi
(p) + \bar{\psi} (-p) \Ld{\bar{\psi} (-p)} S_t \rb\nn\\
&&\quad - \Tr G_t (p) \lb \Ld{\bar{\psi} (-p)} S_t \cdot S_t \Rd{\psi (p)}
+ \Ld{\bar{\psi} (-p)} S_t \Rd{\psi (p)} \rb \Bigg]\,,
\end{eqnarray}
where
\begin{equation}
\lb\begin{array}{c@{~\equiv~}l}
F_t (p) & - \partial_t \ln Z_t (p)\,,\\
G_t (p) & Z_t (p)^2 \partial_t \lb 1/\left(Z_t (p)^2 \bar{A}_t (-p)
    A_t (p)\right)\rb\,. 
\end{array}\right.
\end{equation}

\subsection*{Example 1 (Polchinski)}

We take
\begin{subequations}
\begin{eqnarray}
Z_t (p) &\equiv& \frac{\Kz{p \e^t}}{\Kz{p}} = \frac{K}{K_0}
 \,,\label{diffusion-ZtFermi}\\ 
\frac{1}{\bar{A}_t (-p) A_t (p)} &\equiv& \frac{1}{\fmslash{p} + i m}
K \left( 1 - \frac{K}{K_0}\right)\,.
\end{eqnarray}
\end{subequations}
Then we obtain
\begin{equation}
\lb\begin{array}{c@{~\equiv~}l}
F_t (p) & \Delta (p \e^t/\Lambda_0)/\Kz{p \e^t}\,,\\
G_t (p) & \Delta (p \e^t/\Lambda_0)/\left(\fmslash{p} + i m\right)\,.
\end{array}\right.
\end{equation}

\subsection*{Example 2 (chirality breaking diffusion)}

We keep the same blocking factor $Z_t$, but take the smearing factor
\begin{equation}
\frac{1}{\bar{A}_t A_t} = \frac{1}{i m} K \left( 1 -
    \frac{K}{K_0}\right)\label{diffusion-AtFermi}
\end{equation}
so that
\begin{equation}
G_t (p) = \frac{\Delta (p \e^t/\Lambda_0)}{i m}\,.
\end{equation}
The smearing factor breaks chirality explicitly.
Even if we start from the chiral invariant action
\begin{equation}
S_0 [\psi, \bar{\psi}] = - \int_p \frac{1}{K_0 (p)} \bar{\psi} (-p)
 \fmslash{p} \psi (p)\,, 
\end{equation}
the diffusion process breaks chiral symmetry explicitly, and $S_{t > 0}$
is not manifestly chiral invariant.  But it does inherit the chiral
invariance of the original action $S_0$, as will be shown at the end of
the next subsection.  It is straightforward to find
\begin{equation}
S_t [\psi,\bar{\psi}] = - \int_p \bar{\psi} (-p) D_t (p) \psi (p)\,,
\label{diffusion-freeStFermi}
\end{equation}
where
\begin{equation}
D_t (p) \equiv im \frac{K_0}{K} \frac{\fmslash{p}}{(K_0 - K)
  \fmslash{p} + i m K} \,.\label{diffusion-freeDtFermi}
\end{equation}
This satisfies the initial condition:
\begin{equation}
D_0 (p) = \frac{\fmslash{p}}{K_0}\,.
\end{equation}
Now, we take $\Lambda_0 \to \infty$, while we keep $\Lambda = \Lambda_0
\e^{-t}$ finite.  We obtain the continuum limit:
\begin{equation}
D_\Lambda (p) =  \frac{i m}{\K{p}} \frac{\fmslash{p}}{(1-\K{p})
  \fmslash{p} + im \K{p}}\,.
\end{equation}
This has an interesting property:
\begin{equation}
D_\Lambda (p) \longrightarrow 
\lb\begin{array}{c@{\quad}c}
\frac{i m}{K (p/\Lambda)}& (p \gg \Lambda)\,,\\
\fmslash{p}& (p < \Lambda)\,.\end{array}\right.
\end{equation}
For $p^2 < \Lambda^2$, chiral invariance becomes manifest.

\subsection{WT operator for chiral invariance}

As an application, let us formulate the WT identity for chiral
invariance.  At scale $\Lambda_0$, we define the WT operator
\begin{equation}
\Sigma_0 [\psi,\bar{\psi}] \equiv \int_p \left( S_0 \Rd{\psi (p)} \gamma_5
    \psi (p) + \bar{\psi} (-p) \gamma_5 \Ld{\bar{\psi} (-p)} S_0 \right)
\end{equation}
for chiral invariance (axial invariance to be more precise).  $\Sigma_0
= 0$ implies that the action $S_0$ is invariant under the chiral
transformation:
\begin{equation}
\psi (p) \longrightarrow \gamma_5 \psi (p)\,,\quad
\bar{\psi} (p) \longrightarrow \bar{\psi} (p) \gamma_5\,.
\end{equation}
We wish to study how the chiral symmetry is realized by the Wilson
action $S_t$ with $t > 0$.

A familiar calculation gives
\begin{eqnarray}
    &&\int [d\psi d\bar{\psi}] \Sigma_0 [\psi,\bar{\psi}] \exp
    \left[ S_0 [\psi,\bar{\psi}] + \int 
      (\bar{\xi} \psi + \bar{\psi} \eta)\right]\nn\\
    && = \int [d\psi d\bar{\psi}] \Sigma_t [\psi,\bar{\psi}] \exp
    \left[ S_t
      [\psi, \bar{\psi}] + 
      \int \frac{1}{Z_t} \left(\bar{\xi} \psi + \bar{\psi} \eta
      \right) - \int \frac{1}{Z_t^2} \bar{\xi} \frac{1}{\bar{A}_t A_t}
      \eta \right]\,, 
\end{eqnarray}
where
\begin{eqnarray}
&&\Sigma_t \equiv - \int_p \Tr \gamma_5 \left[ \psi (p) \cdot S_t
\Rd{\psi (p)} + \Ld{\bar{\psi} (-p)} S_t \cdot \bar{\psi} (-p)\right]\nn\\
&&\quad- \int_p \Tr \lb \gamma_5~,~ \frac{1}{\bar{A}_t A_t}\rb \left[
    \Ld{\bar{\psi} (-p)} S_t \cdot 
S_t \Rd{\psi (p)} + \Ld{\bar{\psi} (-p)} S_t \Rd{\psi (p)} \right]\,.
\label{diffusion-SigmatChiral}
\end{eqnarray}
Hence, the chiral invariance of $S_t$ is given by
\begin{equation}
\Sigma_t = 0\,.
\end{equation}

For a free theory\footnote{The quantum master equation and its solutions
have been studied for interacting fermions in
Refs.~\citen{Igarashi:2002ba,Igarashi:2002bs,Igarashi:2001cv}.  See also
Ref. \citen{Ichinose:1999ke}}
\begin{equation}
S_t = - \int_p \bar{\psi} (-p) D_t (p) \psi (p)\,,
\end{equation}
the relation $\Sigma_t = 0$ reduces to the Ginsparg-Wilson
relation\cite{Ginsparg:1981bj}
\begin{equation}
\lb \gamma_5, D_t (p) \rb - D_t (p) \lb \gamma_5, \frac{1}{\bar{A}_t
  A_t} \rb D_t (p) = 0\,.
\end{equation}
ERG gives the following $t$-dependence:
\begin{equation}
D_t (p) = \bar{A}_t A_t - \bar{A}_t A_t Z_t \frac{1}{D_0 + Z_t^2
  \bar{A}_t A_t} Z_t \bar{A}_t A_t\,.
\end{equation}
Substituting this into the Ginsparg-Wilson relation
above, we obtain what is expected:
\begin{equation}
\lb \gamma_5, D_0 \rb = 0\,,
\end{equation}
which is equivalent to
\begin{equation}
\Sigma_0 = 0\,.
\end{equation}

Let us reexamine the second example of the previous subsection, where we
have chosen the blocking factor \bref{diffusion-ZtFermi} and the
chirality breaking smearing factor \bref{diffusion-AtFermi}:
\begin{equation}
Z_t (p) = \frac{K}{K_0},\quad
\frac{1}{\bar{A}_t A_t} = \frac{1}{im} K \left( 1 - \frac{K}{K_0}
\right)\,.
\end{equation}
\bref{diffusion-SigmatChiral} gives
\begin{eqnarray}
\Sigma_t &\equiv& - \int_p \Tr \gamma_5 \Bigg[ \psi (p) \cdot S_t
\Rd{\psi (p)} + \Ld{\bar{\psi} (-p)} S_t \cdot \bar{\psi} (-p)\nn\\
&&\quad + \frac{2 K_0}{im K} (K_0 - K)  \lb \Ld{\bar{\psi} (-p)} S_t \cdot
S_t \Rd{\psi (p)} + \Ld{\bar{\psi} (-p)} S_t \Rd{\psi (p)} \rb
 \Bigg]\,.
\end{eqnarray}
The free action
\begin{equation}
S_t [\psi,\bar{\psi}] = - \int_p \bar{\psi} (-p) i m \frac{K_0}{K}
\frac{\fmslash{p}}{(K_0 - K) \fmslash{p} + i m K} \psi (p)\,,
\end{equation}
given by \bref{diffusion-freeStFermi} and \bref{diffusion-freeDtFermi},
indeed satisfies the WT identity:
\begin{equation}
\Sigma_t = 0\,.
\end{equation}

\newpage

\section{Applications of composite operators\label{appcomp}}

In the main text we have shown how to realize symmetry in terms of
either a WT composite operator $\Sigma_\Lambda$ or a QM composite
operator $\bar{\Sigma}_\Lambda$.  In this appendix we describe two
further applications of composite operators: one to beta functions, and
the other to universality.

\subsection{Beta functions\label{appcomp-beta}}

We recall that the Wilson action of the $\phi^4$ theory in $D=4$ is
defined by the Polchinski equation
\begin{equation}
- \Lambda \frac{\partial}{\partial \Lambda} S_{I,\Lambda} [\phi]
= \int_p \frac{\Delta (p/\Lambda)}{p^2 + m^2} \frac{1}{2} \lb
\frac{\delta S_{I,\Lambda} [\phi]}{\delta \phi (-p)}
\frac{\delta S_{I,\Lambda} [\phi]}{\delta \phi (p)}
+ \frac{\delta^2 S_{I,\Lambda} [\phi]}{\delta \phi (-p) \delta \phi
  (p)} \rb\,,\label{appcomp-beta-Polchinski}
\end{equation}
and the asymptotic behavior
\begin{eqnarray}
S_{I,\Lambda} &\asym& \int d^4 x \, \left[ \left( \Lambda^2 a_2 (\ln
        \Lambda/\mu) + m^2 b_2 (\ln \Lambda/\mu) \right)
    \frac{\phi^2}{2}\right.\nn\\ 
&& \quad \left. + c_2 (\ln \Lambda/\mu) \frac{1}{2}
    \left( \partial_\mu \phi \right)^2 + a_4 (\ln \Lambda/\mu)
    \frac{\phi^4}{4!} \right]\,.\label{appcomp-beta-SILasymp}
\end{eqnarray}
The Wilson action is parametrized not only by $m^2$ which appears in
the differential equation, but also by the three parameters $b_2 (0),
c_2 (0), a_4 (0)$.  Here, for the sake of simplicity, we make a simple
choice\cite{Sonoda:2002pb}:
\begin{equation}
b_2 (0) = c_2 (0) = 0,\quad a_4 (0) = - \lambda\,.\label{appcomp-beta-MS}
\end{equation}
We call this the MS scheme for its similarity to the minimal subtraction
scheme for dimensional regularization.\cite{Hooft:1972fi,Hooft:1973mm}
We immediately notice, though, that the MS scheme depends on the choice
of $\mu$.  The MS scheme (\ref{appcomp-beta-MS}) for a different $\mu$
corresponds to $b_2 (0), c_2 (0), a_4 (0)$ that do not satisfy
(\ref{appcomp-beta-MS}) for the original $\mu$.  It is the purpose of
this appendix to derive the $\mu$ dependence of the Wilson action in the
MS scheme (\ref{appcomp-beta-MS}).\cite{Sonoda:2006ai}(See also
\citen{Hughes:1987rf,Bonini:1996bk} for derivations of beta functions
from the ERG.  Ref.~\citen{Pernici:1998tp} derives the same mass
independent RG equation as \bref{appcomp-beta-MSRG}.)

Since $\mu$ can be regarded as a constant external field, the
derivative of the Wilson action
\begin{equation}
- \mu \frac{\partial \SL}{\partial \mu}
\end{equation}
is a composite operator.  (Here, $m^2$ and $\lambda$ are fixed.) To be
more precise, it is a dimension $4$ composite operator with zero
momentum.  Hence, it must be a linear combination of the three
linearly independent composite operators
\begin{equation}
m^2 \left[ \frac{\phi^2}{2} \right]_\Lambda (0)\,,\quad
\left[ \frac{\phi^4}{4!} \right]_\Lambda (0)\,,\quad
\left[ \frac{1}{2} \left( \partial_\mu \phi \right)^2\right]_\Lambda
(0)\,,\label{appcomp-beta-MSbasis}
\end{equation}
which are defined at the end of \S \ref{comp-asymp}.  The asymptotic
behavior (\ref{appcomp-beta-SILasymp}) implies the following asymptotic
behavior:
\begin{eqnarray}
- \mu \frac{\partial S_\Lambda}{\partial \mu} &\asym&
\int d^4 x\, \left[ \left( \Lambda^2 \dot{a}_2 (\ln \Lambda/\mu) + m^2
        \dot{b}_2 (\ln \Lambda/\mu) \right) \frac{\phi^2}{2}
\right.\nonumber\\
&& \left. \quad + \dot{c}_2 (\ln \Lambda/\mu) \frac{1}{2}
    \left( \partial_\mu \phi \right)^2 + \dot{a}_4 (\ln \Lambda/\mu)
    \frac{\phi^4}{4!} \right]\,,
\end{eqnarray}
where
\begin{equation}
\dot{a}_2 (\ln \Lambda/\mu) \equiv \Lambda \frac{\partial}{\partial
  \Lambda} a_2 (\ln \Lambda/\mu)\,,\,\textrm{etc.}
\end{equation}
Hence, we obtain
\begin{equation}
- \mu \frac{\partial\SL}{\partial\mu} = \dot{b}_2 (0) m^2 \left[
    \frac{\phi^2}{2} \right]_\Lambda (0) + \dot{c}_2 (0) \left[
    \frac{1}{2} \left( \partial_\mu \phi \right)^2\right]_\Lambda (0)
+ \dot{a}_4 (0) \left[ \frac{\phi^4}{4!}\right]_\Lambda (0)\,,
\end{equation}
where the coefficients depend only on $\lambda$.

We now wish to rewrite the above equation as a physically meaningful
equation.  For this purpose, we construct an alternative basis of
dimension $4$ composite operators with zero momentum.

We first consider
\begin{equation}
\Op_\lambda \equiv - \partial_\lambda S_\Lambda = - \partial_\lambda
S_{I,\Lambda} \,.
\end{equation}
This is a composite operator of the type \bref{comp-dSdJ}, since
$\lambda$ can be regarded as a constant external field.  This satisfies
\begin{eqnarray}
    \vev{\Op_\lambda \, \phi (p_1) \cdots \phi (p_n)}^\infty &\equiv&
    \prod_{i=1}^n \frac{1}{\K{p_i}} \cdot \vev{\Op_\lambda \,
      \phi (p_1) \cdots \phi (p_n)}_{S_\Lambda} \nn\\
    &=& - \partial_\lambda \vev{\phi (p_1) \cdots \phi (p_n)}^\infty \,.
\end{eqnarray}

Similarly, we wish to construct a composite operator $\Op_m$ that gives
\begin{equation}
\vev{\Op_m\, \phi (p_1) \cdots \phi (p_n)}^\infty = - \partial_{m^2}
\vev{\phi (p_1) \cdots \phi (p_n)}^\infty 
\end{equation}
for any $n$.  For $n \ne 2$, the $m^2$ derivative of $\SL$ will do:
\begin{eqnarray}
- \partial_{m^2} \vev{\phi (p_1) \cdots \phi (p_n)}^\infty
= \prod_{i=1}^n \frac{1}{\K{p_i}} \cdot \vev{ \left( - \partial_{m^2}
      \SL \right)\, \phi (p_1) \cdots \phi (p_n)}_{\SL}.
\end{eqnarray}
But, for $n=2$, we obtain
\begin{eqnarray}
&&- \partial_{m^2} \vev{\phi (p) \phi (-p)}^\infty \nn\\
&&= - \partial_{m^2} \left( \frac{1}{K^2} \vev{\phi (p) \phi
      (-p)}_{\SL} + \frac{1 - 1/K}{p^2 + m^2} \right)\nn\\
&&= \frac{1}{K^2} \left( \vev{( - \partial_{m^2} \SL)\, \phi (p)
      \phi (-p)}_{\SL} + \frac{K(K-1)}{(p^2 + m^2)^2} \right)\,.
\end{eqnarray}
Hence, we need $\Op'$ that satisfies
\begin{equation}
\lb\begin{array}{r@{~=~}l}
\vev{\Op'\, \phi (p) \phi (-p)}_{\SL} & \frac{K(K-1)}{(p^2 + m^2)^2}\,,\\
\vev{\Op'\, \phi (p_1) \cdots \phi (p_n)}_{\SL} & 0\quad (n \ne 2)\,.
\end{array}\right.
\end{equation}
This is easily constructed as
\begin{equation}
\Op' = \int_p \frac{K(K-1)}{(p^2 + m^2)^2} \frac{1}{2} \lb
\frac{\delta \SL}{\delta \phi (p)} \frac{\delta \SL}{\delta \phi (-p)}
+ \frac{\delta^2 \SL}{\delta \phi (p) \delta \phi (-p)} \rb\,.
\end{equation}
Hence, we obtain
\begin{eqnarray}
    \Op_m &=& - \partial_{m^2} \SL + \Op'\nn\\
&=&  - \partial_{m^2} \SL  - \int_p \frac{K(1-K)}{(p^2 +
      m^2)^2} \frac{1}{2} \lb 
    \frac{\delta \SL}{\delta \phi (p)} \frac{\delta \SL}{\delta \phi (-p)}
    + \frac{\delta^2 \SL}{\delta \phi (p) \delta \phi (-p)} \rb\,.
\end{eqnarray}
It is straightforward to check that this $\Op_m$ satisfies the
ERG differential equation (\ref{comp-compdiffeq}).

The third and last composite operator is defined by
\begin{equation}
\N \equiv - \int_p \K{p} \lb [\phi]_\Lambda (p) \frac{\delta
  \SL}{\delta \phi (p)} + \frac{\delta}{\delta \phi (p)}
[\phi]_\Lambda (p) \rb\,,
\end{equation}
where
\begin{equation}
[\phi]_\Lambda (p) \equiv \frac{1}{K} \phi (p) + \frac{1 - K}{p^2 +
  m^2} \frac{\delta \SL}{\delta \phi (-p)}
\end{equation}
is the composite operator corresponding to $\phi (p)$.  $\N$ is a
composite operator of the type (\ref{comp-dOdphi}), and it has a simple
correlation function
\begin{equation}
\vev{\N\,\phi (p_1) \cdots \phi (p_n)}^\infty = n \vev{\phi (p_1)
  \cdots \phi (p_n)}^\infty
\end{equation}
for any $n$.

Thus, we have obtained an alternative basis of dimension $4$ scalar
composite operators with zero momentum, consisting of
\begin{equation}
\Op_\lambda\,,\quad
m^2 \Op_m\,,\quad
\N\,.
\end{equation}
We can use this basis, instead of the basis consisting of
(\ref{appcomp-beta-MSbasis}), to expand $- \mu
\frac{\partial}{\partial \mu} \SL$:
\begin{equation}
- \mu \frac{\partial S_\Lambda}{\partial \mu} = \beta_m (\lambda) m^2
\Op_m + \beta (\lambda) \Op_\lambda + \gamma (\lambda) \N\,,
\end{equation}
where the coefficients are all functions of $\lambda$ alone, since
they cannot depend on $\ln \Lambda/\mu$.  The physical meaning of this
equation is clear.  It implies, for the correlation functions,
\begin{equation}
\left( - \mu \frac{\partial}{\partial \mu} + \beta_m
    m^2 \partial_{m^2} + \beta \partial_\lambda - n \gamma \right)
\vev{\phi (p_1) \cdots \phi (p_n)}^\infty = 0\,.
\label{appcomp-beta-MSRG}
\end{equation}
This is a mass independent renormalization group equation.
\cite{Hooft:1973mm,PhysRevD.8.3497}

\subsection{Universality\label{appcomp-univ}}

\textbf{Universality} is an important concept in renormalization
theory.  To construct a continuum limit, there are always more than
one way.  The independence of the continuum limit on the particular
method of construction is called universality.  For example, if we use
a lattice to construct a continuum theory, the limit should not depend
on what kind of lattice, whether square or cubic, we use.

In the following we examine universality in two restricted senses.
First, we wish to show that the continuum limit (i.e., the correlation
functions with the suffix $\infty$) does not depend on the particular
asymptotic conditions we use to select a solution of the ERG
differential equation.  Second, we wish to show that the continuum limit
does not depend on the choice of a cutoff function $K$.

\subsection*{Scheme dependence}

Let us compare two solutions of the same Polchinski differential
equation (\ref{appcomp-beta-Polchinski}).  One is $\SL$ satisfying 
the MS condition (\ref{appcomp-beta-MS}), and the other is $\SL +
\delta \SL$ satisfying
\begin{equation}
\lb\begin{array}{c@{~=~}l}
b_2 (0) & \delta b_2\,,\\
c_2 (0) & \delta c_2\,,\\
a_4 (0) & - \lambda + \delta a_4\,,
\end{array}\right.
\end{equation}
where $\delta b_2, \delta c_2, \delta a_4$ are all infinitesimal
constants.  Then, $\delta \SL$ is a composite operator given by
\begin{equation}
\delta \SL = \delta b_2 \, m^2 \left[\frac{\phi^2}{2}\right]_\Lambda (0) +
\delta c_2 \left[\frac{1}{2} \left(\partial_\mu
        \phi\right)^2\right]_\Lambda (0) + \delta a_4 \left[
    \frac{\phi^4}{4!} \right]_\Lambda (0)\,.
\end{equation}
Using the alternative basis, we can rewrite this in the form
\begin{equation}
\delta m^2 \Op_m + \frac{\delta z}{2} \N + \delta \lambda \,\Op_\lambda\,.
\end{equation}
This implies that $\SL + \delta \SL$, which does not satisfy the MS
condition (\ref{appcomp-beta-MS}), gives the same continuum limit as the
Wilson action satisfying (\ref{appcomp-beta-MS}) with squared mass $m^2
+ \delta m^2$ and coupling $\lambda + \delta \lambda$.  The only
difference is in the normalization of fields by a factor $\frac{1}{2}
\delta z$.  Extending this, we can show that the action with any
choice of $b_2 (0), c_2 (0), a_4 (0)$ is equivalent to an action
satisfying the MS condition (\ref{appcomp-beta-MS}) up
to field normalization.

\subsection*{Dependence on the choice of $K$}

In our discussions so far, we have always kept a choice of the cutoff
function $K$.  Physics should not depend on the choice of $K$, and
we will show this in the following.

Let $\SL$ be a solution of the ERG equation with the cutoff function $K
+ \delta K$, infinitesimally different from $K$.  We would like to
construct an equivalent action $\SL'$ that has $K$ as the cutoff
function.  We first recall that the correlation functions in the
continuum limit are given by
\begin{equation}
\lb\begin{array}{r@{~=~}l}
\vev{\phi (p) \phi (-p)}^\infty & \frac{1}{\left(K + \delta
      K\right)^2} \vev{\phi (p) \phi (-p)}_{\SL} +
\frac{1 - 1/(K+\delta K)}{p^2 + m^2}\,,\\
\vev{\phi (p_1) \cdots \phi (p_n)}^\infty & \prod_{i=1}^n
\frac{1}{\left( K + \delta K \right) (p_i)}\,\cdot \vev{\phi (p_1)
  \cdots \phi (p_n)}_{\SL}\,.
\end{array}\right.
\end{equation}
This can be rewritten as
\begin{eqnarray}
\vev{\phi (p) \phi (-p)}^\infty &=& \frac{1}{K^2} \Big\lbrace \vev{\left( 1 -
      \delta K/K\right) \phi (p) \left(1 - \delta K/K\right) \phi
  (-p)}_{\SL}\nn\\
&&\, + \frac{\delta K}{p^2 + m^2}\Big\rbrace
 + \frac{1 - 1/K}{p^2 + m^2}\,,\\
\vev{\phi (p_1) \cdots \phi (p_n)}^\infty &=& \prod_{i=1}^n
\frac{1}{K(p_i)} \nn\\
&&\, \times \vev{\left(1 - \delta K/K\right) \phi (p_1)
  \cdots \left(1 - \delta K/K\right) \phi (p_n)}_{\SL}\,.
\end{eqnarray}

We now construct a new Wilson action by
\begin{eqnarray}
\SL' &\equiv& \SL + \int_p \frac{\delta K}{K} \phi (p) \frac{\delta
  \SL}{\delta \phi (p)}\nn\\
&& + \int_p \frac{\delta K}{p^2 + m^2} \frac{1}{2} \lb \frac{\delta
  \SL}{\delta \phi (p)} \frac{\delta \SL}{\delta \phi (-p)} +
\frac{\delta^2 \SL}{\delta \phi (p) \delta \phi (-p)} \rb\,.
\end{eqnarray}
This has the correlation functions:
\begin{equation}
\lb\begin{array}{r@{~=~}l}
\vev{\phi (p) \phi (-p)}_{\SL'} & \vev{\left(1-\delta K/K\right) \phi
  (p) \left( 1 - \delta K/K\right) \phi (-p)}_{\SL} + \frac{\delta
  K}{p^2 + m^2}\,,\\
\vev{\phi (p_1) \cdots \phi (p_n)}_{\SL'} & \vev{\left( 1 - \delta
      K/K\right) \phi (p_1) \cdots \left( 1 - \delta
      K/K\right) \phi (p_n)}_{\SL}\,,
\end{array}\right.
\end{equation}
so that
\begin{eqnarray}
\vev{\phi (p) \phi (-p)}^\infty &=& \frac{1}{K^2} \vev{\phi (p) \phi
  (-p)}_{\SL'} + \frac{1 - 1/K}{p^2 + m^2}\,,\\
\vev{\phi (p_1) \cdots \phi (p_n)}^\infty &=& \prod_{i=1}^n
\frac{1}{K(p_i)} \cdot \vev{\phi (p_1) \cdots \phi (p_n)}_{\SL'}\,.
\end{eqnarray}
This implies
\begin{enumerate}
\item that $\SL'$ satisfies the Polchinski equation with the cutoff
    function $K$,
\item that $\SL'$ gives the same continuum limit as $\SL$.
\end{enumerate}
Hence, the continuum limit does not depend on the choice of a cutoff
function.  Note that even if $\SL$ obeys the MS condition
\bref{appcomp-beta-MS}, $\SL'$ does not necessarily obey it.

\newpage

\section{ERG differential equations for fixed points\label{WilsonRG}}

In the main text we use the Polchinski differential equation,
(\ref{derivation-polchinskieq}) for $\SIL$ and
(\ref{derivation-polchinskieqfull}) for $\SL$, to construct the
continuum limit of renormalizable theories.  This equation is
\textbf{NOT} what was originally proposed by K.~G.~Wilson for his
non-perturbative studies of field theory \cite{Wilson:1973jj}.  As we
lower the cutoff $\Lambda$, the cutoff function $\K{p}$ keeps changing
its momentum dependence, and hence the Polchinski equation has no fixed
point.  In this appendix we explain how to modify the Polchinki equation
to obtain an ERG differential equation that can have non-trivial fixed
points.

To have a fixed point, we must do two things:
\begin{enumerate}
\item make everything dimensionless by multiplying an appropriate power
      of the cutoff $\Lambda$,
\item rescale the field to normalize the kinetic term.
\end{enumerate}
Let us explain the above one by one:
\begin{enumerate}
\item For the real scalar theory in $D$ dimensions, the Fourier transform
$\phi (p)$ has mass dimension $- \frac{D+2}{2}$.  Using the
      dimensionless ratio
\begin{equation}
\bar{p} \equiv \frac{p}{\Lambda}\,,
\end{equation}
we define the dimensionless field
\begin{equation}
\bar{\phi} (\bar{p}) \equiv \Lambda^{\frac{D+2}{2}} \phi (p)\,.
\end{equation}
We then rewrite the Wilson action
\begin{eqnarray}
\SL [\phi] &=& - \frac{1}{2} \int_p \frac{p^2}{K(p/\Lambda)}\phi (p)
 \phi (-p)  + 
 \sum_{n=1}^\infty \frac{1}{(2n)!} \int_{p_1, \cdots, p_{2n}}  
\phi (p_1) \cdots \phi (p_{2n}) \, \nn\\
&&\,\times  (2 \pi)^D \delta^{(D)} (p_1 + \cdots
p_{2n}) \, \V_{2n} (\Lambda; p_1, \cdots, p_{2n})
\end{eqnarray}
as
\begin{eqnarray}
\bar{S}_t [\bar{\phi}] &\equiv& 
- \frac{1}{2} \int_{\bar{p}} \frac{\bar{p}^2}{K(\bar{p})} \bar{\phi} (-
\bar{p}) \bar{\phi} (\bar{p})
 + \sum_{n=1}^\infty \frac{1}{(2n)!} \int_{\bar{p}_1, \cdots, \bar{p}_{2n}} 
\bar{\phi} (\bar{p}_1) \cdots \bar{\phi} (\bar{p}_{2n}) \nn\\
&&\,\times (2 \pi)^D \delta^{(D)}
(\bar{p}_1 + \cdots \bar{p}_{2n}) \, \bar{\V}_{2n} (t; \bar{p}_1, \cdots,
\bar{p}_{2n}) \,,
\end{eqnarray}
where we define
\begin{equation}
\bar{\V}_{2n} (t; \bar{p}_1, \cdots, \bar{p}_{2n}) \equiv \Lambda^{(n-1)D-2n}
 \, \V_{2n} (\Lambda; p_1, \cdots, p_{2n})
\end{equation}
so that
\begin{equation}
\SL [\phi] = \bar{S}_t [\bar{\phi}]\,.
\end{equation}
The parameter $t$ is a dimensionless parameter such that
\begin{equation}
\Lambda \propto \e^{-t}\,.
\end{equation}
If $\SL [\phi]$ satisfies the Polchinski equation, then $\bar{S}_t$
satisfies
\begin{eqnarray}
\partial_t \bar{S}_t &=& \int_{p} \lb p_\mu \frac{\partial
 \bar{\phi} (p)}{\partial p_\mu} + \left( \frac{D+2}{2} +
\frac{\Delta (p)}{K (p)}\right) \phi (p) \rb \frac{\delta
 \bar{S}_t}{\delta \bar{\phi} (p)}\nn\\
&&\quad + \int_{p} \frac{\Delta (p)}{p^2} \frac{1}{2} \left(
 \frac{\delta \bar{S}_t}{\delta \bar{\phi} (p)}  \frac{\delta
 \bar{S}_t}{\delta \bar{\phi} (-p)} +
\frac{\delta^2 \bar{S}_t}{\delta \bar{\phi} (p) \delta \bar{\phi}
(-p)}\right)\,,
\end{eqnarray}
where we have omitted the bar above the integration variable $\bar{p}$.
\item To obtain a fixed point, we must normalize the field to satisfy 
\begin{equation}
\frac{\partial}{\partial p^2} \bar{\V}_2 (t; p,-p) \Big|_{p^2=0} = 0
\end{equation}
This further modifies the differential equation to
\begin{eqnarray}
&&\partial_t \bar{S}_t = \int_{p} \lb p_\mu \frac{\partial
 \bar{\phi} (p)}{\partial p_\mu} + \left( \frac{D+2}{2} -
						\frac{\eta_t}{2} +
\frac{\Delta (p)}{K (p)}\right) \bar{\phi} (p) \rb \frac{\delta
 \bar{S}_t}{\delta \bar{\phi} (p)}\label{WilsonRG-diffeq}\\
&&\quad + \int_{p} \frac{1}{p^2} \lb \Delta (p)
- \eta_t K(p)(1-K(p))\rb \frac{1}{2} \left(
 \frac{\delta \bar{S}_t}{\delta \bar{\phi} (p)}  \frac{\delta
 \bar{S}_t}{\delta \bar{\phi} (-p)} +
\frac{\delta^2 \bar{S}_t}{\delta \bar{\phi} (p) \delta \bar{\phi}
(-p)}\right)\,, \nn
\end{eqnarray}
where the constant $\eta_t$, \textbf{anomalous dimension}, is determined
as
\begin{equation}
\eta_t = \frac{ - \frac{\partial}{\partial p^2} \int_q \frac{\Delta
	(q)}{q^2} \bar{\V}_4 (t; q,-q, p, -p)\Big|_{p^2=0}}{
1 - \frac{\partial}{\partial p^2} \int_q \frac{K(q)(1-K(q))}{q^2}
 \bar{\V}_4 (t; q, -q, p, -p)\Big|_{p^2=0}}\,.
\end{equation}
\end{enumerate}
The last ERG differential equation (\ref{WilsonRG-diffeq}) has not only
the gaussian fixed point
\begin{equation}
\bar{S}_G [\bar{\phi}] \equiv - \frac{1}{2} \int_p \bar{\phi} (-p)
 \bar{\phi} (p) \frac{p^2}{K(p)}\,,
\end{equation}
but also a non-trivial fixed point, called the Wilson-Fisher fixed
point, if $D < 4$.

In the above we have modified the ERG differential equation not only to
have a fixed point, but also to give simple $t$-dependence to the
correlation functions, similar to (\ref{derivation-vevSBSL}).  We find
the following results:
\begin{subequations}
\begin{eqnarray}
&&\frac{1}{K(p \e^t)^2} \vev{\bar{\phi} (p \e^t) \bar{\phi} (- p
\e^t)}_{\bar{S}_t} + \frac{1 - 1/K(p \e^t)}{p^2 \e^{2t}}\nn\\
&& \quad= \exp \left[ - 2 t + \int^t dt' \,\eta_{t'}\right] \cdot
 (\textrm{$t$-independent})\,, \\
&&\prod_{i=1}^n \frac{1}{K(p+i
\e^t)} \cdot \vev{\bar{\phi} (p_1 \e^t) \cdots \bar{\phi} (p_n
\e^t)}_{\bar{S}_t}\nn\\
&& \quad = \exp \left[ t \left( D - n \frac{D+2}{2}\right) + \frac{n}{2}
		 \int^t dt'\, \eta_{t'} \right] \cdot
(\textrm{$t$-independent}) \,.
\end{eqnarray}
\end{subequations}

The original differential equation, given in \citen{Wilson:1973jj}, is
based upon Wilson's ERG differential equation (Example 1 of
\S\ref{diffusion-derivation}), and is somewhat different from the above:
\begin{eqnarray}
\partial_t \bar{S}_t &=& \int_{\bar{p}} \lb p_\mu \frac{\partial
 \bar{\phi} (p)}{\partial p_\mu} + \left( \frac{D+2}{2} -
						\frac{\eta_t}{2} +
\frac{\Delta (p)}{K (p)}\right) \phi (p) \rb \frac{\delta
 \bar{S}_t}{\delta \bar{\phi} (p)}\label{WilsonRG-originaldiffeq}\\
&&\quad + \int_{p} \lb \frac{\Delta (p)}{K(p)}
+ 1 - \frac{\eta_t}{2} \rb \frac{1}{2} \left(
 \frac{\delta \bar{S}_t}{\delta \bar{\phi} (p)}  \frac{\delta
 \bar{S}_t}{\delta \bar{\phi} (-p)} +
\frac{\delta^2 \bar{S}_t}{\delta \bar{\phi} (p) \delta \bar{\phi}
(-p)}\right)\,. \nn
\end{eqnarray}
The difference in the second line comes from the choice of $G_t$
\bref{diffusion-derivation-WilsonGt}. In \citen{Wilson:1973jj}, the
function $\Delta (p)/K(p)$ is denoted as
\begin{equation}
\rho (p)\,.
\end{equation}

Expanding $\bar{S}_t$ in powers of fields
\begin{equation}
\bar{S}_t = - \sum_{n=1}^\infty \frac{1}{(2n)!} \int \bar{\phi} (p_1)
 \cdots \bar{\phi} (p_{2n}) \, (2\pi)^D \delta^{(D)} (p_1 + \cdots
 p_{2n}) \, u_{2n} (t; p_1, \cdots, p_{2n})\,,
\end{equation}
the anomalous dimension $\eta_t$ is determined by the normalization
condition
\begin{equation}
\frac{\partial}{\partial p^2} u_2 (t; p, -p) \Big|_{p^2=0} = 1
\end{equation}
as
\begin{equation}
\frac{\eta_t}{2} = \frac{ - 2 u_2 (t; 0,0) +
\frac{\partial}{\partial p^2} \int_q \left( \frac{\Delta (q)}{K(q)} +
				      1\right)
\frac{1}{2} u_4 (t; q,-q, p,-p)\Big|_{p^2=0}}{1 - 2 u_2 (t;0,0) +
\frac{\partial}{\partial p^2} \frac{1}{2} \int_q u_4 (t;
q,-q,p,-p)\Big|_{p^2=0}} 
\end{equation}
(\ref{WilsonRG-originaldiffeq}) gives slightly different
$t$-dependence to the two-point correlation function:
\begin{subequations}
\begin{eqnarray}
&&\frac{1}{K(p \e^t)^2} \left( \vev{\bar{\phi} (p \e^t) \bar{\phi} (- p
			 \e^t)}_{\bar{S}_t} - 1\right) \nn\\
&&\quad = \exp \left[ - 2 t + \int^t dt'\, \eta_{t'} \right] \cdot
 (\textrm{$t$-independent})\,, \\
&&\prod_{i=1}^n \frac{1}{K(p_i \e^t)} \cdot \vev{\bar{\phi} (p_1 \e^t)
 \cdots \bar{\phi} (p_{n} \e^t)}_{\bar{S}_t}\nn\\
&&\quad = 
\exp \left[ t \left( D - n
  \frac{D+2}{2} \right) + \frac{n}{2} \int^t dt'\, \eta_{t'}\right] \cdot
(\textrm{$t$-independent})\,.
\end{eqnarray}
\end{subequations}

\newpage

\section{Symmetry of the effective average action\label{appgamma}}

The purpose of this appendix is to rewrite the WT identity and quantum
master equation of the Wilson action as those of the corresponding
effective average action.  We have briefly discussed this in
\S\ref{BV-ST}, where a more functional method is used than the method
here.  Our main tool is the relation between a composite operator and
its 1PI sibling, given by \bref{comp-OO1PI} \& \bref{comp-Phiphi}.

We first consider a generic WT composite operator
\begin{equation}
\Sigma_\Lambda [\phi] \equiv \int_p \K{p} \left[
\Op_\Lambda [\phi] (p) \frac{\delta \SL}{\delta \phi (p)}
+ \frac{\delta}{\delta \phi (p)} \Op_\Lambda [\phi] (p) \right]\,,
\end{equation}
where $\Op_\Lambda [\phi] (p)$ is a composite operator of momentum $p$
giving an infinitesimal transformation of $\phi (p)$.  The
corresponding 1PI composite operator is defined by
\begin{equation}
\Sigma^{1PI}_\Lambda [\Phi] \equiv \Sigma_\Lambda [\phi]\,,
\end{equation}
where
\begin{equation}
\Phi (p) \equiv \frac{1}{\K{p}} \phi (p) + \frac{1 - \K{p}}{p^2 + m^2}
\frac{\delta \SL}{\delta \phi (-p)}\,.
\end{equation}
(In this appendix, we consider only the continuum limit $\Lambda_0 \to
\infty$.)  Since $\Phi$ and $\phi$ coincide asymptotically,
$\Sigma_\Lambda [\phi]$ and $\Sigma_\Lambda^{1PI} [\Phi]$ share the
same asymptotic behavior.

In order to express $\Sigma_\Lambda [\phi]$ in terms of $\Phi$, we
consider the two composite operators $\Op_\Lambda [\phi]$ and $K(p)
\frac{\delta \SL}{\delta \phi (p)}$ one by one.
\begin{itemize}
\item[(i)] $\Op_\Lambda [\phi]$ --- The corresponding 1PI composite
    operator is given by
\begin{equation}
\Op^{1PI}_\Lambda [\Phi] = \Op_\Lambda [\phi]\,.
\end{equation}
\item[(ii)] $K(p) \frac{\delta \SL}{\delta \phi (p)}$ --- Writing
    $\Gamma_\Lambda$ for $\Gamma_{B,\Lambda}$ for short,
    (\ref{derivation-Phiphi}) and (\ref{derivation-phiPhi}) give
\begin{equation}
\Phi (p) = - \frac{1}{K} \frac{1 - K}{p^2 + m^2} \frac{\delta
  \Gamma_\Lambda [\Phi]}{\delta \Phi (-p)} + \frac{1 - K}{p^2 + m^2}
\frac{\delta \SL [\phi]}{\delta \phi (-p)}\,.
\end{equation}
Hence, 
\begin{equation}
\K{p} \frac{\delta \SL}{\delta \phi (-p)}
= - (p^2 + m^2) \Phi (p) + \frac{\delta \Gamma_{I,\Lambda}
  [\Phi]}{\delta \Phi (-p)}\,,
\end{equation}
where the interaction part $\Gamma_{I,\Lambda} [\Phi]$ is defined so
that
\begin{equation}
\Gamma_\Lambda [\Phi] = - \frac{1}{2} \int_p \frac{p^2+m^2}{1 - \K{p}}
\Phi (-p) \Phi (p) + \Gamma_{I, \Lambda} [\Phi]\,.
\end{equation}
\end{itemize}
Thus, we obtain
\begin{eqnarray}
\Sigma^{1PI}_\Lambda  [\Phi] &=& \int_p \left[ \lb - (p^2 + m^2) \Phi
    (-p) + \frac{\delta \Gamma_{I,\Lambda} [\Phi]}{\delta \Phi (p)} \rb
      \Op^{1PI}_\Lambda [\Phi] (p) \right.\nn\\
&&\left.\quad + \K{p} \int_q \frac{\delta \Phi (q)}{\delta \phi (p)}
\frac{\delta \Op^{1PI}_\Lambda [\Phi] (p)}{\delta \Phi (q)} \right]\nn\\
&=& \int_p \left[ \lb - (p^2 + m^2) \Phi
    (-p) + \frac{\delta \Gamma_{I,\Lambda} [\Phi]}{\delta \Phi (p)} \rb
      \Op^{1PI}_\Lambda [\Phi] (p) \right.\nn\\
&&\left.\quad - \int_q \left(\Gamma_\Lambda^{(2)}\right)^{-1}  (q,-p)
R_\Lambda (p) \frac{\delta \Op^{1PI}_\Lambda [\Phi] (p)}{\delta \Phi
(q)} \right] 
\,,
\label{appgamma-Sigma}
\end{eqnarray}
where we have used (\ref{derivation-Gamma2inverse})
\begin{equation}
\frac{\delta \Phi (q)}{\delta \phi (p)} = - \left( \Gamma_\Lambda^{(2)}
\right)^{-1} (q,-p) \frac{p^2 + m^2}{1 - \K{p}}\,,
\end{equation}
and
\begin{equation}
R_\Lambda (p) \equiv \left( p^2 + m^2\right) \frac{\K{p}}{1 - \K{p}}\,.
\end{equation}

For example, the WT identity for QED
\begin{eqnarray}
k_\mu \frac{\delta \SIL}{\delta A_\mu (k)} &=& e \int_p \K{p}
\left( - \SL \Rd{\psi (p)} [\psi] (p-k) + \Tr [\psi] (p-k) \Rd{\psi
      (p)} \right.\nn\\
&& \left.+ [\bar{\psi}] (-p-k) \Ld{\bar{\psi} (-p)} \SL
- \Tr \Ld{\bar{\psi} (-p)} [\bar{\psi}] (-p-k) \right)
\end{eqnarray}
is rewritten as
\begin{eqnarray}
k_\mu \frac{\delta \Gamma_{I,\Lambda}}{\delta \mathcal{A}_\mu (k)}
&=& e \int_p \Bigg[ \, \bar{\Psi} (-p-k) \fmslash{k} \Psi (p)\nn\\
&&\quad - \Gamma_{I,\Lambda} \Rd{\Psi (p+k)} \Psi (p) + \bar{\Psi} (-p-k)
\Ld{\Psi (-p)} \Gamma_{I,\Lambda} \\
&&+ \K{(p+k)} \Tr \Psi (p) \Rd{\psi (p+k)} 
- \K{p} \Tr \Ld{\bar{\psi} (-p)} \bar{\Psi} (-p-k) \Bigg]\,,\nn
\end{eqnarray}
where 
\begin{equation}
\lb\begin{array}{c@{~=~}l}
\mathcal{A}_\mu (k) & [A_\mu]_\Lambda (k)\,,\\
\Psi (p) & [\psi]_\Lambda (p)\,,\\
\bar{\Psi} (-p) & [\bar{\psi}]_\Lambda (-p)\,.
\end{array}\right.
\end{equation}

We next consider the quantum master operator
\begin{equation}
\bar{\Sigma}_\Lambda [\phi,\phi^*] \equiv \int_p \K{p} \left[ \frac{\delta
      \bSL}{\delta \phi (p)} \Ld{\phi^* (-p)} \bSL + \Ld{\phi^* (-p)}
    \frac{\delta \bSL}{\delta \phi (p)}  \right]\,.
\end{equation}
The antifield $\phi^*$ is a classical external field, and it does not
affect the Legendre transformation between $\bSL [\phi,\phi^*]$ and
$\bar{\Gamma}_\Lambda [\Phi,\phi^*]$.  The Legendre transformation is given
by
\begin{equation}
\bar{\Gamma}_\Lambda [\Phi, \phi^*] = \bSL [\phi, \phi^*] + \int_p \frac{p^2
  + m^2}{1 - K} \left( \frac{1}{2 K} \phi (p) \phi (-p) - \Phi (p)
    \phi (-p) \right)\,,
\end{equation}
where
\begin{equation}
\Phi (p) = \frac{1}{K} \phi (p) + \frac{1 - K}{p^2 + m^2} \frac{\delta
  \bSL [\phi, \phi^*]}{\delta \phi (-p)}\,.
\end{equation}
From the general relation (\ref{derivation-deltaGammadeltaW}), we
obtain
\begin{equation}
\Ld{\phi^* (-p)} \bSL [\phi, \phi^*] = \Ld{\phi^* (-p)} \bar{\Gamma}_\Lambda
[\Phi, \phi^*]\,.
\end{equation}
Hence, the 1PI quantum master operator is given by
\begin{eqnarray}
\bar{\Sigma}_\Lambda^{1PI} [\Phi, \phi^*] &=& \int_p \left[
\lb - (p^2 + m^2) \Phi (-p) + \frac{\delta \bar{\Gamma}_{I,\Lambda} [\Phi,
  \phi^*]}{\delta \Phi (p)} \rb \Ld{\phi^* (-p)} \bar{\Gamma}_\Lambda [\Phi,
\phi^*] \right.\nn\\
&&\quad \left.-  \int_q
	 \left(\bar{\Gamma}_\Lambda^{(2)}\right)^{-1} (q,-p)
R_\Lambda (p) \Ld{\phi^* (-p)} \frac{\delta \bar{\Gamma}_\Lambda [\Phi,
\phi^*]}{\delta \Phi (q)}\right]\,.
\label{appgamma-Sigmabar}
\end{eqnarray}
Since
\begin{equation}
\bar{\Sigma}_\Lambda [\phi, \phi^*] = \bar{\Sigma}_\Lambda^{1PI} [\Phi, \phi^*]
\end{equation}
by construction, $\bar{\Sigma}_\Lambda$ and $\bar{\Sigma}_\Lambda^{1PI}$
share the same asymptotic behavior.  For YM theories,
\begin{equation}
\bar{\Sigma}_\Lambda^{1PI} [\Phi, \phi^*] = 0
\end{equation}
is called the \textbf{modified Slavnov-Taylor (ST)
identity}, first obtained by Ellwanger \cite{Ellwanger:1994iz}.

To summarize, for the effective average action, the WT identity
is given by
\begin{equation}
\Sigma^{1PI}_\Lambda [\Phi] = 0\,,\label{appgamma-modifiedWT}
\end{equation}
where $\Sigma^{1PI}_\Lambda [\Phi]$ is given by (\ref{appgamma-Sigma}), and
the quantum master equation is given by the modified ST identity
\begin{equation}
\bar{\Sigma}^{1PI}_\Lambda [\Phi, \phi^*] = 0\,,\label{appgamma-modifiedST}
\end{equation}
where $\bar{\Sigma}^{1PI} [\Phi, \phi^*]$ is given by (\ref{appgamma-Sigmabar}).

In the limit $\Lambda \to 0+$, the effective average action becomes the
effective action (the generating functional of the 1PI correlation
functions):
\begin{equation}
\lb\begin{array}{c@{~=~}l}
\lim_{\Lambda \to 0+} \Gamma_\Lambda [\Phi] & \Gamma [\Phi]\,,\\
\lim_{\Lambda \to 0+} \bar{\Gamma}_\Lambda [\Phi, \phi^*] & \bar{\Gamma} [\Phi,
 \phi^*]\,.
\end{array}\right.
\end{equation}
Since $\K{p}$ vanishes in this limit, the identities
\bref{appgamma-modifiedWT} \& \bref{appgamma-modifiedST} reduce to
\begin{equation}
\lb\begin{array}{l}
\int_p \frac{\delta \Gamma [\Phi]}{\delta \Phi (p)}
 \Op^{1PI}_{\Lambda=0} [\Phi] (p) = 0\,,\\
\int_p \frac{\delta \bar{\Gamma} [\Phi, \phi^*]}{\delta \Phi (p)} 
\Ld{\phi^* (-p)} \bar{\Gamma} [\Phi, \phi^*] = 0\,.
\end{array}\right.
\end{equation}
The latter is the Zinn-Justin equation,\cite{ZinnJustin:1993wc} familiar
from YM theories.

\newpage

\section{O(N) linear sigma model\label{AFexamples-ON}}

In this subsection we apply the antifield formalism to the O(N) linear
sigma model.  The model is a typical example of theories with their
symmetry realized linearly.  In such cases, it should not be surprising
that the WT identity $\Sigma_\Lambda = 0$ or the QME
$\bar{\Sigma}_\Lambda = 0$ for the Wilson action boils down to the
invariance of the action under the na\"ive linear
transformation.\footnote{In preparing this appendix, we have benefited
from the discussions with Drs.~K.~\"Ulker and L.~Akant.}  Though we do
not discuss it here, the Wess-Zumino model with auxiliary fields has its
supersymmetry linearly realized, and the supersymmetry of the Wilson
action is invariant under the na\"ive linear supersymmetry
transformation.\cite{Bonini:1998ec,Pernici:1998ex,Rosten:2008ih}

We consider a theory of $N$ real scalar fields $\phi_i$ invariant
under the following linear O(N) transformation:
\begin{equation}
\delta \phi_i (p) = \xi^a T^a_{ij} \phi_j (p)\,, \quad (i=1,\cdots,N)
\end{equation}
where $\xi^a\,(a=1,\cdots,N(N-1)/2)$ is an infinitesimal constant.
The $N$-by-$N$ matrices $T^a$ are real and antisymmetric, and they
satisfy the commutation relation
\begin{equation}
\left[ T^a, T^b \right] = - f^{abc} T^c\,,
\end{equation}
where the structure constants are real and completely antisymmetric, and
satisfy the Jacobi identity:
\begin{equation}
f^{abd} f^{dce} + f^{bcd} f^{dae} + f^{cad} f^{dbe} = 0\,.
\end{equation}

The Wilson action is given as
\begin{equation}
\SL = \SFL + \SIL\,,
\end{equation}
where the free part is
\begin{equation}
\SFL \equiv \frac{1}{2} \int_p \phi_i (p) \phi_i (-p) \frac{p^2 +
  m^2}{\K{p}}\,.
\end{equation}
Using the composite operator corresponding to the elementary field
\begin{equation}
[\phi_i]_\Lambda (p) \equiv \phi_i (p) + \frac{1-\K{p}}{p^2 + m^2} \frac{\delta
  \SIL}{\delta \phi_i (-p)}\,,
\end{equation}
the WT composite operator is given by
\begin{equation}
\Sigma_\Lambda [\phi] \equiv \int_p \K{p} \left[ \frac{\delta \SL}{\delta
      \phi_i (p)} \xi^a T^a_{ij} 
    [\phi_j]_\Lambda (p) + \frac{\delta}{\delta \phi_i (p)} \left( \xi^a
        T^a_{ij} [\phi_j]_\Lambda (p)\right) \right] \,.
\end{equation}
The WT identity 
\begin{equation}
\Sigma_\Lambda = 0 \label{AFexamples-ON-WT}
\end{equation}
is equivalent to the cutoff independent identities
\begin{equation}
\sum_{j=1}^n \vev{\phi_{i_1} (p_1) \cdots 
\xi^a T^a_{i_j k} \phi_k (p_j) \cdots \phi_{i_n} (p_n)}^\infty
= 0
\end{equation}
for any $n$.  

We now wish to show that the WT identity (\ref{AFexamples-ON-WT}) is
equivalent to the na\"ive identity:
\begin{equation}
    \int_p \frac{\delta \SIL}{\delta \phi_i (p)} \xi^a T^a_{ij} \phi_j (p)
    = 0\,.\label{AFexamples-ON-naive}
\end{equation}
This is done in two steps.
First we show the vanishing of the jacobian:
\begin{eqnarray}
&&\int_p \K{p} \frac{\delta}{\delta \phi_i (p)} \left( \xi^a T^a_{ij}
    [\phi_j]_\Lambda (p) \right) \nn\\
&& = \int_p \K{p} \xi^a T^a_{ij} \left( \delta_{ij} (2\pi)^4 \delta
    (0) + \frac{1-\K{p}}{p^2+m^2} \frac{\delta^2 \SIL}{\delta \phi_i
      (p) \delta \phi_j  (-p)} \right)\nn\\
&& = 0\,.
\end{eqnarray}
This vanishes because of the antisymmetry of $T^a$.  Second we
consider
\begin{eqnarray}
&&\int_p \K{p} \frac{\delta \SL}{\delta \phi_i (p)} \xi^a T^a_{ij}
[\phi_j]_\Lambda (p)\nn\\
&& = \int_p \K{p} \left[ - (p^2+m^2) \frac{1}{\K{p}} \phi_i (-p) +
    \frac{\delta \SIL}{\delta \phi_i (p)} \right]\nn\\
&&\qquad \times\, \xi^a T^a_{ij} \left[ \phi_j (p) + \frac{1 - \K{p}}{p^2
      + m^2} \frac{\delta \SIL}{\delta \phi_j (-p)} \right]\nn\\
&& = \int_p \K{p} \left[ - \frac{1-\K{p}}{\K{p}} \phi_i (-p)
    \frac{\delta \SIL}{\delta \phi_j (-p)} + \frac{\delta \SIL}{\delta
      \phi_i (p)} \phi_j (p) \right] \xi^a T^a_{ij}\nn\\
&& = - \int_p \phi_i (-p) \frac{\delta \SIL}{\delta \phi_j (-p)} \xi^a
T^a_{ij} \,,
\end{eqnarray}
where we have used the antisymmetry of $T^a$ again.  Hence, the WT
identity is equivalent to (\ref{AFexamples-ON-naive}).

We now promote $\xi^a$ to anticommuting ghost fields, constant in space.
We then introduce the fermionic antifields $\phi_i^* (p)$ for $\phi_i
(p)$ and bosonic constant antifields $\xi^{* a}$ for $\xi^a$.  We define
the action by
\begin{eqnarray}
\bSL &\equiv& \SFL + \int_p \phi_i^* (-p) \xi^a T^a_{ij} \phi_j (p)
 + \frac{1}{2} \int_p \phi_i^* (p) \xi^a T^a_{ij} \phi_k^* \xi^b
T^b_{kj} \frac{1 - \K{p}}{p^2 + m^2} \nn\\
&& \, + \SIL \left[ \phi^{sh}_i (p) \right]
 - \frac{1}{2} \xi^{a *} f^{abc} \xi^b \xi^c\,,
\end{eqnarray}
where the interaction action is the same as before, except that the
fields are shifted by antifields:
\begin{equation}
\phi^{sh}_i (p) \equiv \phi_i (p) + \phi_j^* (p) \xi^a
    T^a_{ji} \frac{1-\K{p}}{p^2 + m^2} \,.
\end{equation}
$\bSL$ is constructed so that it satisfies the same Polchinski
differential equation as $\SL$.  The last $\Lambda$-independent term
generates the O(N) transformation of the ghosts:
\begin{equation}
\delta \xi^a = \frac{\partial \bSL}{\partial \xi^{a*}} = - \frac{1}{2}
f^{abc} \xi^b \xi^c\,.
\end{equation}

The QM operator is defined by
\begin{eqnarray}
\bar{\Sigma}_\Lambda [\phi,\phi^*; \xi, \xi^*]
&\equiv& \int_p \K{p} \lb
     \frac{\delta \bSL}{\delta 
      \phi_i (p)} \cdot \Ld{\phi_i^* (-p)} \bSL +
\frac{\delta}{\delta
      \phi_i (p)} \Ld{\phi_i^* (-p)} \bSL \rb\nn\\
&&\quad + \frac{\partial \bSL}{\partial \xi^{c*}}
    \frac{\overrightarrow{\partial}}{\partial \xi^c} \bSL\,.
\end{eqnarray}
We wish to verify that the QME
\begin{equation}
\bar{\Sigma}_\Lambda  [\phi,\phi^*; \xi, \xi^*] = 0
\label{AFexamples-ON-qme}
\end{equation}
is equivalent to the WT identity \bref{AFexamples-ON-WT}, and hence
equivalent to the na\"ive identity \bref{AFexamples-ON-naive}.
Verification is straightforward but takes several steps.
\begin{enumerate}
\item We first compute
\begin{eqnarray}
\Ld{\phi_i^* (-p)} \bSL &=& 
\xi^a T^a_{ij} \Bigg[ \phi_j (p) + \phi_k^* (p) \xi^b T^b_{kj}
\frac{1 - \K{p}}{p^2 + m^2} \nn\\
&&\, + \frac{1-\K{p}}{p^2 + m^2} \frac{\delta \SIL [\phi^{sh}]}{\delta
  \phi^{sh}_j (-p)}\Bigg]\nn\\
&=&\xi^a T^a_{ij} [\phi_j]_\Lambda (p)\Big|_{\phi \to \phi^{sh}}\,.
\end{eqnarray}
Hence,
\begin{equation}
\Ld{\phi_i^* (-p)} \frac{\delta \bSL}{\delta \phi_i (p)} = \xi^a
T^a_{ij} \frac{1-\K{p}}{p^2 + m^2} \frac{\delta^2 \SIL
  [\phi^{sh}]}{\delta \phi_j^{sh} (-p) \delta \phi^{sh}_i (p)}\,,
\end{equation}
and therefore
\begin{equation}
\int_p \K{p} \Ld{\phi_i^* (-p)} \frac{\delta \bSL}{\delta \phi_i (p)}
= 0\label{AFexamples-ON-zerojacobian}
\end{equation}
by the antisymmetry of $T^a$.
\item We then compute
\begin{eqnarray}
\frac{\delta \bSL}{\delta \phi_i (p)} &=& - \frac{1}{\K{p}} (p^2 + m^2)
\phi_i (-p) + \phi_j^* (-p) \xi^a T^a_{ji} + \frac{\delta \SIL
  [\phi^{sh}]}{\delta \phi^{sh}_i (p)}\nn\\
&=& - \frac{1}{\K{p}} (p^2 + m^2) \left( \phi_i (-p) + \phi_j^* (-p)
    \xi^a T^a_{ji} \frac{1-\K{p}}{p^2 + m^2} \right)\nn\\
&&\, + \frac{\delta \SIL [\phi^{sh}]}{\delta \phi^{sh}_i (p)}
 + \frac{1}{\K{p}} \phi_j^* (-p) \xi^a T^a_{ji}\nn\\
&=& \frac{\delta \SL [\phi^{sh}]}{\delta \phi^{sh}_i (p)}
 + \frac{1}{\K{p}} \phi_j^* (-p) \xi^a T^a_{ji}\,.
\end{eqnarray}
\item We therefore obtain
\begin{eqnarray}
&&\int_p \K{p} \frac{\delta \bSL}{\delta \phi_i (p)} \cdot
\Ld{\phi_i^* (-p)} \bSL \nn\\
&&=\int_p \K{p} \left[ \frac{\delta \SL[\phi^{sh}]}{\delta \phi^{sh}_i (p)} + 
\frac{1}{\K{p}} \phi_j^* (-p) \xi^a T^a_{ji} \right] 
\xi^b T^b_{ik} [\phi_k]_\Lambda (p) \Big|_{\phi \to \phi^{sh}}\nn\\
&&= \Sigma_\Lambda \Big|_{\phi \to \phi^{sh}}
+ \int_p \phi_j^* (-p) \xi^a T^{a}_{ji} \xi^b T^b_{jk}
\left[\phi_k\right]_\Lambda (p) \Big|_{\phi \to \phi^{sh}}\,.
\end{eqnarray}
Hence,
\begin{equation}
\bar{\Sigma}_\Lambda = \Sigma_\Lambda \Big|_{\phi \to \phi^{sh}} 
+ \int_p \phi_j^* (-p) \xi^a T^{a}_{ji} \xi^b T^b_{jk}
\left[\phi_k\right]_\Lambda (p) \Big|_{\phi \to \phi^{sh}}
+ \frac{\partial \bSL}{\partial \xi^{c*}}
\frac{\overrightarrow{\partial}}{\partial \xi^c} \bSL\,.
\end{equation}
\item Using
\begin{equation}
\xi^a T^{a}_{ji} \xi^b T^b_{jk} = - \frac{1}{2} f^{abc} \xi^a \xi^b
 T^c_{jk} = \frac{\partial \bSL}{\partial \xi^{c*}} T^c_{jk}\,,
\end{equation}
we obtain
\begin{eqnarray}
&&\int_p \phi_j^* (-p) \xi^a T^a_{ji} \xi^b T^b_{ik}
[\phi_k]_\Lambda (p)\Big|_{\phi \to  \phi^{sh}}\nn\\
&&= \frac{\partial \bSL}{\partial \xi^{c*}} \int_p \phi_i^* (-p) T^c_{ij}
\Bigg[ \phi_j (p) + \phi_k^* (p) \xi^d T^d_{kj} \frac{1-\K{p}}{p^2 +
      m^2} \nn\\
&&\qquad \qquad \qquad + \frac{1-\K{p}}{p^2 + m^2} \frac{\delta \SIL
  [\phi^{sh}]}{\delta \phi^{sh}_j (-p)} \Bigg]\,.
\end{eqnarray}
\item We compute
\begin{eqnarray}
\frac{\overrightarrow{\partial}}{\delta \xi^c} \bSL &=& - \int_p
 \phi_i^* (-p) T^c_{ij} 
\Bigg[ \phi_j (p) + \phi_k^* (p) \xi^d T^d_{kj} \frac{1-\K{p}}{p^2 +
      m^2} \nn\\
&&\qquad + \frac{1-\K{p}}{p^2 + m^2} \frac{\delta \SIL
  [\phi^{sh}]}{\delta \phi^{sh}_j (-p)} \Bigg] + f^{abc}
    \xi^{a*} \xi^b\,.
\end{eqnarray}
Since
\begin{equation}
\frac{\partial \SL}{\partial \xi^{c*}} f^{abc} \xi^{a*} \xi^b
= - \frac{1}{2} f^{cde} \xi^d \xi^e f^{abc} \xi^{a*} \xi^b
= - \frac{1}{2} f^{cde} f^{abc} \xi^d \xi^e \xi^b \xi^{a*}
= 0
\end{equation}
by the Jacobi identity, we obtain
\begin{equation}
\int_p \phi_j^* (-p) \xi^a T^{a}_{ji} \xi^b T^b_{jk}
\left[\phi_k\right]_\Lambda (p) \Big|_{\phi \to \phi^{sh}}
+ \frac{\partial \bSL}{\partial \xi^{c*}}
\frac{\overrightarrow{\partial}}{\partial \xi^c} \bSL
    = 0\,.
\end{equation}
\item  We finally obtain the desired relation
\begin{equation}
\bar{\Sigma}_\Lambda [\phi,\phi^*; \xi, \xi^*]
= \Sigma_\Lambda [\phi] \Big|_{\phi \to \phi^{sh}}\,.
\end{equation}
\end{enumerate}
The last relation is already familiar from QED, for which we have
 derived \bref{AFexamples-QED-SigmabarSigma}, even though the abelian
 gauge symmetry of QED is not strictly linear in the sense used here.

To conclude, we have found that for the O(N) global symmetry, the
following three equations are equivalent:
\begin{subequations}
\begin{eqnarray}
\delta S_\Lambda &\equiv& \int_p \frac{\delta \SL}{\delta \phi_i (p)}
 \xi^a T^a_{ij} \phi_j (p) = 0\,,\\
\Sigma_\Lambda [\phi] &\equiv& \int_p \K{p} \frac{\delta \SL}{\delta
 \phi_i (p)} \xi^a T^a_{ij} [\phi_j]_\Lambda (p) = 0\,,\\
\bar{\Sigma}_\Lambda [\phi,\phi^*; \xi, \xi^*] &\equiv& \int_p \K{p}
 \frac{\delta \bSL}{\delta \phi_i (p)}
\Ld{\phi_i^* (-p)} \bSL  + \frac{\partial \bSL}{\partial \xi^{c*}}
\frac{\overrightarrow{\partial}}{\partial \xi^c} \bSL = 0\,.
\end{eqnarray}
\end{subequations}
The jacobian of the linear transformation is $1$:
\begin{equation}
\int_p \K{p} \xi^a T^a_{ij} \frac{\delta [\phi_j]_\Lambda (p)}{\delta \phi_i
  (p)} = \int_p \K{p} \frac{\delta}{\delta \phi_i (p)}
\Ld{\phi_i^* (-p)} \bSL = 0\,.
\end{equation}

\newpage

\section{Quick summary\label{summary}}

\subsection*{\textbf{ERG}}

\begin{enumerate}
\item A cutoff function $K (p)$ and its log derivative $\Delta (p)
    \equiv - 2 p^2 \frac{d}{dp^2} K(p)$:
\begin{equation}
K(0) = 1,\quad K (p) \stackrel{p^2 \to \infty}{\longrightarrow}
0\,.
\end{equation}
We usually (not always) choose
\begin{equation}
K(p) = 1 \quad (p^2 < 1)\,.
\end{equation}
\item The interaction part $\SIL$ of the Wilson action
satisfies the Polchinski equation
\begin{equation}
- \Lambda \frac{\partial}{\partial \Lambda} S_{I,\Lambda} = \frac{1}{2} \int_p
\frac{\Delta (p/\Lambda)}{p^2 + m^2} \lb \frac{\delta S_{I,\Lambda}}{\delta \phi
  (p)} \frac{\delta S_{I,\Lambda}}{\delta \phi (-p)} + \frac{\delta^2
  S_{I,\Lambda}}{\delta \phi (p) \delta \phi (-p)} \rb\,.
\end{equation}
\item The full Wilson action
\begin{equation}
\SL \equiv - \frac{1}{2} \int_p
  \frac{p^2 + m^2}{\K{p}} \phi (p) \phi (-p) + \SIL
\end{equation}
satisfies the ERG differential equation
\begin{eqnarray}
- \Lambda \frac{\partial}{\partial \Lambda} S_\Lambda &=&
\int_p \frac{\Delta (p/\Lambda)}{p^2 + m^2} \left[
\frac{p^2+m^2}{\K{p}} \phi (p) \frac{\delta S_\Lambda}{\delta \phi
  (p)}\right. \nn\\
&&\quad \left.+
\frac{1}{2} \lb \frac{\delta S_\Lambda}{\delta \phi (p)} \frac{\delta
  S_\Lambda}{\delta \phi (-p)} + \frac{\delta^2 S_\Lambda}{\delta \phi
  (p) \delta \phi (-p)} \rb \right]\,.
\end{eqnarray}
\item Under the initial condition $S_{I, \Lambda_0} = S_{I,B}$, the
    integral solution of the Polchinski equation is given by 
\begin{eqnarray}
&&\exp \left[ S_{I,\Lambda} [\phi] \right] 
= \int [d\phi'] \exp \Big[\\
&&\quad \left. -
    \frac{1}{2} \int_p \frac{p^2+m^2}{\Kz{p}-\K{p}} \phi' (-p) \phi'
    (p) + S_{I,B} [\phi+\phi'] \right]\,.\nn
\end{eqnarray}
\item For renormalizable theories, $S_{I,\Lambda}$ can be determined by
    the Polchinski equation and its asymptotic behavior:
\begin{equation}
S_{I,\Lambda} \asym S_{I,\Lambda}^{asymp}\,.
\end{equation}
For the $\phi^4$ theory in $D=4$, we choose
\begin{eqnarray*}
S_{I,\Lambda}^{asymp} &=& \int d^4 x\, \left[
\left( \Lambda^2 a_2 (\ln \Lambda/\mu) + m^2 b_2 (\ln \Lambda/\mu)
\right) \frac{1}{2} \phi^2 \right.\\
&&\qquad \left.+ c_2 (\ln \Lambda/\mu)
\frac{1}{2} \partial_\mu \phi \partial_\mu \phi + a_4 (\ln
\Lambda/\mu) \frac{1}{4!} \phi^4 \right]\,,
\end{eqnarray*}
where the coefficients $a_2, b_2, c_2, a_4$ are determined by the
      Polchinski equation up to
\begin{equation}
b_2 (0),\quad c_2 (0),\quad a_4 (0)\,,
\end{equation}
which constitute the renormalized parameters of the theory.  A
      particularly simple choice is
\begin{equation}
b_2 (0) = c_2 (0) = 0,\quad a_4 (0) = - \lambda\,.
\end{equation}
\item No loss of information along the ERG flow: the correlation functions
\begin{equation}
\lb \begin{array}{c@{~\equiv~}l}
\vev{\phi (p) \phi (-p)}^\infty & \frac{1}{\K{p}^2} \vev{\phi (p) \phi
  (-p)}_{S_\Lambda} + \frac{1 - 1/\K{p}}{p^2 + m^2}\,,\\
\vev{\phi (p_1) \cdots \phi (p_n)}^\infty & \prod_{i=1}^n
\frac{1}{\K{p_i}} \cdot \vev{\phi (p_1) \cdots \phi (p_n)}_{S_\Lambda}\,,(n>2)
\end{array}\right.
\end{equation}
are independent of $\Lambda$.
\item $\Op_\Lambda$ is a composite operator if $S_\Lambda + \ep
    \Op_\Lambda$ satisfies the ERG differential equation to first
    order in $\ep$, or equivalently if $\Op_\Lambda$ satisfies the
    differential equation
\begin{eqnarray}
- \Lambda \frac{\partial}{\partial \Lambda} \Op &=& \int_p
\frac{\Delta (p/\Lambda)}{p^2 + m^2} \left[ \lb\frac{p^2 + m^2}{\K{p}}
    \phi (p) + \frac{\delta S_\Lambda}{\delta
      \phi (-p)} \rb \frac{\delta}{\delta \phi (p)}\right.\nn\\
&&\qquad\qquad\qquad\left. + \frac{1}{2}
    \frac{\delta^2}{\delta \phi (p) \delta \phi (-p)} \right] \Op_\Lambda\nn\\
&=& \int_p \frac{\Delta (p/\Lambda)}{p^2 + m^2} \left[
\frac{\delta S_{I,\Lambda}}{\delta
      \phi (-p)} \frac{\delta}{\delta \phi (p)} + \frac{1}{2}
    \frac{\delta^2}{\delta \phi (p) \delta \phi (-p)} \right] \Op_\Lambda\,.
\end{eqnarray}
Given an initial condition $\Op_{\Lambda_0} = \Op_B$, the integral
      formula for $\Op_\Lambda$ is
\begin{eqnarray}
&&\Op_\Lambda [\phi] \exp \left[ S_{I,\Lambda} [\phi] \right]
= \int [d\phi'] \Op_B [\phi+\phi'] \\
&&\quad \times \exp \left[ - \frac{1}{2} \int_p
    \frac{p^2 + m^2}{\Kz{p} - \K{p}} \phi' (-p) \phi' (p) + S_{I,B}
    [\phi+\phi'] \right]\,.\nn
\end{eqnarray}
\item The $\Lambda$ independent correlation functions of a composite operator:
\begin{equation}
\vev{\Op_\Lambda \,\phi (p_1) \cdots \phi (p_n)}^\infty = \prod_{i=1}^n
\frac{1}{\K{p_i}} \cdot \vev{\Op \,\phi (p_1) \cdots \phi
(p_n)}_{S_\Lambda}\,. 
\end{equation}
\item Examples of composite operators
\begin{enumerate}
\item $\int_p \K{p} \frac{\delta S_\Lambda}{\delta \phi (-p)}$ (Equation
      of motion)
\begin{equation}
\int_p \K{p} \vev{\frac{\delta S_\Lambda}{\delta \phi (-p)} \phi (p_1) \cdots
  \phi (p_n)}^\infty = - \sum_{i=1}^n \vev{ \cdots \phi (p_i+p)
  \cdots}^\infty\,.
\end{equation}
\item $\frac{\delta S_\Lambda}{\delta J(-p)}$, where $J$ is an external
    source. 
\begin{equation}
\vev{\frac{\delta S_\Lambda}{\delta J(-p)} \phi (p_1) \cdots \phi
  (p_n)}^\infty = \frac{\delta}{\delta J(-p)} \vev{\phi (p_1) \cdots
  \phi (p_n)}^\infty\,.
\end{equation}
\item $[\phi]_\Lambda (p) \equiv \frac{1}{\K{p}} \phi (p) + \frac{1 -
      \K{p}}{p^2 + m^2} \frac{\delta S_\Lambda}{\delta \phi (-p)}$ has the
    asymptotic behavior
\begin{equation}
[\phi]_\Lambda (p) \stackrel{\Lambda \to \infty}{\longrightarrow} \phi (p)\,.
\end{equation}
\item \label{typeA} If $\Op_\Lambda$ is a composite operator, so is
\begin{eqnarray}
A (p) &\equiv& \K{p} \left( \Op_\Lambda \frac{\delta S_\Lambda}{\delta
      \phi (-p)} + \frac{\delta \Op_\Lambda}{\delta \phi (-p)} \right)\nn\\
&=& \exp [- S_\Lambda] \K{p} 
\frac{\delta}{\delta \phi (-p)} \left( \Op_\Lambda\, \exp [S_\Lambda]\right)\,.
\end{eqnarray}
\begin{equation}
\vev{A (p) \phi (p_1) \cdots \phi (p_n)}^\infty =  - \sum_{i=1}^n (2\pi)^D
\delta^{(D)} (p_i + p) \vev{\cdots \widehat{\phi (p_i)} \Op_\Lambda
  \cdots}^\infty\,.
\end{equation}
\item \label{typeB} If $\Op_\Lambda$ is a composite operator, so is
\begin{equation}
B (p) \equiv \frac{\delta}{\delta J(-p)} \Op_\Lambda + \frac{\delta}{\delta
  J(-p)} S_\Lambda\cdot \Op = \exp [-S_\Lambda] \frac{\delta}{\delta
  J(-p)} \left( \Op\, \exp [S_\Lambda]\right)\,.
\end{equation}
\begin{equation}
\vev{B (p) \phi (p_1) \cdots \phi (p_n)}^\infty = \frac{\delta}{\delta
  J(-p)} \vev{\Op\, \phi (p_1) \cdots \phi (p_n)}^\infty\,.
\end{equation} 
\end{enumerate}
\item For the $\phi^4$ theory in $D=4$, the composite operator $[\phi^2/2]
(p)$ is determined by the differential equation and the asymptotic
      behavior
\begin{equation}
[\phi^2/2] (p) \stackrel{\Lambda \to \infty}{\longrightarrow} c(\ln
\Lambda/\mu) \frac{1}{2} \int_q \phi (q) \phi (p-q)\,,
\end{equation}
where $c(0)$ is an arbitrary normalization constant.
\end{enumerate}

\subsection*{\textbf{Ward-Takahashi (WT) identity}}

\begin{enumerate}
\item Under the symmetry transformation of the field $\phi (p)$
\begin{equation}
\delta \phi (p) = \K{p} \Op_\Lambda (p)\,,
\end{equation}
where $\Op_\Lambda (p)$ is a composite operator, the Wilson action
      changes by
\begin{eqnarray}
\Sigma_\Lambda &\equiv& \e^{-\SL} \int_p \K{p} \frac{\delta}{\delta \phi
 (p)} \left( \Op_\Lambda (p) \e^{\SL} \right)\nn\\
&=& \int_p \K{p} \left(
\frac{\delta \SL}{\delta \phi (p)} \Op_\Lambda (p)  
+ \frac{\delta \Op_\Lambda (p)}{\delta \phi (p)}\right)\,.
\end{eqnarray}
The WT composite operator $\Sigma_\Lambda$ is a composite operator of
      type (d).
\item The WT identity is given by
\begin{equation}
\Sigma_\Lambda = 0\,.
\end{equation}
Since $\Sigma_\Lambda$ is a composite operator, 
\begin{equation}
\lim_{\Lambda \to \infty} \Sigma_\Lambda = 0
\end{equation}
guarantees the WT identity for all $\Lambda$.
\item The WT identity is equivalent to
\begin{equation}
\sum_{i=1}^n \vev{\phi (p_1) \cdots \Op (p_i) \cdots \phi (p_n)}^\infty
 = 0\,.
\end{equation}
\end{enumerate}

\subsection*{\textbf{Quantum master equation}}

\begin{enumerate}
\item We generalize the Wilson action $\SL$ to $\bSL$ by introducing an
antifield $\phi^*$, conjugate to $\phi$.  $\phi^*$ has the opposite
statistics to $\phi$.  $\phi^*$ is a classical external source that
generates the  transformation of $\phi$.  
\item The quantum master operator
\begin{eqnarray*}
\bar{\Sigma}_\Lambda &\equiv& \e^{-\bSL} \int_p \K{p}
 \frac{\delta}{\delta \phi (p)} \Ld{\phi^* (-p)} \e^{\bSL}\\
&=& \int_p \K{p}  
\left[
\frac{\delta \bSL}{\delta \phi (p)} \Ld{\phi^* (-p)} \bSL +
\frac{\delta}{\delta \phi (p)} \Ld{\phi^* (-p)} \bSL
\right]
\end{eqnarray*}
is a composite operator of type (d), since $\Ld{\phi^* (-p)} \bSL$ is 
a composite operator of type (e).
\item The quantum master equation
\begin{equation}
\bar{\Sigma}_\Lambda = 0
\end{equation}
is satisfied, if
\begin{equation}
\lim_{\Lambda \to \infty} \bar{\Sigma}_\Lambda = 0\,.
\end{equation}
\item The quantum master equation is equivalent to the WT identity
\begin{equation}
\sum_{i=1}^n \Ld{\phi^* (-p_i)} \vev{ \phi (p_1) \cdots  \widehat{\phi
 (p_i)} \cdots \phi (p_n)}^\infty = 0\,.
\end{equation}
\item The Wilson action $\bSL + \ep \Op_\Lambda$, deformed by the
      composite operator $\Op_\Lambda$, satisfies the quantum master
      equation, if and only if
\begin{equation}
\delta_Q \Op_\Lambda = 0\,,
\end{equation}
where the BRST transformation is defined by
\begin{equation}
\delta_Q \equiv \int_p \K{p} \lb \Ld{\phi^* (-p)} \bSL \cdot \frac{\delta}
{\delta \phi (p)} + \frac{\delta \bSL}{\delta \phi (p)} \Ld{\phi^* (-p)}
+ \Ld{\phi^* (-p)} \frac{\delta}{\delta \phi (p)} \rb\,.
\end{equation}
\item For any composite operator $\Op$, $\delta_Q \Op$ is a composite
    operator if $\bar{\Sigma} = 0$. Then,
\begin{equation}
\vev{\phi (p_1) \cdots \phi (p_n) \delta_Q \Op}^\infty
= - \sum_{i=1}^n \Ld{\phi^* (-p_i)} \vev{ \phi (p_1) \cdots
  \widehat{\phi (p_i)} \cdots \phi (p_n) \Op}^\infty\,.
\end{equation}
For example,
\begin{eqnarray}
\N &\equiv& - \delta_Q \int_p \phi^* (-p) [\phi] (p) \\
&=& \int_p \left[ \phi^* (-p) \Ld{\phi^* (-p)} \bSL - \K{p} \lb
[\phi] (p) \frac{\delta \bSL}{\delta \phi (p)} + \frac{\delta [\phi]
  (p)}{\delta \phi (p)} \rb \right]\nn
\end{eqnarray}
satisfies
\begin{equation}
\vev{\N\, \phi (p_1) \cdots \phi (p_n)}^\infty
= \left( \int_p \phi^* (-p) \Ld{\phi^* (-p)} + n \right) \vev{\phi
  (p_1) \cdots \phi (p_n)}^\infty\,.
\end{equation}
\item The algebraic identity
\begin{equation}
\delta_Q \bar{\Sigma}_\Lambda = 0
\end{equation}
even if $\bar{\Sigma}_\Lambda \ne 0$.
\item The nilpotency $\delta_Q^2=0$ needs the QME $\bar{\Sigma}_\Lambda = 0$.
\end{enumerate}

\subsection*{\textbf{Classical nature of the quantum master equation}}

\begin{enumerate}
\item Loop expansions (bars omitted; $\bSL \to \SL$, etc.)
\begin{equation}
\SL = \sum_{l=0}^\infty S_l (\Lambda),\quad \Sigma_\Lambda = \sum_{l=0}^\infty
 \Sigma_l (\Lambda),\,\cdots
\end{equation}
Loop expansions of the asymptotic parts:
\begin{equation}
\SL^{asymp} = \sum_{l=0}^\infty S_l^{asymp} (\Lambda),\quad
\Sigma_\Lambda^{asymp} = \sum_{l=0}^\infty \Sigma_l^{asymp} (\Lambda),\, \cdots
\end{equation}
\item \textbf{induction hypothesis}: Given $S_{0,\cdots,l-1}$ that
    satisfy $\Sigma_{0,\cdots,l-1} = 0$, we wish to construct $S_l$ so
    that
\begin{equation}
\Sigma_l = 0 \,.
\end{equation}
This condition is equivalent to
\begin{equation}
\Sigma_l^{asymp} = 0\,.
\end{equation}
\item Tree level: we start from the classical action 
    $S_{cl}$ with the classical BRST invariance
\begin{equation}
\int_x \Ld{\phi^* (x)} S_{cl} \cdot \frac{\delta S_{cl}}{\delta \phi
  (x)} = 0\,.
\end{equation}
We then choose
\begin{equation}
S_0^{asymp} = S_{cl}
\end{equation}
so that
\begin{equation}
\Sigma_0^{asymp} = \Sigma_0 = 0\,.
\end{equation}
\item Under the induction hypothesis, the ERG differential equation
      $\Sigma_l$
\begin{equation}
- \Lambda \frac{\partial}{\partial \Lambda} \Sigma_l = \int_p
\frac{\Delta (p/\Lambda)}{p^2 + m^2} \frac{\delta S_0}{\delta \phi
  (-p)} \frac{\delta \Sigma_l}{\delta \phi (p)}
\end{equation}
implies that $\Sigma_l^{asymp}$ is independent of
      $\Lambda$. 
\item $S_l (\Lambda)$ satisfies the differential equation
\begin{equation}
- \Lambda \frac{\partial}{\partial \Lambda} S_l =
\frac{1}{2} \int_p \frac{\Delta (p/\Lambda)}{p^2 + m^2} \left[
\sum_{k=0}^l \frac{\delta S_k}{\delta \phi (-p)} \frac{\delta
  S_{l-k}}{\delta \phi (p)} + \frac{\delta^2 S_{l-1}}{\delta \phi (p)
  \delta \phi (-p)} \right]\,.
\end{equation}
We split $S_l$ into two parts:
\begin{equation}
S_l (\Lambda) = S_{1,l}  + S_{2,l} (\Lambda)
\end{equation}
\begin{enumerate}
\item $S_{1,l}$ is independent of $\Lambda$, not determined by the
      differential equation.
\item $S_{2,l} (\Lambda)$ is a particular solution of the differential
    equation for $S_l$.  
\end{enumerate}
We wish to fine tune $S_{1,l}$ so that $\Sigma_l = 0$.
\item We split $\Sigma_l$ into two parts:
\begin{equation}
\Sigma_l = \Sigma_{1,l} + \Sigma_{2,l}\,,
\end{equation}
where
\begin{eqnarray*}
\Sigma_{1,l} &\equiv& \int_p \K{p} \left[ \Ld{\phi^* (-p)} S_0 \cdot
    \frac{\delta}{\delta \phi (p)} + \frac{\delta S_0}{\delta \phi
      (p)} \Ld{\phi^* (-p)} \right] S_{1,l}\,,\\
\Sigma_{2,l} &\equiv& \int_p \K{p} \left[ \Ld{\phi^* (-p)} S_0 \cdot
    \frac{\delta}{\delta \phi (p)} + \frac{\delta S_0}{\delta \phi
      (p)} \Ld{\phi^* (-p)} \right] S_{2,l}\\
&& + \int_p \K{p} \left[ \sum_{k=1}^{l-1} \Ld{\phi^* (-p)} S_k \cdot
\frac{\delta S_{l-k}}{\delta \phi (p)} +
\Ld{\phi^* (-p)} \frac{\delta}{\delta \phi (p)} S_{l-1} \right]\,.
\end{eqnarray*}
\begin{enumerate}
\item $S_0^{asymp} = S_{cl}$ implies 
$\Sigma^{asymp}_{1,l}$ is $\Lambda$ independent.
\item $\Sigma^{asymp}_l$ is $\Lambda$ independent, and so is
      $\Sigma^{asymp}_{2,l}$.
\end{enumerate}
\item $\Sigma_l = 0$ is equivalent to
\begin{equation}
\Sigma^{asymp}_{1,l}  + \Sigma_{2,l}^{asymp} = 0\,.
\end{equation}
\item The classical BRST transformation is defined by
\begin{equation}
\delta_{cl} \equiv \int_x \left[ \Ld{\phi^* (x)} S_{cl}\cdot
    \frac{\delta}{\delta \phi (x)} + \frac{\delta S_{cl}}{\delta \phi
      (x)} \Ld{\phi^* (x)} \right]\,.
\end{equation}
This is nilpotent:
\begin{equation}
\delta_{cl}^2 = 0\,.
\end{equation}
\item Since
\begin{equation}
\Sigma^{asymp}_{1,l} = \delta_{cl} S_{1,l} \,,
\end{equation}
we obtain
\begin{equation}
\delta_{cl} \Sigma^{asymp}_{1,l} = 0\,.
\end{equation}
\item $\delta_Q \Sigma_\Lambda = 0$ and the induction hypothesis imply
\begin{equation}
\int_p \K{p} \left[ \Ld{\phi^* (-p)} S_0 \cdot \frac{\delta}{\delta
      \phi (p)} + \frac{\delta S_0}{\delta \phi (p)} \Ld{\phi^* (-p)}
\right] \Sigma_l = 0\,.
\end{equation}
Taking the asymptotic form, we obtain
\begin{equation}
\delta_{cl} \Sigma_l^{asymp} = 0\,.
\end{equation}
Hence, $\delta_{cl} \Sigma^{asymp}_{1,l} = 0$ implies
\begin{equation}
\delta_{cl} \Sigma_{2,l}^{asymp} = 0\,.
\end{equation}
\item $\Sigma_l = 0$ is equivalent to
\begin{equation}
\delta_{cl} S_{1,l} = - \Sigma_{2,l}^{asymp}\,.
\end{equation}
\item Thus, we must prove the triviality of the cohomology of
      $\delta_{cl}$:
\begin{equation}
\fbox{$\displaystyle \delta_{cl} \Sigma_{2,l}^{asymp} = 0
 \stackrel{\textrm{?}}{\Longrightarrow} 
\Sigma_{2,l}^{asymp} = \delta_{cl} (- S_{1,l})\,.$}
\end{equation}
\end{enumerate}

\newpage

\begin{thebibliography}{99}
\bibitem{Alford:1994fa}
M.~G.~Alford, \PLB{336,1994,237}.

\bibitem{Aoki:2000wm}
K.-I.~Aoki, \IJMP{B14,2000,1249}.

\bibitem{Aoki:1996gx}
K.-I.~Aoki, {\it Non-perturbative renormalization group approach to dynamical chiral
  symmetry breaking in gauge theories} in *Nagoya 1996, Perspectives of strong coupling gauge theories*, (1996) 171, hep-ph/9706204.

\bibitem{Aoki:1996fn}
K.-I.~Aoki, K.-I.~Morikawa, W.~Souma, J.-I. Sumi, and H. Terao,
\PTP{95,1996,409}.

\bibitem{Aoki:1998um}
K.-I.~Aoki, K.-I.~Morikawa, W.~Souma, J.-I. Sumi, and H. Terao,
\PTP{99,1998,451}.

\bibitem{Arnone:1998zc}
S.~Arnone, C.~Fusi, and K.~Yoshida,
\JHEP{02,1999,022}.

\bibitem{Arnone:2001iy}
S.~Arnone, Y.~A.~Kubyshin, T.~R.~Morris, and J.~F.~Tighe,
\IJMP{A17,2002,2283}.

\bibitem{Arnone:2005fb}
S.~Arnone, T.~R.~Morris, and O.~J.~Rosten.
\JL{Eur. Phys. J., C50,2007,467}.

\bibitem{Bagnuls:2001mw}
C.~Bagnuls and C.~Bervillier, \PRP{348,2001,91}.

\bibitem{Ball:1993zy}
R.~D.~Ball and R.~S.~Thorne, \JL{Ann. Phys.,236,1994,117}.

\bibitem{Ball:1994ji}
R.~D.~Ball, P.~E.~Haagensen, J.~I.~Latorre, and E.~Moreno,
\PLB{347,1995,80}.

\bibitem{Barnich:1994db}
G.~Barnich, F.~Brandt, and M.~Henneaux, \CMP{174,1995,57}.

\bibitem{Barnich:1994mt}
G.~Barnich, F.~Brandt, and M.~Henneaux, \CMP{174,1995,93}.

\bibitem{Barnich:2000zw}
G.~Barnich, F.~Brandt, and M.~Henneaux, \PRP{338,2000,439}.

\bibitem{Barnich:1994ve}
G.~Barnich and M.~Henneaux, \PRL{72,1994,1588}.

\bibitem{Batalin:1981jr}
I.~A.~Batalin and G.~A.~Vilkovisky. \PLB{102,1981,27}.

\bibitem{Becchi:1996an}
C.~Becchi, {\it On the construction of renormalized gauge theories using
  renormalization group techniques}, hep-th/9607188.

\bibitem{Benitez:2009xg}
F.~Benitez, J.-P.~Blaizot, H.~Chat{\' e}, B.~Delamotte, R.~M{\'
	e}ndez-Galain, and N.~Wschebor, arXiv:0901.0128 [cond-mat.stat-mech].

\bibitem{Benitez:2007mk}
F.~Benitez, R.~M{\' e}ndez Galain, and N.~Wschebor, \PRB{77,2008,024431}.

\bibitem{Berges:2000ew}
J.~Berges, N.~Tetradis, and C.~Wetterich, \PRP{363,2002,223}.

\bibitem{Bervillier:2008pt}
C.~Bervillier, B.~Boisseau, and H.~Giacomini, \NPB{801,2008,296}.

\bibitem{Bervillier:2008kh}
C.~Bervillier, B.~Boisseau, and H.~Giacomini, \NPB{789,2008,525}.

\bibitem{Bilal:2007ne}
A.~Bilal, \ANN{323,2008,2311}.

\bibitem{Bonini:1992vh}
M.~Bonini, M.~D'Attanasio, and G.~Marchesini, \NPB{409,1993,441}.

\bibitem{Bonini:1994xj}
M.~Bonini, M.~D'Attanasio, and G.~Marchesini, \PLB{329,1994,249}.

\bibitem{Bonini:1993sj}
M.~Bonini, M.~D'Attanasio, and G.~Marchesini, \NPB{421,1994,429}.

\bibitem{Bonini:1993kt}
M.~Bonini, M.~D'Attanasio, and G.~Marchesini, \NPB{418,1994,81}.

\bibitem{Bonini:1994kp}
M.~Bonini, M.~D'Attanasio, and G.~Marchesini, \NPB{437,1995,163}.

\bibitem{Bonini:1994dz}
M.~Bonini, M.~D'Attanasio, and G.~Marchesini, \PLB{346,1995,87}.

\bibitem{Bonini:1995tx}
M.~Bonini, M.~D'Attanasio, and G.~Marchesini, \NPB{444,1995,602}.

\bibitem{Bonini:1996bk}
M.~Bonini, G.~Marchesini, and M.~Simionato, \NPB{483,1997,475}.

\bibitem{Bonini:2000wr}
M.~Bonini and E.~Tricarico, \NPB{585,2000,253}.

\bibitem{Bonini:2001xm}
M.~Bonini and E.~Tricarico, \NPB{606,2001,231}.

\bibitem{Bonini:1997yv}
M.~Bonini and F.~Vian, \NPB{511,1998,479}.

\bibitem{Bonini:1998ec}
M.~Bonini and F.~Vian, \NPB{532,1998,473}.

\bibitem{Brandt:1996uv}
F.~Brandt, M.~Henneaux, and A.~Wilch, \PLB{387,1996,320}.

\bibitem{Brandt:1997cz}
F.~Brandt, M.~Henneaux, and A.~Wilch, \NPB{510,1998,640}.

\bibitem{DeJonghe:1993zc}
F.~De~Jonghe, hep-th/9403143.

\bibitem{Delamotte:2007pf}
B.~Delamotte, cond-mat/0702365v1 [cond-mat.stat-mech].

\bibitem{Ellwanger:1994iz}
U.~Ellwanger, \PLB{335,1994,364}.

\bibitem{Ellwanger:1995qf}
U.~Ellwanger, M.~Hirsch, and A.~Weber, \JL{Z. Phys. C69,1996,687}.

\bibitem{Fisher:1998kv}
M.~E.~Fisher, \JL{Rev. Mod. Phys., 70, 1998, 653}.

\bibitem{Freire:2000mn}
F.~Freire, D.~F.~Litim, and J.~M.~Pawlowski, \IJMP{A16,2001,2035}.

\bibitem{Freire:1996db}
F.~Freire and C.~Wetterich, \PLB{380,1996,337}.

\bibitem{Freire:2000bq}
F.~Freire, D.~F.~Litim, and J.~M.~Pawlowski, \PLB{495,2000,256}.

\bibitem{Gies:2006wv}
H.~Gies, hep-ph/0611146.

\bibitem{Ginsparg:1981bj}
P.~H.~Ginsparg and K.~G.~Wilson, \PRD{25,1982,2649}.

\bibitem{Golner:fp}
G.~R.~Golner, hep-th/9801124.

\bibitem{BVreview-Gomis}
J.~Gomis, J.~Paris, and S.~Samuel, \PRP{259,1995,1}.

\bibitem{Hasenfratz:1985dm}
A.~Hasenfratz and P.~Hasenfratz, \NPB{270,1986,687}.

\bibitem{Henneaux-Teitelboim}
M.~Henneaux and C.~Teitelboim, \textit{Quantization of Gauge Systems} (Princeton University Press, 1992).

\bibitem{Higashi:2007ie}
T.~Higashi, K.~Higashijima, and E.~Itou, \PTPS{164,2007,103}.

\bibitem{Higashi:2007ax}
T.~Higashi, E.~Itou, and T.~Kugo, \PTP{118,2007,1115}.

\bibitem{Higashijima:2002mh}
K.~Higashijima and E.~Itou, \PTP{108,2002,737}.

\bibitem{Higashijima:2003rp}
K.~Higashijima and E.~Itou, \PTP{110,2003,563}.

\bibitem{Higashijima:2005qk}
K.~Higashijima and E.~Itou, hep-th/0511300v1.

\bibitem{Hughes:1987rf}
J.~Hughes and J.~Liu, \NPB{307,1988,183}.

\bibitem{Ichinose:1999ke}
I.~Ichinose and K.~Nagao, \JL{Chin. J. Phys., 38, 20000, 671}.

\bibitem{Igarashi:2006sc}
Y.~Igarashi, M.~Ishikake, K.~Itoh, H.~Sawanaka, and H.~So, \JP{A39,2006,8023}.

\bibitem{Igarashi:2001jq}
Y.~Igarashi, K.~Itoh, and H.~So, \IJMP{A16,2001,2047}.

\bibitem{Igarashi:2000vf}
Y.~Igarashi, K.~Itoh, and H.~So, \PTP{104,2000,1053}.

\bibitem{Igarashi:1999rm}
Y.~Igarashi, K.~Itoh, and H.~So, \PLB{479,2000,336}.

\bibitem{Igarashi:2001mf}
Y.~Igarashi, K.~Itoh, and H.~So, \PTP{106,2001,149}.

\bibitem{Igarashi:2001cv}
Y.~Igarashi, K.~Itoh, and H.~So, \PLB{526,2002,164}.

\bibitem{Igarashi:2007fw}
Y.~Igarashi, K.~Itoh, and H.~Sonoda, \PTP{118,2007,121}.

\bibitem{Igarashi:2008bb}
Y.~Igarashi, K.~Itoh, and H.~Sonoda, \PTP{120,2008,1017}.

\bibitem{Igarashi:2009ymBV}
Y.~Igarashi, K.~Itoh, and H.~Sonoda, \textit{Quantum master equation for yang-mills theory in the exact
  renormalization group} {\em in preparation}.

\bibitem{Igarashi:2002ba}
Y.~Igarashi, H.~So, and N.~Ukita, \PLB{535,2002,363}.

\bibitem{Igarashi:2002bs}
Y.~Igarashi, H.~So, and N.~Ukita, \NPB{640,2002,95}.

\bibitem{Keller:1991bz}
G.~Keller and C.~Kopper, \CMP{148,1992,445}.

\bibitem{Keller:1992by}
G.~Keller and C.~Kopper, \CMP{153,1993,245}.

\bibitem{Keller:1990ej}
G.~Keller, C.~Kopper, and M.~Salmhofer, \JL{Helv. Phys. Acta, 65, 1992, 32}.

\bibitem{Kugo:1981hm}
T.~Kugo and S.~Uehara, \NPB{197,1982,378}.

\bibitem{Lauscher:2007zz}
O.~Lauscher and M.~Reuter, \textit{Quantum Einstein gravity: Towards an asymptotically safe field
  theory of gravity}, volume 721 (Springer Berlin / Heidelberg, 2007).

\bibitem{Litim:2001up}
D.~F.~Litim, \PRD{64,2001,105007}.

\bibitem{Litim:2008tt}
D.~F.~Litim, arXiv0810.3675 [hep-th].

\bibitem{Litim:1998nf}
D.~F.~Litim and J.~M.~Pawlowski, hep-th/9901063.

\bibitem{Litim:2002hj}
D.~F.~Litim and J~~M.~Pawlowski, \PLB{546,2002,279}.

\bibitem{Luscher:1998pqa}
M. L{\" u}scher, \PLB{428,1998,342}.

\bibitem{Margaritis:1987hv}
A.~Margaritis, G.~Odor, and A.~Patkos, \JL{Z. Phys., C39, 1988, 109}.

\bibitem{Morris:1993qb}
T.~R.~Morris, \IJMP{A9,1994,2411}.

\bibitem{Morris:1998da}
T.~R.~Morris, \PTPS{131,1998,395}.

\bibitem{Morris:1999px}
T.~R.~Morris, \NPB{573,2000,97}.

\bibitem{Morris:2000fs}
T.~R.~Morris, \JHEP{12,2000,012}.

\bibitem{Morris:2005ck}
T.~R.~Morris, \JHEP{07,2005,027}.

\bibitem{Morris:2005tv}
T.~R.~Morris and O.~J.~Rosten, \PRD{73,2006,065003}.

\bibitem{Morris:2006in}
T.~R.~Morris and O.~J.~Rosten, \JP{A39,2006,11657}.

\bibitem{Nicoll:1977hi}
J.~F.~Nicoll and T.~S.~Chang, \PLA{62,1977,287}.

\bibitem{Nicoll:1974zz}
J.~F.~Nicoll, T.~S.~Chang, and H.~E.~Stanley, \PRL{33,1974,540}.

\bibitem{Pawlowski:2005xe}
J.~M.~Pawlowski, \ANN{322,2007,2831}.

\bibitem{Pernici:1998tp}
M.~Pernici and M.~Raciti, \NPB{531,1998,560}.

\bibitem{Pernici:1997ie}
M.~Pernici, M.~Raciti, and F.~Riva, \NPB{520,1998,469}.

\bibitem{Pernici:1998ex}
M.~Pernici, M.~Raciti, and F.~Riva, \PLB{440,1998,305}.

\bibitem{Polchinski:1983gv}
J.~Polchinski, \NPB{231,1984,269}.

\bibitem{Polonyi:2001se}
J.~Polonyi, \JL{Central Eur. J. Phys., 1, 2003, 1}.

\bibitem{Reuter:1993kw}
M.~Reuter and C.~Wetterich, \NPB{417,1994,181}.

\bibitem{Reuter:1994sg}
M.~Reuter and C.~Wetterich, \NPB{427,1994,291}.

\bibitem{Reuter:2007rv}
M.~Reuter and F.~Saueressig, arXiv:0708.1317 [hep-th].

\bibitem{Rosten:2006tk}
O.~J.~Rosten, \IJMP{A21,2006,4627}.

\bibitem{Rosten:2005ka}
O.~J.~Rosten, \JP{A39,2006,8699}.

\bibitem{Rosten:2006qx}
O.~J.~Rosten, \PRD{74,2006,125006}.

\bibitem{Rosten:2008ih}
O.~J.~Rosten, arXiv:0808.2150 [hep-th].

\bibitem{Salmhofer:2001tr}
M.~Salmhofer and C.~Honerkamp, \PTP{105,2001,1}.

\bibitem{Seiberg:1993vc}
N.~Seiberg, \PLB{318,1993,469}.

\bibitem{Shankar:1993pf}
R.~Shankar, \JL{Rev. Mod. Phys., 66, 1994, 129}.

\bibitem{Sonoda:2002pb}
H.~Sonoda, \PRD{67,2003,065011}.

\bibitem{Sonoda:2007dj}
H. Sonoda, \JP{A40,2007,9675}.

\bibitem{Sonoda:2006ai}
H.~Sonoda, \JP{A40,2007,5733}.

\bibitem{Sonoda:2007av}
H.~Sonoda, arXiv:0710.1662.

\bibitem{Sonoda:2008dz}
H.~Sonoda and K.~Ulker, \PTP{120,2008,197}.

\bibitem{Hooft:1973mm}
G.~'t~Hooft, \NPB{61,1973,455}.

\bibitem{Hooft:1972fi}
G.~'t~Hooft and M.~J.~G.~Veltman, \NPB{44,1972,189}.

\bibitem{Tetradis:1993ts}
N.~Tetradis and C.~Wetterich, \NPB{422,1994,541}.

\bibitem{Troost:1989cu}
W.~Troost, P.~van Nieuwenhuizen, and A.~Van~Proeyen, \NPB{333,1990,727}.

\bibitem{Warr:1986we}
B.~J.~Warr, \ANN{183,1988,1}.

\bibitem{Warr:1986ux}
B.~J.~Warr, \ANN{183,1988,59}.

\bibitem{Wegner:1972ih}
F.~J.~Wegner and A.~Houghton, \PRA{8,1973,401}.

\bibitem{PhysRevD.8.3497}
S.~Weinberg, \PRD{8,1973,3497}.

\bibitem{Weinberg:1998uv}
S.~Weinberg, \PRL{80,1998,3702}.

\bibitem{Wess:1973kz}
J.~Wess and B.~Zumino, \PLB{49,1974,52}.

\bibitem{Wetterich:1989xg}
C.~Wetterich, \NPB{352,1991,529}.

\bibitem{Wetterich:1993ne}
C.~Wetterich, \JL{Z. Phys., C60, 1993, 461}.

\bibitem{Wetterich:1992yh}
C.~Wetterich, \PLB{301,1993,90}.

\bibitem{Wilson:1973jj}
K.~G.~Wilson and J.~B.~Kogut, \PRP{12,1974,75}.

\bibitem{ZinnJustin:1993wc}
J.~Zinn-Justin, \textit{Quantum field theory and critical phenomena, Fourth Edition} (Oxford Univ Press, 2002).

\end{thebibliography}

\end{document}